%% file: main.tex
\makeatletter
\def\input@path{{styles/}{}}

\providecommand\p@copyrightTextShortEven{}
\providecommand\p@copyrightTextShortOdd{}
\newcommand{\copyrightTextShortEven}[1]{\renewcommand\p@copyrightTextShortEven{##1}}
\newcommand{\copyrightTextShortOdd}[1]{\renewcommand\p@copyrightTextShortOdd{##1}}
\providecommand\p@copyrightTextShort{}
\providecommand\p@copyrightTextShortEven{}
\providecommand\p@copyrightTextShortOdd{}
\providecommand\p@copyrightTextTitPag{}
\providecommand\p@copyrightTextRunPag{}
\providecommand\p@copyrightTextLong{}

\newcommand{\copyrightTextShort}[1]{%
  \renewcommand\p@copyrightTextShort{#1}%
}

\newcommand{\copyrightTextTitPag}[1]{%
  \renewcommand\p@copyrightTextTitPag{#1}%
}
\newcommand{\copyrightTextRunPag}[1]{%
  \renewcommand\p@copyrightTextRunPag{#1}%
}
\makeatother
\PassOptionsToPackage{dvipsnames,table}{xcolor}
\documentclass{egpubl}
\usepackage{eg2026}

\STAR                   
\usepackage[T1]{fontenc}
\usepackage{cite}  
\BibtexOrBiblatex
\electronicVersion
\PrintedOrElectronic
\ifpdf \usepackage[pdftex]{graphicx} \pdfcompresslevel=9
\else \usepackage[dvips]{graphicx} \fi

\usepackage{siunitx}
\usepackage{t1enc, dfadobe}
\usepackage{egweblnk}
\usepackage{cite}
\usepackage{amsmath}
\usepackage{booktabs}
\usepackage{egweblnk} 
\usepackage{lineno}
\usepackage{booktabs}
\usepackage{graphicx}
\graphicspath{{./}{./figures/}{./figures/parts/}{./figures/bio/}{./icons/}{./icons/custom/}{./icons/table_logos/}}
\usepackage{booktabs}
\usepackage{pgfplotstable}
\usepackage{array}
\usepackage{rotating}
\usepackage{pifont}
\usepackage{stfloats}
\usepackage{tabularx}
\usepackage{booktabs}
\usepackage{enumitem}
\usepackage{pifont}
\usepackage{cleveref}
\usepackage{wrapfig}
\usepackage{fontawesome5}
\usepackage{academicons}
\usepackage{xcolor}
\usepackage{ifthen}
\usepackage{multirow}
\usepackage{tabularx}
\usepackage{multirow}
\usepackage{booktabs}
\usepackage{xcolor}
\usepackage{amssymb}

\input{macros}

\title[STAR]{How to Build Digital Humans? \\ From Priors to Photorealistic Avatars}

\author[W. Zielonka, T. Kirschstein, T. Bolkart, X. Deng, V. Sklyarova, S. Giebenhain, D. Xiang, S. Saito, Y. Liu, M. Niessner, J. Thies]
{\parbox{\textwidth}{\centering 
Wojciech Zielonka$^{1}$$^\dagger$\orcid{0009-0009-6534-8909},
Tobias Kirschstein$^{2}$$^\dagger$\orcid{0009-0002-5308-591X},
Timo Bolkart$^{3}$$^\dagger$\orcid{0000-0002-3829-3924},
Simon Giebenhain$^{2}$\orcid{0000-0002-7588-8767},
Vanessa Sklyarova$^{5, 8}$\orcid{0000-0002-8883-9972},
Xiang Deng$^{4}$\orcid{0009-0005-3047-9950},
Donglai Xiang$^{6}$\orcid{0000-0002-6487-1935},
Shunsuke Saito$^{1}$\orcid{0000-0003-2053-3472},
Yebin Liu$^{4}$,
Matthias Niessner$^{2}$\orcid{0000-0001-6093-5199},
Justus Thies$^{7}$$^\dagger$\orcid{0000-0002-0056-9825}
}
\\
{\parbox{\textwidth}{\centering 
$^1$Meta \hspace{1em}
$^2$Technical University of Munich \hspace{1em}
$^3$Google \hspace{1em}
$^4$Tsinghua University \hspace{1em} \\
$^5$Max Planck Institute for Intelligent Systems \hspace{1em}
$^6$NVIDIA \hspace{1em}
$^7$Technical University of Darmstadt \hspace{1em}
$^8$ETH Zurich \\[0.4em]
\small $^\dagger$ These authors contributed equally.
}}
\\
\\
\parbox{\textwidth}{\centering
    \large 
    \href{https://zielon.github.io/how-to-build-digital-humans/}{https://zielon.github.io/how-to-build-digital-humans/}
}
\vspace{-1.1cm}
}

\begin{document}

\teaser{
 \vspace{-0.1cm}
 \includegraphics[width=1.0\linewidth,trim={0 6.7cm 0 0},clip]{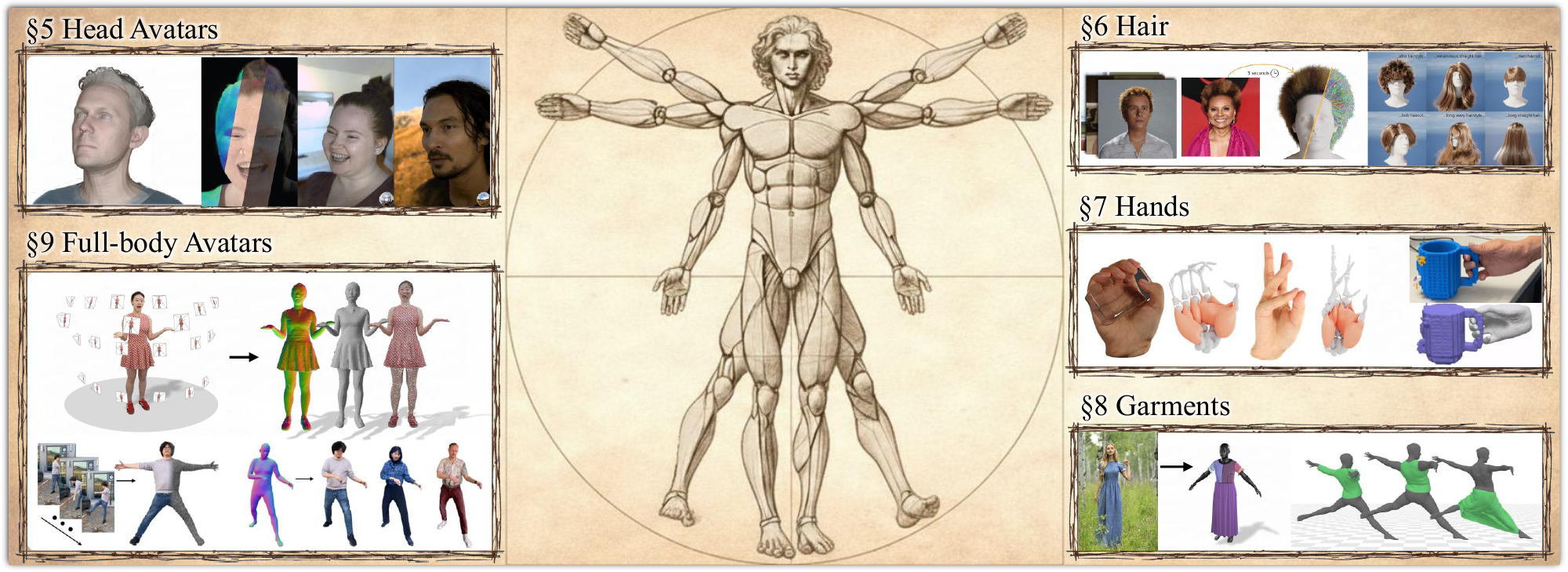}
 \centering
 \caption{Building digital avatars requires considering many components, such as the \textbf{human head, hair, hands, and garments}, and ultimately \textbf{full-body avatars}. Much like da Vinci’s Vitruvian Man symbolizes the ideal proportions and unity of the human form, the construction of digital avatars demands a holistic approach in which each part integrates into a coherent whole. Each of these elements presents distinct challenges and numerous potential design choices across different stages. In this report, we introduce and examine the details of each creation stage, develop a practical taxonomy of the current literature, and discuss future directions in the field of digital avatars. The background image was generated using Google Gemini. \new{\cite{Qian2024gaussianavatars, Saito2024relightable, chen2025taoavatar, Guo2025vid2avatarpro, Radu2025difflocks, Sklyarova2024haar, li2022nimble, fan2024hold, bian2025chatgarment, grigorev2023hood}.}}
\label{fig:teaser}
}

\maketitle

\input{chapters/0_abstract}
\input{chapters/1_introduction}
\input{chapters/3_taxonomy}

\input{chapters/4_fundamentals}

\input{tables/legend.tex}
\input{tables/taxonomy.tex}

\label{sec:heads}
\input{chapters/5_1_faces}
\input{tables/assets.tex}
\input{chapters/5_3_hair}

\label{sec:body}
\input{chapters/5_2_hands}
\input{chapters/5_4_garment}

\input{chapters/5_5_full_body}
\input{chapters/7_future_work}
\newpage
\input{chapters/8_acknowledgment}

\newpage
\input{bios}

\clearpage
\newpage
\appendix
\input{appendix/reighting}
\input{appendix/hair}
\input{appendix/garment}

\input{appendix/datasets}

\newpage
\input{main.bbl}

\end{document}

%% file: macros.tex
\usepackage[dvipsnames]{xcolor}
\usepackage{comment}
\usepackage{lipsum}
\usepackage{tikz}
\usepackage{graphicx}
\usepackage{transparent}
\usepackage{colortbl}
\usepackage{nicematrix}
\usepackage{graphbox}
\usepackage{calc}

\newcommand{\history}[1]{{\color{pink}#1}}

\newcommand{\new}[1]{\leavevmode{\color{black}#1}}

\newcommand{\etal}{\textit{et al.}}

\usepackage[dvipsnames]{xcolor}

\definecolor{cborange}{RGB}{230,159,0}
\definecolor{cbpurple}{RGB}{153,0,153}
\definecolor{cbblue}{RGB}{0,114,178}
\definecolor{cbvermillion}{RGB}{213,94,0}
\definecolor{cbgreen}{RGB}{0,158,115}
\definecolor{cbyellowbrown}{RGB}{204,121,167}
\definecolor{cbpink}{RGB}{255,105,180}
\definecolor{cbgrey}{RGB}{128,128,128}
\definecolor{cbteal}{RGB}{86,180,233}
\definecolor{cbcyan}{RGB}{0,180,180}
\definecolor{cbyellow}{RGB}{255,222,33}

\newcommand{\VS}[1]{{\color{cbpurple}{VS: #1}}} 
\newcommand{\TK}[1]{{\color{cbblue}{TK: #1}}} 
\newcommand{\JT}[1]{{\color{red}{JT: #1}}} 

\newcommand*{\rot}{\rotatebox{90}}
\newcommand{\valsep}{\hspace{1em}}
\newcommand{\val}[2]{\mbox{#2 \ #1}}
\newcommand{\icon}[1]{\includegraphics[height=1.7em,align=c]{icons/#1}}
\newcommand{\crbx}[2]{\scalebox{0.9}{\fcolorbox{#1}{#1}{\makebox[1.5ex][c]{\rule{0pt}{1.5ex}#2}}}}
\newcommand{\crboxbold}[2]{\crbx{#1}{\textcolor{white}{\textbf{\textsf{#2}}}}}
\newcommand{\crboxreg}[2]{\crbx{#1}{\textcolor{black}{\textbf{\textsf{#2}}}}}

\definecolor{bgcolor00}{RGB}{62,253,202}
\definecolor{bgcolor01}{RGB}{251,249,74}
\definecolor{bgcolor02}{RGB}{251,114,238}
\definecolor{bgcolor03}{RGB}{221,209,197}
\definecolor{bgcolor04}{RGB}{94,238,64}
\definecolor{bgcolor05}{RGB}{243,153,92}
\definecolor{bgcolor06}{RGB}{143,212,161}
\definecolor{bgcolor07}{RGB}{155,193,249}
\definecolor{bgcolor08}{RGB}{181,200,70}
\definecolor{bgcolor09}{RGB}{181,133,207}
\definecolor{bgcolor10}{RGB}{85,208,241}
\definecolor{bgcolor11}{RGB}{174,252,231}
\definecolor{bgcolor12}{RGB}{60,245,133}
\definecolor{bgcolor13}{RGB}{232,168,254}
\definecolor{bgcolor14}{RGB}{187,158,137}
\definecolor{bgcolor15}{RGB}{180,255,96}
\definecolor{bgcolor16}{RGB}{233,207,122}
\definecolor{bgcolor17}{RGB}{121,159,210}
\definecolor{bgcolor18}{RGB}{252,159,167}
\definecolor{bgcolor19}{RGB}{77,198,128}
\definecolor{bgcolor20}{RGB}{114,250,251}
\definecolor{bgcolor21}{RGB}{236,254,169}
\definecolor{bgcolor22}{RGB}{174,174,191}
\definecolor{bgcolor23}{RGB}{239,228,251}
\definecolor{bgcolor24}{RGB}{184,241,176}
\definecolor{bgcolor25}{RGB}{125,217,108}
\definecolor{bgcolor26}{RGB}{128,251,191}
\definecolor{bgcolor27}{RGB}{178,208,122}
\definecolor{bgcolor28}{RGB}{122,255,124}
\definecolor{bgcolor29}{RGB}{234,202,62}
\definecolor{bgcolor30}{RGB}{234,143,214}
\definecolor{bgcolor31}{RGB}{85,219,197}
\definecolor{bgcolor32}{RGB}{199,186,231}
\definecolor{bgcolor33}{RGB}{217,181,159}
\definecolor{bgcolor34}{RGB}{198,166,87}
\definecolor{bgcolor35}{RGB}{141,236,70}
\definecolor{bgcolor36}{RGB}{221,132,124}
\definecolor{bgcolor37}{RGB}{184,145,249}
\definecolor{bgcolor38}{RGB}{127,198,209}
\definecolor{bgcolor39}{RGB}{197,236,59}
\definecolor{bgcolor40}{RGB}{165,221,216}
\definecolor{bgcolor41}{RGB}{102,201,67}
\definecolor{bgcolor42}{RGB}{59,240,241}
\definecolor{bgcolor43}{RGB}{218,255,128}
\definecolor{bgcolor44}{RGB}{73,223,96}
\definecolor{bgcolor45}{RGB}{95,226,145}
\definecolor{bgcolor46}{RGB}{215,140,172}
\definecolor{bgcolor47}{RGB}{90,183,216}
\definecolor{bgcolor48}{RGB}{123,181,166}
\definecolor{bgcolor49}{RGB}{226,254,231}
\definecolor{bgcolor50}{RGB}{234,182,205}
\definecolor{bgcolor51}{RGB}{243,218,164}
\definecolor{bgcolor52}{RGB}{61,216,161}
\definecolor{bgcolor53}{RGB}{216,135,85}
\definecolor{bgcolor54}{RGB}{140,199,82}
\definecolor{bgcolor55}{RGB}{251,184,111}
\definecolor{bgcolor56}{RGB}{60,253,89}
\definecolor{bgcolor57}{RGB}{131,223,232}
\definecolor{bgcolor58}{RGB}{150,247,142}
\definecolor{bgcolor59}{RGB}{247,237,119}
\definecolor{bgcolor60}{RGB}{234,201,236}
\definecolor{bgcolor61}{RGB}{107,242,218}
\definecolor{bgcolor62}{RGB}{163,180,148}
\definecolor{bgcolor63}{RGB}{205,226,135}
\definecolor{bgcolor64}{RGB}{230,140,253}
\definecolor{bgcolor65}{RGB}{192,214,253}
\definecolor{bgcolor66}{RGB}{212,201,93}
\definecolor{bgcolor67}{RGB}{221,161,122}
\definecolor{bgcolor68}{RGB}{156,165,223}
\definecolor{bgcolor69}{RGB}{115,183,254}
\definecolor{bgcolor70}{RGB}{193,212,167}
\definecolor{bgcolor71}{RGB}{252,115,112}
\definecolor{bgcolor72}{RGB}{253,236,201}
\definecolor{bgcolor73}{RGB}{211,120,209}
\definecolor{bgcolor74}{RGB}{93,248,168}
\definecolor{bgcolor75}{RGB}{145,247,219}
\definecolor{bgcolor76}{RGB}{253,126,165}
\definecolor{bgcolor77}{RGB}{167,221,86}
\definecolor{bgcolor78}{RGB}{63,198,194}
\definecolor{bgcolor79}{RGB}{168,179,93}
\definecolor{bgcolor80}{RGB}{203,159,215}
\definecolor{bgcolor81}{RGB}{213,225,227}
\definecolor{bgcolor82}{RGB}{140,218,195}
\definecolor{bgcolor83}{RGB}{219,240,192}
\definecolor{bgcolor84}{RGB}{243,102,203}
\definecolor{bgcolor85}{RGB}{98,201,164}
\definecolor{bgcolor86}{RGB}{161,235,115}
\definecolor{bgcolor87}{RGB}{115,191,106}
\definecolor{bgcolor88}{RGB}{254,192,168}
\definecolor{bgcolor89}{RGB}{251,200,207}
\definecolor{bgcolor90}{RGB}{117,202,133}
\definecolor{bgcolor91}{RGB}{201,241,250}
\definecolor{bgcolor92}{RGB}{169,226,148}
\definecolor{bgcolor93}{RGB}{185,181,122}
\definecolor{bgcolor94}{RGB}{196,227,100}
\definecolor{bgcolor95}{RGB}{222,234,83}
\definecolor{bgcolor96}{RGB}{159,198,195}
\definecolor{bgcolor97}{RGB}{193,206,211}
\definecolor{bgcolor98}{RGB}{218,162,59}
\definecolor{bgcolor99}{RGB}{232,182,83}

\definecolor{bgcolordark00}{RGB}{247,85,93}
\definecolor{bgcolordark01}{RGB}{100,60,183}
\definecolor{bgcolordark02}{RGB}{69,136,90}
\definecolor{bgcolordark03}{RGB}{131,73,73}
\definecolor{bgcolordark04}{RGB}{184,107,163}
\definecolor{bgcolordark05}{RGB}{82,112,253}
\definecolor{bgcolordark06}{RGB}{247,62,193}
\definecolor{bgcolordark07}{RGB}{83,132,176}
\definecolor{bgcolordark08}{RGB}{69,197,90}
\definecolor{bgcolordark09}{RGB}{189,103,66}
\definecolor{bgcolordark10}{RGB}{148,87,227}
\definecolor{bgcolordark11}{RGB}{60,79,96}
\definecolor{bgcolordark12}{RGB}{96,182,146}
\definecolor{bgcolordark13}{RGB}{102,94,138}
\definecolor{bgcolordark14}{RGB}{191,62,133}
\definecolor{bgcolordark15}{RGB}{208,60,253}
\definecolor{bgcolordark16}{RGB}{129,131,66}
\definecolor{bgcolordark17}{RGB}{105,151,245}
\definecolor{bgcolordark18}{RGB}{72,60,249}
\definecolor{bgcolordark19}{RGB}{252,72,252}
\definecolor{bgcolordark21}{RGB}{112,94,194}
\definecolor{bgcolordark22}{RGB}{59,60,143}
\definecolor{bgcolordark23}{RGB}{147,123,131}
\definecolor{bgcolordark24}{RGB}{62,99,207}
\definecolor{bgcolordark25}{RGB}{119,152,107}
\definecolor{bgcolordark26}{RGB}{150,126,250}
\definecolor{bgcolordark27}{RGB}{155,75,170}
\definecolor{bgcolordark28}{RGB}{216,106,124}
\definecolor{bgcolordark29}{RGB}{132,66,132}
\definecolor{bgcolordark30}{RGB}{94,107,60}
\definecolor{bgcolordark31}{RGB}{250,71,145}
\definecolor{bgcolordark32}{RGB}{107,167,61}
\definecolor{bgcolordark33}{RGB}{198,60,80}
\definecolor{bgcolordark34}{RGB}{170,145,64}
\definecolor{bgcolordark35}{RGB}{61,163,181}
\definecolor{bgcolordark36}{RGB}{180,121,220}
\definecolor{bgcolordark37}{RGB}{80,63,63}
\definecolor{bgcolordark38}{RGB}{194,128,105}
\definecolor{bgcolordark39}{RGB}{108,84,244}
\definecolor{bgcolordark40}{RGB}{97,62,104}
\definecolor{bgcolordark41}{RGB}{165,94,109}
\definecolor{bgcolordark42}{RGB}{124,153,172}
\definecolor{bgcolordark43}{RGB}{236,120,73}
\definecolor{bgcolordark44}{RGB}{63,92,163}
\definecolor{bgcolordark45}{RGB}{165,67,61}
\definecolor{bgcolordark46}{RGB}{62,138,223}
\definecolor{bgcolordark47}{RGB}{61,132,148}
\definecolor{bgcolordark48}{RGB}{177,59,226}
\definecolor{bgcolordark49}{RGB}{62,165,70}
\definecolor{bgcolordark50}{RGB}{220,98,214}
\definecolor{bgcolordark51}{RGB}{111,131,213}
\definecolor{bgcolordark52}{RGB}{119,106,97}
\definecolor{bgcolordark53}{RGB}{131,118,170}
\definecolor{bgcolordark54}{RGB}{110,182,98}
\definecolor{bgcolordark55}{RGB}{241,60,71}
\definecolor{bgcolordark56}{RGB}{108,128,134}
\definecolor{bgcolordark57}{RGB}{89,158,208}
\definecolor{bgcolordark58}{RGB}{74,157,121}
\definecolor{bgcolordark59}{RGB}{64,106,128}
\definecolor{bgcolordark60}{RGB}{151,66,253}
\definecolor{bgcolordark61}{RGB}{147,117,217}
\definecolor{bgcolordark62}{RGB}{151,140,98}
\definecolor{bgcolordark63}{RGB}{60,63,183}
\definecolor{bgcolordark64}{RGB}{180,90,245}
\definecolor{bgcolordark65}{RGB}{59,167,231}
\definecolor{bgcolordark66}{RGB}{93,105,223}
\definecolor{bgcolordark67}{RGB}{131,63,213}
\definecolor{bgcolordark68}{RGB}{214,87,158}
\definecolor{bgcolordark69}{RGB}{152,99,71}
\definecolor{bgcolordark70}{RGB}{253,108,116}
\definecolor{bgcolordark71}{RGB}{101,196,65}
\definecolor{bgcolordark72}{RGB}{150,145,151}
\definecolor{bgcolordark73}{RGB}{64,91,63}
\definecolor{bgcolordark74}{RGB}{220,92,68}
\definecolor{bgcolordark75}{RGB}{171,93,200}
\definecolor{bgcolordark76}{RGB}{224,66,114}
\definecolor{bgcolordark77}{RGB}{77,79,124}
\definecolor{bgcolordark78}{RGB}{248,91,188}
\definecolor{bgcolordark79}{RGB}{137,95,145}
\definecolor{bgcolordark80}{RGB}{60,119,59}
\definecolor{bgcolordark81}{RGB}{138,102,252}
\definecolor{bgcolordark82}{RGB}{227,73,226}
\definecolor{bgcolordark83}{RGB}{98,62,144}
\definecolor{bgcolordark84}{RGB}{161,65,108}
\definecolor{bgcolordark85}{RGB}{182,137,135}
\definecolor{bgcolordark86}{RGB}{76,69,215}
\definecolor{bgcolordark87}{RGB}{134,146,233}
\definecolor{bgcolordark88}{RGB}{76,140,248}
\definecolor{bgcolordark89}{RGB}{64,91,255}
\definecolor{bgcolordark90}{RGB}{194,77,107}
\definecolor{bgcolordark91}{RGB}{93,94,90}
\definecolor{bgcolordark92}{RGB}{93,162,170}
\definecolor{bgcolordark93}{RGB}{94,114,157}
\definecolor{bgcolordark94}{RGB}{83,159,90}
\definecolor{bgcolordark95}{RGB}{97,143,74}
\definecolor{bgcolordark96}{RGB}{207,97,96}
\definecolor{bgcolordark97}{RGB}{117,126,253}
\definecolor{bgcolordark98}{RGB}{136,86,100}
\definecolor{bgcolordark99}{RGB}{62,114,174}

\definecolor{bgcoloryes}{RGB}{218,223,227}

\newcommand{\crboxUSCHairSalon}{\crboxbold{bgcolordark00}{H}}
\newcommand{\crboxInterHandM}{\crboxbold{bgcolordark01}{IH}}
\newcommand{\crboxFFHQ}{\crboxbold{bgcolordark02}{F}}
\newcommand{\crboxVFHQ}{\crboxbold{bgcolordark03}{V}}
\newcommand{\crboxInternal}{\crboxbold{bgcolordark04}{I}}

\newcommand{\crboxCelebVHQ}{\crboxbold{bgcolordark06}{V}}
\newcommand{\crboxNeRSemble}{\crboxbold{bgcolordark07}{N}}
\newcommand{\crboxMEAD}{\crboxbold{bgcolordark08}{M}}

\newcommand{\crboxAva}{\crboxbold{bgcolordark09}{A}}
\newcommand{\crboxTHuman}{\crboxbold{bgcolordark10}{TH}}
\newcommand{\crboxHumanDiT}{\crboxbold{bgcolordark11}{HD}}
\newcommand{\crboxActorsHQ}{\crboxbold{bgcolordark12}{AH}}
\newcommand{\crboxZJUMoCap}{\crboxbold{bgcolordark13}{Z}}

\newcommand{\crboxClothD}{\crboxbold{bgcolordark15}{CD}}
\newcommand{\crboxNPHM}{\crboxbold{bgcolordark16}{NH}}
\newcommand{\crboxFaceVerse}{\crboxbold{bgcolordark17}{FV}}
\newcommand{\crboxNextDsamples}{\crboxbold{bgcolordark18}{NS}}
\newcommand{\crboxTwindom}{\crboxbold{bgcolordark19}{TW}}
\newcommand{\crboxFaceScape}{\crboxbold{bgcolordark28}{FS}}
\newcommand{\crboxBCNet}{\crboxbold{bgcolordark21}{BC}}
\newcommand{\crboxCTHair}{\crboxbold{bgcolordark22}{CT}}
\newcommand{\crboxSewingPattern}{\crboxbold{bgcolordark23}{SP}}
\newcommand{\crboxPanoHeadsamples}{\crboxbold{bgcolordark24}{PH}}
\newcommand{\crboxAIST}{\crboxbold{bgcolordark65}{AI}}
\newcommand{\crboxSURREAL}{\crboxbold{bgcolordark30}{SR}}
\newcommand{\crboxLPFF}{\crboxbold{bgcolordark25}{LF}}
\newcommand{\crboxRenderPeople}{\crboxbold{bgcolordark29}{RP}}
\newcommand{\crboxCustomHumans}{\crboxbold{bgcolordark30}{CH}}
\newcommand{\crboxKK}{\crboxbold{bgcolordark31}{2K}}
\newcommand{\crboxEnsemble}{\crboxbold{bgcolordark32}{E}}
\newcommand{\crboxMRIHand}{\crboxbold{bgcolordark33}{MH}}

\newcommand{\crboxVoxCeleb}{\crboxreg{bgcolor68}{VC}}

\newcommand{\crboxTrackedDMM}{\crboxreg{bgcolor31}{3}}
\newcommand{\crboxCameras}{\crboxreg{bgcolor33}{C}}
\newcommand{\crboxRegisteredMeshes}{\crboxreg{bgcolor34}{M}}
\newcommand{\crboxTextures}{\crboxreg{bgcolor35}{T}}

\newcommand{\crboxsegmentationmasks}{\crboxreg{bgcolor38}{S}}
\newcommand{\crboxMeshes}{\crboxreg{bgcolor39}{M}}

\newcommand{\crboxKnownLighting}{\crboxreg{bgcolor42}{L}}

\newcommand{\crboxNeuralHaircut}{\crboxreg{bgcolor50}{H}}
\newcommand{\crboxDMM}{\crboxreg{bgcolor51}{3}}
\newcommand{\crboxHandPose}{\crboxreg{bgcolor52}{HP}}
\newcommand{\crboxImageDiffusion}{\crboxreg{bgcolor53}{D}}
\newcommand{\crboxRelighting}{\crboxreg{bgcolor56}{R}}
\newcommand{\crboxGroomCap}{\crboxreg{bgcolor57}{GC}}
\newcommand{\crboxPhysics}{\crboxreg{bgcolor58}{P}}
\newcommand{\crboxContourCraft}{\crboxreg{bgcolor59}{CC}}
\newcommand{\crboxWaveVec}{\crboxreg{bgcolor60}{WV}}
\newcommand{\crboxDeepMVSHair}{\crboxreg{bgcolor61}{DH}}
\newcommand{\crboxSemanticsegmentation}{\crboxreg{bgcolor62}{SS}}
\newcommand{\crboxGarmenttemplates}{\crboxreg{bgcolor63}{GT}}
\newcommand{\crboxDeepFeatures}{\crboxreg{bgcolor64}{DF}}
\newcommand{\crboxPortraitAnimation}{\crboxreg{bgcolor65}{PA}}

\newcommand{\crboxGPTV}{\crboxreg{bgcolor67}{G4}}
\newcommand{\crboxGroundedSAM}{\crboxreg{bgcolor68}{S}}

\newcommand{\crboxMesh}{\crboxreg{bgcolor00}{M}}
\newcommand{\crboxDGS}{\crboxreg{bgcolor01}{G}}
\newcommand{\crboxNeRF}{\crboxreg{bgcolor02}{N}}
\newcommand{\crboxMVP}{\crboxreg{bgcolor03}{MV}}
\newcommand{\crboxNeuralRendering}{\crboxreg{bgcolor04}{NR}}
\newcommand{\crboxSDF}{\crboxreg{bgcolor05}{S}}
\newcommand{\crboxStrands}{\crboxreg{bgcolor05}{ST}}

\newcommand{\icoReal}{\icon{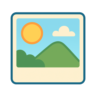}} 
\newcommand{\icoSynth}{\icon{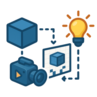}} 
\newcommand{\icoGenerated}{\icon{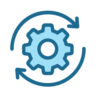}} 
\newcommand{\icoSewingPattern}{\icon{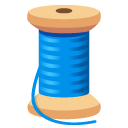}} 

\newcommand{\icoMRI}{\icon{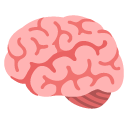}}           
\newcommand{\icoDepth}{\icon{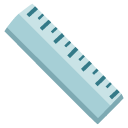}}         

\newcommand{\icoSingleImage}{\icon{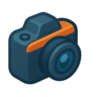}}   
\newcommand{\icoMultiImage}{\icon{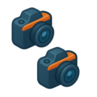}}    
\newcommand{\icoMonoVideo}{\icon{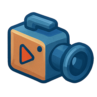}}     
\newcommand{\icoMultiViewVideo}{\icon{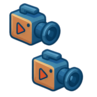}}
\newcommand{\icoOLATImage}{\icon{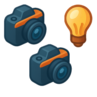}}      
\newcommand{\icoOLATVideo}{\icon{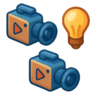}}      
\newcommand{\icoMeshes}{\icon{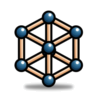}}         

\newcommand{\icoPointCloud}{\icon{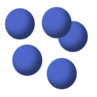}}
\newcommand{\icoStrands}{\icon{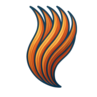}}       
\newcommand{\icoAudio}{\icon{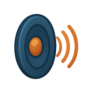}}         
\newcommand{\icoText}{\icon{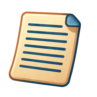}}          

\newcommand{\icoZero}{\icon{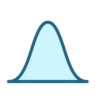}}           
\newcommand{\icoOne}{\icon{custom/icon_single_image.png}}            
\newcommand{\icoFew}{\icon{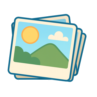}}           

\newcommand{\icoInstant}{\icon{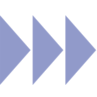}}        
\newcommand{\icoFast}{\icon{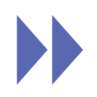}}          
\newcommand{\icoMedium}{\icon{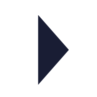}}        
\newcommand{\icoSlow}{\icon{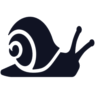}}          

\newcommand{\icoExprThreeDMM}{\icon{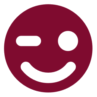}}  
\newcommand{\icoExprGeneral}{\icon{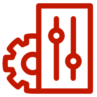}}   
\newcommand{\icoPose}{\icon{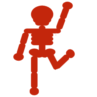}}          
\newcommand{\icoSimulation}{\icon{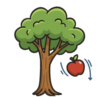}}     
\newcommand{\icoVideo}{\icon{custom/icon_single_video.png}}         

\newcommand{\icoLocalLight}{\icon{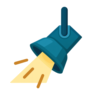}}    
\newcommand{\icoDistantLight}{\icon{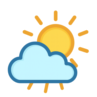}}   

\newcommand{\icoRealtime}{\icon{custom/icon_instant.png}}       
\newcommand{\icoInteractive}{\icon{custom/icon_fast.png}}    
\newcommand{\icoOffline}{\icon{custom/icon_medium.png}}       

\newcommand{\icoYes}{\crboxreg{bgcoloryes}{X}}
\newcommand{\icoNo}{}

\newcommand{\icoFace}{\icon{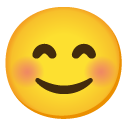}}          
\newcommand{\icoBody}{\icon{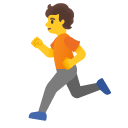}}          
\newcommand{\icoHair}{\icon{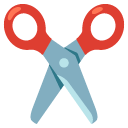}}           
\newcommand{\icoHand}{\icon{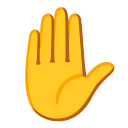}}           
\newcommand{\icoGarment}{\icon{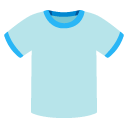}}       



%% file: chapters/0_abstract.tex
\begin{abstract}
This state-of-the-art report provides an overview of controllable 3D human avatar creation. We describe current 3D avatar systems, which typically consist of three stages: (i) learning priors of human appearance and motion, (ii) creating a personalized avatar, and (iii) animating the avatar. To limit the scope, we focus on the prior learning and avatar creation stages. We define current avatar representations and introduce a taxonomy that categorizes existing work along multiple axes, including body regions and employed priors. We review methods for full-body and head avatars, as well as layered representations that decompose the body into components such as hands, hair, and garments. Finally, we outline common underlying principles, reference key literature for newcomers, and discuss open challenges and future research directions.
\end{abstract}

%% file: chapters/1_introduction.tex
\section{Introduction}
\label{sec:intro}
For decades, the vision of constructing a photo-realistic digital representation of a person has driven research at the intersection of Computer Graphics, Computer Vision, and Machine Learning. Richard Feynman famously stated, “What I cannot create, I do not understand,” highlighting the epistemic value of creation as a means of understanding human appearance, behavior, and motion. 

\new{
But why do digital humans matter?
There are already several industries that depend on digital humans, especially in the entertainment field (gaming, VFX), but also in the e-commerce field (virtual try-on).
As interactions shift into virtual and hybrid environments (VR/AR, telepresence, online collaboration), people need human-centered representations that preserve identity, expression, and social nuance.
\emph{Digital humans become the bridge between physical and virtual communication, not only between humans, but also between humans and machines.}
%
%
}
%
%
%
%
%
%

%
A central question that is asked by many researchers, given the recent developments, is: \textit{Do we need 3D avatars in a world which is increasingly dominated by 2D representations?}
With the rise of image and video diffusion models, animatable portraits~\cite{xu2024vasa1, hu2023animateanyone}, and stylized video avatars~\cite{SVP:ECCV:24}, 2D avatars have demonstrated their value.
%
%
However, 3D avatars come with the promise of spatial interpretability and the ability to interact naturally within spatial environments, such as augmented and virtual reality, teleconferencing, and embodied AI. 
These capabilities render 3D avatars indispensable in scenarios where 2D representations fall short due to their inherent limitations in viewpoint consistency, representation compactness, render efficiency, and dynamic context adaptation.

\paragraph*{Scope.}
In this report, we will review the state-of-the-art in building 3D digital humans.
However, we will also discuss the nuanced relationship between 2D and 3D digital humans.
%
%
Our main goal is to develop a taxonomy for the field of digital humans, which will be especially helpful for new researchers in the field.
This report covers both human head avatars and full-body avatars.
We also present state-of-the-art methods for individual components such as hair, hands, and garments.
We emphasize topics related to building universal priors for avatars and analyze different representations used for tasks in both 2D and 3D. 
%
\new{Note that we assume avatar motion and control signals are given, and concentrate on techniques for photorealistic appearance modeling and synthesis.}
%
%

\paragraph*{Selection scheme.}
This report aims to provide a comprehensive overview of the current research field, list key literature for readers exploring specific areas, and help standardize notations and definitions.
The report includes papers published in the proceedings of major computer vision and computer graphics conferences from 2021 to 2025, as well as preprints available on arXiv.
The papers were carefully selected to fit the scope of this survey and to provide readers with an overview of a broad range of techniques.
Readers are encouraged to consult the cited publications for deeper insights.
%


\paragraph*{Structure of the report.}
After positioning the report in relation to existing surveys in \Cref{sec:related_surveys}, we introduce a formal definition of avatars in \Cref{sec:avatar_definition}. 
In \Cref{sec:fundamentals}, we introduce the fundamental components employed by modern methods to build prior models that can generalize across multiple actors, leveraging generative frameworks and statistical analysis.
In \Crefrange{sec:faces}{sec:full_body}, the main chapters, the state-of-the-art methods for different body parts are presented in detail.
As photo-realistic avatars also come with certain risks of misuse, we discuss social and ethical implications in \Cref{sec:ethics}.
\new{The report concludes with a discussion on the future of digital avatars in \Cref{sec:future_work}.}
Specifically, we formulate important research questions that are still open, such as improved efficiency, behavior modeling, spatial-aware avatars, or spatial audio.
%

\new{Overall, the report follows the digital human pipeline from representation and priors, through per-region modeling of heads, hair, hands, and garments, to full-body integration. As such, our goal is to paint a holistic picture of the current state of digital humans.}
%

%
%

%


\section{Related Surveys}
\label{sec:related_surveys}
%
Modeling the appearance and motion of 3D human avatars has attracted considerable research in the areas of Computer Graphics, Computer Vision, and Machine Learning.
%
%
%
%
The focus of this state-of-the-art report lies on reconstructing controllable human avatars leveraging a diverse set of 2D and 3D priors.
%
One of the most prominent priors in the field of digital humans are morphable models. Here, we want to highlight the excellent report by Egger \etal~\cite{Egger20203d}.
However, general 2D priors for natural image and video generation are also widely used.
We refer to the state-of-the-art reports on controllable video generation~\cite{Ma2025diff_star}, as well as the report on diffusion models for visual computing~\cite{po2024state}.
%
%
We do not review literature in the field of motion capturing and motion synthesis.
We want to direct the interested reader to the report on face tracking by Zollhöfer \etal~\cite{Zollhofer2018state}, the 3D face reconstruction survey of Morales \etal~\cite{Morales2021survey}, and a report on non-rigid 3D reconstruction by Tretsch \etal~\cite{TretschkNonRigidSurvey2023}.
Modeling the appearance of digital humans relies on different 3D representations that we will briefly introduce in \Cref{sec:fundamentals}.
For an in-depth review of current representations and the corresponding rendering schemes, we refer the reader to the reports on neural rendering by Tewari \etal~\cite{Tewari2020state,Tewari2022advances}.
The survey on 3D human avatar modeling of Wang \etal~\cite{wang2024survey} is closely related to our report. However, they focus on general avatar reconstruction, pixel-aligned representations, and generative modeling, e.g., with text-based control.
\new{
While the work by Gu \etal~\cite{gu20253dhumans} provides a detailed survey of neural field-based techniques for reconstructing 3D human avatars, and Wang \etal~\cite{wang2024survey} offers a thorough review of interaction design paradigms for 3D virtual humans, our survey aims to take a broader perspective. We cover a wider range of topics across the construction of digital humans, including foundational reconstruction and synthesis methods as well as specialized methods for human heads, hair, garments, hands, and full body avatars, thereby providing a more comprehensive overview of how digital humans are built and represented. In particular, our report includes a review of commonly used human priors as well as a unified framework of avatar systems facilitating classification of existing methods.
}

%% file: chapters/3_taxonomy.tex
\begin{figure*}[tb]
    \centering
    \includegraphics[width=1.0\linewidth]{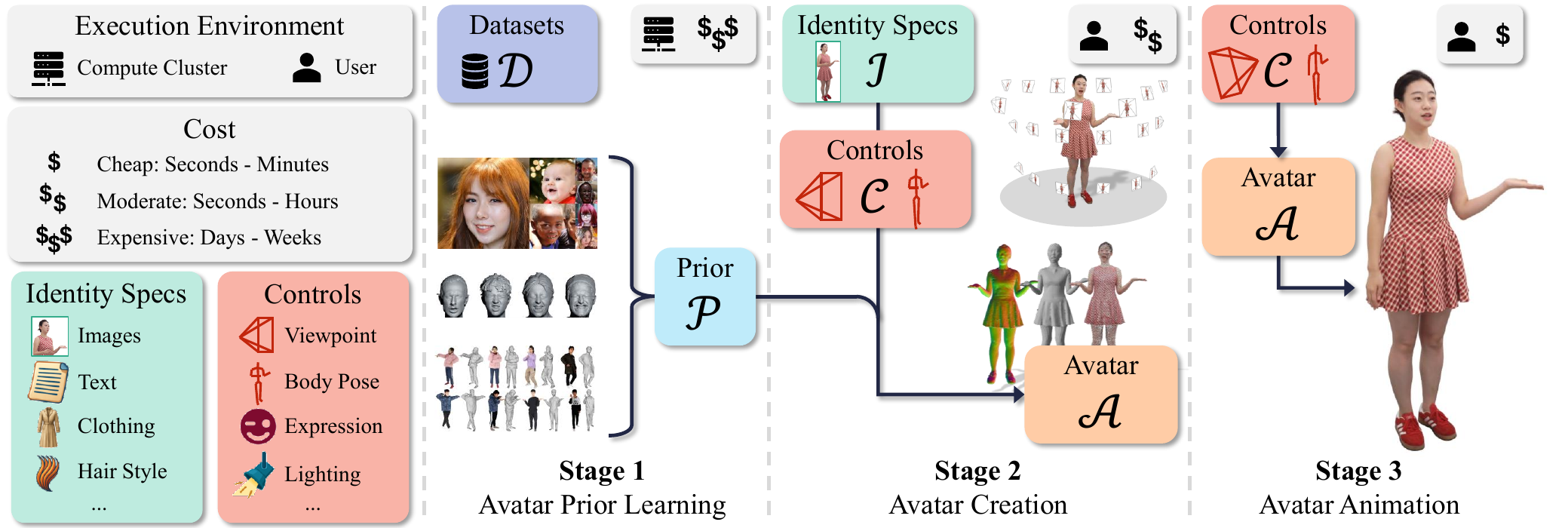} 
    \caption{A Common Framework of Avatar Systems. Images adapted from FFHQ~\cite{karras2021alias}, NPHM~\cite{giebenhain2023nphm}, THuman~\cite{yu2021function4d}, and TaoAvatar~\cite{chen2025taoavatar}.}
    \label{fig:avatar_system}
    \vspace{-5mm}
    \label{fig:taxonomy}
\end{figure*}

\section{Definition of an Avatar System}
\label{sec:avatar_definition}
\new{Given the breadth of approaches covered in this report, we propose a framework for categorizing different avatar systems.} 
We define an \textit{avatar} as a function $\mathcal{A}(\cdot)$ that represents the appearance of a person by taking a control signal $\mathcal{C}$ as input to synthesize an image $I$.
\paragraph*{Control signals.}
The control signal $\mathcal{C}$ consists of animation controls $\mathcal{M}$ to animate the facial expression or body pose, optional lighting $\mathcal{L}$, and camera controls $\pi$ that specify the viewpoint:
\[\mathcal{C} = \{\mathcal{M}[,\mathcal{L}], \pi\}.\]
%
%
%
\new{In our framework, control signals $\mathcal{C}$ are restricted to modifying non–identity-related attributes of the avatar, such as pose or rendering viewpoint. By contrast, inputs that would alter the avatar’s identity, such as editing instructions or alternative reference images, are incorporated during the avatar creation process which we will describe below. The animation controls $\mathcal{M}$ may include low-level specifications, such as per-frame body joint angles, as well as high-level cues, such as emotional states.}
\paragraph*{2D and 3D Avatars.}
The \textit{avatar} can be based on an explicit or an implicit image formation process.
A \textit{2D avatar} $\mathcal{A}_{2D}$ directly synthesizes the final image $I$ given all control signals $\mathcal{C}$:
\[I = \mathcal{A}_{2D}(\mathcal{C}).\]
Note that in our definition, a \textit{2D avatar} is still controllable by the camera viewpoint $\pi$, which creates an apparent 3D effect without true 3D modeling.
In contrast, a \textit{3D avatar} $\mathcal{A}_{3D}$ factorizes the image formation process into two parts: (i) Creating a 3D representation $f_{3D}(\mathcal{M}[,\mathcal{L}])$ given the animation control signals $\mathcal{M}$ and optional lighting $\mathcal{L}$, and (ii) rendering the created 3D representation from camera $\pi$ using a renderer $\mathcal{R}$:
\[I = \mathcal{A}_{3D}(\mathcal{C}) = \mathcal{R}(f_{3D}(\mathcal{M}[,\mathcal{L}]), \pi).\]
As such, a \textit{3D avatar} can be seen as a special case of a \textit{2D avatar} where the use of a 3D representation in the pipeline is enforced. \\
An \textit{avatar system} produces avatars matching specific identity inputs $\mathcal{I}$, such as a person’s images or a textual description.
%
Current avatar systems consist of (1) an optional avatar prior learning stage, (2) the avatar creation stage, and (3) an avatar animation stage.
These stages differ by how costly they are and by whom they are executed:
While learning a prior $\mathcal{P}$ on a dataset $\mathcal{D}$ may be very expensive and is only executed once by the avatar system's builder, the creation stage is executed by the end user for each desired avatar, and the animation stage is run for each final frame, ideally in real-time. A schematic overview of the different stages in an avatar system is depicted in~\Cref{fig:avatar_system}.
%

\paragraph*{Stage 1 (optional) - avatar prior learning.}
The prior learning stage enables under-constrained avatar creation tasks that are unsolvable by optimization alone, for example, single-image avatar generation.
A prior $\mathcal{P}$ is distilled by training a generalized model on diverse human data, requiring a large-scale dataset $\mathcal{D}$:
\begin{equation}
    \mathcal{D} \rightarrow \mathcal{P}.
\end{equation}
%
%
Publications vary widely in the types and modalities of data they use (see~\Cref{tab:avatar_pipeline_taxonomy} - Prior Stage).
%

\paragraph*{Stage 2 - avatar creation.}
In the creation stage, the actual avatar $\mathcal{A}$ is created from identity specifications $\mathcal{I}$, corresponding control signals $\mathcal{C}$, and (optional) human priors $\mathcal{P}$. For example, this could involve optimizing for an animatable 3D representation given videos of a person, or \new{generating} an avatar from text:
\begin{equation}
    \{\mathcal{I}, \mathcal{C}, \mathcal{P}\} \rightarrow \mathcal{A}.
\end{equation}
\new{The resulting avatar $\mathcal{A}$ is intended to represent the person specified by the identity parameters $\mathcal{I}$. This formulation encompasses both (i) unimodal approaches, in which an avatar is generated from images or a textual description, as well as (ii) multimodal and iterative creation processes, where editing instructions modify an existing avatar or additional specifications such as clothing materials further refine its appearance.}
To make avatar creation tractable, many systems use existing priors such as the geometry of a 3D morphable model. Additionally, the input requirements can vary drastically, ranging from a single image to multi-view video recordings with expensive mesh registrations (see~\Cref{tab:avatar_pipeline_taxonomy} - Creation Stage). These requirements determine in which scenarios an avatar system can be deployed.
%
%

\paragraph*{Stage 3 - avatar animation.}
The animation stage enables animation via control signals $\mathcal{C}$ and produces the final image $I$:
\begin{equation}
    I = \mathcal{A}(\mathcal{C}).
\end{equation}
The animation stage is mainly dependent on the animation signal, but may also incorporate additional controls such as lighting (see~\Cref{tab:avatar_pipeline_taxonomy} - Animation Stage).
The animation speed is crucial in this stage, as the avatar should ideally be animated in real-time on an end-user device such as a smartphone or VR headset.

\new{Note that while this report discusses all three stages, it does not have a focus on how the control and animation signal $\mathcal{C}$ is obtained or generated. For face and head tracking that is used as control and animation signal in a series of methods, we refer to the report of Zollhöfer \etal~\cite{Zollhofer2018state}. In addition, we give a short overview of hair and garment animation in \Cref{appendix:hair,appendix:garment}.}

\history{}\new{With this framework in place, we next introduce the fundamental building blocks, 3D representations, human priors, animation controls, and material models, that underlie the methods reviewed in the subsequent chapters.}

%% file: chapters/4_fundamentals.tex
\section{Fundamentals}
\label{sec:fundamentals}


This section covers the fundamental concepts for creating 3D digital avatars \new{that recur across all body regions and creation scenarios}, following the progression from the avatar's representation and creation to its animation and realistic appearance.
%
This addresses four fundamental questions:
\begin{itemize}
\item \textbf{What is it made of?} We begin with an overview of core 3D representations such as meshes, point clouds, and neural fields, which form the avatar's underlying structure.
\item \textbf{How do we build it?} Next, we explore types of human 2D and 3D priors -- statistical models crucial for generating plausible digital humans, especially from limited and incomplete data.
\item \textbf{How does it move?} We then discuss the control signals and animation of digital humans.
\item \textbf{How does it look?} The section concludes with an examination of commonly used material and lighting models, which determine the avatar's photorealistic appearance.
\end{itemize}
%

\subsection{What is it made of? -- 3D Representations}

For the digitization of a human, we need a digital representation that stores information about the appearance of the human, including its geometry and material properties.
%
These representations can be completely implicit and not interpretable by a human (e.g., the latent codes of some diffusion model), but they can also be human-interpretable, e.g., a mesh, and, thus, compatible with classical computer graphics concepts and pipelines.
%
\new{Compatibility with classical graphics pipelines enables not only explicit camera control, but for example also grants access to established geometry processing tools that allow for editing and animation.}
%
The reader is advised to consult the state-of-the-art report on 3D neural rendering~\cite{Tewari2022advances} where a detailed introduction to the different 3D representation types is given.
In the context of digital humans, we predominantly see 3D representations that either define the surface (explicitly with a mesh or implicitly with a signed distance function (SDF)) or the volume (occupancy, optical densities, radiance fields).
For the surface and volume representations, different types of rendering schemes can be used.
Every surface that can be indexed can be rendered in a forward process, where the surface is projected onto the image plane. Techniques like rasterization are extremely efficient, as modern GPUs have dedicated hardware units for it.
Implicit representation, as well as volume information, needs to be gathered along rays that originate at a camera position and are cast through a pixel into the scene.
This process is called ray casting, or ray-tracing if additional rays are shot to evaluate global illumination effects.
Both ray casting and rasterization allow for explicit control of the camera.
Surface and volume representations can be stored in discrete, e.g., points, triangles, voxel grids, or continuous formats. 
%
In the past few years, continuous functions were widely used, especially in the form of neural networks, as they can be seen as universal function approximators that can store 3D content information.
A very prominent example is NeRF (neural radiance fields)~\cite{Mildenhall2020nerf}, a volumetric representation that uses a multi-layer perceptron (MLP) as a function approximator to store optical density and radiance information at given points in 3D space.
With a differentiable volumetric integration scheme using ray casting, this representation can be used for reconstructing and storing a digital model of a scene or object from a set of input images.
The model can then be used for novel view synthesis by changing the camera pose.
Recently, many approaches have been using 3D Gaussian primitives~\cite{Kerbl20233d}. These are radial basis functions (RBF)~\cite{rbf} that span a smooth function across a 3D space.
It can be seen as a hybrid between an explicit representation (discrete point samples) and a smooth continuous function (Gaussian kernel) attached to it.
The radial basis functions are often initialized on the surface by leveraging sparse point clouds obtained \new{via classical structure from motion~\cite{schoenberger2016sfm,schoenberger2016mvs}, learning-based dense point-cloud estimations~\cite{dust3r_cvpr24,mast3r_eccv24,wang2025vggt},} or from a reconstructed mesh (e.g., from FLAME~\cite{Li2017flame}).
Using a volumetric rendering scheme that leverages rasterization of the primitives, it can be implemented in a very efficient way, and, similar to NeRF, allows for novel view synthesis once the representation is optimized.
For digital humans, motions and surface deformations are a key component.
In this context, explicit representations such as the Gaussian primitives have an advantage over implicit structures such as NeRF.
For 3D Gaussians, deformation fields can be applied in a forward process, while implicit representations need a backward process.
%
%
%
%
\new{
Such deformation fields for modeling the human motion can be based on a classical motion prior like a Blendshape-based morphable model~\cite{Li2017flame,Blanz1999}, a motion proxy, e.g., cage-based~\cite{cage-based-deformation-survey} or skeleton-based deformation (often based on linear blend skinning (LBS)~\cite{loper2015smpl}), physics-based simulation (especially, in the context of hair and garment), or learned representations.
}
It is important to understand that these different representations allow for various controls of the avatar.
\new{For example, learned deformation field representations like HyperNeRF~\cite{Park2021hypernerf} or Nerfies~\cite{Park2021nerfies} are able to model the deformations in an input sequence conditioned on a time stamp, which enables free viewpoint video playback of the sequence.}
However, these methods do not allow for generating novel motion. 
\new{
In contrast, morphable models~\cite{Li2017flame,Blanz1999} and skeleton-based deformations~\cite{loper2015smpl} can be controlled by the underlying motion parameters -- they can be seen as generative models that could generate an infinite amount of novel deformation fields.
}
%


\subsection{How do we build it? -- Human Priors}

\begin{figure}[tb]
    \centering
    \includegraphics[width=\columnwidth]{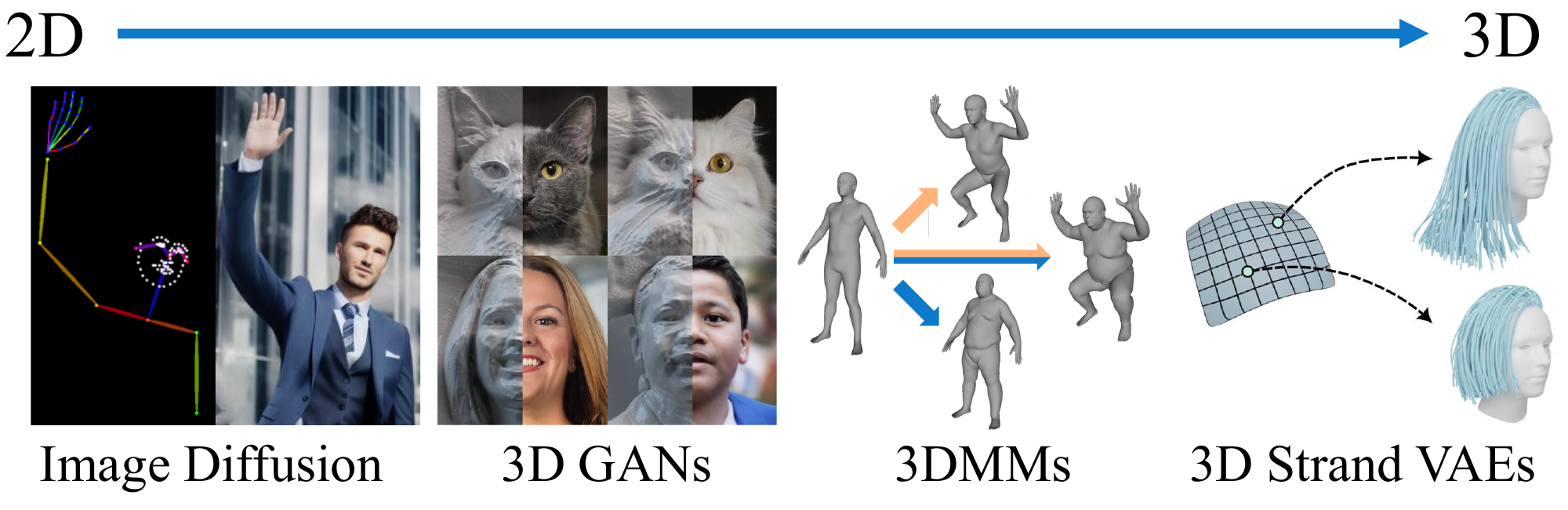}
    \caption{
        Types of 2D and 3D priors commonly used for digital humans. Images adapted from Controlnet~\cite{zhang2023adding}, EG3D~\cite{chan2022eg3d}, SMPL~\cite{loper2015smpl}, and Perm~\cite{He2025perm}.
    }
    \vspace{-5mm}
    \label{fig:priors}
\end{figure}

\new{While the 3D representations discussed above define what an avatar is made of, they do not by themselves resolve the ambiguities inherent in avatar creation. Many avatar creation scenarios are heavily under-constrained, such as creating an avatar from a text prompt or a few photos.} In these cases, the provided identity specification is not sufficient to describe the desired avatar and therefore a prior model encoding generic beliefs about humans has to be employed to hallucinate or fill in unseen regions. Such human prior models can be quite diverse in their design (see \Cref{fig:priors}). Some may only provide guidance on typical 3D human surface geometry while others focus on human appearance in 2D. All such priors are learned from large-scale datasets of human examples.

\subsubsection{What is a Human Prior?}

In the context of this report, a human prior models the distribution of human shape, appearance, and motion in 2D or 3D.
Typical examples include 3D morphable models (3DMMs) that provide animatable geometry, e.g., by rigging 3D Gaussians to a 3DMM mesh~\cite{Qian2024gaussianavatars}. Other approaches leverage image diffusion model priors to regularize unseen viewpoints~\cite{tang2025gaf, Taubner2025cap4d} or employ 3D GANs for avatar reconstruction from sparse inputs~\cite{tran2024voodooxp}.



\subsubsection{3D Morphable Models (3DMMs)}
\label{sec:3dmm}
Pioneered by Blanz and Vetter~\cite{Blanz1999}, there is a long-standing tradition of building mesh-based 3D parametric models for bodies and body parts
~\cite{loper2015smpl, pavlakos2019expressive,Li2017flame, paysan2009bfm, romero2017mano, potamias2023handy}.
These are typically obtained by first registering a generic human template mesh to a dataset of 3D human scans and then performing dimensionality reduction techniques on the registered meshes. The result is a (linear) parametric 3D geometry model with low-dimensional controls for aspects such as body shape, skeletal articulation, facial expression, or head pose. For more details on 3DMMs, see the survey by Egger \etal~\cite{Egger20203d}. Recent works have explored different 3D representations to build 3DMMs~\cite{Giebenhain2024mononphm, giebenhain2023nphm, yenamandra2021i3dmm, zheng2022imface}. 
In the context of human avatars, 3DMMs are tremendously useful because they offer a coarse geometry proxy and disentangled control over human identity and articulation. A common approach is to rig 3D primitives, such as 3D Gaussians, to an existing animatable 3DMM~\cite{Qian2024gaussianavatars,karunratanakul2023harp}. As evident from~\Cref{tab:taxonomy}, the most prominent control signals for animating a human avatar are still facial expression or body pose codes from a 3DMM.

\subsubsection{3D Generative Adversarial Networks (3D GANs)}
With sufficient observations of a person, such as multi-view studio captures, avatar creation can be treated as a pure optimization problem.
However, more challenging scenarios, such as single- or few-image reconstruction settings, require additional prior knowledge about the typical appearance of 3D humans to plausibly generate regions of the avatar that have never been observed. 
One such prior is 3D GANs that learn a distribution of realistic 3D humans from a large 2D image dataset~\cite{chan2022eg3d, abdal2024gsm}. Once trained, a 3D GAN allows sampling random 3D humans from the learned distribution. The 3D prior learned by 3D GANs is particularly useful in underconstrained reconstruction scenarios, as the model's generative capabilities can fill in unseen regions and plausibly complete the final avatar. For example, several 3D portrait animation methods build on top of pre-trained 3D GANs to enable creating a 3D avatar from just a single image~\cite{tran2024voodooxp, ma2023otavatar} or use 3D GANs as a synthetic multi-view data creation engine~\cite{Deng2024portrait4d, ye2024real3d}.

\subsubsection{2D Image Generation Priors}
Similar to 3D GANs, pre-trained 2D image generation models help in under-constrained avatar creation tasks. However, additional measures are necessary to harness the learned 2D human appearance prior to 3D completion. Two major questions arise: (i) How to extract 3D knowledge from the 2D image generation model, and (ii) how to ensure that its output is relevant to the target avatar. Regarding (i), common approaches use the image generation model either as a critic, e.g., via Score Distillation Sampling~\cite{poole2023dreamfusion, zhou2024headstudio}, or as a data generation engine that synthesizes additional images from various viewpoints, poses, or facial expressions for subsequent avatar reconstruction~\cite{Taubner2025cap4d}. \new{Another approach is to use a pre-trained image generation model as a neural renderer to directly generate the pixels of the final avatar~\cite{kirschstein2024diffusionavatars, ding2023diffusionrig, svdp}.}
Regarding (ii), a typical solution conditions image generation on coarse 3DMM renderings or on facial landmarks of the target~\cite{tang2025gaf, Taubner2025cap4d, zhou2024headstudio, wu2025animportrait3d}. This way, the viewpoint and other aspects are controlled in the generated images, ensuring that they provide constraints for avatar reconstruction.

\new{\subsubsection{Video Generation Priors}}
\new{Similar to 2D image priors, open-source video generation models~\cite{hong2022cogvideo, blattmann2023stable, wan2025wan, hacohen2024ltx, kong2024hunyuanvideo} can be a powerful tool to create avatars in under-constrained scenarios. For example, such models can be used to generate additional data of a person, which can then be distilled into an animatable 3D avatar~\cite{zhou2025zero1toA, taubner2025mvp4d}. Moreover, video diffusion models with animation control \cite{cheng2025wananimate, hu2023animateanyone, lin2025omnihuman1, ma2024followyouremoji} are capable of producing highly realistic portrait animations, but typically lack precise control over camera motion. To overcome this limitation, several recent works have investigated methods for jointly controlling camera trajectories and human motion in generated videos~\cite{li2025tokenmotion, li2025adaviewplanner, cao2025uni3c, liang2025realismotion}. Collectively, these advances suggest that controllable video generation models may emerge as a viable alternative to fully 3D-based avatar modeling.}

\subsection{How does it move? -- Animation Control}

\new{Beyond representations and priors,} a key aspect of digital humans is their animatability, where the animation can be driven by diverse control signals, as seen in~\Cref{tab:taxonomy}.

\paragraph*{3DMM expression~\scalebox{0.6}{\icoExprThreeDMM} and pose~\scalebox{0.75}{\icoPose}.}
The predominant approaches rely on low-dimensional pose and expression parameters from mesh-based parametric 3D morphable models (3DMMs) of the body \cite{loper2015smpl, pavlakos2019expressive}, head \cite{Li2017flame}, or hands \cite{romero2017mano}.
These 3DMM parameters can be integrated into the avatar's architecture in several ways.
Some methods directly apply the model’s Linear Blend Skinning (LBS) pose articulation and expression blendshapes to deform the avatar’s core representation.
This technique is used across avatar types, including full-body methods where latent codes are anchored to and deformed by a SMPL model \cite{peng2021neural}, head avatars where 3D Gaussian splats are rigged to a FLAME head mesh \cite{Qian2024gaussianavatars}, and hand \cite{karunratanakul2023harp} and clothing methods \cite{Feng2022scarf}, where the 3DMM’s articulation drives the deformation of the final geometry.
More indirectly, some methods use the 3DMM for 2D-based neural rendering by rasterizing feature maps attached to the 3DMM’s body \cite{yu2025humanram} or head \cite{kirschstein2024diffusionavatars} mesh surface from a target viewpoint, which is then input to an image generator.
At a higher abstraction level, some techniques bypass the 3DMM’s explicit geometry entirely and instead directly condition the avatar’s neural representation on the 3DMM’s expression parameters \cite{Gafni2021dynamic}.
%
%
One key advantage of using a generic 3DMM is the ability to drive the avatar with off-the-shelf tools that reconstruct body and face parameters from images \cite{feng2021pixie, Feng2021deca}.
\new{Furthermore, these models can be animated via high-level signals such as audio~\scalebox{0.8}{\icoAudio} involving music \cite{tseng2023edge} or speech \cite{chhatre2024amuse, aneja2024gaussianspeech} modulated by categorical emotions \cite{danecek2023emote,peng2023emotalk}, or through textual descriptions}
\cite{Petrovich2023tmr}.

\paragraph*{General expression~\scalebox{0.75}{\icoExprGeneral}.}
To capture fine-grained, subject-specific details beyond the linear basis of generic 3DMMs, many methods learn custom neural embeddings.
This involves creating a personalized expression latent space for heads, for instance by modeling the face as a set of deformable volumetric primitives controlled by a latent expression code \cite{Lombardi21mvp}, learning neural deformations on top of a SMPL body mesh \cite{liu2021neural}, or modeling custom hand articulations on top of an LBS-articulated mesh \cite{Iwase2023relightablehands} or an articulated MANO skeleton \cite{corona2022lisa}.
While this produces higher-fidelity results, it requires a dedicated encoder to map driving inputs (e.g., multi-view images~\scalebox{0.8}{\icoMultiImage} \cite{teotia2024gaussianheads}, monocular videos~\scalebox{0.8}{\icoVideo}, or speech~\scalebox{0.8}{\icoAudio}) to this custom expression space, which is trained jointly with the avatar \cite{Remelli2022drivable} or as a separately fine-tuned regressor \cite{Zielonka2025gem}.

\paragraph*{Simulation~\scalebox{0.75}{\icoSimulation}.}
For avatars with secondary elements like clothing and hair, a common two-stage control mechanism uses the primary body motion (e.g., specified by SMPL pose parameters), to inform a secondary simulation.
\new{
This secondary step can involve physics-based simulation using 3DMM meshes as colliders for simulating clothing layers~\cite{zheng2024physavatar, xiang2022DressingAvatars} or hair dynamics~\cite{Zakharov2024haircut}
(see \Cref{appendix:hair,appendix:garment}).}

\paragraph*{Parametric model-free control.}
Finally, some methods bypass parametric models entirely, opting for direct frame-based or audio-based control.
These approaches learn a direct mapping from a driving video for reenactment \cite{delagorce2025volume, tran2024voodooxp} or from speech \new{with control signals such as emotion or eye gaze} to synthesize expressive motion without any explicit 3D model \cite{xu2024vasa1}.

\subsection{How does it look? -- Material and Relighting}
\new{The final fundamental aspect concerns appearance. }A key requirement for digital avatars is consistent visual accuracy under varying lighting conditions.
This depends on modeling both surface material properties and scene illumination.
One of the main challenges lies in properly modeling the materials of digital avatars.
%
For example, human skin is translucent and is traditionally modeled using subsurface scattering~\cite{hanrahan1993reflection, wann2001practical}. Garments present a similarly complex challenge due to their anisotropic structures~\cite{ward1992measuring, ashikhmin2000anisotropic}, while hair is particularly difficult to model because of its fine geometry and complex light interactions~\cite{marschner2003light, zinke2008dual}.

This report considers two illumination configurations supported by the selected relighting methods: \textbf{distant} and \textbf{point} light sources. 

\noindent
\textbf{Distant light}~\scalebox{0.9}{\icoDistantLight}~assumes the light source to be located infinitely far away from the surface. Consequently, the incoming direction and intensity are constant over the entire object, and the illumination depends only on the incident direction~$\boldsymbol{\omega}_i$ (if self-shadowing and other effects are ignored). This formulation models sunlight as a directional source or environment illumination represented by a high-dynamic-range (HDR) environment map~\cite{debevec2000acquiring}:
\begin{equation}
L_i(\mathbf{x}, \boldsymbol{\omega}_i) = L_i(\boldsymbol{\omega}_i).
\end{equation}
For Lambertian surfaces, illumination is efficiently approximated with spherical harmonics~\cite{ramamoorthi2001efficient}:
\begin{equation}
L_i(\boldsymbol{\omega}_i)
\approx
\sum_{\ell=0}^{L} \sum_{m=-\ell}^{\ell}
c_{\ell m}\, Y_{\ell m}(\boldsymbol{\omega}_i),
\end{equation}
where the incoming direction~$\boldsymbol{\omega}_i$ is used to evaluate the spherical harmonic basis functions~$Y_{\ell m}$.

\noindent
\textbf{Local light}~\scalebox{0.9}{\icoLocalLight}~represents one or more light sources positioned at a finite distance from the surface. Consequently, both the direction and intensity of the incoming light vary with the surface position~$\mathbf{x}$. The illumination in this case is expressed as:
\begin{equation}
L_i(\mathbf{x}, \boldsymbol{\omega}_i)
= \frac{I(\mathbf{p})}{\|\mathbf{p} - \mathbf{x}\|^2}\, V(\mathbf{x}, \mathbf{p}),
\end{equation}
where~$\mathbf{p}$ denotes the light position,~$I(\mathbf{p})$ its emitted intensity, and~$V(\mathbf{x}, \mathbf{p})$ a visibility term indicating whether the light source is visible from the surface point~$\mathbf{x}$. This configuration models localized illumination, such as a lamp, flash, or other nearby light sources. Further details on material representation and relighting implementations are provided in the appendix.

\history{}\new{With these fundamentals established, i.e., 3D representations, human priors, animation controls, and material models, the following sections will discuss the state-of-the-art methods for individual body regions 
to full-body avatars, beginning with human head avatars.
}

%% file: tables/legend.tex
\begin{table*}[th!]
  \centering
  \small
  \setlength{\tabcolsep}{4pt}
  \renewcommand{\arraystretch}{1.5}
  \begin{NiceTabularX}{\linewidth}{lX}[colortbl-like]
   \toprule
\Block[fill=ForestGreen!10]{1-2}{\textbf{Prior Stage}} & \\
   \midrule
   Datasets &
\val{2K2K}{\scalebox{0.9}{\crboxKK}}\valsep\val{AIST}{\scalebox{0.9}{\crboxAIST}}\valsep\val{ActorsHQ}{\scalebox{0.9}{\crboxActorsHQ}}\valsep\val{Ava256}{\scalebox{0.9}{\crboxAva}}\valsep\val{BCNet}{\scalebox{0.9}{\crboxBCNet}}\valsep\val{CT2Hair}{\scalebox{0.9}{\crboxCTHair}}\valsep\val{CelebV-HQ}{\scalebox{0.9}{\crboxCelebVHQ}}\valsep\val{Cloth3D}{\scalebox{0.9}{\crboxClothD}}\valsep\val{CustomHumans}{\scalebox{0.9}{\crboxCustomHumans}}\valsep\val{Ensemble}{\scalebox{0.9}{\crboxEnsemble}}\valsep\val{FFHQ}{\scalebox{0.9}{\crboxFFHQ}}\valsep\val{FaceScape}{\scalebox{0.9}{\crboxFaceScape}}\valsep\val{FaceVerse}{\scalebox{0.9}{\crboxFaceVerse}}\valsep\val{Human4DiT}{\scalebox{0.9}{\crboxHumanDiT}}\valsep\val{InterHand2.6M}{\scalebox{0.9}{\crboxInterHandM}}\valsep\val{Internal}{\scalebox{0.9}{\crboxInternal}}\valsep\val{LPFF}{\scalebox{0.9}{\crboxLPFF}}\valsep\val{MEAD}{\scalebox{0.9}{\crboxMEAD}}\valsep\val{MRI-Hand}{\scalebox{0.9}{\crboxMRIHand}}\valsep\val{NPHM}{\scalebox{0.9}{\crboxNPHM}}\valsep\val{NeRSemble}{\scalebox{0.9}{\crboxNeRSemble}}\valsep\val{Next3D samples}{\scalebox{0.9}{\crboxNextDsamples}}\valsep\val{PanoHead samples}{\scalebox{0.9}{\crboxPanoHeadsamples}}\valsep\val{RenderPeople}{\scalebox{0.9}{\crboxRenderPeople}}\valsep\val{SURREAL}{\scalebox{0.9}{\crboxSURREAL}}\valsep\val{Sewing Pattern}{\scalebox{0.9}{\crboxSewingPattern}}\valsep\val{THuman2.0}{\scalebox{0.9}{\crboxTHuman}}\valsep\val{Twindom}{\scalebox{0.9}{\crboxTwindom}}\valsep\val{USC-HairSalon}{\scalebox{0.9}{\crboxUSCHairSalon}}\valsep\val{VFHQ}{\scalebox{0.9}{\crboxVFHQ}}\valsep\val{VoxCeleb2}{\scalebox{0.9}{\crboxVoxCeleb}}\valsep\val{ZJU-MoCap}{\scalebox{0.9}{\crboxZJUMoCap}}
\\
\rowcolor{green!04}
   Data Type &
\val{Real}{\scalebox{0.9}{\icoReal}}\valsep\val{Synthetic}{\scalebox{0.9}{\icoSynth}}\valsep\val{Generated (e.g., by a pre-trained generative model)}{\scalebox{0.9}{\icoGenerated}}
\\
   Data Modality &
\val{3D Strands}{\scalebox{0.9}{\icoStrands}}\valsep\val{MRI}{\scalebox{0.9}{\icoMRI}}\valsep\val{Meshes}{\scalebox{0.9}{\icoMeshes}}\valsep\val{Mono video}{\scalebox{0.9}{\icoMonoVideo}}\valsep\val{Multi-view image}{\scalebox{0.9}{\icoMultiImage}}\valsep\val{Multi-view video}{\scalebox{0.9}{\icoMultiViewVideo}}\valsep\val{OLAT multi-view image}{\scalebox{0.9}{\icoOLATImage}}\valsep\val{OLAT multi-view video}{\scalebox{0.9}{\icoOLATVideo}}\valsep\val{Sewing Pattern}{\scalebox{0.9}{\icoSewingPattern}}\valsep\val{Single image}{\scalebox{0.9}{\icoSingleImage}}
\\
   \midrule
\Block[fill=RoyalBlue!10]{1-2}{\textbf{Creation Stage}} & \\
   \midrule
   Needed Assets &
\val{Cameras}{\scalebox{0.9}{\crboxCameras}}\valsep\val{Hand Pose}{\scalebox{0.9}{\crboxHandPose}}\valsep\val{Known Lighting}{\scalebox{0.9}{\crboxKnownLighting}}\valsep\val{Meshes}{\scalebox{0.9}{\crboxMeshes}}\valsep\val{Registered Meshes}{\scalebox{0.9}{\crboxRegisteredMeshes}}\valsep\val{Textures}{\scalebox{0.9}{\crboxTextures}}\valsep\val{Tracked 3DMM}{\scalebox{0.9}{\crboxTrackedDMM}}\valsep\val{segmentation masks}{\scalebox{0.9}{\crboxsegmentationmasks}}
\\
\rowcolor{blue!04}
   Input &
\val{Text}{\scalebox{0.9}{\icoText}}\valsep\val{Audio}{\scalebox{0.9}{\icoAudio}}\valsep\val{Zero}{\scalebox{0.9}{\icoZero}}\valsep\val{One}{\scalebox{0.9}{\icoOne}}\valsep\val{Few}{\scalebox{0.9}{\icoFew}}\valsep\val{Mono video}{\scalebox{0.9}{\icoMonoVideo}}\valsep\val{Multi-view images}{\scalebox{0.9}{\icoMultiImage}}\valsep\val{Multi-view video}{\scalebox{0.9}{\icoMultiViewVideo}}\valsep\val{OLAT multi-view images}{\scalebox{0.9}{\icoOLATImage}}\valsep\val{OLAT multi-view video}{\scalebox{0.9}{\icoOLATVideo}}\valsep\val{Depth}{\scalebox{0.9}{\icoDepth}}\valsep\val{3D density volume}{\scalebox{0.9}{\icoMRI}}\valsep\val{Point cloud}{\scalebox{0.9}{\icoPointCloud}}
\\
   Additional Priors &
\val{3DMM}{\scalebox{0.9}{\crboxDMM}}\valsep\val{ContourCraft}{\scalebox{0.9}{\crboxContourCraft}}\valsep\val{DeepFeatures}{\scalebox{0.9}{\crboxDeepFeatures}}\valsep\val{DeepMVSHair}{\scalebox{0.9}{\crboxDeepMVSHair}}\valsep\val{GPT-4V}{\scalebox{0.9}{\crboxGPTV}}\valsep\val{Garment templates}{\scalebox{0.9}{\crboxGarmenttemplates}}\valsep\val{GroomCap}{\scalebox{0.9}{\crboxGroomCap}}\valsep\val{Grounded-SAM}{\scalebox{0.9}{\crboxGroundedSAM}}\valsep\val{Image Diffusion}{\scalebox{0.9}{\crboxImageDiffusion}}\valsep\val{NeuralHaircut}{\scalebox{0.9}{\crboxNeuralHaircut}}\valsep\val{Physics}{\scalebox{0.9}{\crboxPhysics}}\valsep\val{PortraitAnimation}{\scalebox{0.9}{\crboxPortraitAnimation}}\valsep\val{Relighting}{\scalebox{0.9}{\crboxRelighting}}\valsep\val{Semantic segmentation}{\scalebox{0.9}{\crboxSemanticsegmentation}}\valsep\val{Wave2Vec}{\scalebox{0.9}{\crboxWaveVec}}
\\
\rowcolor{blue!04}
   Creation Speed &
\val{Slow (>6 hours)}{\scalebox{0.9}{\icoSlow}}\valsep\val{Medium (<6 hours)}{\scalebox{0.9}{\icoMedium}}\valsep\val{Fast (<30 minutes)}{\scalebox{0.9}{\icoFast}}\valsep\val{Instant (<1 minute)}{\scalebox{0.9}{\icoInstant}}
\\
   \midrule
\Block[fill=BurntOrange!10]{1-2}{\textbf{Animation Stage}} & \\
   \midrule
   Animation Signal &
\val{3DMM expr}{\scalebox{0.9}{\icoExprThreeDMM}}\valsep\val{Audio}{\scalebox{0.9}{\icoAudio}}\valsep\val{General expr}{\scalebox{0.9}{\icoExprGeneral}}\valsep\val{Multi-view image}{\scalebox{0.9}{\icoMultiImage}}\valsep\val{Pose}{\scalebox{0.9}{\icoPose}}\valsep\val{Video}{\scalebox{0.9}{\icoVideo}}
\\
\rowcolor{orange!04}
   Lighting Control &
\val{Distant Light}{\scalebox{0.9}{\icoDistantLight}}\valsep\val{Local Light}{\scalebox{0.9}{\icoLocalLight}}
\\
   Animation Speed &
\val{Offline (<1 fps)}{\scalebox{0.9}{\icoOffline}}\valsep\val{Interactive (>5 fps)}{\scalebox{0.9}{\icoInteractive}}\valsep\val{Real-time (>30 fps)}{\scalebox{0.9}{\icoRealtime}}
\\
\rowcolor{orange!04}
   Image Synthesis &
\val{3DGS}{\scalebox{0.9}{\crboxDGS}}\valsep\val{MVP}{\scalebox{0.9}{\crboxMVP}}\valsep\val{Mesh}{\scalebox{0.9}{\crboxMesh}}\valsep\val{NeRF}{\scalebox{0.9}{\crboxNeRF}}\valsep\val{Neural Rendering}{\scalebox{0.9}{\crboxNeuralRendering}}\valsep\val{SDF}{\scalebox{0.9}{\crboxSDF}}\valsep\val{Strands}{\scalebox{0.9}{\crboxStrands}}
\\
   \bottomrule
  \end{NiceTabularX}
  \caption{Avatar pipeline taxonomy legend of the three stages: prior learning, avatar creation and avatar animation.}
  \label{tab:avatar_pipeline_taxonomy}
\end{table*}

%% file: tables/taxonomy.tex
\begin{table*}[th!]
  \centering
  \footnotesize
  \setlength{\tabcolsep}{2pt}
  \resizebox{\textwidth}{!}{%
  \begin{NiceTabular}{|l|l|>{\centering\arraybackslash}p{1.1cm}>{\centering\arraybackslash}p{1.8cm}>{\centering\arraybackslash}p{1.1cm}>{\centering\arraybackslash}p{1.5cm}|>{\centering\arraybackslash}p{1.1cm}>{\centering\arraybackslash}p{1.1cm}>{\centering\arraybackslash}p{1.1cm}>{\centering\arraybackslash}p{1.1cm}>{\centering\arraybackslash}p{0.8cm}|>{\centering\arraybackslash}p{0.8cm}>{\centering\arraybackslash}p{0.8cm}>{\centering\arraybackslash}p{0.8cm}>{\centering\arraybackslash}p{0.8cm}>{\centering\arraybackslash}p{0.8cm}>{\centering\arraybackslash}p{1.1cm}|}[colortbl-like]
   \toprule
\Block[B]{2-2}{\includegraphics[height=1.9cm]{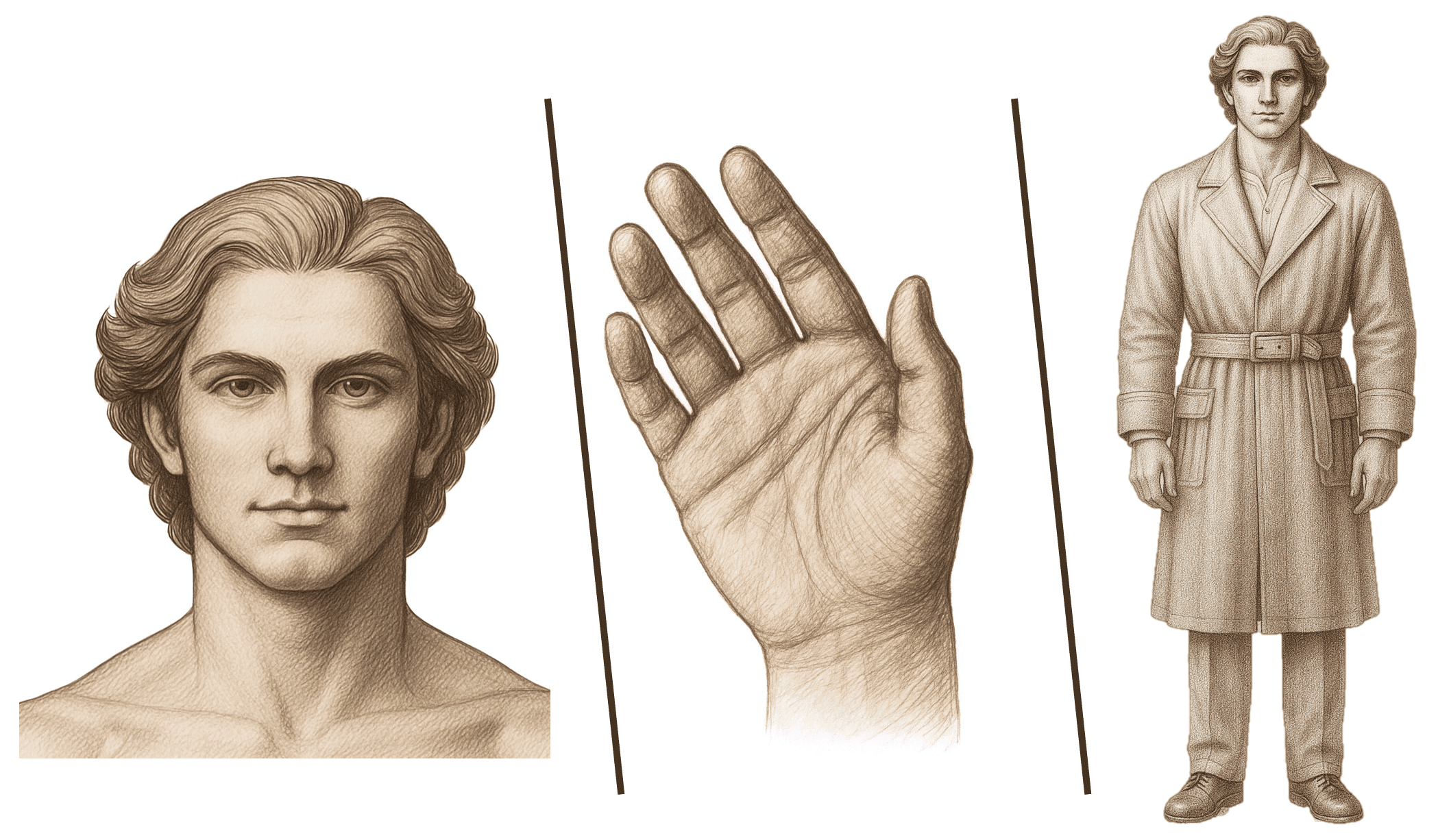}\\\\ \textbf{Digital Human Avatars}} & & \Block[fill=ForestGreen!10]{1-4}{\textbf{Avatar Prior}} &&& & \Block[fill=RoyalBlue!10]{1-5}{\textbf{Avatar Creation}} &&&& & \Block[fill=BurntOrange!10]{1-6}{\textbf{Avatar Animation}} &&&&& \\
    & & \rot{\cellcolor{ForestGreen!10}\textbf{Prior Dataset Size}} & \rot{\cellcolor{ForestGreen!10}\textbf{Datasets}} & \rot{\cellcolor{ForestGreen!10}\textbf{Data Type}} & \rot{\cellcolor{ForestGreen!10}\textbf{Data Modality}} & \rot{\cellcolor{RoyalBlue!10}\textbf{Needed Assets}} & \rot{\cellcolor{RoyalBlue!10}\textbf{Input}} & \rot{\cellcolor{RoyalBlue!10}\textbf{Additional Priors}} & \rot{\cellcolor{RoyalBlue!10}\textbf{Req. Optimization}} & \rot{\cellcolor{RoyalBlue!10}\textbf{Creation Speed}} & \rot{\cellcolor{BurntOrange!10}\textbf{Animation Signal}} & \rot{\cellcolor{BurntOrange!10}\textbf{Lighting Control}} & \rot{\cellcolor{BurntOrange!10}\textbf{Animation Speed}} & \rot{\cellcolor{BurntOrange!10}\textbf{Image Synthesis}} & \rot{\cellcolor{BurntOrange!10}\textbf{Image Refinement}} & \rot{\cellcolor{BurntOrange!10}\textbf{Contents}} \\
    \midrule
    \Block{20-1}{\rotate Face}
 & \cellcolor{blue!04}RGCA~\cite{Saito2024relightable} & \cellcolor{blue!04} & \cellcolor{blue!04} & \cellcolor{blue!04} & \cellcolor{blue!04} & \cellcolor{blue!04}\crboxCameras \crboxRegisteredMeshes \crboxTextures & \cellcolor{blue!04}\icoOLATVideo & \cellcolor{blue!04}\crboxDMM & \cellcolor{blue!04}\icoYes & \cellcolor{blue!04}\icoSlow & \cellcolor{blue!04}\icoExprGeneral & \cellcolor{blue!04}\icoLocalLight \icoDistantLight & \cellcolor{blue!04}\icoRealtime & \cellcolor{blue!04}\crboxDGS & \cellcolor{blue!04}\icoNo & \cellcolor{blue!04}\icoFace \\
 & GaussianAvatars~\cite{Qian2024gaussianavatars} &  &  &  &  & \crboxCameras \crboxTrackedDMM & \icoMultiViewVideo & \crboxDMM & \icoYes & \icoSlow & \icoExprThreeDMM &  & \icoRealtime & \crboxDGS & \icoNo & \icoFace \\
 & \cellcolor{blue!04}GaussianHeads~\cite{teotia2024gaussianheads} & \cellcolor{blue!04} & \cellcolor{blue!04} & \cellcolor{blue!04} & \cellcolor{blue!04} & \cellcolor{blue!04}\crboxCameras \crboxTrackedDMM & \cellcolor{blue!04}\icoMultiViewVideo & \cellcolor{blue!04}\crboxDMM & \cellcolor{blue!04}\icoYes & \cellcolor{blue!04}\icoSlow & \cellcolor{blue!04}\icoMultiImage & \cellcolor{blue!04} & \cellcolor{blue!04}\icoRealtime & \cellcolor{blue!04}\crboxDGS & \cellcolor{blue!04}\icoNo & \cellcolor{blue!04}\icoFace \\
 & Neural Head Avatars~\cite{Grassal2022nha} &  &  &  &  & \crboxCameras \crboxTrackedDMM & \icoMonoVideo & \crboxDMM & \icoYes & \icoSlow & \icoExprThreeDMM &  & \icoInteractive & \crboxMesh & \icoNo & \icoFace \\
 & \cellcolor{blue!04}NeRFace~\cite{Gafni2021dynamic} & \cellcolor{blue!04} & \cellcolor{blue!04} & \cellcolor{blue!04} & \cellcolor{blue!04} & \cellcolor{blue!04}\crboxCameras \crboxTrackedDMM & \cellcolor{blue!04}\icoMonoVideo & \cellcolor{blue!04} & \cellcolor{blue!04}\icoYes & \cellcolor{blue!04}\icoSlow & \cellcolor{blue!04}\icoExprThreeDMM & \cellcolor{blue!04} & \cellcolor{blue!04}\icoOffline & \cellcolor{blue!04}\crboxNeRF & \cellcolor{blue!04}\icoNo & \cellcolor{blue!04}\icoFace \\
 & INSTA~\cite{Zielonka2023insta} &  &  &  &  & \crboxCameras \crboxTrackedDMM & \icoMonoVideo & \crboxDMM & \icoYes & \icoFast & \icoExprThreeDMM &  & \icoRealtime & \crboxNeRF & \icoNo & \icoFace \\
 & \cellcolor{blue!04}URAvatar~\cite{Li2024urvatar} & \cellcolor{blue!04}345 & \cellcolor{blue!04}\crboxInternal & \cellcolor{blue!04}\icoReal & \cellcolor{blue!04}\icoOLATVideo & \cellcolor{blue!04}\crboxCameras \crboxRegisteredMeshes \crboxTextures & \cellcolor{blue!04}\icoMonoVideo & \cellcolor{blue!04}\crboxRelighting & \cellcolor{blue!04}\icoYes & \cellcolor{blue!04}\icoSlow & \cellcolor{blue!04}\icoExprGeneral & \cellcolor{blue!04}\icoLocalLight \icoDistantLight & \cellcolor{blue!04}\icoRealtime & \cellcolor{blue!04}\crboxDGS & \cellcolor{blue!04}\icoNo & \cellcolor{blue!04}\icoFace \\
 & HairCUP~\cite{kim2025haircup} & 252 & \crboxInternal & \icoReal & \icoOLATVideo & \crboxCameras \crboxRegisteredMeshes \crboxTextures & \icoMonoVideo & \crboxImageDiffusion & \icoYes & \icoMedium & \icoExprGeneral & \icoDistantLight & \icoRealtime & \crboxDGS & \icoNo & \icoFace \icoHair \\
 & \cellcolor{blue!04}Avat3r~\cite{kirschstein2025avat3r} & \cellcolor{blue!04}256 & \cellcolor{blue!04}\crboxAva & \cellcolor{blue!04}\icoReal & \cellcolor{blue!04}\icoMultiViewVideo & \cellcolor{blue!04}\crboxCameras & \cellcolor{blue!04}\icoFew & \cellcolor{blue!04}\crboxDeepFeatures & \cellcolor{blue!04}\icoNo & \cellcolor{blue!04}\icoInstant & \cellcolor{blue!04}\icoExprGeneral & \cellcolor{blue!04} & \cellcolor{blue!04}\icoRealtime & \cellcolor{blue!04}\crboxDGS & \cellcolor{blue!04}\icoNo & \cellcolor{blue!04}\icoFace \\
 & InvertAvatar~\cite{Zhao2024invertavatar} & 77500 & \crboxFFHQ \crboxNextDsamples \crboxVFHQ & \icoGenerated \icoReal & \icoSingleImage \icoMonoVideo & \crboxCameras \crboxTrackedDMM & \icoFew \icoOne & \crboxDMM & \icoNo & \icoInstant & \icoExprThreeDMM &  & \icoRealtime & \crboxNeRF & \icoYes & \icoFace \\
 & \cellcolor{blue!04}Cap4D~\cite{Taubner2025cap4d} & \cellcolor{blue!04}6317 & \cellcolor{blue!04}\crboxAva \crboxMEAD \crboxNeRSemble \crboxVFHQ & \cellcolor{blue!04}\icoReal & \cellcolor{blue!04}\icoMeshes \icoMultiViewVideo & \cellcolor{blue!04}\crboxCameras \crboxTrackedDMM & \cellcolor{blue!04}\icoFew \icoOne & \cellcolor{blue!04}\crboxDMM \crboxImageDiffusion & \cellcolor{blue!04}\icoYes & \cellcolor{blue!04}\icoSlow & \cellcolor{blue!04}\icoExprThreeDMM & \cellcolor{blue!04} & \cellcolor{blue!04}\icoRealtime & \cellcolor{blue!04}\crboxDGS & \cellcolor{blue!04}\icoNo & \cellcolor{blue!04}\icoFace \\
 & HeadNeRF~\cite{hong2022headnerf} & 4543 & \crboxFFHQ \crboxFaceScape \crboxInternal & \icoReal & \icoSingleImage \icoMultiImage \icoOLATImage & \crboxCameras & \icoOne \icoFew \icoMonoVideo & \crboxDMM & \icoYes & \icoFast & \icoExprGeneral & \icoDistantLight & \icoRealtime & \crboxNeRF & \icoYes & \icoFace \\
 & \cellcolor{blue!04}GPHM~\cite{Xu2024gphm} & \cellcolor{blue!04}15625 & \cellcolor{blue!04}\crboxFaceVerse \crboxNPHM \crboxNeRSemble \crboxVFHQ & \cellcolor{blue!04}\icoReal \icoGenerated & \cellcolor{blue!04}\icoMultiViewVideo \icoMonoVideo & \cellcolor{blue!04}\crboxTrackedDMM & \cellcolor{blue!04}\icoOne \icoFew \icoMonoVideo & \cellcolor{blue!04}\crboxPortraitAnimation & \cellcolor{blue!04}\icoYes & \cellcolor{blue!04}\icoMedium & \cellcolor{blue!04}\icoExprGeneral & \cellcolor{blue!04} & \cellcolor{blue!04}\icoRealtime & \cellcolor{blue!04}\crboxDGS & \cellcolor{blue!04}\icoYes & \cellcolor{blue!04}\icoFace \\
 & GAGAvatar~\cite{Chu2024generalizable} & 7228 & \crboxVFHQ & \icoReal & \icoMonoVideo & \crboxCameras \crboxTrackedDMM & \icoOne & \crboxDMM \crboxDeepFeatures & \icoNo & \icoInstant & \icoExprThreeDMM &  & \icoRealtime & \crboxDGS & \icoYes & \icoFace \\
 & \cellcolor{blue!04}VOODOOXP~\cite{tran2024voodooxp} & \cellcolor{blue!04}103220 & \cellcolor{blue!04}\crboxCelebVHQ \crboxFFHQ \crboxNeRSemble & \cellcolor{blue!04}\icoReal & \cellcolor{blue!04}\icoSingleImage \icoMonoVideo & \cellcolor{blue!04} & \cellcolor{blue!04}\icoOne & \cellcolor{blue!04} & \cellcolor{blue!04}\icoNo & \cellcolor{blue!04}\icoInstant & \cellcolor{blue!04}\icoVideo & \cellcolor{blue!04} & \cellcolor{blue!04}\icoRealtime & \cellcolor{blue!04}\crboxNeRF & \cellcolor{blue!04}\icoYes & \cellcolor{blue!04}\icoFace \\
 & VASA1~\cite{xu2024vasa1} & 9500 & \crboxInternal \crboxVoxCeleb & \icoReal & \icoMonoVideo &  & \icoOne & \crboxPortraitAnimation \crboxWaveVec & \icoNo & \icoInstant & \icoAudio &  & \icoRealtime & \crboxNeuralRendering & \icoNo & \icoFace \\
 & \cellcolor{blue!04}GaussianSpeech~\cite{aneja2024gaussianspeech} & \cellcolor{blue!04} & \cellcolor{blue!04} & \cellcolor{blue!04} & \cellcolor{blue!04} & \cellcolor{blue!04}\crboxCameras \crboxTrackedDMM & \cellcolor{blue!04}\icoMultiViewVideo \icoAudio & \cellcolor{blue!04}\crboxDMM \crboxWaveVec & \cellcolor{blue!04}\icoYes & \cellcolor{blue!04}\icoSlow & \cellcolor{blue!04}\icoAudio & \cellcolor{blue!04} & \cellcolor{blue!04}\icoRealtime & \cellcolor{blue!04}\crboxDGS & \cellcolor{blue!04}\icoNo & \cellcolor{blue!04}\icoFace \\
 & HeadStudio~\cite{zhou2024headstudio} &  &  &  &  &  & \icoText & \crboxDMM \crboxImageDiffusion & \icoYes & \icoMedium & \icoExprThreeDMM &  & \icoRealtime & \crboxDGS & \icoNo & \icoFace \\
 & \cellcolor{blue!04}AnimPortrait3D~\cite{wu2025animportrait3d} & \cellcolor{blue!04}89590 & \cellcolor{blue!04}\crboxFFHQ \crboxLPFF & \cellcolor{blue!04}\icoReal & \cellcolor{blue!04}\icoSingleImage & \cellcolor{blue!04} & \cellcolor{blue!04}\icoText & \cellcolor{blue!04}\crboxDMM \crboxImageDiffusion & \cellcolor{blue!04}\icoYes & \cellcolor{blue!04}\icoMedium & \cellcolor{blue!04}\icoExprThreeDMM & \cellcolor{blue!04} & \cellcolor{blue!04}\icoRealtime & \cellcolor{blue!04}\crboxDGS & \cellcolor{blue!04}\icoNo & \cellcolor{blue!04}\icoFace \\
 & Next3D~\cite{Sun2023next3d} & 70000 & \crboxFFHQ & \icoReal & \icoSingleImage &  & \icoZero & \crboxDMM & \icoNo & \icoFast & \icoExprThreeDMM &  & \icoRealtime & \crboxNeRF & \icoYes & \icoFace \\
    \midrule
    \Block{23-1}{\rotate Full-body}
 & \cellcolor{blue!04}RFGCA~\cite{wang2025relightable} & \cellcolor{blue!04} & \cellcolor{blue!04} & \cellcolor{blue!04} & \cellcolor{blue!04} & \cellcolor{blue!04}\crboxKnownLighting \crboxTrackedDMM & \cellcolor{blue!04}\icoOLATVideo & \cellcolor{blue!04}\crboxDMM & \cellcolor{blue!04}\icoYes & \cellcolor{blue!04}\icoSlow & \cellcolor{blue!04}\icoExprGeneral \icoPose & \cellcolor{blue!04}\icoLocalLight \icoDistantLight & \cellcolor{blue!04}\icoRealtime & \cellcolor{blue!04}\crboxDGS \crboxNeuralRendering & \cellcolor{blue!04}\icoYes & \cellcolor{blue!04}\icoBody \\
 & HuGS~\cite{moreau2024human} &  &  &  &  & \crboxCameras \crboxTrackedDMM & \icoMultiViewVideo & \crboxDMM & \icoYes & \icoSlow & \icoPose &  & \icoRealtime & \crboxDGS & \icoNo & \icoBody \\
 & \cellcolor{blue!04}Driv. Vol. Avatars~\cite{Remelli2022drivable} & \cellcolor{blue!04} & \cellcolor{blue!04} & \cellcolor{blue!04} & \cellcolor{blue!04} & \cellcolor{blue!04}\crboxTrackedDMM & \cellcolor{blue!04}\icoMultiViewVideo & \cellcolor{blue!04}\crboxDMM & \cellcolor{blue!04}\icoYes & \cellcolor{blue!04}\icoSlow & \cellcolor{blue!04}\icoPose \icoVideo & \cellcolor{blue!04} & \cellcolor{blue!04}\icoRealtime & \cellcolor{blue!04}\crboxMVP & \cellcolor{blue!04}\icoNo & \cellcolor{blue!04}\icoBody \\
 & Driv. Aware Avatars~\cite{Bagautdinov2021driving} &  &  &  &  & \crboxTextures \crboxTrackedDMM & \icoMultiViewVideo & \crboxDMM & \icoNo & \icoSlow & \icoPose &  & \icoRealtime & \crboxMesh & \icoNo & \icoBody \\
 & \cellcolor{blue!04}Neural Body~\cite{peng2021neural} & \cellcolor{blue!04} & \cellcolor{blue!04} & \cellcolor{blue!04} & \cellcolor{blue!04} & \cellcolor{blue!04}\crboxTrackedDMM & \cellcolor{blue!04}\icoMultiViewVideo & \cellcolor{blue!04}\crboxDMM & \cellcolor{blue!04}\icoNo & \cellcolor{blue!04}\icoSlow & \cellcolor{blue!04}\icoPose & \cellcolor{blue!04} & \cellcolor{blue!04}\icoOffline & \cellcolor{blue!04}\crboxNeRF & \cellcolor{blue!04}\icoNo & \cellcolor{blue!04}\icoBody \\
 & Anim. Gaussians~\cite{Li2024animatable} &  &  &  &  & \crboxTrackedDMM & \icoMultiViewVideo & \crboxDMM & \icoYes & \icoSlow & \icoPose &  & \icoInteractive & \crboxDGS & \icoNo & \icoBody \\
 & \cellcolor{blue!04}TaoAvatar~\cite{chen2025taoavatar} & \cellcolor{blue!04} & \cellcolor{blue!04} & \cellcolor{blue!04} & \cellcolor{blue!04} & \cellcolor{blue!04}\crboxCameras \crboxTrackedDMM & \cellcolor{blue!04}\icoMultiViewVideo & \cellcolor{blue!04}\crboxDMM & \cellcolor{blue!04}\icoYes & \cellcolor{blue!04}\icoSlow & \cellcolor{blue!04}\icoPose & \cellcolor{blue!04} & \cellcolor{blue!04}\icoRealtime & \cellcolor{blue!04}\crboxDGS & \cellcolor{blue!04}\icoNo & \cellcolor{blue!04}\icoBody \\
 & AvatarRex~\cite{zheng2023avatarrex} &  &  &  &  & \crboxTrackedDMM & \icoMultiViewVideo & \crboxDMM & \icoYes & \icoSlow & \icoPose &  & \icoInteractive & \crboxNeRF & \icoNo & \icoBody \\
 & \cellcolor{blue!04}D3GA~\cite{Zielonka2023d3ga} & \cellcolor{blue!04} & \cellcolor{blue!04} & \cellcolor{blue!04} & \cellcolor{blue!04} & \cellcolor{blue!04}\crboxTrackedDMM & \cellcolor{blue!04}\icoMultiViewVideo & \cellcolor{blue!04}\crboxDMM & \cellcolor{blue!04}\icoNo & \cellcolor{blue!04}\icoSlow & \cellcolor{blue!04}\icoPose & \cellcolor{blue!04} & \cellcolor{blue!04}\icoRealtime & \cellcolor{blue!04}\crboxDGS & \cellcolor{blue!04}\icoNo & \cellcolor{blue!04}\icoBody \\
 & GPS-Gaussian~\cite{Zheng2024gps} & 1926 & \crboxInternal \crboxTHuman & \icoReal & \icoMeshes &  & \icoMultiViewVideo &  & \icoYes & \icoSlow & \icoPose &  & \icoInteractive & \crboxDGS & \icoNo & \icoBody \\
 & \cellcolor{blue!04}Instant Avatar~\cite{jiang2023instantavatar} & \cellcolor{blue!04} & \cellcolor{blue!04} & \cellcolor{blue!04} & \cellcolor{blue!04} & \cellcolor{blue!04}\crboxTrackedDMM & \cellcolor{blue!04}\icoMonoVideo & \cellcolor{blue!04} & \cellcolor{blue!04}\icoYes & \cellcolor{blue!04}\icoInstant & \cellcolor{blue!04}\icoPose & \cellcolor{blue!04} & \cellcolor{blue!04}\icoInteractive & \cellcolor{blue!04}\crboxNeRF & \cellcolor{blue!04}\icoNo & \cellcolor{blue!04}\icoBody \\
 & Vid2Avatar-Pro~\cite{Guo2025vid2avatarpro} & 1000 & \crboxInternal & \icoReal & \icoMultiViewVideo & \crboxTrackedDMM & \icoMonoVideo & \crboxDMM & \icoYes & \icoMedium & \icoPose &  & \icoRealtime & \crboxDGS & \icoNo & \icoBody \\
 & \cellcolor{blue!04}Intrinsicavatar~\cite{wang2024intrinsicavatar} & \cellcolor{blue!04} & \cellcolor{blue!04} & \cellcolor{blue!04} & \cellcolor{blue!04} & \cellcolor{blue!04}\crboxCameras \crboxTrackedDMM & \cellcolor{blue!04}\icoMonoVideo & \cellcolor{blue!04}\crboxDMM & \cellcolor{blue!04}\icoYes & \cellcolor{blue!04}\icoMedium & \cellcolor{blue!04}\icoPose & \cellcolor{blue!04}\icoDistantLight & \cellcolor{blue!04}\icoOffline & \cellcolor{blue!04}\crboxNeRF & \cellcolor{blue!04}\icoNo & \cellcolor{blue!04}\icoBody \\
 & Relightable avatar~\cite{xu2024relightable} &  &  &  &  & \crboxCameras \crboxTrackedDMM & \icoMonoVideo & \crboxDMM & \icoYes & \icoSlow & \icoPose & \icoDistantLight & \icoOffline & \crboxSDF & \icoNo & \icoBody \\
 & \cellcolor{blue!04}RANA~\cite{iqbal2023rana} & \cellcolor{blue!04}400 & \cellcolor{blue!04}\crboxRenderPeople & \cellcolor{blue!04}\icoReal & \cellcolor{blue!04}\icoMeshes & \cellcolor{blue!04}\crboxCameras \crboxTextures \crboxTrackedDMM & \cellcolor{blue!04}\icoMonoVideo & \cellcolor{blue!04}\crboxDMM & \cellcolor{blue!04}\icoYes & \cellcolor{blue!04}\icoMedium & \cellcolor{blue!04}\icoPose & \cellcolor{blue!04}\icoDistantLight & \cellcolor{blue!04}\icoRealtime & \cellcolor{blue!04}\crboxMesh & \cellcolor{blue!04}\icoNo & \cellcolor{blue!04}\icoBody \\
 & Fresa~\cite{Wang2025fresa} & 1100 & \crboxInternal & \icoReal & \icoMultiViewVideo \icoMeshes & \crboxTrackedDMM & \icoFew &  & \icoNo & \icoInstant & \icoPose &  & \icoRealtime & \crboxSDF & \icoNo & \icoBody \\
 & \cellcolor{blue!04}PuzzleAvatar~\cite{xiu2024puzzleavatar} & \cellcolor{blue!04}525 & \cellcolor{blue!04}\crboxTHuman & \cellcolor{blue!04}\icoReal & \cellcolor{blue!04}\icoMeshes & \cellcolor{blue!04} & \cellcolor{blue!04}\icoFew & \cellcolor{blue!04}\crboxDMM \crboxGPTV \crboxGroundedSAM \crboxImageDiffusion & \cellcolor{blue!04}\icoYes & \cellcolor{blue!04}\icoMedium & \cellcolor{blue!04}\icoPose & \cellcolor{blue!04} & \cellcolor{blue!04}\icoRealtime & \cellcolor{blue!04}\crboxMesh & \cellcolor{blue!04}\icoNo & \cellcolor{blue!04}\icoBody \\
 & HumanRAM~\cite{yu2025humanram} & 7100 & \crboxActorsHQ \crboxHumanDiT \crboxTHuman \crboxZJUMoCap & \icoReal & \icoMultiViewVideo \icoMeshes & \crboxTrackedDMM & \icoOne \icoFew &  & \icoNo & \icoInstant & \icoPose &  &  & \crboxMesh \crboxNeuralRendering & \icoYes & \icoBody \\
 & \cellcolor{blue!04}ICON~\cite{Xiu2022icon} & \cellcolor{blue!04}450 & \cellcolor{blue!04}\crboxRenderPeople & \cellcolor{blue!04}\icoReal \icoGenerated & \cellcolor{blue!04}\icoSingleImage \icoMeshes & \cellcolor{blue!04}\crboxTrackedDMM & \cellcolor{blue!04}\icoOne & \cellcolor{blue!04}\crboxDMM & \cellcolor{blue!04}\icoNo & \cellcolor{blue!04}\icoInstant & \cellcolor{blue!04}\icoPose & \cellcolor{blue!04} & \cellcolor{blue!04}\icoOffline & \cellcolor{blue!04}\crboxSDF & \cellcolor{blue!04}\icoNo & \cellcolor{blue!04}\icoBody \\
 & PERSONA~\cite{sim2025persona} &  &  &  &  & \crboxTrackedDMM & \icoOne & \crboxDMM \crboxImageDiffusion & \icoYes & \icoMedium & \icoPose &  & \icoRealtime & \crboxDGS & \icoNo & \icoBody \\
 & \cellcolor{blue!04}AniGS~\cite{qiu2025anigs} & \cellcolor{blue!04}6124 & \cellcolor{blue!04}\crboxKK \crboxCustomHumans \crboxRenderPeople \crboxTHuman \crboxTwindom & \cellcolor{blue!04}\icoReal \icoGenerated \icoSynth & \cellcolor{blue!04}\icoMonoVideo & \cellcolor{blue!04}\crboxTrackedDMM & \cellcolor{blue!04}\icoOne & \cellcolor{blue!04}\crboxDMM & \cellcolor{blue!04}\icoYes & \cellcolor{blue!04}\icoFast & \cellcolor{blue!04}\icoPose & \cellcolor{blue!04} & \cellcolor{blue!04}\icoRealtime & \cellcolor{blue!04}\crboxDGS & \cellcolor{blue!04}\icoNo & \cellcolor{blue!04}\icoBody \\
 & IDOL~\cite{zhuang2025idol} & 100000 & \crboxEnsemble & \icoReal \icoSynth \icoGenerated & \icoMultiViewVideo \icoMeshes & \crboxTrackedDMM & \icoOne & \crboxDMM & \icoNo & \icoInstant & \icoPose &  & \icoRealtime & \crboxDGS & \icoNo & \icoBody \\
 & \cellcolor{blue!04}GNaRF~\cite{bergman2022gnarf} & \cellcolor{blue!04}175 & \cellcolor{blue!04}\crboxAIST \crboxSURREAL & \cellcolor{blue!04}\icoReal & \cellcolor{blue!04}\icoMonoVideo \icoMultiViewVideo & \cellcolor{blue!04} & \cellcolor{blue!04}\icoZero & \cellcolor{blue!04} & \cellcolor{blue!04}\icoNo & \cellcolor{blue!04}\icoInstant & \cellcolor{blue!04}\icoPose & \cellcolor{blue!04} & \cellcolor{blue!04}\icoRealtime & \cellcolor{blue!04}\crboxNeRF & \cellcolor{blue!04}\icoYes & \cellcolor{blue!04}\icoBody \\
    \midrule
    \Block{4-1}{\rotate Hands}
 & RelightableHands~\cite{Iwase2023relightablehands} & 2 & \crboxInternal & \icoReal & \icoMultiViewVideo & \crboxTrackedDMM & \icoOLATVideo & \crboxDMM & \icoNo &  & \icoPose & \icoLocalLight \icoDistantLight & \icoRealtime & \crboxMVP & \icoNo & \icoHand \\
 & \cellcolor{blue!04}Hand Avatar~\cite{chen2023handavatar} & \cellcolor{blue!04}27 & \cellcolor{blue!04}\crboxInterHandM & \cellcolor{blue!04}\icoReal & \cellcolor{blue!04}\icoMultiViewVideo & \cellcolor{blue!04}\crboxTrackedDMM & \cellcolor{blue!04}\icoMonoVideo & \cellcolor{blue!04}\crboxDMM & \cellcolor{blue!04}\icoYes & \cellcolor{blue!04}\icoMedium & \cellcolor{blue!04}\icoPose & \cellcolor{blue!04}\icoDistantLight & \cellcolor{blue!04}\icoOffline & \cellcolor{blue!04}\crboxNeRF & \cellcolor{blue!04}\icoNo & \cellcolor{blue!04}\icoHand \\
 & URHand~\cite{Chen2024urhand} & 93 & \crboxInternal & \icoReal & \icoMultiViewVideo & \crboxHandPose \crboxMeshes \crboxTextures & \icoMonoVideo \icoDepth & \crboxDMM & \icoYes &  & \icoPose & \icoLocalLight \icoDistantLight & \icoRealtime & \crboxMesh & \icoNo & \icoHand \\
 & \cellcolor{blue!04}Nimble~\cite{li2022nimble} & \cellcolor{blue!04}35 /20 & \cellcolor{blue!04}\crboxMRIHand & \cellcolor{blue!04}\icoReal & \cellcolor{blue!04}\icoMRI \icoMeshes \icoOLATVideo & \cellcolor{blue!04}\crboxTrackedDMM & \cellcolor{blue!04}\icoOne \icoDepth & \cellcolor{blue!04}\crboxDMM & \cellcolor{blue!04}\icoYes & \cellcolor{blue!04}\icoFast & \cellcolor{blue!04}\icoPose & \cellcolor{blue!04}\icoLocalLight \icoDistantLight & \cellcolor{blue!04}\icoOffline & \cellcolor{blue!04}\crboxMesh & \cellcolor{blue!04}\icoNo & \cellcolor{blue!04}\icoHand \\
  \bottomrule
  \end{NiceTabular}
  }
  \caption{Digital Humans Taxonomy: We show representative state-of-the-art methods for human faces, bodies and hands.}
  \label{tab:taxonomy}
\end{table*}

%% file: chapters/5_1_faces.tex

\pagebreak
\section{Human Head Avatars} 
\label{sec:faces}


\begin{wrapfigure}{r}{0.2\columnwidth} 
    \centering 
    \vspace{-1.0cm}
    \hspace*{-0.5cm}
    \includegraphics[width=0.25\columnwidth]{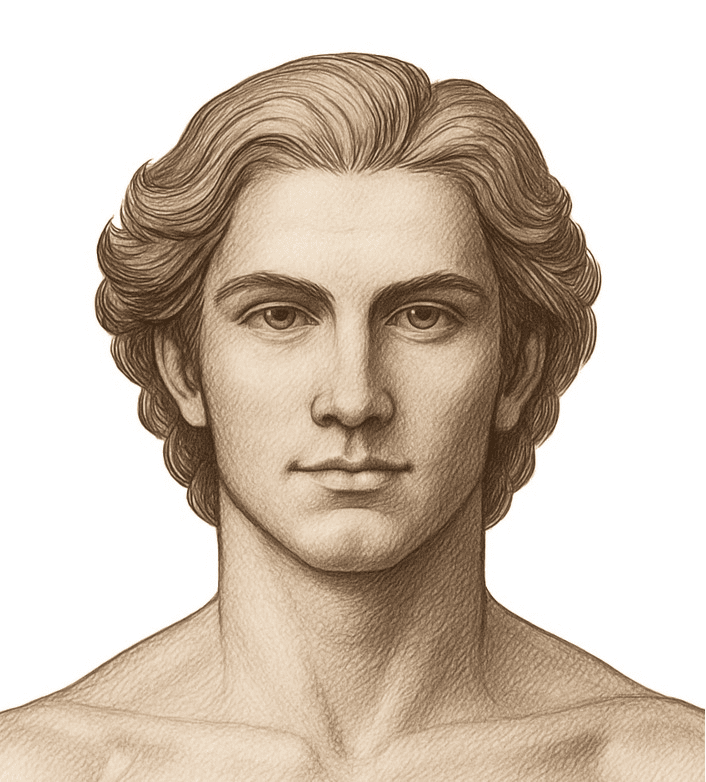} 
    \vspace{-0.6cm}
\end{wrapfigure}
Human head avatars play a crucial role in building a believable digital replica of a person.
%
The human face is a key identifier of a human conveying personality, intent, and social cues.
The nuanced modeling and synthesis of facial expressions and accurate likeness foster trust, empathy, and effective communication.
The level of motion and appearance quality of human head avatars can differ drastically depending on the available data.
The film industry is using sophisticated studios to record multi-view, multi-modal data to recover high-quality, relightable and editable avatars.
However, such data is not available for end users who might want to have an avatar for a teleconference at home.
Therefore, there is intense research in building avatar systems that work with \new{sparse observations of a person}, requiring powerful priors to compensate for the lack of data.
We review this transition from high-quality to few-shot reconstruction methods.
%

\subsection{Studio-level 3D Head Avatars}
This category aims at creating the highest-quality virtual avatars, at the cost of requiring expensive capturing set-ups to gather enough identity observations $\mathcal{I}$, i.e., extensive calibrated and time-synchronized multi-view videos that can take hours to capture.
Consequently, $\mathcal{I}$ provides dense observations of the person, w.r.t. both view angles and facial expression.
\paragraph*{Learned expression space.}
While early approaches were still mesh-based~\cite{lombardi2018deep}, the field quickly evolved to volumetric representations, such as NeuralVolumes~\cite{lombardi2019neuvol} and Mixtures of Volumetric Primitives (MVP)~\cite{Lombardi21mvp}. Upon the invention of 3DGS~\cite{Kerbl20233d}, 3DGS-based avatars were quickly proposed~\cite{Saito2024relightable,teotia2024gaussianheads}. All these methods learn expression codes that act as motion control $\mathcal{M}$ in an end-to-end manner, yielding high flexibility, but rendering cross-reenactment difficult.
GEM~\cite{Zielonka2025gem} compresses a studio-level head avatar into a personalized low-dimensional eigenbasis. At inference time, for self- and cross-reenactment tasks, the coefficients for the eigenbasis can be regressed using a neural network, enabling real-time performance, and abandoning the need for 3DMM tracking.

\paragraph*{3DMM-based expression space.}
A related line of work derives motion control $\mathcal{M}$ from 3DMM expression codes~\cite{Zielonka2023insta}. GaussianAvatars~\cite{Qian2024gaussianavatars} attach 3DGS-primitives to the mesh faces of FLAME~\cite{Li2017flame}. GaussianHeadAvatars~\cite{Xu2024gaussian} increase the dynamics by proposing MLP-based neural deformations conditioned on $\mathcal{M}$. NPGA~\cite{giebenhain2024npga} focus on higher-fidelity motion control, by replacing the mesh-based 3DMM with a neural-3DMM, i.e., MonoNPHM~\cite{Giebenhain2024mononphm}, and by introducing per-primitive features that allow for primitive-specific expression behavior.


\begin{figure*}[t!]
    \centering
    \includegraphics[width=\linewidth]{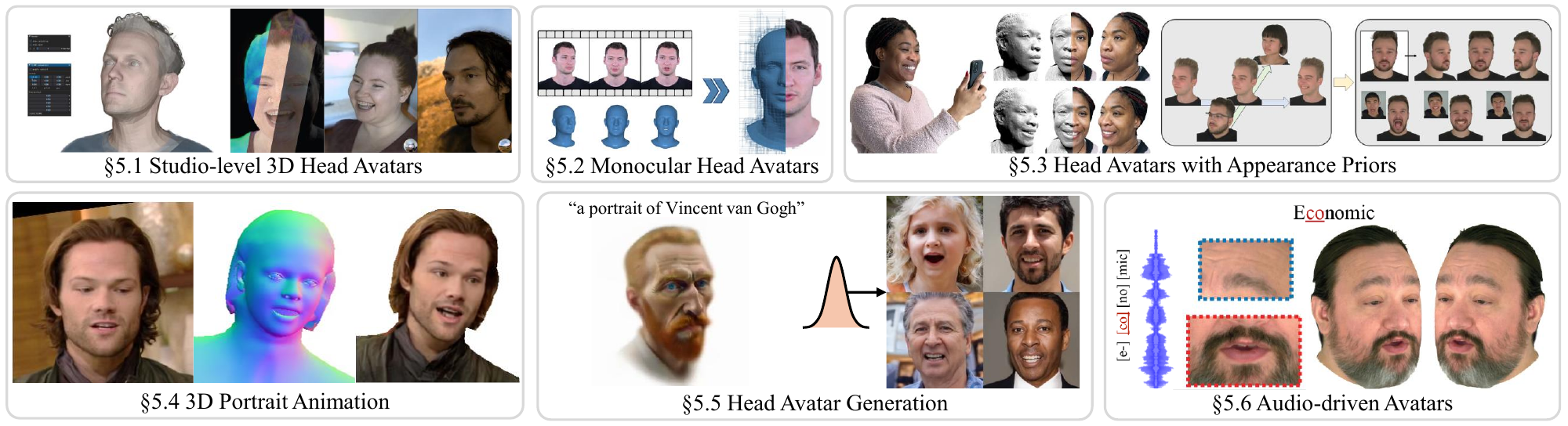}
    \caption{Overview of different research fields in 3D Head Avatars. Representative images are taken from GaussianAvatars~\cite{Qian2024gaussianavatars}, RGCA~\cite{Saito2024relightable}, INSTA~\cite{Zielonka2023insta}, Cao~\etal~\cite{cao2022authentic}, GPHM~\cite{Xu2024gphm}, ROME~\cite{khakhulin2022rome}, HeadStudio~\cite{zhou2024headstudio}, Next3D~\cite{Sun2023next3d}, and GaussianSpeech~\cite{aneja2024gaussianspeech}.}
\end{figure*}

\paragraph*{Relightable studio avatars.}
Relightability is another core quality for the application of avatars in virtual environments. To this end, elaborate light-stage capture set-ups have been built that can control the lighting position, direction, and intensity for each timestep. 
RGCA~\cite{Saito2024relightable} proposes a relightable formulation for 3DGS, based on pre-computed radiance transfer. The follow-up BecomingLit~\cite{schmidt2025becominglit} proposed an alternative shading formulation, which requires less diverse lighting conditions during avatar creation. 
Finally, DiffRelight~\cite{he2024diffrelight} disentangles relighting from the avatar's facial animation by exploiting the image prior of a 2D image diffusion model, and fine-tuning it for relighting uniformly lit 3DGS-based avatars. The diffusion model learns complex light transport effects, such as light transmission through translucent body parts, such as ears, which are difficult to model with graphics-based shading formulations. While DiffRelight suffers from flickering between timesteps, a video-diffusion backbone could ensure temporally smooth results in the future.  
The follow-up work LuxPostFacto~\cite{mei2025lux} employs a video diffusion model to overcome temporal inconsistencies, and focuses on relighting arbitrary portrait videos, by first learning how to delight videos on studio light stage data, and then augmenting their relighting training data with in-the-wild videos with pseudo ground truth albedo.
\new{3DPR~\cite{rao20253dpr} addresses the complementary problem of single-image 3D portrait relighting by leveraging generative priors to synthesize plausible relighting from a single photograph.}


\subsection{Monocular Head Avatars without Appearance Priors}
Carefully captured monocular videos can also be used to construct head avatars without relying on more priors than in the studio setting, as long as the video is easy to track and showcases diverse head rotations, which essentially imitate multi-view observations. 
Across recent years, several methods have been proposed that explored the most promising 3D representations of their time. While very early work, like Face2Face~\cite{Thies2016face} used mesh-based representations, more recent approaches have explored NeRFs~\cite{Gafni2021dynamic}, meshes in combination with deferred neural rendering~\cite{Grassal2022nha}, InstantNGP~\cite{Zielonka2023insta}, SDFs~\cite{Zheng2022avatar}, point clouds~\cite{zheng2023pointavatar}, and 3DGS~\cite{Xiang2024flashavatar, Shao2024splattingavatar}.
Compared to studio-grade avatars, such methods are mainly limited due to underlying face tracking, which is complicated due to depth ambiguity, and often struggles to accurately capture extreme expressions and head rotations. Therefore, such avatars often overfit to frontal views and simple facial expressions; thus, they typically exhibit significantly noticeable artifacts when facing stronger viewpoint changes and facial expressions.
In an attempt to alleviate the need for accurate 3DMM tracking during the avatar creation stage, GAN-Avatar~\cite{kabadayi24ganavatar} personalizes the 3D-aware GAN EG3D~\cite{chan2022eg3d} on person-specific data, only requiring a collection of images and camera parameters. Animation is enabled by learning a mapping from 3DMM codes to the GAN latent space.

\subsection{Monocular Head Avatars with \new{Appearance} Priors}
\new{While the methods above rely on carefully captured monocular videos, many practical scenarios provide even less data.} As observations $\mathcal{I}$ become sparse, such as casually captured or in-the-wild monocular videos, or when provided with a few casual selfie images, the need for a strong prior $\mathcal{P}$ increases.
\paragraph*{3D prior + fine-tuning.}
A popular approach for a learning-based 3D prior is Universal Prior Models (UPMs)~\cite{cao2022authentic,Li2024urvatar,mendiratta2025grmm}, which are UNets, operating in UV-space to transfer coarse UV-space geometry and texture from a tracked mesh to high-fidelity volumetric representation, such as MVP~\cite{cao2022authentic} or Gaussian primitives~\cite{Li2024urvatar, Zielonka2025synshot}. While UPMs produce high-quality results once a tracked mesh has been obtained, their avatar creation stage commonly includes fine-tuning of the encoder and decoder networks based on $\mathcal{I}$, which can take up to several hours~\cite{cao2022authentic,Li2024urvatar} or several minutes~\cite{Zielonka2025synshot}.
In general, since UPMs operate in UV-space, they are reliant on high-quality mesh-based tracking, which is a challenging task on its own. 


\paragraph*{Photo-realistic 3D head models.}
\new{
Pioneered by HeadNeRF~\cite{hong2022headnerf}, some works aim at building a parametric 3D head model with an additional photorealistic appearance space. To achieve this, one has to go beyond the geometrical perspective of classical 3DMMs~\cite{paysan2009bfm, Li2017flame} and train a morphable model on a large dataset of human videos. Instead of meshes or SDFs, as common for 3DMMs, photo-realistic 3D head models therefore rely on NeRF~\cite{hong2022headnerf, yang2024vrmm, zhuang2022mofanerf, yu2024one2avatar} or 3DGS~\cite{Xheng2024headgap, Xu2024gphm, mendiratta2025grmm}. The most common architectural choice is an autodecoder~\cite{park2019deepsdf} that optimizes separate codebooks: one for each person, facial expression, or light configuration. The disentanglement of these factors often follows from the use of an (OLAT) multi-view image dataset where the same person, expression, and potentially lighting configuration is seen from multiple cameras. Once trained, the parametric head model can be fitted to any number of observations, similar to classical 3DMMs. As such, photo-realistic 3D head models are very versatile and can be used in many scenarios, from creating an avatar with a single image to multi-view video fitting. On the downside, the creation involves potentially costly fitting and assumes that the target identity actually exists in the model's latent space. An assumption that barely holds given the small identity coverage of existing multi-view datasets. While HeadNeRF~\cite{hong2022headnerf} finds that including monocular images from the FFHQ dataset~\cite{Karras2019style} in their Autodecoder training leads to better generalizability, One2Avatar~\cite{yu2024one2avatar} observes that such an approach leads to limited 3D completeness in the final avatars. In general, how to train parametric 3D models on partial data, such as monocular videos or single images, remains an open problem.}

\paragraph*{Feed-forward 3D prior.}
Following the recent success of multi-view transformers for 3D reconstruction~\cite{wang2024dust3r, hong2024lrm, wang2025vggt}, Avat3r~\cite{kirschstein2025avat3r} proposes a formulation that lifts four input images into a set of canonical Gaussian primitives using a transformer-based architecture similar to GS-LRM~\cite{Zhang2024gslrm}. Expression control is achieved using a cross-attention module that attends to latent expression codes. 
As such, Avat3r targets avatar creation with minimum requirements on observed images $\mathcal{I}$ and minimal compute budget during the avatar creation stage, which merely involves a single forward pass of the transformer.
It is worth noting that the output is not a structured representation
(e.g., a mesh or primitives attached to it), 
but pixel-aligned 3D Gaussian primitives.

\paragraph*{Image-based priors.}
Inspired by the recent success of image diffusion and multi-view diffusion models, CAP4D~\cite{Taubner2025cap4d} and GAF~\cite{tang2025gaf} propose a 3DMM-conditioned multi-view diffusion model, which learns to generate high-quality 3D consistent images, based on the prompted camera view and 3DMM expression. In this way, the diffusion model is tasked with generating novel views from unseen camera angles and unseen facial expressions, which are included in the optimization of a 3DGS-based avatar.

\subsection{3D Portrait Animation}

Portrait animation methods further reduce the required identity specification $\mathcal{I}$ to a single image. As such, the problem becomes heavily under-constrained, and consequently, portrait animation methods are heavily prior-reliant and learning-based.

\paragraph*{Feed-forward prediction.}
A straightforward approach to portrait animation is to train a feed-forward network to predict 3D Gaussian primitives, similar to Avat3r~\cite{kirschstein2025avat3r}, but based on a \emph{single} input image instead. Notable methods like GAGavatar~\cite{Chu2024generalizable} and LAM~\cite{he2025lam} predict Gaussian parameters, which are attached to a 3DMM mesh, allowing for animation of these Gaussians, but limiting the representational capacity w.r.t. expressions to that of the underlying 3DMM. While LAM purely relies on mesh-attached Gaussians, GAGavatar uses a combination of mesh-attached Gaussians and a set of pixel-aligned Gaussians, which are non-animatable.
VoluMe~\cite{delagorce2025volume}, in contrast, purely relies on pixel-aligned Gaussians, but predicts two Gaussians per pixel, to accommodate for occluded regions. VoluMe cannot be animated via some abstract expression conditioning, but only using images from the same person, since each input image is directly transferred to 3D.
While this leaves the method without a proper avatar animation stage, it demonstrates that their avatar creation stage is efficient enough for real-time performance and is temporally consistent.
Earlier works~\new{\cite{ki2024export3d, yu2023nofa, ye2024real3d, tran2024voodoo3d, tran2024voodooxp}} adopted the efficient TriPlane representation~\new{\cite{chan2022eg3d}} for fast volumetric rendering, which is crucial when training on large video datasets. A technique is to predict two TriPlanes from the input image: one that contains the canonical 3D face of the person, and a second that contains the 3D motion of the driver~\cite{ye2024real3d}.  

\paragraph*{3D GANs.}
Yet another approach to portrait animation is 3D GANs, with added control for facial expressions.
3D GANS~\cite{chan2022eg3d,kirschstein2024gghead,barthel2025cgsgan} have the unique ability to learn a distribution over 3D heads from large 2D image collections, without the need for costly 3D datasets, which usually cover a much narrower data distribution. However, such generative models, by default, cannot disentangle identity and expression variation.
Popular approaches to add expression control into 3D GANs are generating features in UV space to inherit the underlying animation capabilities from a 3DMM, as explored in Next3D~\cite{Sun2023next3d}, or training a neural network to canonicalize the generated triplanes and learning to add expressions to the triplanes using a separate network, as proposed in VOODOOXP~\cite{tran2024voodooxp}. A single input image can be inverted using pivotal tuning~\cite{Roich2022pivotal}, resulting in a latent that can be decoded into a 3D head avatar. Another approach to add disentangled expression control to an existing 3D GAN is via latent space exploration. Here, the assumption is that the 3D GAN's latent space contains vectors representing the same identity with different facial expressions. Disentangled animation can therefore be achieved by training a separate network that maps identity and expression codes into the latent space of the 3D GAN~\cite{ma2023otavatar}. 

\paragraph*{Image-based prior + optimization.}
There exist several approaches that exploit the image prior of existing image/video diffusion models to hallucinate novel views, conditioned on a single image. The synthesized novel views can be baked into a 3DGS representation to obtain a 3D avatar. CAP4D~\cite{Taubner2025cap4d} and GAF~\cite{tang2025gaf} also support the single-image setting, while Zero-1-to-A~\cite{zhou2025zero1toA} specializes in portrait animation, by fine-tuning a video diffusion model to generate novel trajectories, with potentially higher 3D consistency.

\paragraph*{2D and 2.5D avatars.}
A core question in the community is whether the avatars necessarily have to follow some underlying 3D representation, whether they can be purely 2D, or some mixture which has some 3D components. While recent trends in video and world generation 
suggest that a surprising amount of consistency can be achieved by scaling up 2D generative models, the specific requirements for head avatars in teleconferencing systems, such as real-time performance, long-term consistency, support for multiple avatars, and consumer-grade edge devices, add more complexity to the debate.
Face-Vid2Vid~\cite{wang2021facevid2vid} and LivePortait~\cite{guo2024liveportrait}, for example, choose to learn in a 2.5D space, using a simple orthographic projection and learned implicit 3D keypoints, which condition a warping in a 3D feature volume. MegaPortraits~\cite{Drobyshev2022megaportraits} alleviates the bottleneck of using sparse keypoints to condition the warping by replacing them with learned latent codes.
Another line of work attempts to exploit the excellent image generation properties of diffusion models. A common choice to condition the diffusion process on the desired facial expression and camera view are renderings from a 3DMM, for example, see~\cite{ding2023diffusionrig,prinzler2024joker,Taubner2025cap4d,tang2025gaf}.
As such, these approaches abandon all sorts of internal 3D representation or rendering mechanisms. Instead of relying on human-crafted 3D inductive biases, they embrace a more data-driven approach.
Alternatively, some works decide to condition on landmarks~\cite{ma2024followyouremoji,wang2025fantasytalking}, or learnable 3DMM-attached features~\cite{learn2control}, rather than rendered 3DMM meshes.
An even more extreme version is video-driven portrait animation models, which condition video diffusion models on a source image $\mathcal{I}$, and use a driving video as motion control $\mathcal{M}$ 
\cite{cheng2025wananimate,chen2025hunyuanvideoavatarhighfidelityaudiodrivenhuman,svdp,sang2025lynxhighfidelitypersonalizedvideo}. 
As a consequence, camera and motion control in the animation stage cannot be directly modified. 

\subsection{Head Avatar Generation}

\paragraph*{Text to 3D head avatars.}
By further reducing the information content in $\mathcal{I}$, we arrive at the task of creating 3D head avatars conditioned merely on text.
%
While native 3D generative models text-conditioned head avatars are still rare, especially when requiring animation controllability, several optimization-based approaches exist~\cite{aneja2023clipface,zhang2024teca}. 
For example, HeadStudio~\cite{zhou2024headstudio} uses score distillation sampling (SDS) to optimize for Gaussian primitives rigged to the FLAME mesh. Similarly, AnimPortrait3D~\cite{wu2025animportrait3d} use Interval Score Matching (ISM) and SDEdit (SDE) based on a 3DMM conditioned 2D diffusion model, inspired by Joker~\cite{prinzler2024joker}.

An alternative for text-conditioned head avatar creation is to employ a text-conditioned 2D generation model, such as FLUX~\cite{flux2024}, to create an image based on a textual description and subsequently feed the image to one of the above-mentioned portrait animation methods, in order to create an avatar.

\paragraph*{Unconditional generation.}
\new{Finally, some works aim at learning the entire distribution of 3D head avatars, enabling unconditional generation without any identity specification.} Compared to the previous sections, this area is slightly underexplored.
Controllable 3D GANs, such as Next3D~\cite{Sun2023next3d}, GNARF~\cite {bergman2022gnarf}, or GAIA~\cite{yu2025gaia}, are one of the few approaches for unconditional avatar generation. To achieve this, a common approach is to start with an existing static 3D GAN such as EG3D~\cite{chan2022eg3d} and incorporate the motion prior from a 3DMM such as FLAME~\cite{Li2017flame} to make it animatable. RodinHD~\cite{Zhang2024rodinhd} is a latent diffusion model to generate 3D heads, but does not support animation capabilities. 
However, similar to text-conditioned generation, one can employ an unconditional image generator, followed by a portrait animation method.

\subsection{Audio-Driven Avatars}
Many applications of head avatars are centered around speech. Thus, there is an increased interest in the physical plausibility of mouth movement during talking and proper lip synchronization. 
GaussianSpeech~\cite{aneja2024gaussianspeech} builds studio-grade 3DGS-based avatars, which are animated using audio signals.
VASA-1~\cite{xu2024vasa1}, on the other hand, is a latent diffusion model that operates in the latent space of Megaportraits~\cite{Drobyshev2022megaportraits}. The identity component of the avatar is obtained using the Megaportraits encoder, and the generation task boils down to the generation of a sequence of expression codes, conditioned on an audio sequence as motion control $\mathcal{M}$. 
UniGAHA~\cite{teotia2025audiodrivenuniversalgaussianhead}, similarly, employs a latent diffusion model to map an audio sequence to the latent expression space of a UPM.

%% file: tables/assets.tex
\begin{table*}[th!]
  \centering
  \footnotesize
  \setlength{\tabcolsep}{2pt}
  \resizebox{\textwidth}{!}{%
  \begin{NiceTabular}{|l|l|>{\centering\arraybackslash}p{1.1cm}>{\centering\arraybackslash}p{1.1cm}>{\centering\arraybackslash}p{1.1cm}>{\centering\arraybackslash}p{1.5cm}|>{\centering\arraybackslash}p{1.1cm}>{\centering\arraybackslash}p{1.1cm}>{\centering\arraybackslash}p{1.1cm}>{\centering\arraybackslash}p{1.1cm}>{\centering\arraybackslash}p{1.1cm}>{\centering\arraybackslash}p{1.1cm}>{\centering\arraybackslash}p{1.1cm}>{\centering\arraybackslash}p{1.1cm}|}[colortbl-like]
   \toprule
\Block[B]{2-2}{\includegraphics[height=1.9cm]{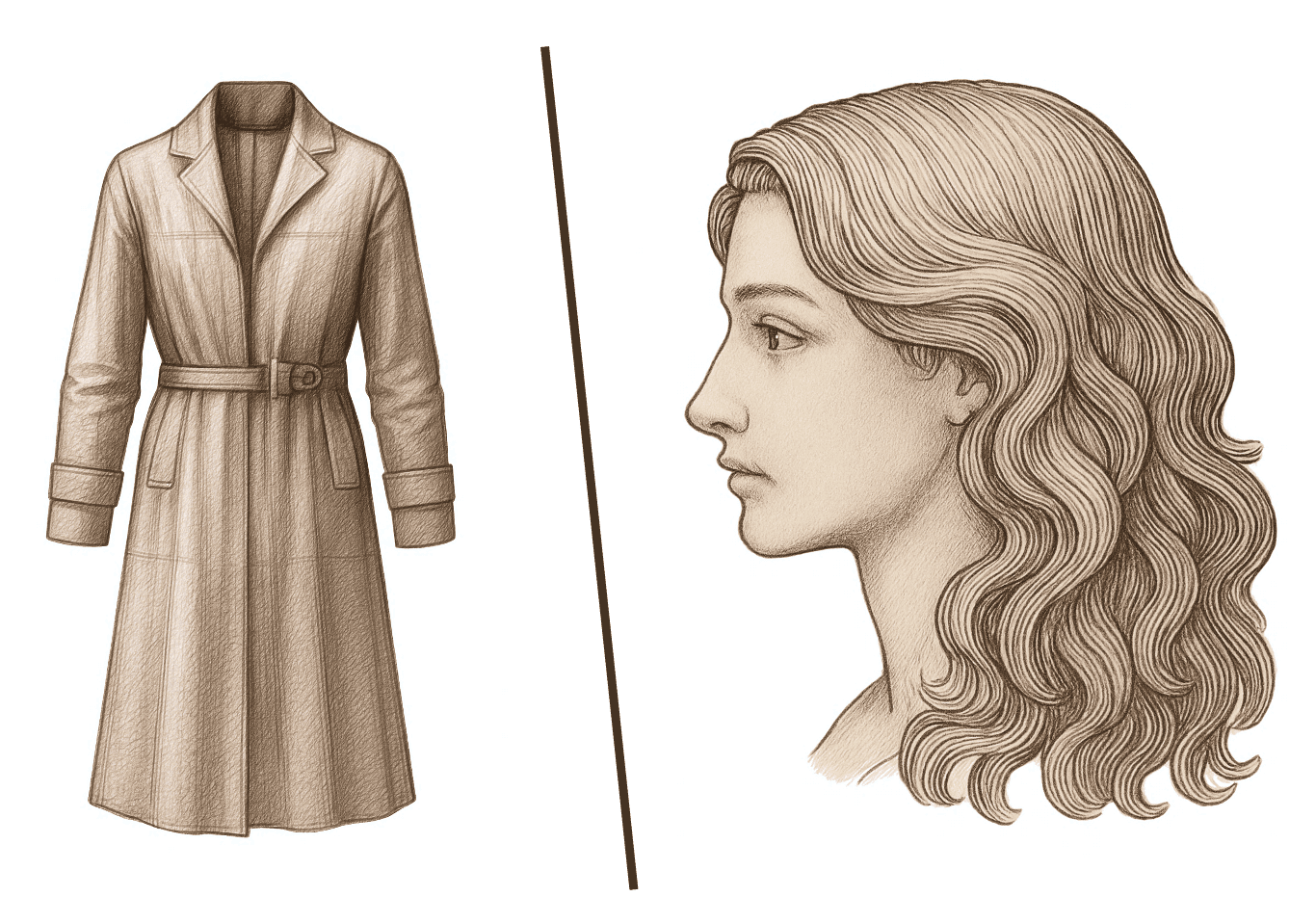}\\\\ \textbf{Digital Human Assets}} & & \Block[fill=ForestGreen!10]{1-4}{\textbf{Assets Prior}} &&& & \Block[fill=RoyalBlue!10]{1-8}{\textbf{Assets Creation}} &&&&&&& \\
    & & \rot{\cellcolor{ForestGreen!10}\textbf{Prior Dataset Size}} & \rot{\cellcolor{ForestGreen!10}\textbf{Datasets}} & \rot{\cellcolor{ForestGreen!10}\textbf{Data Type}} & \rot{\cellcolor{ForestGreen!10}\textbf{Data Modality}} & \rot{\cellcolor{RoyalBlue!10}\textbf{Needed Assets}} & \rot{\cellcolor{RoyalBlue!10}\textbf{Input}} & \rot{\cellcolor{RoyalBlue!10}\textbf{Additional Priors}} & \rot{\cellcolor{RoyalBlue!10}\textbf{Creation Speed}} & \rot{\cellcolor{RoyalBlue!10}\textbf{Representation}} & \rot{\cellcolor{RoyalBlue!10}\textbf{Simulation Ready}} & \rot{\cellcolor{RoyalBlue!10}\textbf{Lighting Control}} & \rot{\cellcolor{RoyalBlue!10}\textbf{Contents}} \\
    \midrule
    \Block{10-1}{\rotate Hair}
 & \cellcolor{blue!04}CT2Hair~\cite{shen2023CT2Hair} & \cellcolor{blue!04} & \cellcolor{blue!04} & \cellcolor{blue!04} & \cellcolor{blue!04} & \cellcolor{blue!04}\crboxTrackedDMM & \cellcolor{blue!04}\icoMRI & \cellcolor{blue!04} & \cellcolor{blue!04}\icoSlow & \cellcolor{blue!04}\crboxStrands & \cellcolor{blue!04}\icoYes & \cellcolor{blue!04} & \cellcolor{blue!04}\icoHair \\
 & NeuralStrands~\cite{Radu2022neuralstrands} &  & \crboxInternal & \icoSynth & \icoStrands & \crboxTrackedDMM & \icoMultiImage &  & \icoSlow & \crboxStrands & \icoYes &  & \icoHair \\
 & \cellcolor{blue!04}GroomCap~\cite{Zhou2024groomcap} & \cellcolor{blue!04} & \cellcolor{blue!04} & \cellcolor{blue!04} & \cellcolor{blue!04} & \cellcolor{blue!04} & \cellcolor{blue!04}\icoMultiImage \icoMonoVideo & \cellcolor{blue!04} & \cellcolor{blue!04}\icoSlow & \cellcolor{blue!04}\crboxMesh \crboxStrands & \cellcolor{blue!04}\icoYes & \cellcolor{blue!04} & \cellcolor{blue!04}\icoHair \\
 & Neural Haircut ~\cite{Sklyarova2023neural} & 343 & \crboxUSCHairSalon & \icoSynth & \icoStrands & \crboxTrackedDMM & \icoMonoVideo \icoMultiImage &  & \icoSlow & \crboxStrands & \icoNo &  & \icoHair \\
 & \cellcolor{blue!04}Monohair~\cite{Wu2024monohair} & \cellcolor{blue!04}2 744 & \cellcolor{blue!04}\crboxUSCHairSalon & \cellcolor{blue!04}\icoSynth & \cellcolor{blue!04}\icoStrands & \cellcolor{blue!04} & \cellcolor{blue!04}\icoMonoVideo \icoMultiImage & \cellcolor{blue!04}\crboxDeepMVSHair & \cellcolor{blue!04}\icoMedium & \cellcolor{blue!04}\crboxStrands & \cellcolor{blue!04}\icoNo & \cellcolor{blue!04} & \cellcolor{blue!04}\icoHair \\
 & GroomLight~\cite{Zheng2025GroomLight} &  &  &  &  &  & \icoOLATImage & \crboxGroomCap & \icoSlow & \crboxDGS \crboxStrands & \icoYes & \icoDistantLight & \icoHair \\
 & \cellcolor{blue!04}DiffLocks~\cite{Radu2025difflocks} & \cellcolor{blue!04}40 000 & \cellcolor{blue!04}\crboxInternal & \cellcolor{blue!04}\icoSynth & \cellcolor{blue!04}\icoStrands & \cellcolor{blue!04} & \cellcolor{blue!04}\icoOne & \cellcolor{blue!04} & \cellcolor{blue!04}\icoFast & \cellcolor{blue!04}\crboxStrands & \cellcolor{blue!04}\icoYes & \cellcolor{blue!04} & \cellcolor{blue!04}\icoHair \\
 & HAAR~\cite{Sklyarova2024haar} & 9825 & \crboxCTHair \crboxInternal \crboxUSCHairSalon & \icoSynth \icoGenerated & \icoStrands & \crboxTrackedDMM & \icoText \icoZero & \crboxNeuralHaircut & \icoInstant & \crboxStrands & \icoYes &  & \icoHair \\
 & \cellcolor{blue!04}3DGH~\cite{he2025head} & \cellcolor{blue!04}70000 & \cellcolor{blue!04}\crboxPanoHeadsamples & \cellcolor{blue!04}\icoReal & \cellcolor{blue!04} & \cellcolor{blue!04} & \cellcolor{blue!04}\icoZero & \cellcolor{blue!04} & \cellcolor{blue!04} & \cellcolor{blue!04}\crboxDGS & \cellcolor{blue!04}\icoNo & \cellcolor{blue!04} & \cellcolor{blue!04}\icoHair \\
 & Doubly~\cite{Chen2025doubly} & 658 & \crboxUSCHairSalon & \icoGenerated \icoSynth & \icoStrands &  & \icoZero &  & \icoInstant & \crboxStrands & \icoYes &  & \icoHair \\
    \midrule
    \Block{6-1}{\rotate Garment}
 & \cellcolor{blue!04}Neural Tailor~\cite{korosteleva2022neuraltailor} & \cellcolor{blue!04}22 000 & \cellcolor{blue!04}\crboxSewingPattern & \cellcolor{blue!04}\icoSynth & \cellcolor{blue!04}\icoMeshes \icoSewingPattern & \cellcolor{blue!04} & \cellcolor{blue!04}\icoPointCloud & \cellcolor{blue!04} & \cellcolor{blue!04}\icoInstant & \cellcolor{blue!04}\crboxMesh & \cellcolor{blue!04}\icoYes & \cellcolor{blue!04} & \cellcolor{blue!04}\icoGarment \\
 & Rec~\cite{qiu2023rec} &  &  &  &  & \crboxTrackedDMM & \icoMonoVideo & \crboxGarmenttemplates \crboxSemanticsegmentation & \icoSlow & \crboxSDF & \icoNo &  & \icoGarment \\
 & \cellcolor{blue!04}BCNet~\cite{jiang2020bcnet} & \cellcolor{blue!04}48 000 & \cellcolor{blue!04}\crboxBCNet & \cellcolor{blue!04}\icoSynth & \cellcolor{blue!04}\icoSingleImage \icoMeshes \icoSewingPattern & \cellcolor{blue!04}\crboxTrackedDMM & \cellcolor{blue!04}\icoOne & \cellcolor{blue!04} & \cellcolor{blue!04}\icoInstant & \cellcolor{blue!04}\crboxMesh & \cellcolor{blue!04}\icoNo & \cellcolor{blue!04} & \cellcolor{blue!04}\icoGarment \\
 & SCARF~\cite{Feng2022scarf} &  &  &  &  & \crboxTrackedDMM \crboxsegmentationmasks & \icoMonoVideo & \crboxDMM & \icoSlow & \crboxMesh \crboxNeRF &  &  & \icoGarment \\
 & \cellcolor{blue!04}Gaussian Garments~\cite{Rong2024gaussgarment} & \cellcolor{blue!04} & \cellcolor{blue!04} & \cellcolor{blue!04} & \cellcolor{blue!04} & \cellcolor{blue!04}\crboxRegisteredMeshes & \cellcolor{blue!04}\icoMultiViewVideo & \cellcolor{blue!04}\crboxDMM \crboxContourCraft \crboxPhysics & \cellcolor{blue!04}\icoSlow & \cellcolor{blue!04}\crboxDGS \crboxMesh & \cellcolor{blue!04} & \cellcolor{blue!04} & \cellcolor{blue!04}\icoGarment \\
 & SimAvatar~\cite{li2025simavatar} & 20 000 & \crboxClothD \crboxSewingPattern & \icoSynth & \icoMeshes &  & \icoText & \crboxDMM & \icoMedium & \crboxDGS &  &  & \icoGarment \\
  \bottomrule
  \end{NiceTabular}
  }
  \caption{Digital Human Assets Taxonomy: We show representative state-of-the-art methods for human avatar assets like hair and garments.}
  \label{tab:assets}
\end{table*}

%% file: chapters/5_3_hair.tex


\section{Hair}
\label{sec:hair}

\begin{wrapfigure}{r}{0.2\columnwidth} 
\centering 
\vspace{-0.4cm}
\hspace*{-1cm}
\includegraphics[width=0.25\columnwidth]{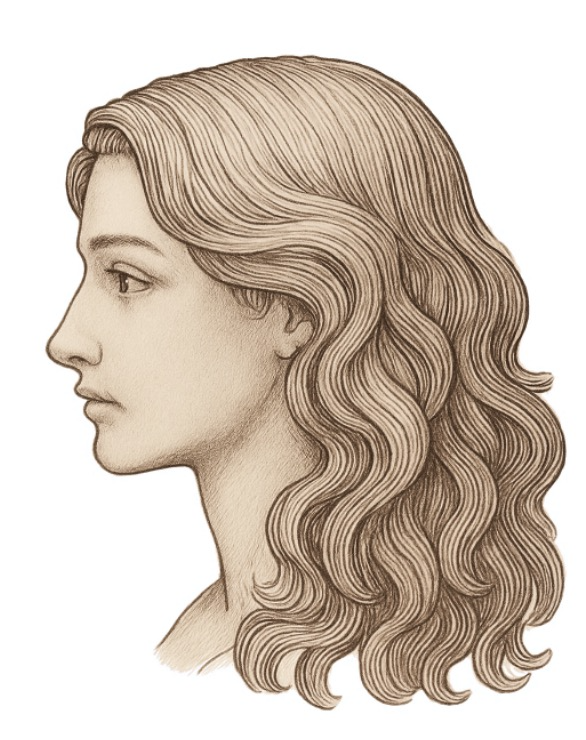} 
\vspace{-0.5cm}
\end{wrapfigure}
\new{
Most 3D head avatar methods mentioned in the previous section treat the head as a single, monolithic representation. However, the head is composed of structurally distinct elements, most notably hair, whose geometry, dynamics, and appearance differ fundamentally from those of the face.
}
%
%
While the head can be approximated as a smooth, largely rigid geometry, hair consists of thousands of thin, flexible fibers with complex appearance and dynamic behavior. A unified representation conflates these characteristics, limiting generalization, reducing rendering fidelity, and restricting editability. Disentangled representations, on the other hand, model hair as a separate layer from the head~\cite{kim2025haircup, liu2025lucas, HAIRFREE:NEURIPS:25, he2025head}. Although this approach introduces additional complexity compared to embedding hair directly on the head, it offers key advantages, including accurate geometric modeling, independent control of hair and head, smooth interpolation, and compositional flexibility for layered avatar construction.

In the earliest days of computer graphics, hairstyles were created almost entirely by hand, using textured polygonal strips (hair cards) or a few manually drawn spline curves duplicated to suggest a full head of hair~\cite{maya, Blender}.
However, automatically reconstructing realistic hairstyles from observations is also highly challenging.
A typical hairstyle consists of 100,000–200,000 individual strands, yet only a small fraction are directly visible due to severe self-occlusion.
Since a strand is essentially a 3D curve, recovering the hidden internal structure is particularly difficult.
Even with multi-view data, ambiguities persist, as different internal arrangements can produce nearly identical external appearances. Moreover, hairstyles must capture both global coherence, such as consistent flow directions, and local irregularities like curls, wisps, and frizz to achieve realism and physical plausibility.

\new{
In the following sections, we provide a detailed discussion of hair reconstruction and generation methods.
For hair animation and simulation, as well as physics-based rendering and relighting of hair, we refer the reader to \Cref{appendix:hair}.
}

\subsection{Hair Reconstruction}

We review the state-of-the-art hair reconstruction methods by input modality, progressing from costly multi-view systems toward affordable, consumer-level phone captures.
A common strategy in image-based approaches is to optimize or regress a 3D hairstyle such that its projection matches the observed 2D orientation features in the image (often using Gabor filters). 
Regardless of the input modality, most hair reconstruction approaches can be grouped into two main categories: (i) \textit{hair growing} methods and (ii) direct \textit{hair map} prediction / optimization.
Both types typically begin by estimating a 3D volumetric representation of the hairstyle along with an orientation field.
\textit{Hair growing} methods~\cite{Luo2013structure, Chai2013dynamic, Zhang2017data, Zhang2018modeling, Shunsuke2018hair, Zhang2019hair, Nam2019strandaccurate, Wu2022neuralhdhair, Kuang2022deepmvshair, zheng2023hairstep, Wu2024monohair} employ a hair growing procedure by sampling root positions on the scalp and tracing strands along the orientation field. 
In contrast, methods such as~\cite{Zhou2018hairnet, Radu2022neuralstrands, Sklyarova2023neural, Zhou2023groomgen, Sklyarova2024haar, Zakharov2024haircut, Luo2024gaussianhair, Takimoto2024drhair, Radu2025difflocks} directly predict or optimize a (latent) \textit{hair map} based on orientation fields, volumetric or silhouette information, to recover the final strand-level geometry.

\subsubsection{\new{Hair Reconstruction from} Multi-view Data}


%
Even with multi-view calibrated data, state-of-the-art methods rely on strong priors to capture the internal structure of hair.
%
%
Zhang \etal~\cite{Zhang2017data} propose a four-view image-based method, reconstructing realistic 3D hair by combining database retrieval with texture-derived 3D direction fields, even when the input views come from unpaired hairstyles. DeepMVSHair~\cite{Kuang2022deepmvshair} introduces a deep implicit formulation, representing hair as a continuous 3D growth direction field. It predicts occupancy and direction at any query point from multi-view features, using a view-aware transformer to aggregate information and preserve high-frequency details. A hair-growing and deformation stage further refines strand geometry. Building on the hairstyle prior of DeepMVSHair for capturing accurate internal geometry, MonoHair~\cite{Wu2024monohair} reconstructs external hair geometry from monocular video using Patch-based Multi-View Optimization (PMVO), then applies a hair growth algorithm to resolve directional ambiguities.

Nam \etal~\cite{Nam2019strandaccurate} combine a multi-view stereo method with a novel cost function, clustering, and strand growing. While their method achieves accurate reconstructions, the strands are often disconnected from the scalp and hairstyles may contain holes, limiting applicability in downstream simulation. Neural Strands~\cite{Radu2022neuralstrands} extends this line of work to full hairstyle reconstruction by optimizing geometry in a latent hair-map space aligned with a scalp UV map. At each iteration, strand roots are sampled and decoded using a pretrained strand prior, then projected back into 2D space via soft rasterization of lines. Geometry is supervised using RGB images, Gabor features, and 3D strand priors from Nam \etal~\cite{Nam2019strandaccurate}.

Since calibrated multi-view capture is impractical in consumer applications, recent research has shifted toward monocular 360° video under natural conditions.
Neural Haircut~\cite{Sklyarova2023neural} introduces a two-stage pipeline for strand-based hair reconstruction from monocular video captured with a smartphone.
It directly optimizes for a latent hair map using a diffusion-based hairstyle prior, which is trained on synthetic data.
To establish data terms of the optimization, Neural Haircut uses a Structure-from-Motion (SfM)-based 3D geometry estimation of the visible surface and surface orientation fields extracted from the input images.
To improve internal strand flow, Dr.Hair~\cite{Takimoto2024drhair} introduces a Laplace-based initialization, connectivity-aware reparameterization, and an enhanced renderer.
Gaussian Haircut~\cite{Zakharov2024haircut}, GaussianHair~\cite{Luo2024gaussianhair}, and HairGS~\cite{pan2025hairgs} propose to represent strands as sequences of cylindrical 3D Gaussian primitives anchored at segment midpoints, which enables efficient rasterization and improves reconstruction quality. To address motion blur during phone capture and inaccuracies in camera poses estimated by COLMAP~\cite{schoenberger2016sfm,schoenberger2016mvs}, Gaussian Haircut~\cite{Zakharov2024haircut} further proposes camera pose finetuning, leading to improved reconstruction quality.
In contrast to these prior-driven approaches at the strand or hairstyle level, GroomCap~\cite{Zhou2024groomcap} learns a neural implicit representation for 3D occupancy and orientations using a volumetric 3D orientation rendering algorithm, coupled with 2D orientation distribution supervision. 
Similar to~\cite{Zakharov2024haircut, Luo2024gaussianhair}, they use a 
3D Gaussian primitives-based representation
enabling direct photometric supervision from input images. 
It is worth noting that these methods can only handle static geometries.
There are only a few methods like Hu \etal~\cite{Hu2017simulation} that tackle both reconstruction and the estimation of physical hair properties, thereby bridging hairstyle modeling with dynamic simulation.

\subsubsection{\new{Hair Reconstruction from a} Single Image}

Reconstructing strand-based hairstyles from a single image is inherently ill-posed, as only the frontal view is observed while multiple plausible back views may exist. Furthermore, Gabor features capture local hair orientations but remain undirected, leaving directional ambiguities unresolved. To address these challenges, early methods~\cite{Chai2012single, Chai2013dynamic, Hu2014robust, Hu2015single} incorporate user-provided strokes as additional constraints.
Methods like~\cite{Chai2016autohair, Hu2017avatar, Zhang2018modeling} improve automation by retrieving hairstyles from a dataset, which are then refined through optimization. Recent approaches, such as Hairmony~\cite{Meishvili2024hairmony} employ learned hairstyle classifiers to guide reconstruction.
Building on these approaches, recent single-view methods leverage learned hairstyle priors. These priors are trained on synthetic datasets of realistic strand-based hair. To mitigate the domain gap (sim-to-real), methods use orientation maps as input~\cite{Zhou2018hairnet, Shunsuke2018hair, Zhang2019hair, Wu2022neuralhdhair, zheng2023hairstep}.
Hairstep~\cite{zheng2023hairstep} additionally proposes dense direction maps labeled on a small dataset.
In DiffLocks~\cite{Radu2025difflocks}, procedurally generated curly hairstyles are used to train a diffusion model, conditioned on DINOv2~\cite{Oquab2023DINOv2LR} features extracted from input images, to synthesize realistic strand-based reconstructions. 
In contrast to these complex prior models, there are also classical PCA-based hair model priors as used in Perm~\cite{He2025perm}, representing both coarse and detailed hair geometry.
It reconstructs hairstyles from single images via inversion in PCA space.
Building on this idea, Im2Haircut~\cite{sklyarova2025im2haircut} trains a transformer-based prior to predict the PCA map from an input image, followed by inversion in the prior space, improving efficiency and accuracy.
The methods above aim to develop general solutions for a wide range of hairstyles; however, others focus on specific styles, such as braids or curls, using specialized priors or procedural rules.
This reflects a trade-off between versatility and precision: general methods handle diverse hair but may struggle with complex structures, whereas specialized approaches achieve high fidelity for targeted hairstyles.
For example, Hu \etal~\cite{Hu2014capturing} reconstruct braided hairstyles from a single RGB-D camera using compact procedural braid models. 
%
Luo \etal~\cite{Luo2013structure} reconstruct coherent hair wisps from still images by detecting local structures, reasoning about connectivity and direction, and synthesizing strands robust to occlusions, suitable for animation. Chai \etal~\cite{Chai2015high} further reconstruct high-quality 3D hair from minimal input by combining depth cues, a 3D helical hair prior, and Shape-from-Shading normals.

\subsubsection{\new{Hair Reconstruction from CT Data}}  
Unlike the image-based methods that we discussed above, which can only capture the visible outer surface, computed tomography (CT) scans enable reconstruction of internal hair structures. CT2Hair~\cite{shen2023CT2Hair} is the first method to leverage CT scans to generate high-fidelity 3D hair models from real-world hair wigs. The approach employs a coarse-to-fine process: guide strands with 3D orientation fields are first recovered, then dense strands are populated via neural interpolation and refined to match the density volumes.
While this is an interesting experiment, CT scans for avatar reconstruction are not feasible in practice.
%

\begin{figure}[t]
    \centering
    \includegraphics[width=\columnwidth]{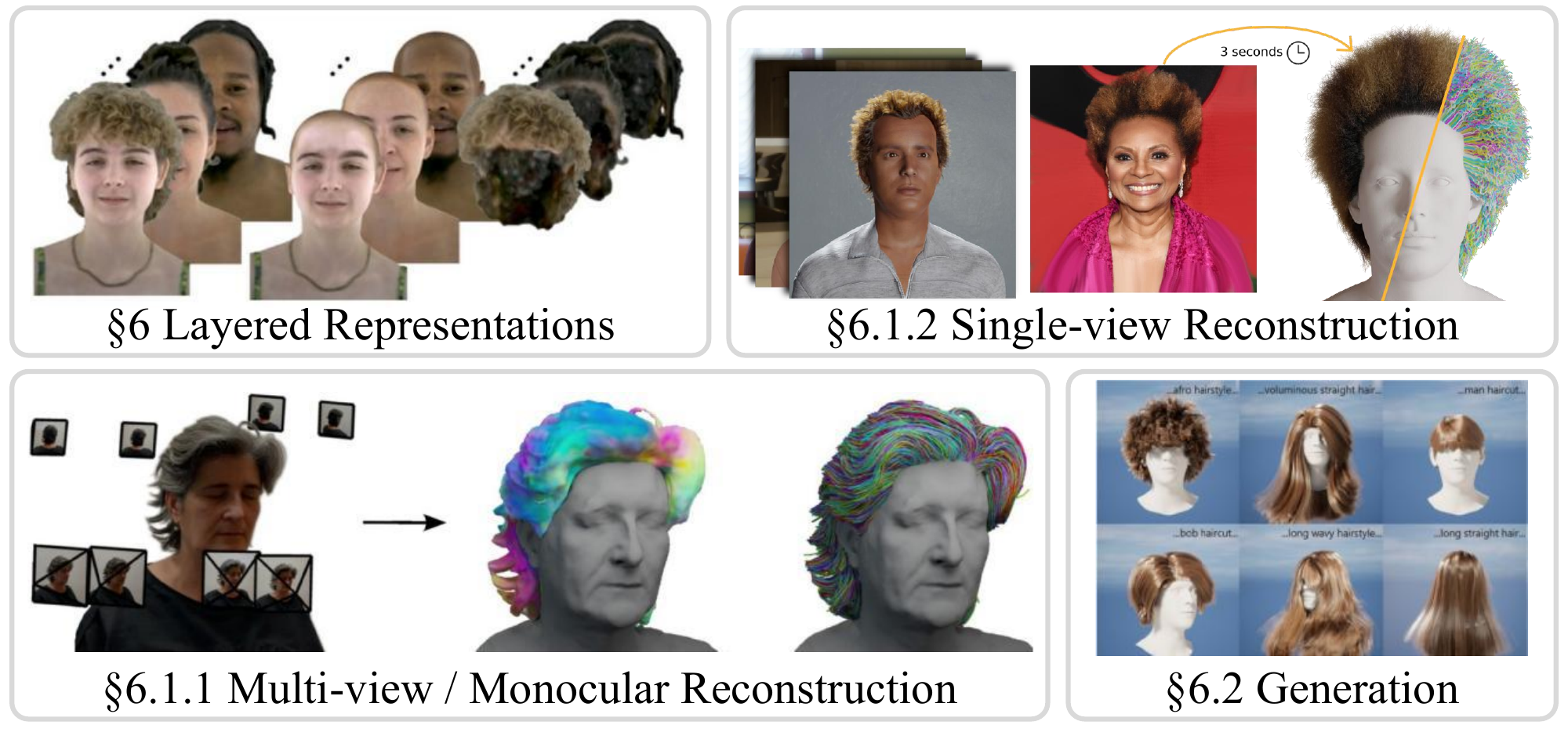}
    \caption{ 
    Hair reconstruction methods. Representative images are taken from HairCUP~\cite{kim2025haircup}, Difflocks~\cite{Radu2025difflocks}, Neural Haircut~\cite{Sklyarova2023neural},  HAAR~\cite{Sklyarova2024haar}.
    }
    \vspace{-3mm}
\end{figure}

\subsection{Hair Generation}

Most hair capture techniques focus on recreating a single subject, whereas generative hair models aim to represent and manipulate a wide range of hairstyles, but have so far been less studied.

\subsubsection{\new{Unconditional 3D Hair Generation}} 
GroomGen~\cite{Zhou2023groomgen} is the first generative model for dense, strand-level hair geometry, employing hierarchical latent spaces to represent individual strands, sparse guide hairs, and full hairstyles. Trained on an augmented, artist-designed strand-based dataset, the model can generate realistic and diverse hairstyles at inference. 
Perm~\cite{He2025perm} introduces a PCA-based representation for hairstyles. Unlike prior work that jointly models global structure and local curls, Perm disentangles them in the frequency domain, separating hair geometry into low-frequency guide textures and high-frequency residual textures. These components are parameterized with separate generative models, allowing Perm not only to reconstruct hairstyles accurately but also to generate novel hairstyles, perform hairstyle editing, and serve as a prior for downstream tasks.
Another hierarchical strand-level approach is proposed by Chen \etal~\cite{Chen2025doubly}, which first generates coarse guide hairs and then dense strands with high-frequency details. By leveraging a discrete cosine transform to separate global structure from local curls and k-medoids clustering to select guide strands, this method preserves hairstyle characteristics and supports off-grid modeling for more realistic dense strand generation.

Extending from hairstyle-level to full-head generation, 3DGH~\cite{he2025head} models 3D human heads with composable hair and face components. Hair is represented using deformable 3D Gaussians within a template-based Gaussian splatting framework, capturing geometric variations across hairstyles. A dual-generator 3D GAN with cross-attention models hair-face correlations, enabling high-quality 3D hairstyle synthesis and composable hairstyle editing, outperforming prior 3D GAN approaches.

\subsubsection{\new{Text-based 3D Hair Generation}} 
HAAR~\cite{Sklyarova2024haar} proposes the first strand-based generative model that produces 3D hairstyles from textual descriptions. It leverages VQA-annotated~\cite{liu2024improvedllava, liu2023llava} synthetic hair models to train a latent diffusion model in a common hairstyle UV space, enabling realistic and diverse hairstyle generation. Later, StrandHead~\cite{Sun2025strandhead} extends this approach to model disentangled full head avatars. Instead of relying on large-scale paired data, they propose to distill knowledge from a pretrained 2D diffusion model to generate an avatar.

\subsubsection{\new{2D Hair Appearance Generation with 3D Control}} 
\new{
ControlHair~\cite{lin2025controlhairphysicallybasedvideodiffusion} combines 3D hair reconstruction and physics-based animation with a conditional video diffusion model.
Specifically, an appearance prior in terms of a video diffusion model is learned on real videos by fine-tuning a ControlNet~\cite{zhang2023adding}-like architecture based on WAN2.1~\cite{wan2025wan}.
As a control signal, it leverages a reference image of the subject and orientation maps that can be extracted from original videos during train-time~\cite{zheng2023hairstep}.
At test-time these orientation maps are provided by a classical computer graphics rendering scheme that leverages 3D hair strands that are recovered from the reference image by DiffLocks~\cite{Radu2025difflocks} and animated with Blender.
While the usage of such a learned hair rendering prior results in promising results, it heavily relies on the accuracy of the 3D hair strand reconstruction and the simulation.
Automatically recovering physics parameters of a hairstyle from a reference image that can be used for simulation is an open challenge, and thus, ControlHair requires manual fine-tuning of the physics parameters in Blender.
PhysAnimator~\cite{xie2025physanimatorphysicsguidedgenerativecartoon} is a related method that follows mainly the same concept, but uses a mesh-based representation to model dynamics and contour-lines as guiding signal of the video diffusion model.
It is more general than ControlHair, but lacks strand-level control and is only evaluated on cartoon animation.
}

\subsection{Facial Hair Reconstruction}

Recently, several works have focused on modeling beards, eyebrows, and eyelashes using strand-based representations. Beeler \etal~\cite{Beeler2012coupled} present a method that reconstructs various facial-hair styles along with the underlying skin surface. Using multi-view images, it detects and traces hairs for beards and eyebrows and lifts them into 3D to produce strand-based geometry, while also handling hair occlusion to generate realistic skin geometry underneath.
Li \etal~\cite{Li2024strand} further propose dense facial hair reconstruction and tracking. By combining line-based multi-view stereo and line segment matching, it recovers a dense hair point cloud, extracts strands via a forward Euler method, and tracks them over time with space-time optimization. The refined skin mesh produces highly accurate and complete facial hair models suitable for dynamic facial performances. EMS~\cite{Li2023ems} reconstructs 3D eyebrows from a single image by modeling them as fiber curves. It uses three modules: RootFinder for root locations, OriPredictor for 3D growth directions, and FiberEnder to stop growth. Trained on a synthetic 3D eyebrow dataset, EMS handles a wide range of eyebrow styles and lengths.
Kerbiriou \etal~\cite{Kerbiriou2024reconstruction} introduce the first data-driven semantic model of eyelashes. Using multi-view 3D reconstruction, they build the first 3D eyelash dataset and extract semantic features to describe shape and variation, enabling intuitive and realistic eyelash synthesis for avatars and synthetic data generation.
%


%% file: chapters/5_2_hands.tex
\section{Hands}
\label{sec:hands}


\begin{wrapfigure}{r}{0.2\columnwidth} 
\centering 
\vspace{-0.4cm}
\hspace*{-1cm}
\includegraphics[width=0.25\columnwidth]{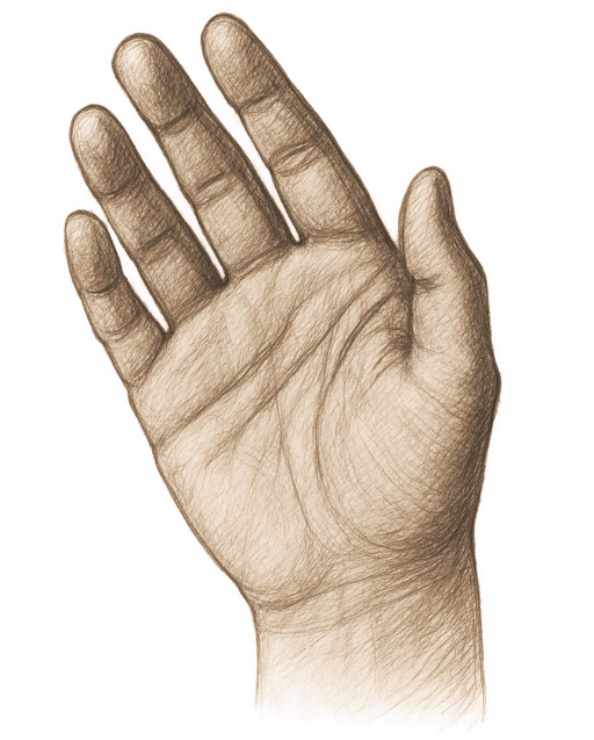} 
\vspace{-0.5cm}
\end{wrapfigure}
Although hands are part of the body, we pay special attention to hand modeling, as hands play a central role in everyday interactions with hands, other body parts, and objects.
%
In addition, hand modeling is uniquely challenging because it needs to support large articulations.
Changes in poses lead to drastic geometry and appearance change due to large self-contact and shadow casting.
In this section, we review the literature on hand priors and hand modeling, and discuss hand avatar control.

\begin{figure}[t]
    \centering
    \includegraphics[width=\columnwidth]{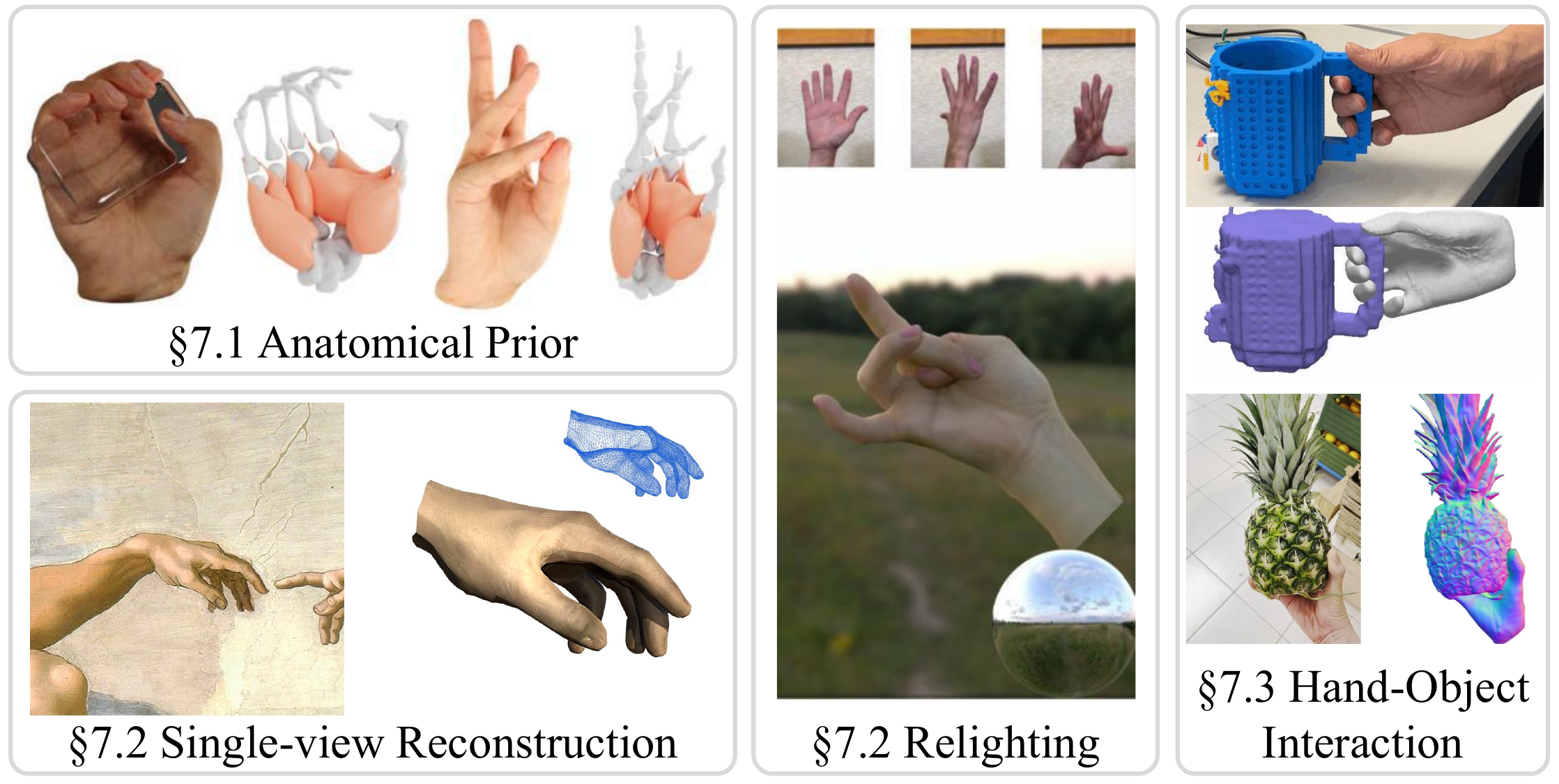}
    \caption{ 
    Hand reconstruction methods. Representative images are taken from NIMBLE~\cite{li2022nimble}, Handy~\cite{potamias2023handy}, URHand~\cite{Chen2024urhand}, and HOLD~\cite{fan2024hold}, respectively. 
    }
    \vspace{-3mm}
\end{figure}

\subsection{Hand Priors}
\label{sec:hand_prior}


Due to the large articulation of hands, various priors are proposed for hand modeling and tracking. Earlier work utilizes a collection of primitives to represent hands~\cite{rehg1994visual,oikonomidis2011efficient,tagliasacchi2015robust}. Sridhar \etal~instead use a collection of 3D Gaussians~\cite{sridhar2013interactive,sridhar2015fast}. Sphere-meshes~\cite{oikonomidis2011full,tkach2016sphere} are also used to simplify the shape representation of hands for tracking. 
Schmidt \etal~\cite{schmidt2014dart} combine voxelized primitives and signed distance function (SDF) to further improve collision handling. 
While these representations are useful to handle heavy collision and occlusion in an efficient manner, their geometric fidelity remains limited. 

A triangular mesh has gained the most popularity for shape prior due to its efficiency and fidelity. Similar to face 3DMMs~\cite{Blanz1999}, Khamis \etal~\cite{khamis2015learning} apply PCA to the scans of 50 people using a depth sensor to obtain a linear identity shape space. However, pose-dependent deformation is limited to linear blend skinning (LBS), leading to limited fidelity. 
Later, in the same spirit as SMPL~\cite{loper2015smpl}, MANO~\cite{romero2017mano} learns identity linear shape space and pose-dependent corrective space independently using PCA from approximately 1000 scans consisting of up to 51 poses from 31 subjects.
Non-linear variants of identity and pose correctives have also been explored. Dkulon \etal~\cite{dkulon2019rec} use graph convolution to represent correctives. However, this work models identity shape and pose-dependent deformations into a single latent space, not supporting disentangled control.
DeepHandMesh ~\cite{moon2020deephandmesh} present a self-supervised method to learn non-linear identity and pose correctives parameterized by MLPs. The model is learned from 3D scans obtained from InterHand2.6M~\cite{moon2020interhand2}, showing superior details compared to the MANO model. 
UHM~\cite{moon2024authentic} additionally incorporates image-based supervision through appearance modeling. 
Incorporating anatomical priors has also been explored in several works~\cite{li2022nimble,wang2019hand,zheng2022simulation}. These approaches use magnetic resonance imaging (MRI) to obtain underlying tissues and bones. These anatomical priors and physics-based volumetric prior~\cite{smith2020constraining} are highly effective at handling large deformations and self-contacts of hands.

Since hands share a common topology across identities, building an appearance prior on UV space is a common choice. 
HTML~\cite{qian2020html} builds a PCA linear appearance space of textures on UV space under a uniformly lit environment from 51 subjects.
NIMBLE~\cite{li2022nimble} learns PCA linear appearance space parameterized by PBR reflectance maps (i.e., surface normal, albedo, and specular map) from 38 scanned subjects. 
Despite being compact, a linear basis tends to blur the final outputs.
Handy~\cite{potamias2023handy} learns a generative model of hand geometry and appearance using StyleGAN3~\cite{karras2021alias} from approximately 1200 high-quality dome-captured data. The expressiveness of StyleGAN enables more photorealistic texture synthesis.
URHand~\cite{Chen2024urhand} builds a relightable appearance prior on UV space from 93 identities with approximately 42,000 frames of time-multiplexed illuminations for each identity. URHand does not contain a latent bottleneck unlike the aforementioned approaches, and instead directly translates a mean texture map into relightable features. This also supports more detailed appearance.
Unlike UV space, directly representing geometry and appearance in 3D space using Neural Fields~\cite{xie2022neural} such as NeRF is also a promising direction due to its expressiveness and photorealism. 
LISA~\cite{corona2022lisa} learns a prior model of part-based SDF from InterHand2.6M in an end-to-end manner.
OHTA~\cite{zheng2024ohta} presents a NeRF-based prior model that learns albedo, shadow, and opacity fields from InterHand2.6M~\cite{moon2020interhand2} dataset. The fields are conditioned by a global latent code, and can be quickly personalized via inversion.
GaussianHand~\cite{huang2024learning} adopts 3D Gaussian Splatting~\cite{Kerbl20233d} for geometry and appearance representations, and builds a prior in UV-aligned feature space from InterHand2.6M.





\subsection{Hand Avatar Creation}

Hand avatar creation has been explored with various input setups. 
On one hand, multi-view dome capture data is used to push the envelope of fidelity in hand avatars. These approaches are typically person-specific and have minimal use of priors and preprocessing, with a training time longer than a day. 
RelightableHands~\cite{Iwase2023relightablehands} builds a person-specific relightable hand model learned from multi-view dome data with time-multiplexed illumination patterns. RelightableHands directly learns a mapping from given lights to final appearance, achieving photorealistic relighting with complex global light transport effects such as subsurface scattering and non-local shadows.
LiveHand~\cite{mundra2023livehand} and HandNeRF~\cite{guo2023handnerf} learn a person-specific non-relightable hand model represented by NeRF from multi-view dome data. LiveHand parameterizes NeRF using UV coordinates on MANO and height fields, following Neural Actor~\cite{liu2021neural}, whereas HandNeRF parameterizes it in a canonical space and deforms it via LBS-guided deformation fields.

On the other hand, if the input is a monocular video, the problem often becomes ill-posed due to limited observations. To address this, leveraging prior models discussed in \Cref{sec:hand_prior} is critical. 
HARP~\cite{karunratanakul2023harp} uses meshes as a shape representation and introduces tailored loss functions with various shape regularization terms to stabilize the optimization process. 
HandAvatar~\cite{chen2023handavatar} learns a photorealistic hand avatar represented by part-wise occupancy fields together with shading and albedo appearance fields. The key to the successful optimization is to condition the neural fields with a high-resolution version of the MANO model, MANO-HD.
UHM~\cite{moon2024authentic} and URHand~\cite{Chen2024urhand} both constrain the geometry with a global latent code as an information bottleneck, allowing us to quickly optimize via inversion. 

%
Similarly, having a compact latent code for geometry and appearance enables hand avatar creation from a single image. Given a single image, we can run inversion to obtain a corresponding latent code by minimizing image-based losses~\cite{qian2020html, corona2022lisa, zheng2024ohta, huang2024learning}. Handy~\cite{potamias2023handy} uses similar loss functions to train a regressor to enable feed-forward reconstruction of a personalized hand avatar.




\new{\subsection{Hand-Object Interaction}
Recent advancements have shifted toward the holistic modeling of hands in context, specifically their interactions with hands, objects and other body parts~\cite{shimada2023decaf,fan2024hold,chen2025interactavatar,ohkawa2025generative}.}
HOLD~\cite{fan2024hold} addresses the joint reconstruction of a hand avatar and an object from monocular video, demonstrating that simultaneous optimization reduces depth and scale ambiguities inherent in isolated modeling. 
\new{Beyond object manipulation, Decaf~\cite{shimada2023decaf} explores complex face-hand interactions by explicitly learning contact regions and local deformations from monocular images.
While Decaf relies on an iterative optimization-in-the-loop to resolve final geometries, InteractAvatar~\cite{chen2025interactavatar} advances the state-of-the-art by moving toward a feed-forward approach.
It achieves high-fidelity results by jointly regressing 3D Gaussian correctives that capture the fine-grained geometric and photometric nuances induced by face-hand contact.}

\subsection{Hand Avatar Control}

Unlike hair or clothing, hand shape and appearance are mostly deterministic based on input poses. Therefore, the aforementioned hand avatar models are all solely driven by hand pose parameters. For relightable hand models~\cite{karunratanakul2023harp,li2022nimble,Iwase2023relightablehands,Chen2024urhand}, we also take a target illumination as an input condition.

\subsection{2D Generative Models for Hands}

\new{Early diffusion-based animation models typically suffer from precisely reconstructing hand shapes~\cite{Ma2025diff_star,hu2024animate}. Several works focus on specifically addressing the hand quality issues of image/video diffusion models by introducing region-aware losses~\cite{fu2024adaptive} or 3D hand embeddings~\cite{narasimhaswamy2024handiffuser}. FoundHand~\cite{chen2025foundhand} achieves precise pose and camera control of hand images with an image-to-image diffusion model.}

%% file: chapters/5_4_garment.tex
\section{Garments}
\label{sec:garment}
%
%
%
%
\begin{wrapfigure}{r}{0.2\columnwidth} 
\centering 
\vspace{-0.4cm}
\hspace*{-0.75cm}
\includegraphics[width=0.25\columnwidth]{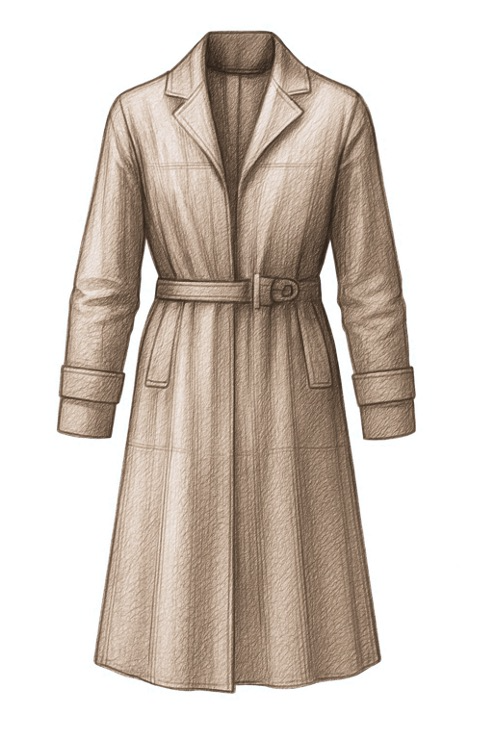} 
\vspace{-0.5cm}
\end{wrapfigure}
Garments play an important role in building immersive human avatars.
Subtle movements in body pose can have a large influence on the motion and appearance of the garment, e.g., wrinkles get visible or disappear during animation.
In contrast to faces, or minimally clothed human bodies, there are no general morphable models that could be used to represent a large variety of clothing types and their deformations. Challenges like topological changes need to be handled, as well as 'free-form' deformations.
In the following, we organize the work based on the garment-related aspects being addressed, such as its role in the avatar, reconstruction, generation, or animation (see \new{\Cref{appendix:garment}}).



\subsection{Avatars with Separate Clothing}






Among the existing work on full-body digital avatars, several methods model garments in different layers from the underlying bodies. Such a design choice is possible with different representations, including meshes~\cite{xiang2021ClothingCodecAvatar,xiang2022DressingAvatars,Xiang2023drivable}, NeRF~\cite{Feng2022scarf}, 3DGS~\cite{lin2024layga,Zielonka2023d3ga}, or even hybrid representations \cite{zheng2024physavatar}. Avatars with disentangled clothing are inherently more complicated to build compared to those with body and clothing in the same layer, but they offer several key benefits.

First, treating clothing separately alleviates the difficulty of modeling loose-fitting garments such as skirts and dresses~\cite{Feng2022scarf,lin2024layga}. These garments deviate significantly from the underlying body shape, and are thus hard to represent by just offsets from commonly-used full-body models such as SMPL \cite{loper2015smpl} and SMPL-X \cite{pavlakos2019expressive}. Second, this design choice allows the methods to focus on the sophisticated temporal dynamics of the garments beyond skinning techniques that rely primarily on body pose information. Some examples include learning clothing dynamics in the latent space in an autoregressive manner \cite{xiang2021ClothingCodecAvatar}, incorporating physics-based cloth simulation \cite{xiang2022DressingAvatars,zheng2024physavatar}, or learning clothing deformation from video input to bridge the gap between animation and reconstruction \cite{Xiang2023drivable}.

\subsection{Garment Reconstruction}

\label{sec:garment-reconstruction}

By garment reconstruction or capture, we refer to methods that convert visual data of garments into 3D animatable assets, including geometry and appearance. Depending on the input, methods can be divided into multi-view reconstruction, which focuses on quality, and monocular reconstruction, which favors applicability. In another dimension, image-based approaches recover static garment assets, and video-based reconstruction focuses on dynamic formation or physical parameters. We distinguish these reconstruction methods from text-conditioned approaches, which we refer to as ``generation'' of garments and cover mostly in \Cref{sec:garment-generation}.


\subsubsection{Multi-View Reconstruction \new{of Garments}}

Garment reconstruction usually leverages high-quality surface output from standard multi-view reconstruction algorithms such as Multi-View Stereo (MVS)~\cite{furukawa2009accurate} or Photometric Stereo~\cite{woodham1980photometric}. Early works like Markerless Garment Capture~\cite{bradley2008markerless} and Clothcap~\cite{pons2017clothcap,zhang2017detailed} focus on segmenting out the garment region from the whole scanned geometry and registering the garment sequence with a template of consistent topology. These segmentation and registration steps turn out to be quite useful for creating the training data for the following works on clothed avatars \cite{xiang2021ClothingCodecAvatar,xiang2022DressingAvatars} or monocular garment reconstruction \cite{bhatnagar2019multi}. To further boost the accuracy of registration, some approaches~\cite{scholz2005garment,halimi2022pattern} print color-coded patterns on the fabric to enable reliable feature detection, providing high-quality datasets for learning-based registration approaches~\cite{guo2024diffusion}.

Some recent work addresses the problem of simulation-ready garment reconstruction, i.e., assets that can be readily consumed by physics-based simulators (including neural simulators). This line of work focuses on two challenges. First, as almost all simulators rely on template meshes for garments, other 3D representations that are used for rendering photorealistic appearances, especially 3DGS, still need to be coupled with meshes for simulation. PhysAvatar~\cite{zheng2024physavatar}, PGC~\cite{Guo2025pgc}, and Gaussian Garments~\cite{Rong2024gaussgarment} achieve this purpose by embedding Gaussians within the mesh structure, synergizing the physical fidelity of garment dynamics with high-quality appearance from 3DGS.

Second, the dynamic behavior of garments is subject to the influence of physical parameters, including the stretching and bending stiffness of cloth, the sewing pattern or template mesh in the undeformed configuration, as well as the underlying body shape/pose. Compared with work that directly measures forces and deformation in a laboratory setting \cite{wang2011data,miguel2012data,clyde2017modeling,zhang2024estimating}, it is inherently more challenging to estimate these parameters from just visual input, yet more relevant to our discussion on avatars. The basic paradigm is to optimize the estimated parameters by comparing the simulation output with some visual measurement. For example, DiffAvatar \cite{li2024diffavatar} estimates body and clothes parameters through a differentiable eXtended Position-Based Dynamics (XPBD) simulator~\cite{stuyck2023diffxpbd}. Physavatar \cite{zheng2024physavatar} utilizes a finite-difference method to compute parameter gradients that optimize a Codimensional-IPC simulation \cite{li2021codimensional}. CaPhy~\cite{su2023caphy} and Gaussian Garments~\cite{Rong2024gaussgarment}, on the other hand, estimate parameters for neural simulators, namely, SNUG~\cite{santesteban2022snug} and ContourCraft~\cite{Grigorev2024contourcraft}. Despite such progress, this problem is still challenging. For example, the underlying body pose has a huge influence on the garment dynamics. An optimization that takes into account body pose has a number of Degrees of Freedom (DoF) and requires careful initialization and regularization \cite{li2024diffavatar}. In addition, the simulation of cloth on the human body is contact-rich, for which the gradient estimation can be noisy~\cite{zheng2024physavatar}.

\begin{figure}[t!]
    \centering
    \includegraphics[clip, trim={0 10.3cm 17cm 0},width=\columnwidth]{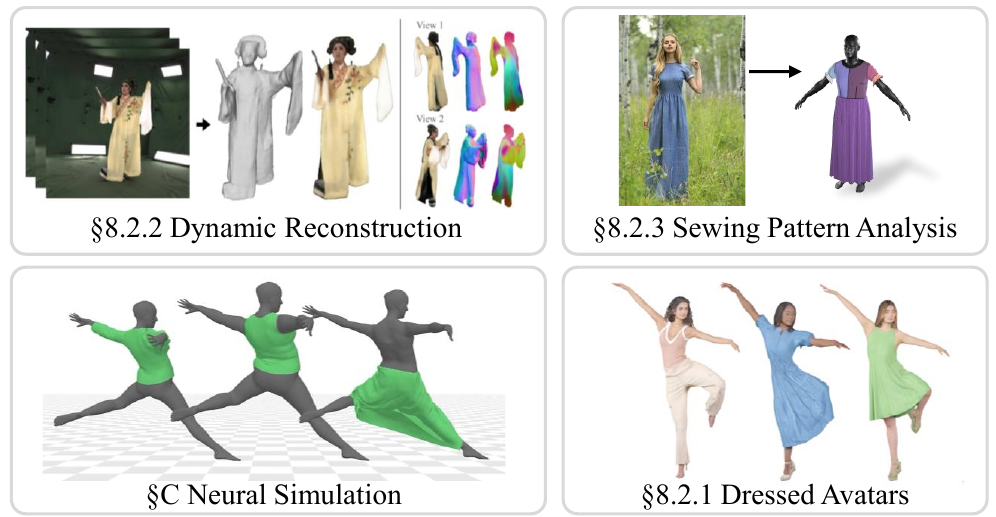}
    \caption{Garment reconstruction methods. Representative images are taken from DressRecon~\cite{tan2025dressrecon}, ChatGarment~\cite{bian2025chatgarment}, HOOD~\cite{grigorev2023hood} and PhysAvatar~\cite{zheng2024physavatar}, respectively. 
    }
    \vspace{-3mm}
\end{figure}

\subsubsection{Monocular Reconstruction \new{of Garments}}

Accurate reconstruction of separated garments from a single RGB image is a highly challenging problem. Due to the ill-posed nature of this problem, almost all methods in this area rely on some \textit{garment templates}, and learn a neural network that maps single images to the samples or deformation of the templates. For example, Multi-Garment Net~\cite{bhatnagar2019multi} and BCNet~\cite{jiang2020bcnet} build PCA subspaces of garments and learn networks to reconstruct PCA coefficients. ClothWild~\cite{moon20223d} leverages SMPLicit~\cite{corona2021smplicit}, a multi-category generative model for implicit garment geometry, to regularize the estimated garment shape. DeClotH~\cite{nam2025decloth} further refines the output geometry and texture from ClothWild~\cite{moon20223d} using an SDS optimization~\cite{poole2023dreamfusion}. ReEF~\cite{zhu2022registering} leverages the surface prediction from a PIFu-style implicit network~\cite{Saito2020pifuhd}, but still registers clothing templates to the predicted surface for high-quality, complete garment shape reconstruction. Garment3DGen~\cite{sarafianos2025garment3dgen} first reconstructs a watertight target garment geometry using an LRM-style~\cite{hong2024lrm} image-to-3D network, and then deforms a garment template to the target using Jacobian Fields~\cite{aigerman2022neural} to obtain high-quality, simulation-ready mesh assets. Generally speaking, this area is relatively under-explored, possibly due to its inherent difficulty in achieving high-quality output that warrants disentangled predictions and a lack of large-scale, realistic training data.

In addition, several methods have been proposed to reconstruct garments that move separately from the underlying body, given a monocular RGB video as input. These methods usually model the underlying body with SMPL/SMPL-X, but separately track the garment deformation. This problem is less ill-posed than single-image estimation, but still highly challenging when realistic clothing dynamics are desired. MonoClothCap~\cite{xiang2020monoclothcap} fits a sequence of clothing shapes in PCA subspaces by differentiable rendering and comparing against the input monocular video. Similar to the idea of monocular dynamic NeRF~\cite{Park2021nerfies}, several methods~\cite{qiu2023rec,guo2024reloo,chen2025d,tan2025dressrecon} simultaneously optimize for a canonical garment shape and its per-frame deformation to perform volumetric differentiable rendering against the input video. In particular, Rec-MV~\cite{qiu2023rec} and D$^3$-Human~\cite{chen2025d} address the issue of representing open garment surface, by tracking the boundary curve, and proposing a human manifold Signed Distance Field (hmSDF), respectively. ReLoo~\cite{guo2024reloo} and DressRecon~\cite{tan2025dressrecon} focus on the challenge of tracking highly dynamic clothing by adopting the bag-of-bones deformation model~\cite{pan2022predicting,yang2022banmo}.

\subsubsection{Sewing Pattern Reconstruction/Generation}

A notable trend in garment reconstruction and generation is that, instead of operating in the ambient 3D geometry space, predictions can be made in the 2D space of sewing patterns. Roughly speaking, sewing patterns are 2D templates of garments, defined by outlines and markings, that guide how fabric should be cut and assembled to make the garments. In Computer Graphics (CG) and Computer-Aided Design (CAD), sewing patterns are represented by panels of cloth defined by a set of 2D planar curves, and stitching marks that connect the panels together, forming a compact representation that allows neural networks to make predictions in. For example, NeuralTailor~\cite{korosteleva2022neuraltailor} proposes to regress sewing patterns from point clouds of the input garments, and Neural Sewing Machine~\cite{chen2022structure} trains an autoencoder for sewing patterns that enables monocular garment reconstruction by an additional CNN that maps images to sewing pattern latents. DMap~\cite{li2025dmap} treats the 2D space of sewing pattern space as a unified UV parameterization of garments, which bridges pixel coordinates and 3D garment geometry for accurate monocular reconstruction.

With the advancement of transformer models, techniques have been developed to tokenize sewing patterns, allowing foundational models to learn their correlation with other modalities such as text and images. Such techniques thus enable reconstruction of sewing patterns from a single image, as in SewFormer~\cite{liu2023sewformer}, and generation of sewing patterns from text description, as in DressCode~\cite{he2024dresscode}, both through the transformer architecture. AIpparel~\cite{nakayama2025aipparel} and ChatGarment~\cite{bian2025chatgarment} develop a multi-modal foundational model for sewing patterns that can handle multiple garment-related tasks, such as text/image-to-garments and text-based editing. Another advantage of working with sewing patterns is that they can be easily used for cloth simulators as a rest shape template. For example, Dress-1-to-3~\cite{li2025dress1to3} employs a differentiable cloth simulator to further optimize the sewing patterns predicted by SewFormer~\cite{liu2023sewformer}, and improves the alignment between the draped garments and the input image.

\subsection{Garment Generation}

\label{sec:garment-generation}

In this section, we discuss recent methods that generate 3D garments, including geometry and texture, from text descriptions. Here, we focus on methods where the generated garments are disentangled from the underlying body. Early methods such as CAPE~\cite{Ma2020cape} and SMPLicit~\cite{corona2021smplicit} build generative models (e.g., GANs) that map latent distributions to scanned and registered clothing geometry. As 3D generation methods based on Score Distillation Sampling (SDS) \cite{poole2023dreamfusion} gain popularity, which optimize 3D representations while requiring only a general 2D image diffusion model, similar techniques have been applied to 3D clothed body generation. Wang and colleagues~\cite{wang2024disentangled} generate garment geometry and texture in an offset layer from the underlying SMPL body shape. DAGSM~\cite{zhuang2025dagsm} generates separate body and garments in two different layers in 2DGS representation, from which garment meshes can be further extracted.

Similar to the trend in general 3D generation that shifts from general 2D-based SDS to 3D-specific priors, recent work starts to leverage garment-specific priors trained on 3D garment data. GarmentDreamer~\cite{li2025garmentdreamer} and SimAvatar~\cite{li2025simavatar} train specific diffusion models on garment data for effective generation of base geometry, and then refine the associated 3DGS appearance with SDS. LayerAvatar~\cite{zhang2025disentangled} proposes a diffusion model that generates multi-layer body and garment components as 3DGS embedded in disentangled UV regions. Compared with full-body avatars where garments and body are generated in a single layer~\cite{kolotouros2023dreamhuman}, these disentangled avatars enjoy the capability of garment editing and transfer~\cite{zhang2025disentangled}, as well as realistic physical dynamics from cloth simulators~\cite{li2025garmentdreamer,li2025simavatar}.
%


%% file: chapters/5_5_full_body.tex
\section{Full-body Avatars}
\label{sec:full_body}


\begin{wrapfigure}{r}{0.15\columnwidth}
    \centering 
    \vspace{-0.8cm}
    \hspace*{-1cm}
    \includegraphics[width=0.275\columnwidth]{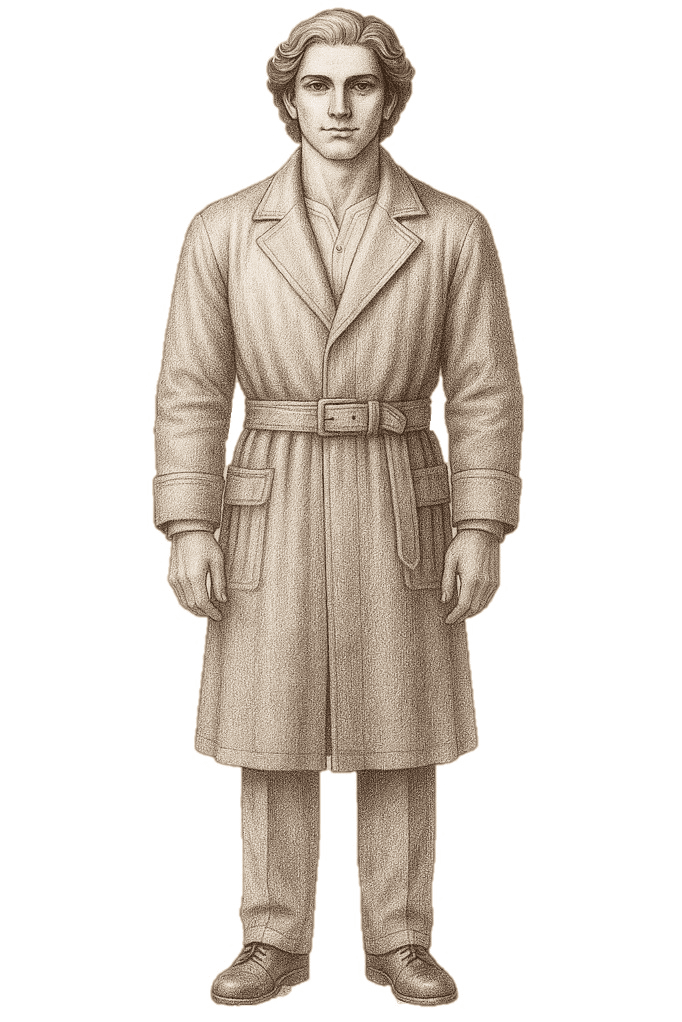}
    \vspace{-0.4cm}
\end{wrapfigure}
This section traces the evolution of full-body avatar creation along a clear data-efficiency trajectory, beginning with constrained 3D scans, advancing to dynamic multi-view setups, and ultimately to monocular, few-shot, and one-shot methods that enable scalable and personalized digital humans by leveraging powerful generative human priors.
Each stage reflects advances in neural modeling and learning-based priors that compensate for reduced input data. This progression highlights the field’s shift toward real-world applicability and rapid personalization, ultimately emphasizing the transition from 3D to 2D approaches.

\subsection{\new{Full-body Avatars from }3D Scan Data} 
\label{sec:3d_scan}

3D scan-based human avatar modeling aims to reconstruct high-fidelity, animatable digital humans from raw 3D scan data, tackling core challenges such as articulated motion, non-rigid deformations, and pose generalization, while preserving fine geometric and textural details. Early advances in this domain laid the foundation by integrating neural implicit surfaces with linear blend skinning to enable pose-independent deformation modeling \cite{Chen2021snarf, ma2021power, mihajlovic2021leap}, developing template-free, weakly supervised canonicalization strategies \cite{Saito2021scanimate,tiwari2021neural,qian2022unif,lin2022learning}, and introducing locally articulated representations to better capture complex clothing dynamics \cite{Ma2021scale,zhang2023closet}. Subsequent work improved physical realism by incorporating physics-based priors into the deformation framework \cite{su2023caphy}, while others extended full-body modeling to encompass expressive facial and hand articulations through part-aware skinning and pose-conditioned texture networks \cite{shen2023x}. More recently, the focus has shifted toward disentangled and reusable avatar representations to enable controllable reanimation and editing \cite{kim2024gala, ho2023learning,bai2022autoavatar}. Concurrently, generative frameworks have advanced multi-subject avatar synthesis with stochastic clothing variations \cite{Chen2022gdna}, and enabled the direct generation of explicit, textured meshes from 2D normal maps, yielding high-fidelity avatars well-suited for real-time rasterization \cite{zhang2023getavatar}.

\begin{figure*}[t!]
    \centering
    \includegraphics[width=\linewidth]{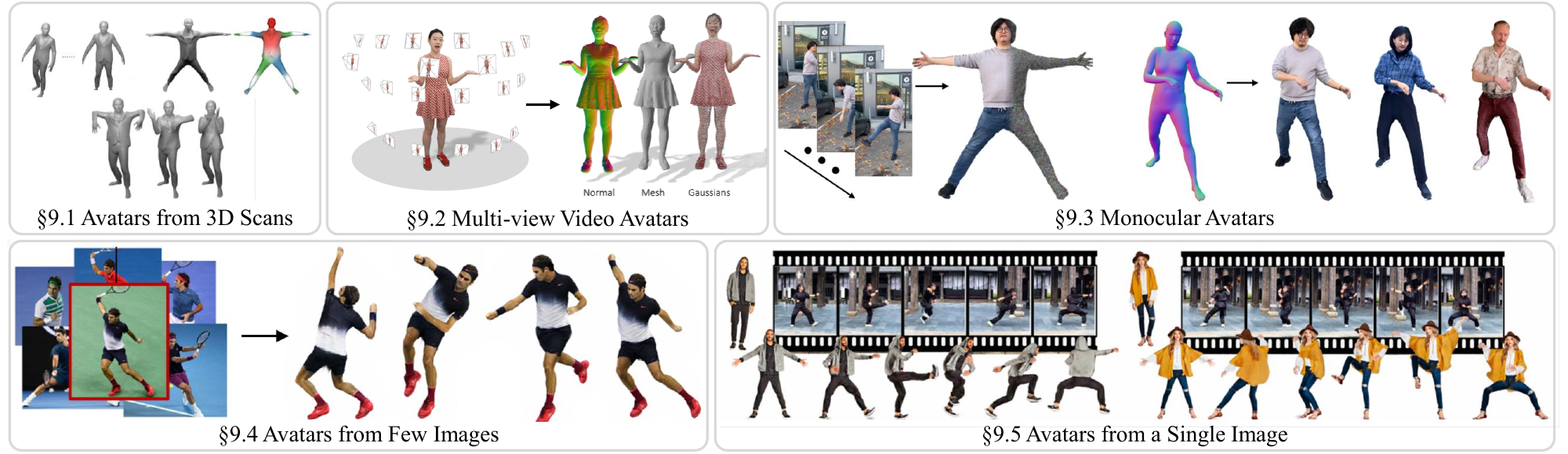}
    \caption{Overview of 3D full-body avatar research. Representative images are taken from SNARF~\cite{Chen2021snarf}, TaoAvatar~\cite{chen2025taoavatar}, Vid2Avatar-Pro~\cite{Guo2025vid2avatarpro}, PersonNeRF~\cite{weng2023personnerf}, and AniGS~\cite{qiu2025anigs}.
    }
\end{figure*}


\subsection{\new{Full-body Avatars from }Multi-View Video}
\label{sec:multi_view_video}
\new{Moving beyond the requirement of explicit 3D scans, }multi-view video enables high-fidelity human avatar reconstruction by providing dense visual observations across both space and time.
Recent research has shifted from purely implicit fields toward hybrid and explicit representations, with the aim of improving fidelity, efficiency, and controllability. This section reviews key advances in representation modeling, 3D Gaussian Splatting, and relighting techniques for achieving photorealistic and animatable avatars.

\subsubsection{From Implicit Fields to Hybrid Representations}

Early approaches to high-fidelity human avatar reconstruction from multi-view video employed encoder-decoder architectures to learn disentangled representations of pose and appearance, enabling view-consistent novel pose synthesis~\cite{Bagautdinov2021driving, Remelli2022drivable}. These methods laid the foundation for pose-controllable avatar modeling but often suffered from limited view consistency and texture blur under large pose deviations.
Neural radiance fields (NeRF)~\cite{Mildenhall2020nerf} marked a significant shift, enabling high-fidelity view synthesis through implicit volumetric rendering. Applied to human avatars, NeRF-based methods learn continuous radiance functions conditioned on pose, appearance, and spatial coordinates, achieving photorealistic novel view synthesis from multi-view video inputs~\cite{peng2021animatable, peng2021neural, zhou2024animatable,su2021anerf}, which typically use parametric body models (e.g., SMPL-X~\cite{pavlakos2019expressive}) to define canonical spaces via deformation fields. 
Despite their visual quality, these implicit representations suffer from high computational cost, slow rendering, and limited geometric controllability.
To address these limitations, recent works have transitioned toward hybrid representations that combine explicit geometric priors with neural rendering, moving beyond deformation-driven canonical spaces to structured 3D layouts. These methods tightly integrate parametric models or template meshes with local radiance fields defined on surfaces or volumes, enabling more precise geometric control and efficient rendering. Examples include surface-aligned radiance volumes~\cite{shao2022doublefield}, mesh-anchored light fields~\cite{zheng2022structured, kwon2023deliffas}, \new{real-time deep dynamic character models~\cite{habermann2021real, habermann2023hdhumans}, tri-plane representations for controllable synthesis~\cite{zhu2024trihuman},} and signed distance function-based reconstructions~\cite{chen2024meshavatar}. By decoupling geometry and appearance, hybrid schemes also enable efficient texture editing and improved training stability~\cite{zheng2023avatarrex, li2023posevocab}. Temporal coherence is further enhanced via spatio-temporal diffusion models conditioned on pose sequences~\cite{jin2025diffuman4d}, marking a shift toward more structured, animatable, and scalable avatar modeling.
This evolution from fully implicit to hybrid representations has paved the way for fully explicit and highly efficient frameworks, like 3D Gaussian Splatting.


\subsubsection{3D Gaussian Splatting \new{for Full-body Avatars}}


3D Gaussian Splatting (3DGS)~\cite{Kerbl20233d} has introduced a favorable trade-off between rendering quality, computational efficiency, and controllability, offering faster inference than neural radiance fields with comparable image quality and explicit scene representation.
Early efforts adapted 3DGS to animatable human avatars by modeling pose-dependent clothing deformations through template-based 2D parameterization and StyleGAN-driven CNNs~\cite{Li2024animatable}, while HuGS~\cite{moreau2024human} improved visual fidelity via motion-aware Gaussian updates and hierarchical body-part modeling. \new{ASH~\cite{Pang2024ash} further demonstrated efficient and photoreal human rendering through animatable Gaussian splats.} Subsequent work has significantly advanced controllability by proposing a two-layer Gaussian representation to disentangle facial expressions from body motion~\cite{junkawitsch2025eva, Zielonka2023d3ga}, a principle extended by LayGA~\cite{lin2024layga} to separate clothing from the body for animatable virtual try-on using layered Gaussian representations with geometry-aware supervision.
Further enhancing generalization and speed, Zhan \etal~\cite{zhan2025real} integrate spatially distributed MLPs with position-based interpolation and learned basis vectors for Gaussian offset prediction, enabling real-time, generalizable avatar rendering. 
To improve deployment efficiency, SqueezeMe~\cite{iandola2025squeezeme} distills neural pose correctives into lightweight linear layers and shares deformations across neighboring Gaussians, achieving real-time performance for multiple avatars on mobile VR hardware.
At the system level, TaoAvatar~\cite{chen2025taoavatar} enables real-time conversational interaction, while Tele-Aloha~\cite{tu2024tele} demonstrates low-latency streaming of Gaussian-based avatars using lightweight hardware.
Meanwhile, extending 3DGS to multi-person scenarios, GeoAvatar~\cite{lee2025geoavatar} introduces a geometrically-consistent 3DGS framework that preserves spatial relationships and prevents inter-penetration through monocular priors and surface ordering.
On the physical realism front, PhysAvatar~\cite{zheng2024physavatar} integrates inverse physics with mesh-aligned 4D Gaussians and differentiable rendering to estimate fabric material properties from multi-view video, enabling physically plausible cloth dynamics under novel motions.
\new{Complementary efforts extend these ideas to sparse-view and egocentric capture setups~\cite{shetty2024holoported, sun2024metacap, chen2024egoavatar, sun2025dut}.}
While these advances achieve high-fidelity geometry and dynamics, true realism requires consistent appearance under varying illumination. Relighting avatars plausibly in unseen environments is thus essential for their deployment in dynamic VR or AR applications.


\subsubsection{Photorealistic Relighting \new{of Full-body Avatars}}
The evolution of relightable human avatars began with Guo \etal~\cite{guo2019relightables}, which used a programmable lighting system to capture reflectance maps for high-fidelity image-based relighting.
This was followed by learning-based approaches such as Relighting4D~\cite{chen2022relighting4d}, which introduced a self-supervised neural decomposition of appearance into intrinsic components for relighting from monocular videos.
Extensions like IntrinsicAvatar~\cite{wang2024intrinsicavatar} and RANA~\cite{iqbal2023rana} further leverage ray tracing for visibility estimation, promoting disentangled representation of geometry, appearance, and illumination.
More recently, methods such as Xu \etal~\cite{xu2024relightable} and Carbonera \etal~\cite{carbonera2024relightable} advanced intrinsic decomposition using 3D-to-UV mapping and distance queries, paving the way for structured representations like Wang \etal~\cite{wang2025relightable}, which model pose-dependent lighting via zonal harmonics and shadow networks in a canonical Gaussian framework.
These are complemented by Li \etal~\cite{li2023animatable}, which use template-guided 2D parameterization for CNN-driven appearance synthesis, and Bolanos \etal~\cite{bolanos2024gaussian}, which enables efficient shadow computation via closed-form occlusion integrals, collectively advancing high-fidelity, generalizable, and efficient relighting.
This progression reflects a clear trend toward physically consistent, controllable, and scalable relighting through increasingly explicit modeling of light transport and material appearance.










\subsection{\new{Full-body Avatars from }Monocular Video}
\label{sec:monocular_video}

\new{While multi-view setups achieve high fidelity, they remain impractical for most end users. }Monocular video provides a far more accessible yet challenging input modality for human avatar creation, demanding high fidelity, efficiency, and robustness under real-world conditions. Recent advances leverage neural and explicit 3D representations to achieve scalable, high-quality reconstruction and animation from casual video. This section reviews progress in appearance modeling and generalization, highlighting the co-evolution of accuracy and practicality.

\subsubsection{High-Fidelity Appearance Modeling}

Early approaches to monocular human avatar creation employed conditional generative models to synthesize images directly from pose inputs, yielding plausible but often blurry outputs with limited fine detail and temporal stability~\cite{liu2019neural, liu2020Neural}. To overcome these limitations, later methods have embraced advances in neural and hybrid 3D representations, including implicit fields~\cite{Mildenhall2020nerf} or neural textures~\cite{thies2019deferred}, enabling view-consistent, photorealistic rendering of high-frequency details~\cite{jiang2023instantavatar, Guo2023vid2avatar, yoon2022learning, Feng2022scarf, Su2023npc, Xu2021hnerf}.
These approaches typically build upon parametric body templates such as SMPL-X~\cite{pavlakos2019expressive} to achieve stable, animatable reconstructions with pose-aware deformations~\cite{alldieck2018detailed, alldieck2018video, burov2021dynamic}. In contrast, template-free methods leverage canonical point clouds to allow greater topological flexibility and more accurate reconstruction of loose clothing and complex garments~\cite{su2023iccv}. Temporal coherence and motion plausibility are further enhanced through equivariant feature representations and frame prediction strategies~\cite{kappel2021high, zhuo2022fast, yoon2022learning, chan2019everybody}, effectively reducing flickering and rendering artifacts in dynamic sequences.


The integration of 3D Gaussian Splatting has shown significant promise for monocular human avatars by offering high rendering fidelity with low computational overhead, though it faces inherent challenges, including severe depth ambiguity, sparse spatial observations, and temporal flickering under self-occlusion.
To address these, recent methods leverage geometric priors from parametric models for robust 3D initialization~\cite{hu2024gauhuman, hu2024gaussianavatar}, design pose-aware deformation mechanisms via learnable skinning fields or direct mesh-Gaussian coupling, and anchor Gaussians to deformable meshes to enhance geometric accuracy and generalization~\cite{Shao2024splattingavatar, paudel2024ihuman, wen2024gomavatar}, thereby achieving high-fidelity, temporally coherent renderings at interactive rates.


To enhance expressiveness, part-aware modeling and dedicated modules for facial and hand articulation enable fine-grained control over local geometry and dynamic details~\cite{deng2024ram, hu2024expressive}. For appearance editing and relighting, IntrinsicAvatar~\cite{wang2024intrinsicavatar} introduces a Monte Carlo rendering framework for physically accurate decomposition of albedo and illumination, while RANA~\cite{iqbal2023rana} enables relightable avatar synthesis from a single video through disentangled modeling of geometry, texture, and lighting, augmented with synthetic pretraining to improve generalization under real-world conditions.


\subsubsection{Efficiency and Generalization to the Wild}


Recent advances in monocular human avatar creation have prioritized training efficiency and robust generalization to uncontrolled, real-world environments. A central step has been the development of efficient neural representations that reduce training time from hours to minutes, achieved through strategies such as part-based decomposition with shared priors~\cite{geng2023learning} or optimized neural rendering pipelines that accelerate convergence without sacrificing visual quality~\cite{jiang2023instantavatar}.
Building on this progress, 3D Gaussian Splatting (3DGS) enables real-time, high-fidelity human avatar rendering from monocular video, with improved articulation stability achieved by anchoring Gaussians to deformable template meshes~\cite{hu2024gauhuman,paudel2024ihuman}.

To enhance robustness in unconstrained settings, recent methods adopt holistic scene-aware modeling by jointly reconstructing humans and their surroundings using neural scene representations. These approaches disentangle foreground avatars from dynamic backgrounds via dual-field architectures~\cite{Guo2023vid2avatar}, and extend to multi-person scenarios through layered representations and hybrid 2D/3D supervision~\cite{jiang2024multiply}. Moreover, to address the challenge of limited pose coverage in casual video, techniques such as Vid2Avatar-Pro leverage universal priors distilled from large-scale multi-view datasets, enabling high-quality, animatable reconstructions even under sparse or repetitive motion~\cite{Guo2025vid2avatarpro}.










\subsection{\new{Full-body Avatars from Few-Shot Images}}
\label{sec:few_shot}
%
%
\new{Reducing the input requirement even further,} reconstructing animatable human avatars from just a few unconstrained images remains highly under-constrained due to limited pose and appearance variation, prompting recent methods to rely on strong learned priors for generalization.
Early approaches use structural and articulatory priors to stabilize reconstruction such as meta-learned initialization of neural SDFs~\cite{Wang2021metaavatar} or keypoint-guided spatial encodings~\cite{Mihajlovic2022keypointnerf}, thus allowing coherent geometry with few input images and avoiding per-subject fine-tuning.
To improve reconstruction completeness under extreme data scarcity, HaveFun~\cite{yang2024have} employs a two-phase optimization that separates identity alignment from appearance completion, enabling robust avatar modeling from very few unconstrained images.
For animatability, later works establish canonical spaces by predicting explicit skinning weights for deformation transfer~\cite{shin2024canonicalfusion}, or modeling pose-dependent deformations implicitly via shared motion fields conditioned on identity-specific latent codes~\cite{weng2023personnerf}.
A new paradigm leverages priors from large-scale multimodal pretraining. PuzzleAvatar~\cite{xiu2024puzzleavatar} fine-tunes vision-language models to extract disentangled semantic tokens for reconstruction and editing without pose estimation. FRESA~\cite{Wang2025fresa} learns a universal generative prior over large-scale 3D human data, enabling zero-shot, feedforward avatar creation with 3D-consistent canonicalization.
TAGA~\cite{zhai2025taga} complements these by learning a self-supervised plausibility prior that corrects implausible skinnings without dense annotations.
%
%
%
\subsection{\new{Full-body Avatars from }Single Image}
\label{sec:single_image}

Reconstructing a detailed and animatable 3D human avatar from a single image remains a highly ill-posed problem, necessitating strong geometric and appearance priors. The field has evolved along two primary axes: optimization- and regression-based methods that progressively refine 3D geometry through differentiable rendering and structured priors, and emerging generative approaches, particularly diffusion-based models, that synthesize plausible geometry and dynamics by leveraging large-scale data. These paradigms represent complementary strategies for balancing fidelity, generalization, and controllability in avatar creation.
\subsubsection{Optimization- or Regression-Based Methods}

%
Early one-shot human avatar reconstructions focused on translating 2D visual inputs into 3D geometry using UV-space representations for mesh reconstruction, advancing beyond basic image-to-image translation tasks \cite{alldieck2019tex2shape, smith2019facsimile, alldieck2019learning}. To improve anatomical accuracy and interpretability, later methods incorporated auxiliary supervision signals, such as anthropometric measurements and semantic attributes \cite{choutas2022accurate}, along with spatial layout and depth priors to enhance multi-person pose and shape estimation \cite{sun2022putting}.
Innovative optimization strategies, like alternating between neural prediction and gradient-based parameter refinement, have enabled high-fidelity reconstructions under limited supervision \cite{kolotouros2019learning}. Multi-task learning paradigms also enriched feature semantics by integrating segmentation and keypoint detection tasks \cite{varol2018bodynet}. To address occlusion and ambiguity, part-aware attention mechanisms dynamically reweight visible body regions \cite{kocabas2021pare}, while probabilistic approaches model uncertainty in body configurations \cite{zhang2023probabilistic, kolotouros2021probabilistic}.
The introduction of neural implicit representations marked a significant advancement, enabling high-resolution joint geometry and texture recovery from single images \cite{Saito2019pifu, Saito2020pifuhd}. ICON improved pose generalization by conditioning local features on surface normals and refining predictions through a feedback loop \cite{Xiu2022icon}, while ECON enhanced structural coherence by fusing normal map predictions with template registration \cite{Xiu2023econ}. 

\subsubsection{Avatar Generation Based on Generative Models}

\new{Generative 3D human modeling initially relied on GAN-based frameworks trained on 2D image collections, using compositional NeRFs~\cite{hong2023eva3d}, articulated tri-plane representations~\cite{dong2023ag3d}, multi-part models for expressive face and hand control~\cite{xu2023xagen}, and unpaired learning for clothed mesh generation~\cite{sanyal2024sculpt}. Subsequent works transitioned to diffusion-based pipelines, operating on structured latent spaces defined on the body manifold~\cite{hu2024structldm} or UV-encoded 3D Gaussians~\cite{yang2024e3gen}.}

\paragraph*{Human video diffusion models.}
Diffusion models trained on large-scale data have revolutionized one-shot human avatar synthesis, enabling high-fidelity and temporally coherent video generation~\cite{blattmann2023align, blattmann2023stable, zhang2023adding, chen2023primdiffusion}. Building on this, UNet-based approaches improve quality and consistency through structured motion conditioning, including 2D skeletons~\cite{hu2024animate}, SMPL parameters~\cite{zhu2024champ}, or confidence-aware pose guidance with progressive latent fusion~\cite{zhang2025mimicmotion}.
\new{Further refining controllability, recent works decouple motion representations via token disentanglement~\cite{li2025tokenmotion} or structure-appearance factorization~\cite{wang2025mosa}.}
More recently, diffusion transformers (DiTs) have advanced the state-of-the-art by factorizing spatial, temporal, and multi-view dependencies through hierarchical attention, enabling controllable and view-consistent synthesis. Building on this framework, Pippo \etal~\cite{kant2025pippo} achieve fully one-shot generation from a single image without relying on parametric models or camera calibration, leveraging large-scale pretraining and attention biasing.
To address data scarcity, SIGMAN~\cite{yang2025sigman} introduces HGS-1M, a large-scale 3D Gaussian avatar dataset, and proposes a UV-structured VAE with a DiT-based generator for end-to-end synthesis. IDOL~\cite{zhuang2025idol} directly outputs animatable 3D Gaussians from a single image using a disentangled transformer and synthetic HuGe100K data.
\new{Preserving subject identity without sacrificing motion dynamics is addressed through hybrid preference optimization~\cite{li2025magicid}, reward-supervised identity disentanglement~\cite{li2025personalvideo}, and multimodal feature fusion~\cite{wei2025echovideo}.}

Concurrently, there is growing interest in scalable, calibration- and pose-free 3D reconstruction. PF-LHM~\cite{qiu2025pf} enables reconstruction without pose or camera calibration using a Point-Image Transformer and 3D Gaussian Splatting, while PARTE~\cite{nam2025parte} improves texture fidelity in novel views by leveraging 3D part segmentation as a structural prior.
To achieve 4D consistency, recent methods incorporate spatiotemporal modeling into diffusion-based frameworks, including interspatial attention~\cite{shao2024isa4d}, synchronized two-stream optimization~\cite{wang20254real}, and geometrically guided video diffusion~\cite{lu2025gas}.
To support large viewpoint changes and diverse driving signals, methods correlate source poses with multiple references for detail recovery~\cite{hong2025free} and unify audio, video, and multimodal inputs within scalable DiT architectures for rich, controllable animation~\cite{lin2025omnihuman1}. \new{This multimodal paradigm has rapidly expanded through dual-modality transformers for joint video and motion generation~\cite{yang2025echomotion}, unified world models integrating speech, text, and video~\cite{xie2025x, pang2025mavid}, cognitive simulation via large language models for emotionally expressive behavior~\cite{jiang2025omnihuman}, and streaming diffusion architectures for low-latency interactive generation~\cite{wang2026flowact}.}
Beyond single-character animation, emerging frameworks model interactions through occlusion-aware representations~\cite{pang2025manivideo}, region-aware diffusion for contact modeling~\cite{lin2025interanimate}, environment masking~\cite{hu2025animate}, and depth-guided 3D layout decomposition for interactive editing~\cite{men2025mimo}, \new{multi-identity animation via structural video diffusion~\cite{wang2025multi}, and affordance-aware scene population~\cite{shan2025populate},} signaling a shift toward physically plausible, context-aware avatars.

\paragraph*{Human avatar via auxiliary view synthesis.}
Recent advances in one-shot 3D avatar generation leverage 2D diffusion models to synthesize auxiliary views or textures from a single in-the-wild image, which are then lifted into 3D for reconstruction. This paradigm effectively mitigates monocular ambiguity while preserving high fidelity and enabling animatability.
Early works such as DINAR~\cite{svitov2023dinar} and FAMOUS~\cite{hema2024famous} employ diffusion-based texture synthesis to handle occlusions: DINAR operates in a latent neural texture space, while FAMOUS leverages fashion datasets to generate plausible back views. AvatarPopUp~\cite{kolotouros2024instant} decouples 2D view synthesis from 3D reconstruction, achieving multi-view generation in under two seconds and significantly improving scalability.
More recent methods extend diffusion-based synthesis to enhance geometric accuracy, identity preservation, and pose controllability. AniGS~\cite{qiu2025anigs} combines transformer-based video diffusion with Gaussian Splatting for dynamic view synthesis and real-time animation. AdaHuman~\cite{huang2025adahuman} introduces a pose-conditioned 3D joint diffusion model to improve pose awareness, while PSHuman~\cite{li2025pshuman} enforces anatomical consistency through cross-scale diffusion under SMPL-X guidance. MoGA~\cite{dong2025moga} fits a generative 3D Gaussian avatar to synthesized multi-view images by optimizing latent codes, achieving high-fidelity reconstruction. PERSONA~\cite{sim2025persona} further improves identity consistency by combining diffusion-based data augmentation with geometry-weighted optimization during fitting.

%% file: chapters/7_future_work.tex
\section{Ethical and Social Implications}
\label{sec:ethics}
\new{As shown in the preceding sections, the state-of-the-art now enables highly realistic digital humans to be created from increasingly sparse inputs.} 
Although these technologies promise broad social and economic benefits, they also raise ethical concerns about potential misuse.

\paragraph*{Misinformation.}
The realism of digital humans allows for powerful forms of misinformation. 
In particular, AI-generated audiovisual content can convincingly depict people saying or doing things they never did.
%
Such manipulated or created media can be more persuasive than text alone and can spread rapidly, particularly in political or crisis contexts. 
Early deepfakes already demonstrated misuse risks~\cite{Thies2016face}; with increased realism, distinguishing genuine from synthetic content has become extremely difficult for human observers.

\paragraph*{Trust in visual data.}
Beyond deliberate misinformation, digital humans and deepfakes pose a broader challenge: they can erode public trust in visual and auditory evidence. 
Visual and audio media have long been trusted as reliable documentation in journalism, courts, and public discourse.
Now that such media can be synthetically generated or altered with high realism, their value as evidence is increasingly questioned. 
For instance, footage from surveillance cameras may be dismissed as fake even though it is authentic. 
This growing uncertainty undermines confidence in digital media as a whole, where the mere possibility of manipulation is enough to cause harm even in cases where no manipulation has occurred.
%
%
%
%

\paragraph*{Identity theft.}
Another potential danger is the ability to impersonate people in online conversations or videos.
In particular, \textit{stealing or mimicking} someone's face, voice, or mannerisms can be used to deceive others. 
A recent real-world incident illustrates this threat: In Hong Kong, fraudsters used a deepfake video call to convincingly impersonate a senior executive and authorize substantial money transfers~\cite{guardian2024}.
%
%
Beyond financial crimes, spoofing biometric identifiers such as facial or voice recognition presents growing security risks as generative tools become more accessible.
%
%
We expect these challenges to become more prominent as real-time generation methods become more widespread, thus requiring more advanced verification techniques.

\paragraph*{Social challenge of remote work / communication.}
The growing prevalence of remote interaction and avatar-mediated communications poses risks to the quality of human connection.
%
%
Synthetic avatars could be misused in remote work environments to impersonate colleagues, create fraudulent meetings, or manipulate behavior. 
The psychological effects include reduced trust, anxiety, or paranoia about whether one is communicating with a real counterpart. 
There is a risk of increased surveillance or monitoring under the guise of authentication, which may affect worker autonomy. 

\subsection{Forgery Detection of Synthetic Media}
While there are certain challenges, there have also been countermeasures to verify and validate the authenticity of visual content, in particular in the human context.

\paragraph*{Active defense.}
Proactive protection mechanisms embed authenticity at capture time, for example, through secure cameras, cryptographic signatures, or digital watermarking. For instance, the ``Secure Digital Camera'' concept \cite{blythe2004secure} integrates biometric identifiers into captured data, while more recent proposals (e.g., Content Authenticity Initiative) attach provenance metadata directly to files. Although significant progress has been made in watermarking and signature schemes~\cite{zhang2024dual,neekhara2022facesigns}, such methods are rarely implemented in consumer hardware.

\paragraph*{Method-specific detection.}
Supervised approaches learn to recognize artifacts of a particular synthesis model. Classic examples include FaceForensics++~\cite{rossler2019faceforensics++} or the Photoshop-based dataset~\cite{wang2019detecting}. 
Newer works extend this to diffusion-based or neural-rendered content such as DiffusionFake~\cite{chen2024diffusionfake} or DigiFakeAV~\cite{liu2025beyond}, which expose generator-specific spectral and temporal patterns. These models achieve high accuracy but generalize poorly to unseen forgery types.

\paragraph*{Generalized detection approaches.}
Manipulation-independent techniques instead assess the physical and statistical plausibility of visual data, testing consistency across lighting, geometry, motion, and temporal coherence. 
ForensicTransfer \cite{cozzolino2018forensictransfer} demonstrates how domain-adaptive features can transfer across unseen manipulation methods. 
ID-Reveal \cite{cozzolino2020idreveal} shifted detection from artifact patterns to identity-based representations, improving robustness to diverse generators. 
More recently, Zero-Shot Detection \cite{cozzolino2024zeroshotdetectionaigeneratedimages} shows that models trained only on real data can identify synthetic content as statistical outliers. While such self-supervised approaches generalize (e.g., to diffusion-based or neural-rendered media), achieving reliability against fully realistic and adversarial content remains an open challenge, underscoring the need for hybrid forensic and provenance-based solutions.


\section{Conclusion}
\label{sec:future_work}


\new{
This report has reviewed the construction of digital humans through the lens of three key stages: prior learning, personalized avatar creation, and animation.
We have reviewed 3D-based avatar methods as well as hybrid 2D-3D models that allow for different types of input controls.
We did not cover the generation of such control signals, but gave reference to related reports in this area.
In the introduction, we raised the question whether we need 3D avatars in a world which is increasingly dominated by 2D representations.
There is no obvious answer to this and it highly depends on the use case and available data.
In the following, we want to further discuss this, in the context of what is missing and how 2D and 3D can help each other.
}

\subsection{What is still missing?}
\paragraph*{Holistic solutions of creation and animation.}
Currently, avatar creation and animation are typically decoupled into independent stages.
While motion generation can leverage dynamical priors and constraints to enforce physical realism, for example, by ensuring that grasps do not float or that walking motions avoid ground penetration, retargeting these motions to specific 3D avatars introduces new challenges.
Discrepancies in body shape and proportions often result in physical inconsistencies such as self-intersections or implausible contact.
Recent academic efforts to address these retargeting issues remain limited \new{and typically focus on mesh-based characters in traditional computer graphics pipelines~\cite{villegas2021contact, lee2024learning}}.
A promising direction for future work is to explore the co-synthesis of appearance and motion through video generation models~\cite{ressler2025dismo, kansy2025reenact}.
Integrating the generation of the avatar’s visual appearance with its dynamic behavior within a unified framework could improve physical coherence and reduce artifacts that arise when these stages are treated separately.

\paragraph*{Avatar agency.}
An important aspect of building truly digital humans, beyond motion generation and appearance, is their agency. 
Unlike traditional animation, there is no external motion signal; the avatars act and behave autonomously, similar to real humans. 
This means that the avatars can infer the context of a conversation, provide non-verbal cues, and mimic emotions and behaviors. 
They can interact with and respond to the environment, as well as to other participants within the scene. 
This capability can be viewed as a step toward the Turing test for avatars.
\new{Achieving realistic avatar agency is therefore highly relevant even beyond the digital realm, e.g., in the context of embodied agents~\cite{fung2025embodied}.}
\new{However, }current research efforts in this direction remain limited~\cite{agrawal2025seamless, ki2026avatar, zhu2025infp}, largely due to data scarcity and the inherent difficulty of modeling and training multimodal avatars. 
\new{A convincing avatar system must not only observe its environment with a variety of sensors, but also produce a human-like reaction that is appropriate in the social context, all while operating under a strict real-time constraint.}
%
In our view, future work on digital humans will increasingly focus on reasoning and agency, as the areas of appearance and motion control continue to reach a higher level of maturity.
%

\paragraph*{General 3D avatar prior.}
Building truly generative 3D human priors that can be leveraged across various downstream applications, such as avatar inversion from a single or a few input images, is highly challenging, especially when the target reconstruction involves a full 3D avatar.
There are several popular approaches:
\begin{itemize}
  \item \textbf{Training on internet-scale datasets}: Leveraging massive and diverse web-scale datasets to improve the generalization and realism of digital humans (e.g., Sapiens~\cite{Khirodkar2024sapiens}).
  \item \textbf{Using synthetic datasets}: Exploiting high-quality synthetic datasets to enhance generalization (e.g., David~\cite{Saleh2025david}, SynShot~\cite{Zielonka2025synshot}).
  \item \textbf{Scaling multi-view dataset capture}: Designing systems and pipelines capable of capturing high-quality multi-view data at scale (e.g., NeRSemble~\cite{Kirschstein2023nersemble}, Seamless~\cite{seamless2023}, DNA-Rendering~\cite{cheng2023dna}).
  \item \textbf{Scaling model capacity to represent everyone}: Expanding model expressivity to accurately capture the full diversity of human appearance, behavior, and motion across populations (e.g., LHM~\cite{qiu2025lhm}, AniGS~\cite{qiu2025anigs}).
\end{itemize}
While several works focus on these aspects, one also has to critically ask whether a general avatar prior that can represent any person and asset (hairstyle, garment) can actually be learned.
The closest ``general'' priors that we have today are generative image and video models, as well as large language models that can reason about certain aspects of human behavior.
An open question is whether explicitly human-specific priors will remain necessary, given the universality and representational power of foundational models, particularly recent video diffusion systems such as Veo~[2,3] (Google), Sora~2 (OpenAI), or Genie~[2,3]~\cite{parker2024genie}, which already model human appearance and motion in a largely domain-agnostic manner.

\paragraph*{Hybrid solutions of 2D and 3D avatars.}
While 2D video generation, exemplified by models such as Sora~2 or Veo~3, demonstrates impressive visual fidelity, it remains limited in controllability, spatiotemporal consistency, and rendering efficiency compared to 3D-based approaches. 
\new{Recently, adding more fine-grained camera and animation control to video generation models has garnered traction~\cite{li2025tokenmotion, li2025adaviewplanner, cao2025uni3c, liang2025realismotion}. These endeavors demonstrate the potential of 2D-video based approaches to match existing studio-level 3D avatars in controllability and consistency, while also inheriting useful knowledge about human behavior from their video prior. These approaches lend themselves especially well to more open-ended use-cases such as interactive avatar conversations. Nevertheless, achieving perfect view-consistency with video models at a small real-time budget remains elusive.  }
In the future, 3D avatar priors could serve as conditioning signals or structural guides to enhance 2D video generation. 
More importantly, 3D digital humans remain indispensable in applications such as VR/AR, the metaverse, 3D virtual try-on, holographic conferencing, and gaming, where unified 3D representations enable multi-user consistency, real-time rendering, and precise geometric fidelity (e.g., accurate garment measurements). 
Although 3D avatar technology still faces substantial challenges, we anticipate a synergistic evolution of 2D and 3D paradigms, ultimately converging into a unified framework for digital humans.
%


\paragraph*{Spatial awareness.}
While substantial progress has been made in modeling human–object and human–scene interactions, spatial awareness remains underexplored, particularly in the context of multi-person interaction. 
\new{Key challenges involve accurate physical simulation, responsiveness to auditory cues, integration of haptic feedback, and even the generation of environments with which agents can meaningfully interact. Consequently, spatial awareness spans} not only human behavior modeling, geometry, and appearance, but also spatial cues and the mechanisms by which humans or virtual agents perceive, interpret, and respond to them. \new{In the long term, advances in spatial awareness may enable avatars to serve as effective learning partners for physical agents.}
%


\paragraph*{Avatar benchmarks.}
\new{Standardized evaluation protocols are essential for fair, transparent, and reproducible assessment of a method’s capabilities. However, 3D avatar creation research still lacks widely established benchmarks. Instead, many subfields have converged on de facto evaluation protocols that have emerged organically and are propagated from paper to paper. For instance, evaluations in 3D portrait animation are commonly conducted on the test splits of the VFHQ~\cite{xie2022vfhq} or HDTF~\cite{zhang2021hdtf} datasets. In monocular full-body avatar creation, performance is often assessed on selected subjects from ZJU-MoCap~\cite{peng2021neural} or People-Snapshot~\cite{alldieck2018video}. Similarly, for 3D hand avatar reconstruction, subsets of the InterHand2.6M dataset~\cite{moon2020interhand2} have emerged as the default evaluation setting. However, such loosely defined evaluation protocols always carry the risk of subtle evaluation mismatches that can ultimately render reported numbers incomparable across works. A notable exception is the NeRsemble benchmark~\footnote{\url{https://kaldir.vc.cit.tum.de/nersemble_benchmark/}} which provides a hidden test set and a public leaderboard for two well-defined tasks: Dynamic novel view synthesis of heads and FLAME-driven monocular 3D avatar reconstruction. Overall, however, further efforts are required to unify evaluations across avatar subfields to ensure steady progress. This challenge becomes more pronounced as the capabilities of avatar systems grow. For example, it is less obvious how to evaluate a controllable video generation model that can synthesize avatars from text descriptions and place them into realistically rendered, artificial environments. Developing evaluation protocols that scale with increasingly expressive avatar systems is therefore an open and pressing challenge for the community.}

\subsection{Beyond 3D Digital Humans}
In the future, we hope to see insights from the field of 3D digital humans translated into the physical world. 
Knowledge about motion and human interaction with objects and scenes could be leveraged to enable humanoid robots to perform analogous actions. 
Modeling the geometry and physical properties of humans, including the structure of body parts and associated assets (e.g., garments), will play a critical role in facilitating safe and effective human–robot interaction—for instance, when assisting elderly individuals with daily tasks such as dressing. 
Although most current digital avatar methods do not aim to recover precise physical attributes, such as metrical dimensions or material properties, these capabilities will be essential in the future.

\history{}\new{As the field matures across the full pipeline, from representations and priors to per-region modeling and full-body integration, we expect the convergence of appearance, motion, and agency to ultimately yield digital humans that are indistinguishable from, and as capable as, their physical counterparts.}

%% file: chapters/8_acknowledgment.tex
\section*{Acknowledgment}

Justus Thies is supported by the ERC Starting Grant 101162081 "LeMo" and the DFG Excellence Strategy (EXC-3057). Tobias Kirschstein is supported by the ERC Consolidator Grant Gen3D (101171131). Vanessa Sklyarova is supported by the Max Planck ETH Center for Learning Systems.

%% file: bios.tex
\section*{Contributor Biographies}

\noindent
\begin{minipage}[t]{0.25\columnwidth}
\centering
\vspace{0pt}
\includegraphics[width=0.9\linewidth]{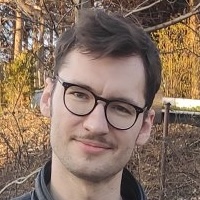}
\end{minipage}%
\hfill
\begin{minipage}[t]{0.73\columnwidth}
\vspace{0pt}
\textbf{Wojciech Zielonka}
is a Research Scientist at Meta in Pittsburgh. He earned his PhD in 2025 from the Technical University of Darmstadt. Before that, he conducted his doctoral research at the Max Planck Institute for Intelligent Systems (2021--2025). He holds a master's degree in Computer Science from the Technical University of Munich (2018--2021). His research focuses on digital avatars, neural rendering, and human-agent interaction, with particular interests in conversational agents, as well as relighting, hair, and garment modeling.
\\
Webpage: \href{https://zielon.github.io/}{https://zielon.github.io/}
\end{minipage}

\bigskip

\noindent
\begin{minipage}[t]{0.25\columnwidth}
\centering
\vspace{0pt}
\includegraphics[width=0.9\linewidth]{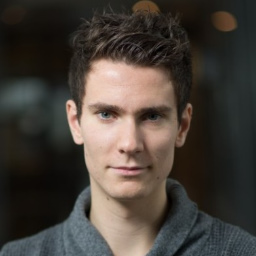}
\end{minipage}%
\hfill
\begin{minipage}[t]{0.73\columnwidth}
\vspace{0pt}
\textbf{Tobias Kirschstein}
is a PhD student at the Technical University of Munich. He is the builder and maintainer of the NeRSemble dataset, and his research is primarily concerned with photo-realistic 3D head avatars. He received his MSc degree in Computer Science at the Technical University of Munich and was a research scientist intern in the Codec Avatars team of Meta's Reality Lab in Pittsburgh.
\\
Webpage: \href{https://tobias-kirschstein.github.io/}{https://tobias-kirschstein.github.io/}
\end{minipage}

\bigskip

\noindent
\begin{minipage}[t]{0.25\columnwidth}
\centering
\vspace{0pt}
\includegraphics[width=0.9\linewidth]{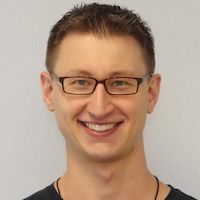}
\end{minipage}%
\hfill
\begin{minipage}[t]{0.73\columnwidth}
\vspace{0pt}
\textbf{Timo Bolkart}
is a Senior Research Scientist at Google in Z\"{u}rich. Previously, he was a Research Scientist at the Max Planck Institute (MPI) for Intelligent Systems in T\"{u}bingen (2018--2023), while also a Visiting Academic at Amazon (2021--2023). He joined the MPI as a Postdoctoral Researcher in 2016, after receiving his doctoral degree from Saarland University. During his time at the MPI, he built the widely used parametric head and body models FLAME and SMPL-X. His research centers on digital humans, including the capture and reconstruction of 3D faces and bodies from images and videos, the modeling of their 3D shape, non-rigid deformations, and photorealistic appearance, and their animation with speech.
\\
Webpage: \href{https://sites.google.com/site/bolkartt}{https://sites.google.com/site/bolkartt}
\end{minipage}

\bigskip

\noindent
\begin{minipage}[t]{0.25\columnwidth}
\centering
\vspace{0pt}
\includegraphics[width=0.9\linewidth]{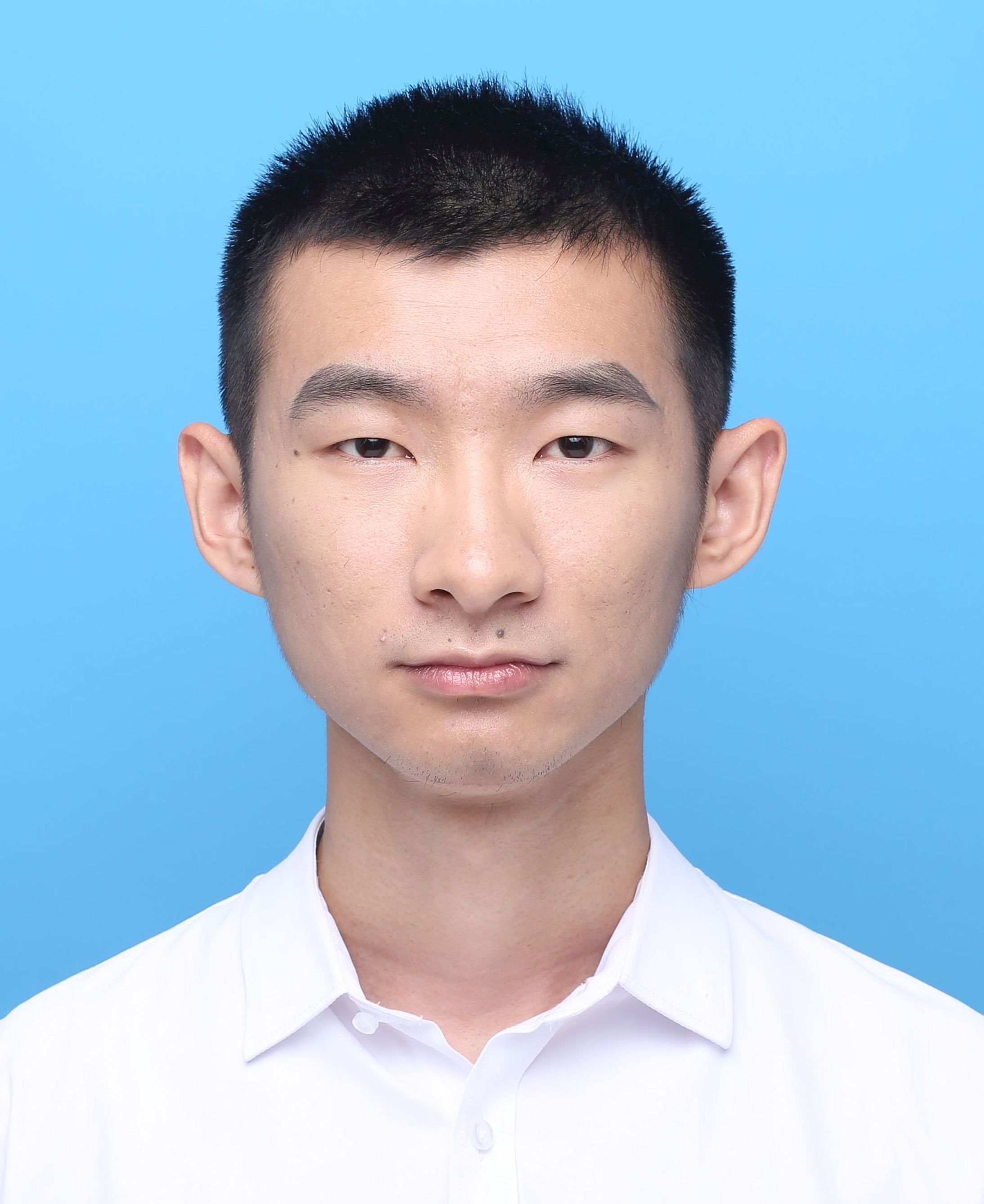}
\end{minipage}%
\hfill
\begin{minipage}[t]{0.73\columnwidth}
\vspace{0pt}
\textbf{Xiang Deng}
received the B.E. degree from the Institute of Computer and Science, Beijing Jiaotong University, China, in 2020. He is currently the third year Ph.D. student from the Automation Department, Tsinghua University, Beijing, China. His research areas include human motion generation and human video generation.
\\
Webpage: \href{https://scholar.google.com/citations?hl=zh-CN&user=X0U6Yj0zg7gC}{https://scholar.google.com/deng}
\end{minipage}

\bigskip

\noindent
\begin{minipage}[t]{0.25\columnwidth}
\centering
\vspace{0pt}
\includegraphics[width=0.9\linewidth]{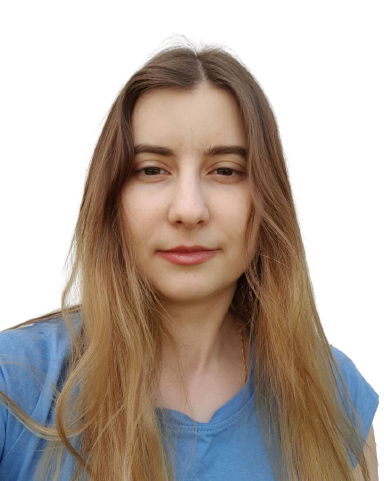}
\end{minipage}%
\hfill
\begin{minipage}[t]{0.73\columnwidth}
\vspace{0pt}
\textbf{Vanessa Sklyarova}
is a Ph.D. student at the Max Planck ETH Center for Learning Systems (CLS), conducting research at the Max Planck Institute for Intelligent Systems and ETH Zurich since 2023. Previously, she was a Research Engineer at Samsung AI Center from 2021 to 2023. She received her M.Sc. in Computer Science from a joint program between the Skolkovo Institute of Science and Technology and the Moscow Institute of Physics and Technology. Her research lies at the intersection of computer vision and computer graphics, focusing on realistic digital humans, neural rendering, and hair reconstruction.
\\
Webpage: \href{https://vanessik.github.io/}{https://vanessik.github.io/}
\end{minipage}

\bigskip

\noindent
\begin{minipage}[t]{0.25\columnwidth}
\centering
\vspace{0pt}
\includegraphics[width=0.9\linewidth]{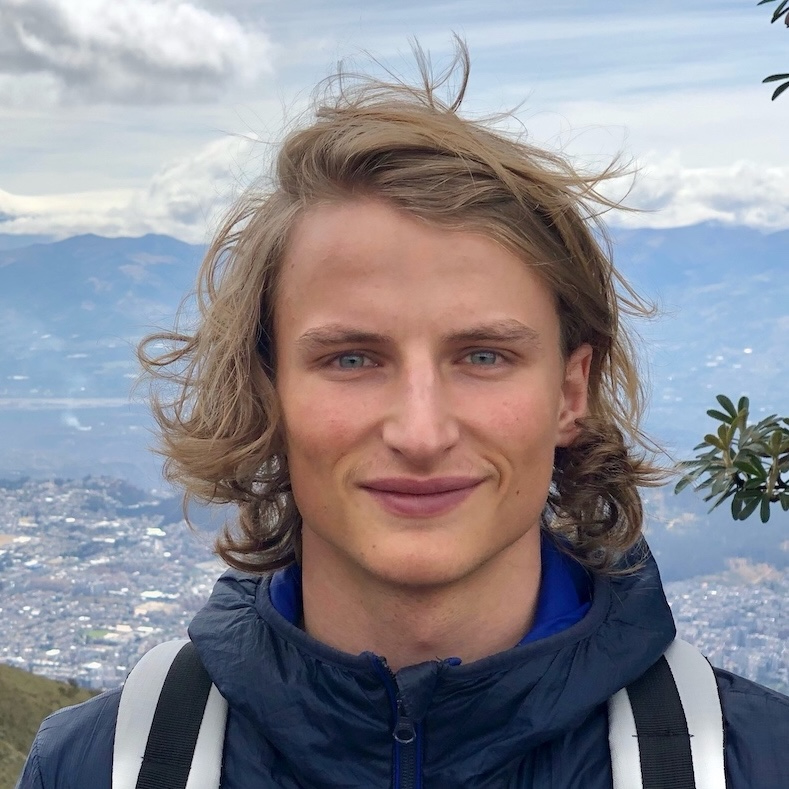}
\end{minipage}%
\hfill
\begin{minipage}[t]{0.73\columnwidth}
\vspace{0pt}
\textbf{Simon Giebenhain}
is a PhD student at the Technical University of Munich. He is the builder and maintainer of the NPHM dataset, and his research is primarily concerned with 3D reconstruction and tracking of human faces, in order to achieve more lifelike expression control for photo-realistic 3D head avatars. He received his MSc degree in Computer Science from the University of Konstanz.
\\
Webpage: \href{https://simongiebenhain.github.io/}{https://simongiebenhain.github.io/}
\end{minipage}

\bigskip

\noindent
\begin{minipage}[t]{0.25\columnwidth}
\centering
\vspace{0pt}
\includegraphics[width=0.9\linewidth]{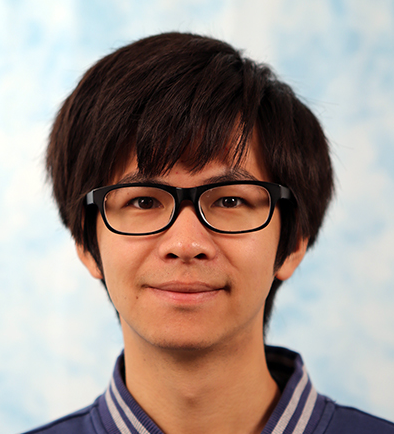}
\end{minipage}%
\hfill
\begin{minipage}[t]{0.73\columnwidth}
\vspace{0pt}
\textbf{Donglai Xiang} is a Research Scientist at NVIDIA Spatial Intelligence Lab (SIL). He obtained his PhD (2019--2023) and MS (2017--2019) degrees from Carnegie Mellon University. His research interest spans computer graphics and computer vision, with a focus on physics-based simulation, geometry processing, dynamic reconstruction and generation. His research work has been published in venues such as SIGGRAPH (Asia), CVPR, ICCV, and received the best paper honorable mention award at 3DV 2020. He contributed to Kaolin and Edify 3D at NVIDIA and Clothed Codec Avatars at Meta.
\\
Webpage: \href{https://xiangdonglai.github.io/}{https://xiangdonglai.github.io/}
\end{minipage}

\bigskip

\noindent
\begin{minipage}[t]{0.25\columnwidth}
\centering
\vspace{0pt}
\includegraphics[width=0.9\linewidth]{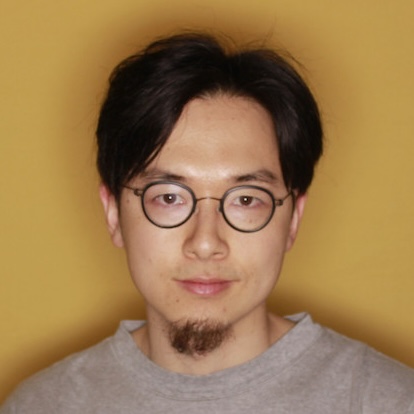}
\end{minipage}%
\hfill
\begin{minipage}[t]{0.73\columnwidth}
\vspace{0pt}
\textbf{Shunsuke Saito} is a Director, Research Scientist at Meta Codec Avatars Lab in Pittsburgh. He obtained his PhD degree at the University of Southern California. His research lies in the intersection of computer graphics, computer vision and machine learning, especially centered around digital human, 3D reconstruction, and performance capture. His work has been published in SIGGRAPH, SIGGRAPH Asia, NeurIPS, ECCV, ICCV and CVPR, three of which have been nominated for CVPR Best Paper Award (2019, 2021) and ECCV Best Paper Award (2024). His real-time volumetric teleportation work also won the Best in Show award at SIGGRAPH 2020 Real-time Live!.
\\
Webpage: \href{https://shunsukesaito.github.io/}{https://shunsukesaito.github.io/}
\end{minipage}

\bigskip

\noindent
\begin{minipage}[t]{0.25\columnwidth}
\centering
\vspace{0pt}
\includegraphics[width=0.9\linewidth]{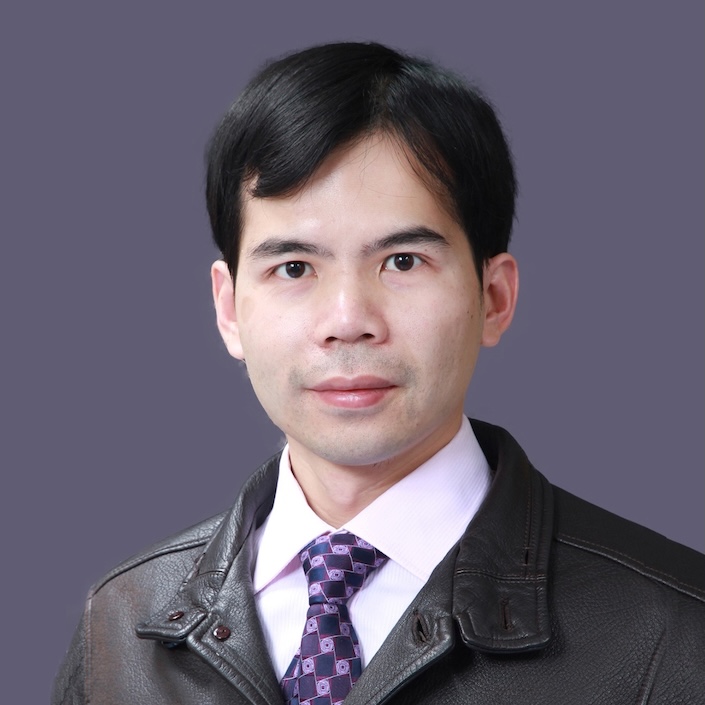}
\end{minipage}%
\hfill
\begin{minipage}[t]{0.73\columnwidth}
\vspace{0pt}
\textbf{Yebin Liu} is a Professor at the Department of Automation, Tsinghua University, Beijing, China, where he leads Digital Media Lab. He obtained his PhD degree from Tsinghua University. Prof. Liu's research lies at the intersection of computer vision, computer graphics, and machine learning, where he is particularly interested in 3D reconstruction, neural rendering, human performance capture, and computational photography. His work has been published at leading venues including CVPR, ICCV, ECCV, SIGGRAPH, SIGGRAPH Asia, and TPAMI. He has received multiple awards including the NSFC Excellent Young Scientist Fund and several best paper awards at international conferences.
\\
Webpage: \href{https://www.liuyebin.com/}{https://www.liuyebin.com/}
\end{minipage}

\bigskip

\noindent
\begin{minipage}[t]{0.25\columnwidth}
\centering
\vspace{0pt}
\includegraphics[width=0.9\linewidth]{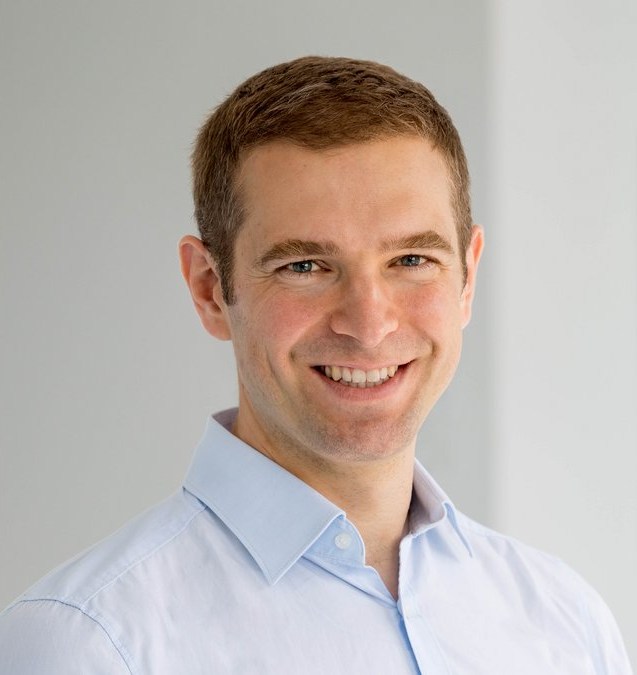}
\end{minipage}%
\hfill
\begin{minipage}[t]{0.73\columnwidth}
\vspace{0pt}
\textbf{Matthias Nie{\ss}ner}
is a Professor at the Technical University of Munich, where he leads the Visual Computing \& AI Lab. Before, he was a Visiting Assistant Professor at Stanford University. Prof. Nie{\ss}ner's research lies at the intersection of computer vision, graphics, and machine learning, where he is particularly interested in cutting-edge techniques for 3D generative AI, AI-driven video synthesis, spatial AI, and photorealistic AI Avatars. 
In total, he has published over 300 academic publications
; several of these works won best paper awards, including at SIGCHI'14, HPG'15, SPG'18, and the SIGGRAPH'16 Emerging Technologies Award for the best Live Demo.
In addition to his academic impact, Prof. Nie{\ss}ner is a co-founder of Synthesia Inc., a unicorn startup dedicated to democratize synthetic media generation with cutting-edge AI-driven video synthesis technology. Prof. Nie{\ss}ner is also co-founder and CEO of SpAItial AI, a brand-new venture that is at the forefront of generative AI and Spatial Foundation Models.
\\
Webpage: \href{https://niessnerlab.org/publications.html}{https://niessnerlab.org/publications.html}
\end{minipage}

\bigskip

\noindent
\begin{minipage}[t]{0.25\columnwidth}
\centering
\vspace{0pt}
\includegraphics[width=0.9\linewidth]{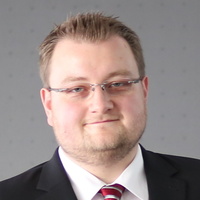}
\end{minipage}%
\hfill
\begin{minipage}[t]{0.73\columnwidth}
\vspace{0pt}
\textbf{Justus Thies}
is a full professor for 3D Graphics \& Vision at the Technical University of Darmstadt. 
He received his doctoral degree from the University of Erlangen-Nuremberg in 2017, was as a postdoc at TU Munich (2017--2021), and a Max Planck Independent Research Group Leader in Tübingen (2021--2024). He has a focus on motion capturing of facial performances, human bodies, and general non-rigid objects, as well as AI-based techniques that allow for photorealistic image and video synthesis. He was awarded with the German Pattern Recognition Award 2024, the Eurographics Young Researcher Award 2024, and an ERC Starting Grant 2024.
\\
Webpage: \href{https://justusthies.github.io/}{https://justusthies.github.io/}
\end{minipage}

%% file: appendix/reighting.tex
\definecolor{StrongRed}{RGB}{220,0,0}
\definecolor{StrongGreen}{RGB}{0,150,0}
\definecolor{StrongBlue}{RGB}{0,0,200}
\definecolor{StrongBrown}{RGB}{160,82,45}
\definecolor{StrongMagenta}{RGB}{200,0,150}

\newcommand{\Lo}{\textcolor{StrongRed}{L_o}}
\newcommand{\Le}{\textcolor{StrongGreen}{L_e}}
\newcommand{\fr}{\textcolor{StrongBlue}{f_r}}
\newcommand{\f}{\textcolor{StrongBlue}{f}}
\newcommand{\Li}{\textcolor{StrongBrown}{L_i}}
\newcommand{\ndotl}{\textcolor{StrongMagenta}{|\mathbf{n}\!\cdot\!\mathbf{l}|}}

\newcommand{\x}{\mathbf{x}}
\newcommand{\vdir}{\mathbf{v}}
\newcommand{\ldir}{\mathbf{l}}
\newcommand{\nrm}{\mathbf{n}}
\newcommand{\domega}{\,d\ldir}
\newcommand{\Omegap}{\Omega^+}

\begin{figure}[b]
    \centering
    \includegraphics[width=1.0\columnwidth]{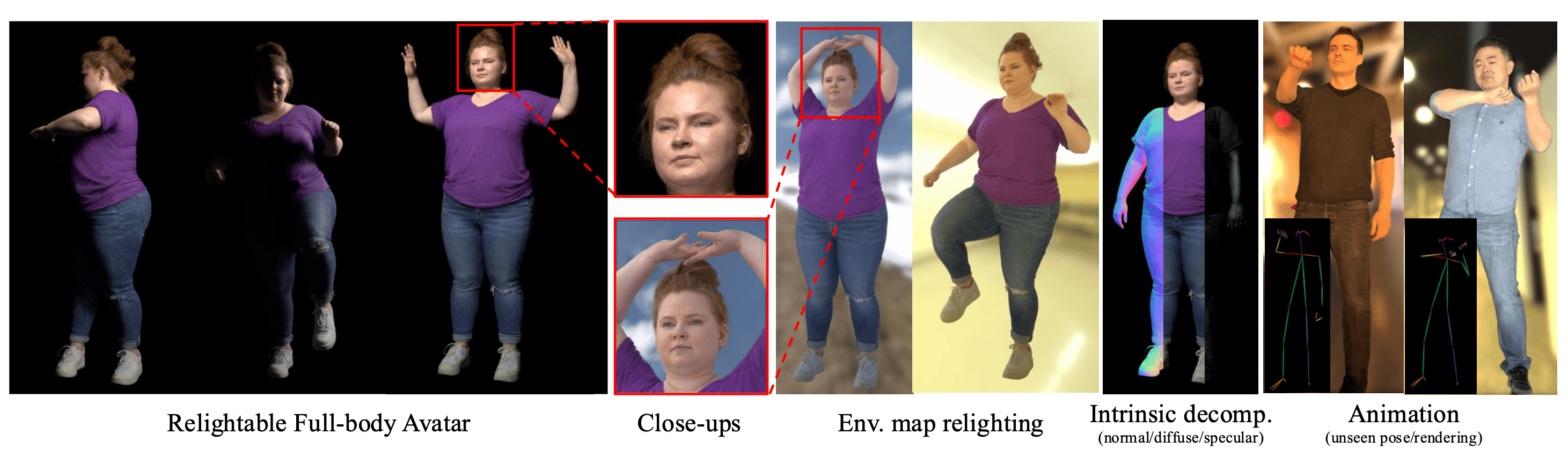} 
    \caption{Relighting requires proper disentangling of material properties such as normals, diffuse, and specular from illumination, often by using OLAT~\cite{debevec2000acquiring} captures. Image from RFGCA~\cite{wang2025relightable}.}
    \label{fig:relighting}
\end{figure}

\section{Physically Based Rendering}
In the Computer Graphics field, illumination and material effects are often modeled with the following rendering equation~\cite{kajiya1986rendering}:
\begin{equation}
\underbrace{\Lo(\x,\vdir;\,\textcolor{Sepia}{\Li})}_{\textcolor{StrongRed}{\textbf{Radiance}}}
=
\int_{\Omega}
\underbrace{\f(\x,\ldir,\vdir)}_{\textcolor{StrongBlue}{\textbf{Material}}}\;
\underbrace{\Li(\x,\ldir)}_{\textcolor{StrongBrown}{\textbf{Illumination}}}\;
\underbrace{\ndotl}_{\textcolor{StrongMagenta}{\textbf{Cosine law}}}
\,\domega.
\label{eq:rendering}
\end{equation}
It consists of several parts, which we will discuss in the following. Please note that we explicitly consider the relighting version of Equation~\ref{eq:rendering} and omit the emission term $\Le(\x,\vdir)$.

\noindent
\textcolor{StrongRed}{\textbf{Radiance}} measures the light energy density traveling in a given direction $\vdir$. It is the final output of the rendering equation and is perceived by the human visual system as color, based on the three types of cones in the retina~\cite{stockman1999cone}. In practice, this process is formalized using color matching functions such as CIE 1931 XYZ~\cite{wyszecki2000color}, which can then be converted into RGB values.

\noindent
\textbf{\textcolor{StrongBlue}{Material}} is represented by the BRDF (Bidirectional Reflectance Distribution Function) ~\cite{united1977geometrical}, which describes how light interacts with a surface \cite{pharr2023physically}, or BSDF (Bidrectional Scattering Distribution Function), which additionally incorporate transmittance for translucent materials. Intuitively, it tells us how much of the incoming light from a source is reflected or transmitted toward the viewer. It can be modeled using parametric microfacet models like Cook–Torrance~\cite{cook1982reflectance} or neural networks~\cite {sztrajman2021neural}.

\noindent
\textcolor{StrongBrown}{\textbf{Illumination}} at a point is described by the distribution of incident radiance over all directions. It describes how much light is incoming from the scene and environment, and can be represented by HDR maps~\cite{debevec2008rendering}, a collection of point lights~\cite{debevec2000acquiring}, or Spherical Harmonics~\cite{ramamoorthi2001efficient}.

\noindent
\textcolor{StrongMagenta}{\textbf{Cosine law}}, also referred to as Lambert's cosine law~\cite{lambert1760photometria}, expresses that the contribution of incoming radiance to a surface point is proportional to the cosine of the angle between the surface normal $\nrm$ and the incident light direction $\ldir$, ensuring physically consistent energy weighting.

\subsection{From Lambertian Material to Relightable Avatars}

Over the years, the traditional physically based rendering, as formulated in Equation~\ref{eq:rendering}, has been adapted to benefit from neural representations, enhancing quality and simplifying modeling through learning-based approaches in the context of digital humans. In the following sections, we present the evolution from SH Illumination~\cite{ramamoorthi2001efficient, sloan2002precomputed} to relightable 3DGS avatars~\cite{Saito2024relightable, wang2025relightable, schmidt2025becominglit}.


\noindent
\textbf{Lambertian surfaces.}
When assuming a pure Lambertian surface, for which the BRDF is $\fr(\x,\ldir,\vdir) = \rho(\x)/\pi$, the rendering equation is view-independent:
\begin{equation}
\Lo(\mathbf{x}) = \tfrac{\rho(x)}{\pi} \int_{\Omega} \Li(\ldir)\,(n \cdot \ldir)\, d\omega_{\ldir}.
\end{equation}
In the seminal work of Ramamoorthi \etal~\cite{ramamoorthi2001efficient}, the incoming illumination $\Li$ is represented by Spherical Harmonics (SH)~\cite{colton1998inverse, muller2006spherical}, effectively encoding the irradiance integral into only 9 coefficients.
This efficient approximation was quickly adapted to model face illumination by Blanz and Vetter~\cite{Blanz1999} and later by Face2Face~\cite{Thies2016face} for real-time face tracking applications. 
The equation is denoted as:
\begin{equation}
\Lo(\mathbf{x}) \approx \rho(\mathbf{x}) \sum_{l=0}^2 \sum_{m=-l}^{l} c_{lm} \, Y_l^m(\mathbf{n}(\mathbf{x})),
\label{eq:sh_classic}
\end{equation}
where the sum is truncated to the first three bands ($l=0,1,2$) of Spherical Harmonics, which were found to be sufficient to approximate diffuse irradiance.
In this line of work~\cite{Blanz1999, Thies2016face, Zielonka2022mica}, the albedo $\rho(\mathbf{x})$ and geometry are optimizable parameters derived from a statistical face model (3DMM Sec~\ref{sec:3dmm}), while the SH coefficients $c_{lm}$ are optimized to represent the scene illumination.
The term $\mathbf{n}(\mathbf{x})$ denotes the surface normal at point $\mathbf{x}$, which accounts for the orientation of the face with respect to the incoming light.

\noindent
\textbf{View-dependent effects.}
Kerbl~\etal~\cite{Kerbl20233d} introduced a modified approximation of Eq.~\ref{eq:rendering} that explicitly accounts for the view direction, where each primitive is represented as a small volumetric blob with its own set of SH coefficients:
\begin{equation}
\Lo(\mathbf{x}, \mathbf{v}) \approx \sum_{l=0}^L \sum_{m=-l}^{l} c_{lm} \, Y_l^m(\mathbf{v}).
\label{eq:sh_3dgs}
\end{equation}
The major difference compared to Ramamoorthi \etal~\cite{ramamoorthi2001efficient} is that in 3DGS the spherical harmonics encode the \emph{outgoing radiance} (directional appearance), whereas in Eq.~\ref{eq:sh_classic} they encode the \emph{incident illumination}. In this sense, 3DGS uses Eq.~\ref{eq:sh_3dgs} as a surrogate representation of the full rendering equation (Eq.~\ref{eq:rendering}), effectively baking material properties and illumination into the SH coefficients. For further discussion on this interpretation and its deviations from principled volumetric rendering, we refer the reader to the analysis by Celarek~\etal~\cite{celarek2025does}.

\noindent
\textbf{Relighting.}
The 3DGS~\cite{Kerbl20233d} formulation does not support relighting, i.e., the ability to apply arbitrary scene illumination for novel environment rendering. To enable this capability, several modifications to Eq.~\ref{eq:sh_3dgs} are required. RGCA~\cite{Saito2024relightable} achieve this by explicitly modeling the diffuse component (Eq.~\ref{eq:rgca_diffuse_sh}) and the specular component (Eq.~\ref{eq:rgca_specular_sg}) of the rendering equation. In particular, in Eq.~\ref{eq:rgca_diffuse_sh}, novel illumination (e.g., a new HDR environment map denoted as $\Li$) can be incorporated via the coefficients $L_{lm}$, which are convolved with the learned radiance transfer functions $d^{\,k}_{lm}$:
\begin{equation}
L^{\mathrm{D}}_k
\;\approx\;
\rho_k \sum_{l=0}^{L_d}\sum_{m=-l}^{l} \textcolor{StrongBrown}{L_{lm}}\, d^{\,k}_{lm}.
\label{eq:rgca_diffuse_sh}
\end{equation}
To properly learn the radiance transfer functions, RGCA requires \emph{one-light-at-a-time} (OLAT) captures~\cite{debevec2000acquiring}, which provide controlled illumination and allow disentangling of material properties ($\rho_k, d^{\,k}_{lm}$) from illumination coefficients ($L_{lm}$). This disentanglement is essential for enabling relighting under novel environments. OLAT can be understood as analogous to PCA~\cite{abdi2010principal} or Fourier bases~\cite{baron2003analytical}: the system probes the face with calibrated white light from individual basis (lights) directions and measures its response. Once these bases are constructed, any novel illumination can be synthesized by projecting an HDR environment map ($\Li$) onto either linear~\cite{ramamoorthi2002analytic, debevec2000acquiring} or neural~\cite{Lombardi21mvp, wang2025relightable, li2024uravatar, Iwase2023relightablehands} OLAT bases.

Since the diffuse part only models low-frequency lighting effects such as smooth shading, RGCA~\cite{Saito2024relightable} incorporates a specular component to capture high-frequency, view-dependent effects in human appearance, such as skin oiliness, eyes, and hair highlights, by using:
\begin{equation}
L^{\mathrm{S}}_k(\mathbf{v})
=
v_k(\mathbf{v}) \int_{S^2} \Li(\mathbf{l})\,
G_s\!\big(\mathbf{l};\, q_k(\mathbf{v}),\, \sigma_k\big)\, d\mathbf{l},
\label{eq:rgca_specular_sg}
\end{equation}
where $G_s(\cdot)$ denotes a spherical Gaussian (SG) kernel~\cite{zhang2021physg, wang2009all} that compactly approximates specular BRDF lobes and allows a closed-form convolution with environment illumination. Here, $v_k(\mathbf{v})$ denotes the learned view-dependent visibility term, $q_k(\mathbf{v})$ is the reflection direction for Gaussian $k$ and view $\mathbf{v}$, and $\sigma_k$ controls the sharpness of the specular lobe. In practice, the integral in Eq.~\ref{eq:rgca_specular_sg} is efficiently evaluated for any HDR environment maps by prefiltering them into a mipmap representation~\cite{wang2009all}, requiring only a single texture lookup per Gaussian~\cite{Saito2024relightable}. An alternative design, as proposed in BecomingLit~\cite{schmidt2025becominglit}, is to use a Cook–Torrance microfacet model~\cite{cook1982reflectance} and split-sum approximation~\cite{lagarde2014moving} with a two-lobe Blinn–Phong distribution~\cite{riviere2020single}, where only the linear roughness is optimized per Gaussian while the Fresnel and geometry terms remain analytic, simplifying the specular part.

\noindent
The final neural rendering equation is given by the sum over all Gaussians $k$:
\begin{equation}
\Lo(\x,\vdir;\,\Li)\;\approx\;
\sum_{k}\alpha_k(\x)\Big[\,L^{\mathrm{D}}_k \;+\; L^{\mathrm{S}}_k(\vdir)\,\Big].
\label{eq:rgca_combined}
\end{equation}
This produces the final radiance as a blend over all face primitives, where $\alpha_k$ is the weight for each Gaussian after volumetric integration. 
Other methods~\cite{Lombardi21mvp, li2023megane, wang2025relightable, Li2024urvatar, Chen2024urhand} follow a similar approach. However, the major differences arise in how neural components are integrated into the rendering equation, whether based on NeRF, MVP, or 3DGS formulations, and which approximations are used for the diffuse and specular components of Eq.~\ref{eq:rendering}. In addition to 3D space, there are several methods that operate only in 2D space, performing single-image relighting~\cite{jin2024neural_graffer, shu2017neural, ranjan2023facelit, ji2022geometry, kanamori2019relighting}. In terms of digital humans, these approaches aim to modify the illumination of a portrait or full-body image directly in the image domain, enabling realistic appearance changes without requiring explicit geometry or material reconstruction.

%% file: appendix/hair.tex
\section{Hair Rendering and Animation}
\label{appendix:hair}
\subsection{\thesection.1.~Rendering and Relighting} 

Hair rendering is challenging because thousands of fine fibers interact with light through reflection, transmission, and internal scattering, producing glints, highlights, and subtle color shifts. Existing methods can be broadly grouped into physics-based~\cite{marschner2003light, zinke2008dual, Zinke2009practical, Sadeghi2010artist, Xu2011interactive, Deon2011energy, Deon2014fiber, Chiang2015practical, Khungurn2017azimuthal, Huang2022microfacet, Yuksel2006rendering, Yuksel2008deep} that describe light transport through and within strands using physical understanding of the world, neural-based approaches~\cite{Wang2022hvh, Wang2023neuwigs, Radu2022neuralstrands, Sklyarova2023neural, Takimoto2024drhair, Zakharov2024haircut, Luo2024gaussianhair, Zhou2024groomcap} that try to learn everything from data and hybrid methods~\cite{Sun2021humanhair, Zheng2025GroomLight}, that utilize physics-based prior for rendering and relighting.

\paragraph*{Physics-based rendering.}

Early approaches such as the Kajiya–Kay model~\cite{Kajiya1989Rendering} treated hair as a collection of oriented cylinders with a simple reflection model. This enabled efficient rendering but could not reproduce physically accurate scattering effects. Physics-based models for hair rendering later advanced through specialized BSDFs that capture light transport in fibers. A major breakthrough came with Marschner \etal~\cite{marschner2003light}, who introduced a multi-lobe scattering BSDF for hair fibers. Their model captures surface reflection (R), direct transmission (TT), and transmission with one internal reflection (TRT), which together explain characteristic phenomena such as specular glints, wavelength-dependent color shifts, and the softness of highlights observed in real hair.  Building on this foundation, subsequent works improved realism and usability in several ways: Yuksel \etal explored the effects of global illumination~\cite{Yuksel2006rendering} and shadowing from semi-transparent objects~\cite{Yuksel2008deep}; Zinke \etal~\cite{zinke2008dual,Zinke2009practical} introduced approximations for multiple scattering between fibers; Deon \etal developed energy-conserving and non-separable fiber scattering models accounting for multiple internal reflections, non-Gaussian longitudinal scattering, azimuthal roughness, caustics, and angle-dependent specular cone shifts~\cite{Deon2011energy,Deon2014fiber}; Chiang \etal proposed a novel, energy-conserving BFSDF for efficient, expressive, and easily controllable hair and fur~\cite{Chiang2015practical}; Khungurn \etal studied scattering from elliptical hair fibers, highlighting forward-specular effects from small deviations in cross-section~\cite{Khungurn2017azimuthal}; Huang \etal introduced the first microfacet-based, non-separable hair BSDF supporting tilted cuticle scales, elliptical cross-sections, and accurate importance sampling while explaining glint-like forward scattering~\cite{Huang2022microfacet};  finally, efforts to combine realism with usability led to art-directable hair shading~\cite{Sadeghi2010artist} and interactive GPU-based rendering using 1D circular Gaussian models for real-time scattering, dynamic editing, and hair eccentricity under complex lighting~\cite{Xu2011interactive}.

Hair rendering and relighting are implemented in real-time engines, production renderers, and research frameworks. Real-time engines like Unreal Engine~\cite{unrealengine} and Unity~\cite{Haas2014unity} offer GPU-accelerated strand- or shell-based hair with interactive editing and approximate lighting. Production renderers such as Arnold (Maya)~\cite{Georgiev2018arnold, maya}, RenderMan~\cite{Christensen2018RenderMan}, and Blender (Cycles)~\cite{Blender} provide physically accurate multi-lobe scattering and full environment illumination. Research frameworks like Mitsuba3~\cite{Wenzel2022Mitsuba3},  and NVIDIA OptiX~\cite{Parker2010Optix} enable custom BSDFs, differentiable rendering, and multi-scattering experiments.

\paragraph*{Neural-based methods.} 

Neural-based approaches learn hair appearance from data by projecting 3D hair geometry into 2D space and supervising through photometric losses.
HVH~\cite{Wang2022hvh} and NeuWigs~\cite{Wang2023neuwigs} propose to attach unstructured volumetric primitives to hair geometry and use Volumetric Raymarching for rendering. Later, 
NeuralStrands~\cite{Radu2022neuralstrands} show first differentiable rasterization approach for 3D strands, by first projecting them into 2D descriptor map using Line Rasterization and then use rendering network to obtain rgba image. To improve gradient propagation Neural Haircut~\cite{Sklyarova2023neural} and Dr.Hair~\cite{Takimoto2024drhair} instead convert strands into mesh quads and use mesh-based rasterization of quads. Gaussian Haircut~\cite{Zakharov2024haircut}, GaussianHair~\cite{Luo2024gaussianhair}, and GroomCap~\cite{Zhou2024groomcap} propose to use Gaussian Splatting rasterization by attaching gaussians in the middle of each strand segment for differentiable rasterization.

\paragraph*{Hybrid methods.}

Hybrid approaches have emerged that combine prior knowledge from physical hair models with data-driven learning.
Sun \etal~\cite{Sun2021humanhair} perform inverse rendering of hair reflectance properties, estimating parameters such as longitudinal roughness and single-strand color. Their method assumes a fixed azimuthal roughness and leverages images captured from multiple viewpoints under varying illumination for optimization. GroomLight~\cite{Zheng2025GroomLight} extends this idea by inverting the material parameters of a hair BSDF from multi-view OLAT (one-light-at-a-time) images. The optimized parameters include the absorption coefficients, longitudinal and azimuthal roughness, and the interior and exterior indices of refraction. To go beyond the limitations of the analytical model, GroomLight further introduces an implicit light-aware residual representation that captures subtle view- and illumination-dependent appearance details not explained by the parametric BSDF.

\subsection{\thesection.2.~Simulations} 

Hair motion is a key component in character animation, games, and virtual avatars, but remains challenging due to the large number of strands, complex physical properties, and frequent collisions. Since simulating every strand is computationally prohibitive, most methods simulate a subset of guide hairs and interpolate the rest~\cite{Lyu2022real}. While interpolation enables fast real-time performance, it often introduces artifacts; to address this, Hsu \etal~\cite{Hsu2024real} propose a physics-driven interpolation scheme that leverages guide hair dynamics. Reduced models also struggle with hair-solid interactions, where motion coherence breaks down; Chai \etal~\cite{Chai2017adaptive} alleviates this by adaptively selecting guide hairs and correcting collisions, enabling real-time simulation of over 150K strands with complex hair-solid contacts. Another key challenge is the ``sagging effect,'' where hairstyles lose shape after initialization; Hsu \etal~\cite{Hsu2023sag} introduce a four-stage sag-free initialization framework that stabilizes quasistatic configurations for strand-based dynamics.

Hair dynamics can be broadly categorized into physics-based simulators~\cite{Bertails2006super,Herrera2024augmented, daviet2023interactive, Huang2023towards}, data-driven neural approaches~\cite{Yang2019dynamic, Wang2022hvh, Wang2023neuwigs, Liao2025hhavatar, Zhou2023groomgen}, self-supervised methods~\cite{Stuyck2024quaffure} that incorporate physics-based constraints as losses, hybrid approaches~\cite{Lyu2022real}, which use a physics simulator to generate guiding strands and then interpolate the full hairstyle using neural networks. The following sections provide a detailed description of each category.

\paragraph*{Physics-based simulation.} 
Physics-based Simulators explicitly model hair dynamics using physical laws, providing stability and interpretability. Hair is typically represented as strands, rods, or volumes, with different physical models chosen to balance realism and computational efficiency. The most widely used physical models for hair simulation include bone-based simulation, super helices~\cite{Bertails2006super}, mass-spring models~\cite{Herrera2024augmented}, and discrete elastic rods~\cite{daviet2023interactive, Huang2023towards}.
Early hair simulation methods employed a rigid skeleton with bones, where hair strands are approximated by rigid segments attached to a character’s mesh and physics constraints drive their motion. This approach is efficient and suitable for real-time applications, though it cannot capture complex strand interactions and requires careful bone design. 
Therefore, later works introduced more physically accurate models: super helices~\cite{Bertails2006super} represent the curliness and coiling of hair using helical shapes, making wavy and curly hairstyles more realistic; mass-spring models~\cite{Herrera2024augmented} treat strands as particles connected by springs, balancing simplicity and real-time performance; and discrete elastic rods~\cite{daviet2023interactive, Huang2023towards} model strands as elastic rods capturing both bending and twisting, producing high-fidelity hair motion at higher computational cost.
Today, physical simulation techniques are widely integrated into \textit{graphics engines} such as Maya~\cite{maya}, Unreal Engine~\cite{unrealengine}, Blender~\cite{Blender}, and Digital Salon~\cite{He2024digitalsalon}, allowing artists to model, simulate, and render hairstyles with greater realism.

\paragraph*{Data-driven methods.}

Instead of relying on physical priors, several approaches learn hair dynamics directly from data, using monocular video~\cite{Yang2019dynamic}, multi-view capture~\cite{Yamaguchi2009video, Zexiang2014dynamic, Wang2022hvh, Wang2023neuwigs, Liao2025hhavatar}, or data generated by a physics simulator~\cite{Zhou2023groomgen, wang2025dgh}. These methods require large datasets and often struggle with generalization compared to physics-based simulators.
Early approaches rely on classical optimization methods and do not use neural networks. Yamaguchi \etal~\cite{Yamaguchi2009video} reconstruct realistic dynamic hair from synchronized multi-camera footage with strobe lighting, modeling motion by growing strands segment by segment from root to tip while enforcing orientation consistency and smooth temporal behavior. \cite{Zexiang2014dynamic} uses a motion-path analysis algorithm to track local hair motion across frames, and enforces spatial–temporal coherence through a spacetime optimization framework solved iteratively. 
Later, neural network–based approaches were introduced. Yang \etal~\cite{Yang2019dynamic} capture hair dynamics from monocular video using HairSpatNet to estimate 3D occupancy and orientation and HairTempNet to predict motion across frames. HairSpatNet is trained on both static and dynamic synthetic data, while HairTempNet uses only dynamic hair simulated with a mass-spring model. 
In HVH~\cite{Wang2022hvh} hair is represented volumetrically using thousands of primitives and learns dynamics from multi-view data. The method works in two stages: first, guide hairs are tracked with multi-view optical flow and per-frame reconstruction; then, sparse strands are expanded into volumetric primitives, which are optimized with multi-view RGB and optical flow using volumetric ray marching. 
NeuWigs~\cite{Wang2023neuwigs} is using differentiable volumetric ray marching to first train an autoencoder to compress hair and head appearance from multi-view RGB images. In a second stage, it introduces a temporal transfer module (T2M) that models hair dynamics in the compressed space, enabling hair animation driven indirectly by head motion and head-relative gravity. 
GroomGen~\cite{Zhou2023groomgen} learns a quasi-static neural simulator from Houdini data, enabling real-time animation of diverse hairstyles. Driven by head poses, it supports up to 3000 strands running interactively. \new{DGH~\cite{wang2025dgh} further improve hair dynamics driven by head motions, while respecting upper-body collision. Method models dynamic hair appearance to achieve photorealistic novel-view synthesis.}
In contrast to these methods, HHAvatar~\cite{Liao2025hhavatar} models head avatars with a layered representation, where hair is represented by unstructured Gaussians. Hair dynamics are captured with an MLP that simulates non-rigid deformations, with motion strength depending on the distance between hair and scalp. The deformation network is conditioned on past hair positions along with current and recent head poses.

\paragraph*{Self-supervised methods.}

Quaffure~\cite{Stuyck2024quaffure} introduces the first real-time quasi-static neural hair simulator trained in a self-supervised manner, eliminating the need for collected data. The method reformulates Cosserat rods for strand representation and leverages physics-based losses—such as elastic potential and penetration—for supervision. At inference, it can animate a wide variety of hairstyles conditioned on body shape and pose. 

\paragraph*{Hybrid simulators.}

Hybrid hair simulators leverage prior knowledge of hair physics by simulating a small number of guide strands with a physical simulator and then training a neural interpolator to compute dynamic weights for interpolating the full hairstyle~\cite{Lyu2022real}.

%% file: appendix/garment.tex
\section{Garment Animation}
\label{appendix:garment}

In this section, we discuss approaches to garment animation, which can be divided into two major categories. One category focuses on producing animation in the 3D space for garment assets that have been either manually created by digital artists, or automatically reconstructed/generated by algorithms discussed in Sec.~\ref{sec:garment-reconstruction} and \ref{sec:garment-generation}. These methods, further divided into physics-based, data-driven, and hybrid methods, are addressed in Sec.~\ref{appendix:garment}.1. The other category directly generates images and videos of the animated garments in the pixel space, bypassing the process of animating, or even generating, garment assets in 3D, which we discuss in Sec.~\ref{appendix:garment}.2.

\subsection{\thesection.1.~3D Garment Animation}

\paragraph{Physics-based simulation.}

Physics-based cloth simulation lays the foundation of garment animation in movies, games, and virtual reality. Modern cloth simulators take form in the seminal work of Baraff and Witkin~\cite {baraff2023large}, which proposes a formulation based on the implicit Euler integrator, allowing simulations to run in relatively large time steps while remaining stable. Since then, cloth simulation has been extensively studied and improved in all kinds of aspects. For example, the efficiency of cloth simulation benefits from various types of new solvers, including Position-Based Dynamics (PBD)~\cite{muller2007position}, XPBD~\cite{Macklin2016XPBD}, Projective Dynamics~\cite{Bouaziz2014ProjectiveDynamics}, and Vertex Block Descent (VBD)~\cite{Chen2024VertexBlockDescent}. We have also seen significant improvement in collision handling for cloth in Codimensional-IPC~\cite{li2021codimensional} and followup works~\cite{li2023subspace,wang2023fast}, which utilize log-barrier functions together with Continuous Collision Detection (CCD) to ensure penetration-free simulation. A comprehensive review of cloth simulation methods is beyond the scope of this report, which we leave for future work.

\paragraph{Data-driven animation.}

Physics-based simulation can produce realistic garment motion, but can be slow for accurate, high-resolution results. In addition, it is nontrivial to set up the simulation correctly and keep the simulation stable in the presence of complicated body-cloth collisions. In light of these difficulties, some methods have been proposed to learn clothing deformation with neural networks from simulation results. The basic formulation is to train a network that takes body pose, shape, and a garment template as input, and predicts a garment motion sequence~\cite{santesteban2019learning,gundogdu2019garnet,patel2020tailornet,pan2022predicting}. To deal with garments of arbitrary 3D topology, convolutional networks on meshes or graphs become a natural choice in these methods~\cite{chentanez2020cloth,vidaurre2020fully}. Some approaches~\cite{bertiche2021deepsd,zhang2022motion} make predictions in the canonical space to leverage the skinning function, so that the networks only need to learn the residual deformation. LayersNet~\cite{shao2023towards} extends this formulation to handle multi-layer garments. As a powerful generative modeling formulation, diffusion models have also been applied to predict detailed wrinkle on top of base geometry predictions~\cite{vidaurre2025diffusedwrinkles} in data-driven cloth animation.

\paragraph{Hybrid neural simulator.}

A particularly interesting line of work explores leveraging physics as a source of self-supervision for garment animation. Instead of training the neural networks on pre-generated simulated data, some work has demonstrated the possibility of supervising the prediction with physics-inspired losses. These losses have similar or the same definitions as the optimization objectives in the implicit time stepping in cloth simulation, such as strain energy, bending energy, and collision penalty, thus effectively bridging the areas of physical and neural simulators. This line of work starts with the draping of clothes in (quasi-)static scenes~\cite{bertiche2021pbns,de2023drapenet,kairanda2024neuralclothsim}, and then includes dynamic garment movement as well~\cite{santesteban2022snug,bertiche2022neural}. HOOD~\cite{grigorev2023hood} trains a unified neural network for various garment types, and ContourCraft~\cite{Grigorev2024contourcraft} handles the collision between multiple layers of garments. EUNet~\cite{shao2024learning} proposes to further learn the constitutive model of the cloth, which defines the mathematical formulation of the strain energy, from an existing cloth trajectory, and then apply this learned energy as part of the physics-inspired loss to train a neural cloth simulator. Interestingly, these methods enjoy both the benefits of physical realism from a physics-informed formulation and the fast inference speed of neural networks.

\subsection{\thesection.2.~Pixel-Space Garment Animation}

In this section, we discuss methods that directly generate images or videos of garment animation, which have demonstrated potential for virtual try-on applications. These methods usually assume a pair of images of a subject and a garment as input, and utilize deep neural networks to generate images of the subject wearing the garment \cite{han2018viton,wang2018toward,choi2021viton}. Some methods \cite{dong2019towards,albahar2021pose} allows the generated image to be controlled by a target 2D pose or motion sequence, thus truly enabling ``animation'' of the garments in the synthesized output. While early work often adopts direct supervision or adversarial training~\cite{lewis2021tryongan}, more recent work ~\cite{zhu2023tryondiffusion,zhu2024m,li2025anydressing} leverages the powerful formulation of diffusion models ~\cite{ho2020denoising} and DiT~\cite{peebles2023scalable}, achieving impressive realism and high resolution output. Beyond static images, generating videos of garments poses a particular challenge, due to the high level of temporal consistency required by garment animation, especially loose-fitting clothing that interacts with the human body and undergoes complicated dynamics \cite{he2024wildvidfit,karras2024fashion,chen2025virtual,li2025pursuing}. As video foundation models advance rapidly, it remains an open and intriguing question whether scaling laws alone will enable them to capture garment physics, or whether 3D and physics-based approaches will play a complementary role by offering controllability, consistency, and physical accuracy.

%% file: appendix/datasets.tex
\section{Digital Human Datasets}

Table \ref{tab:datasets} provides an overview of widely used digital human datasets, covering various body parts (face, body, hands, garments, and hair), data types (real, synthetic, or generated), and modalities (images, videos, meshes, or strands). It summarizes dataset scale, camera configurations, and calibration availability, offering a unified reference for benchmarking methods on animatable human modeling and reconstruction.


\include{tables/datasets.tex}

%% file: tables/datasets.tex
\begin{table*}[th!]
  \centering
  \footnotesize
  \setlength{\tabcolsep}{2pt}
  \begin{NiceTabular}{l>{\centering\arraybackslash}p{1.1cm}>{\centering\arraybackslash}p{1.1cm}>{\centering\arraybackslash}p{1.1cm}>{\centering\arraybackslash}p{1.5cm}>{\centering\arraybackslash}p{1.1cm}>{\centering\arraybackslash}p{1.1cm}}[colortbl-like]
    \toprule
    \textbf{Dataset} & \rot{\cellcolor{ForestGreen!10}\textbf{Body Parts}} & \rot{\cellcolor{ForestGreen!10}\textbf{Data Type}} & \rot{\cellcolor{ForestGreen!10}\textbf{Data Modality}} & \rot{\cellcolor{ForestGreen!10}\textbf{Size}} & \rot{\cellcolor{ForestGreen!10}\textbf{\# Cameras}} & \rot{\cellcolor{ForestGreen!10}\textbf{ Is Calibrated}} \\
    \midrule
    \midrule
{\crboxNeRSemble} NeRSemble~\cite{Kirschstein2023nersemble} & \cellcolor{blue!04}\icoFace & \cellcolor{blue!04}\icoReal & \cellcolor{blue!04}\icoMultiViewVideo & \cellcolor{blue!04}425 & \cellcolor{blue!04}16 & \cellcolor{blue!04}\icoYes \\
{\crboxAva} Ava256~\cite{Martinez2024codec} & \icoFace & \icoReal & \icoMultiViewVideo & 256 & 80 & \icoYes \\
{\crboxFaceScape} FaceScape~\cite{Zhu2023facescape} & \cellcolor{blue!04}\icoFace & \cellcolor{blue!04}\icoReal & \cellcolor{blue!04}\icoMultiImage \icoMeshes & \cellcolor{blue!04}359 & \cellcolor{blue!04}60 & \cellcolor{blue!04}\icoYes \\
{\crboxNPHM} NPHM~\cite{giebenhain2023nphm} & \icoFace & \icoReal & \icoMeshes & 477 &  &  \\
{\crboxFaceVerse} FaceVerse~\cite{wang2022faceverse} & \cellcolor{blue!04}\icoFace & \cellcolor{blue!04}\icoReal & \cellcolor{blue!04}\icoMeshes & \cellcolor{blue!04}128 & \cellcolor{blue!04} & \cellcolor{blue!04} \\
{\crboxMEAD} MEAD~\cite{wang2020mead} & \icoFace & \icoReal & \icoMultiViewVideo & 60 & 7 & \icoNo \\
{\crboxSURREAL} SURREAL~\cite{varol2017surreal} & \cellcolor{blue!04}\icoBody & \cellcolor{blue!04}\icoSynth & \cellcolor{blue!04}\icoMonoVideo & \cellcolor{blue!04}145 & \cellcolor{blue!04}1 & \cellcolor{blue!04} \\
{\crboxCelebVHQ} CelebV-HQ~\cite{zhu2022celebvhq} & \icoFace & \icoReal & \icoMonoVideo & 15653 & 1 & \icoNo \\
{\crboxVoxCeleb} VoxCeleb2~\cite{nagrani2017voxceleb} & \cellcolor{blue!04}\icoFace & \cellcolor{blue!04}\icoReal & \cellcolor{blue!04}\icoMonoVideo & \cellcolor{blue!04}6000 & \cellcolor{blue!04}1 & \cellcolor{blue!04}\icoNo \\
{\crboxVFHQ} VFHQ~\cite{xie2022vfhq} & \icoFace & \icoReal & \icoMonoVideo & 7228 & 1 & \icoNo \\
{\crboxFFHQ} FFHQ~\cite{Karras2019style} & \cellcolor{blue!04}\icoFace & \cellcolor{blue!04}\icoReal & \cellcolor{blue!04}\icoSingleImage & \cellcolor{blue!04}70000 & \cellcolor{blue!04}1 & \cellcolor{blue!04}\icoNo \\
{\crboxActorsHQ} ActorsHQ~\cite{Icsik2023humanrf} & \icoBody \icoGarment & \icoReal & \icoMultiViewVideo & 8 & 160 & \icoYes \\
{\crboxTHuman} THuman2.0~\cite{yu2021function4d} & \cellcolor{blue!04}\icoBody & \cellcolor{blue!04}\icoReal & \cellcolor{blue!04}\icoMeshes & \cellcolor{blue!04}500 & \cellcolor{blue!04}128 & \cellcolor{blue!04}\icoYes \\
{\crboxZJUMoCap} ZJU-MoCap~\cite{peng2021neural} & \icoBody & \icoReal & \icoMultiViewVideo & 9 & 21 & \icoYes \\
{\crboxHumanDiT} Human4DiT~\cite{shao2024human4dit} & \cellcolor{blue!04}\icoBody & \cellcolor{blue!04}\icoReal & \cellcolor{blue!04}\icoMeshes & \cellcolor{blue!04}5000 & \cellcolor{blue!04}180 & \cellcolor{blue!04}\icoYes \\
{\crboxClothD} Cloth3D~\cite{bertiche2020cloth3d} & \icoGarment & \icoSynth & \icoMeshes & 11300 &  &  \\
{\crboxSewingPattern} Sewing Pattern~\cite{korosteleva2021sewingpatterns} & \cellcolor{blue!04}\icoGarment & \cellcolor{blue!04}\icoSynth & \cellcolor{blue!04}\icoMeshes & \cellcolor{blue!04}23500 & \cellcolor{blue!04} & \cellcolor{blue!04} \\
{\crboxBCNet} BCNet~\cite{jiang2020bcnet} & \icoGarment & \icoSynth & \icoMeshes & 54467 &  &  \\
{\crboxInterHandM} InterHand2.6M~\cite{moon2020interhand2} & \cellcolor{blue!04}\icoHand & \cellcolor{blue!04}\icoReal & \cellcolor{blue!04}\icoMultiViewVideo & \cellcolor{blue!04}27 & \cellcolor{blue!04}140 & \cellcolor{blue!04}\icoYes \\
{\crboxUSCHairSalon} USC-HairSalon~\cite{Hu2015single} & \icoHair & \icoSynth & \icoStrands & 343 & - &  \\
{\crboxCTHair} CT2Hair~\cite{shen2023CT2Hair} & \cellcolor{blue!04}\icoHair & \cellcolor{blue!04}\icoGenerated & \cellcolor{blue!04}\icoStrands & \cellcolor{blue!04}10 & \cellcolor{blue!04}- & \cellcolor{blue!04} \\
  \bottomrule
  \end{NiceTabular}
  \caption{Overview of commonly used digital human datasets, detailing covered body parts (face, body, hands, garments, hair), data types (real, synthetic, or generated), and modalities (images, videos, meshes, strands).}
  \label{tab:datasets}
\end{table*}

%% file: main.bbl
\newcommand{\etalchar}[1]{$^{#1}$}

%% file: main.bbl
\begin{thebibliography}{\uppercase{APMTM19}}

\bibitem[AAA{\etalchar{*}}25]{agrawal2025seamless}
\textsc{Agrawal V., Akinyemi A., Alvero K., Behrooz M., Buffalini J., Carlucci F.~M., Chen J., Chen J., Chen Z., Cheng S., et~al.}:
\newblock Seamless interaction: Dyadic audiovisual motion modeling and large-scale dataset.
\newblock \emph{arXiv preprint arXiv:2506.22554} (2025).

\bibitem[AGK{\etalchar{*}}22]{aigerman2022neural}
\textsc{Aigerman N., Gupta K., Kim V.~G., Chaudhuri S., Saito J., Groueix T.}:
\newblock Neural jacobian fields: learning intrinsic mappings of arbitrary meshes.
\newblock \emph{Transactions on Graphics, (Proc. SIGGRAPH) 41}, 4 (2022), 1--17.

\bibitem[ALY{\etalchar{*}}21]{albahar2021pose}
\textsc{AlBahar B., Lu J., Yang J., Shu Z., Shechtman E., Huang J.-B.}:
\newblock Pose with style: Detail-preserving pose-guided image synthesis with conditional stylegan.
\newblock \emph{Transactions on Graphics (TOG) 40}, 6 (2021), 1--11.

\bibitem[AMB{\etalchar{*}}19]{alldieck2019learning}
\textsc{Alldieck T., Magnor M., Bhatnagar B.~L., Theobalt C., Pons-Moll G.}:
\newblock Learning to reconstruct people in clothing from a single rgb camera.
\newblock In \emph{Conference on Computer Vision and Pattern Recognition (CVPR)} (2019), pp.~1175--1186.

\bibitem[AMX{\etalchar{*}}18a]{alldieck2018detailed}
\textsc{Alldieck T., Magnor M., Xu W., Theobalt C., Pons-Moll G.}:
\newblock Detailed human avatars from monocular video.
\newblock In \emph{3DV} (2018), pp.~98--109.

\bibitem[AMX{\etalchar{*}}18b]{alldieck2018video}
\textsc{Alldieck T., Magnor M., Xu W., Theobalt C., Pons-Moll G.}:
\newblock Video based reconstruction of 3d people models.
\newblock In \emph{Conference on Computer Vision and Pattern Recognition (CVPR)} (2018), pp.~8387--8397.

\bibitem[APMTM19]{alldieck2019tex2shape}
\textsc{Alldieck T., Pons-Moll G., Theobalt C., Magnor M.}:
\newblock Tex2shape: Detailed full human body geometry from a single image.
\newblock In \emph{International Conference on Computer Vision (ICCV)} (2019), pp.~2293--2303.

\bibitem[AS00]{ashikhmin2000anisotropic}
\textsc{Ashikhmin M., Shirley P.}:
\newblock An anisotropic phong brdf model.
\newblock \emph{Journal of graphics tools 5}, 2 (2000), 25--32.

\bibitem[ASK{\etalchar{*}}24]{aneja2024gaussianspeech}
\textsc{Aneja S., Sevastopolsky A., Kirschstein T., Thies J., Dai A., Nießner M.}:
\newblock Gaussianspeech: Audio-driven gaussian avatars, 2024.

\bibitem[ATDN23]{aneja2023clipface}
\textsc{Aneja S., Thies J., Dai A., Nie{\ss}ner M.}:
\newblock Clipface: Text-guided editing of textured 3d morphable models.
\newblock In \emph{SIGGRAPH Conference Papers (SA)} (2023).

\bibitem[AW10]{abdi2010principal}
\textsc{Abdi H., Williams L.~J.}:
\newblock Principal component analysis.
\newblock \emph{Wiley interdisciplinary reviews: computational statistics 2}, 4 (2010), 433--459.

\bibitem[AYS{\etalchar{*}}24]{abdal2024gsm}
\textsc{Abdal R., Yifan W., Shi Z., Xu Y., Po R., Kuang Z., Chen Q., Yeung D.-Y., Wetzstein G.}:
\newblock Gaussian shell maps for efficient 3d human generation.
\newblock In \emph{Conference on Computer Vision and Pattern Recognition (CVPR)} (2024), pp.~9441--9451.

\bibitem[BAC{\etalchar{*}}06]{Bertails2006super}
\textsc{Bertails F., Audoly B., Cani M.-P., Querleux B., Leroy F., L\'{e}v\^{e}que J.-L.}:
\newblock Super-helices for predicting the dynamics of natural hair.
\newblock \emph{Transactions on Graphics, (Proc. SIGGRAPH) 25}, 3 (2006), 1180--1187.

\bibitem[BBN{\etalchar{*}}12]{Beeler2012coupled}
\textsc{Beeler T., Bickel B., Noris G., Beardsley P., Marschner S., Sumner R.~W., Gross M.}:
\newblock Coupled 3d reconstruction of sparse facial hair and skin.
\newblock \emph{Transactions on Graphics, (Proc. SIGGRAPH) 31}, 4 (2012).

\bibitem[BBR{\etalchar{*}}22]{bai2022autoavatar}
\textsc{Bai Z., Bagautdinov T., Romero J., Zollh{\"o}fer M., Tan P., Saito S.}:
\newblock Autoavatar: Autoregressive neural fields for dynamic avatar modeling.
\newblock In \emph{European Conference on Computer Vision (ECCV)} (2022), Springer, pp.~222--239.

\bibitem[BCM{\etalchar{*}}23]{seamless2023}
\textsc{Barrault L., Chung Y.-A., Meglioli M.~C., Dale D., Dong N., Duppenthaler M., Duquenne P.-A., Ellis B., Elsahar H., Haaheim J., et~al.}:
\newblock Seamless: Multilingual expressive and streaming speech translation.
\newblock \emph{arXiv} (2023).

\bibitem[BDK{\etalchar{*}}23]{blattmann2023stable}
\textsc{Blattmann A., Dockhorn T., Kulal S., Mendelevitch D., Kilian M., Lorenz D., Levi Y., English Z., Voleti V., Letts A., et~al.}:
\newblock Stable video diffusion: Scaling latent video diffusion models to large datasets.
\newblock \emph{arXiv} (2023).

\bibitem[BF{\etalchar{*}}03]{baron2003analytical}
\textsc{Baron~Fourier J. B.~J., et~al.}:
\newblock \emph{The analytical theory of heat}.
\newblock Courier Corporation, 2003.

\bibitem[BF04]{blythe2004secure}
\textsc{Blythe P., Fridrich J.}:
\newblock Secure digital camera.
\newblock \emph{Digital Investigation} (2004).

\bibitem[BKY{\etalchar{*}}22]{bergman2022gnarf}
\textsc{Bergman A., Kellnhofer P., Yifan W., Chan E., Lindell D., Wetzstein G.}:
\newblock Generative neural articulated radiance fields.
\newblock \emph{Advances in Neural Information Processing Systems (NeurIPS) 35} (2022), 19900--19916.

\bibitem[BME20]{bertiche2020cloth3d}
\textsc{Bertiche H., Madadi M., Escalera S.}:
\newblock Cloth3d: clothed 3d humans.
\newblock In \emph{European Conference on Computer Vision (ECCV)} (2020), Springer, pp.~344--359.

\bibitem[BME21]{bertiche2021pbns}
\textsc{Bertiche H., Madadi M., Escalera S.}:
\newblock Pbns: physically based neural simulation for unsupervised garment pose space deformation.
\newblock \emph{Transactions on Graphics, (Proc. SIGGRAPH Asia) 40}, 6 (2021), 1--14.

\bibitem[BME22]{bertiche2022neural}
\textsc{Bertiche H., Madadi M., Escalera S.}:
\newblock Neural cloth simulation.
\newblock \emph{Transactions on Graphics, (Proc. SIGGRAPH) 41}, 6 (2022), 1--14.

\bibitem[BMH{\etalchar{*}}25]{barthel2025cgsgan}
\textsc{Barthel F., Morgenstern W., Hinzer P., Hilsmann A., Eisert P.}:
\newblock Cgs-gan: 3d consistent gaussian splatting gans for high resolution human head synthesis, 2025.

\bibitem[BML{\etalchar{*}}14]{Bouaziz2014ProjectiveDynamics}
\textsc{Bouaziz S., Martin S., Liu T., Kavan L., Pauly M.}:
\newblock Projective dynamics: Fusing constraint projections for fast simulation.
\newblock \emph{Transactions on Graphics, (Proc. SIGGRAPH) 33}, 4 (2014), 154:1--154:11.

\bibitem[BMTE21]{bertiche2021deepsd}
\textsc{Bertiche H., Madadi M., Tylson E., Escalera S.}:
\newblock Deepsd: Automatic deep skinning and pose space deformation for 3d garment animation.
\newblock In \emph{Conference on Computer Vision and Pattern Recognition (CVPR)} (2021), pp.~5471--5480.

\bibitem[BNT21]{burov2021dynamic}
\textsc{Burov A., Nie{\ss}ner M., Thies J.}:
\newblock Dynamic surface function networks for clothed human bodies.
\newblock In \emph{International Conference on Computer Vision (ICCV)} (2021), pp.~10754--10764.

\bibitem[BPS{\etalchar{*}}08]{bradley2008markerless}
\textsc{Bradley D., Popa T., Sheffer A., Heidrich W., Boubekeur T.}:
\newblock Markerless garment capture.
\newblock \emph{Transactions on Graphics, (Proc. SIGGRAPH)} (2008), 1--9.

\bibitem[BRL{\etalchar{*}}23]{blattmann2023align}
\textsc{Blattmann A., Rombach R., Ling H., Dockhorn T., Kim S.~W., Fidler S., Kreis K.}:
\newblock Align your latents: High-resolution video synthesis with latent diffusion models.
\newblock In \emph{Conference on Computer Vision and Pattern Recognition (CVPR)} (2023), pp.~22563--22575.

\bibitem[BSR24]{bolanos2024gaussian}
\textsc{Bolanos L., Su S.-Y., Rhodin H.}:
\newblock Gaussian shadow casting for neural characters.
\newblock In \emph{Conference on Computer Vision and Pattern Recognition (CVPR)} (2024), pp.~20997--21006.

\bibitem[BTTPM19]{bhatnagar2019multi}
\textsc{Bhatnagar B.~L., Tiwari G., Theobalt C., Pons-Moll G.}:
\newblock Multi-garment net: Learning to dress 3d people from images.
\newblock In \emph{International Conference on Computer Vision (ICCV)} (2019), pp.~5420--5430.

\bibitem[BV99]{Blanz1999}
\textsc{Blanz V., Vetter T.}:
\newblock A morphable model for the synthesis of 3d faces.
\newblock In \emph{Proc. SIGGRAPH} (1999), ACM Press/Addison-Wesley Publishing Co., pp.~187--194.

\bibitem[BW98]{baraff2023large}
\textsc{Baraff D., Witkin A.}:
\newblock Large steps in cloth simulation.
\newblock In \emph{SIGGRAPH Conference Papers (SA)}. 1998, pp.~43--54.

\bibitem[BWS{\etalchar{*}}21]{Bagautdinov2021driving}
\textsc{Bagautdinov T., Wu C., Simon T., Prada F., Shiratori T., Wei S.-E., Xu W., Sheikh Y., Saragih J.}:
\newblock Driving-signal aware full-body avatars.
\newblock \emph{Transactions on Graphics, (Proc. SIGGRAPH) 40}, 4 (2021), 1--17.

\bibitem[BXX{\etalchar{*}}25]{bian2025chatgarment}
\textsc{Bian S., Xu C., Xiu Y., Grigorev A., Liu Z., Lu C., Black M.~J., Feng Y.}:
\newblock Chatgarment: Garment estimation, generation and editing via large language models.
\newblock In \emph{Conference on Computer Vision and Pattern Recognition (CVPR)} (2025), pp.~2924--2934.

\bibitem[CBC{\etalchar{*}}01]{rbf}
\textsc{Carr J.~C., Beatson R.~K., Cherrie J.~B., Mitchell T.~J., Fright W.~R., McCallum B.~C., Evans T.~R.}:
\newblock Reconstruction and representation of 3d objects with radial basis functions.
\newblock In \emph{Proceedings of the 28th annual conference on Computer graphics and interactive techniques} (2001), pp.~67--76.

\bibitem[CBTB15]{Chiang2015practical}
\textsc{Chiang M. J.-Y., Bitterli B., Tappan C., Burley B.}:
\newblock A practical and controllable hair and fur model for production path tracing.
\newblock In \emph{SIGGRAPH Conference Papers (SA)} (2015).

\bibitem[CBVW25]{chen2025virtual}
\textsc{Chen J.-K., Bansal A., Vo M.~P., Wang Y.-X.}:
\newblock Virtual fitting room: Generating arbitrarily long videos of virtual try-on from a single image.
\newblock In \emph{Advances in Neural Information Processing Systems (NeurIPS)} (2025).

\bibitem[CCF{\etalchar{*}}23]{cheng2023dna}
\textsc{Cheng W., Chen R., Fan S., Yin W., Chen K., Cai Z., Wang J., Gao Y., Yu Z., Lin Z., et~al.}:
\newblock Dna-rendering: A diverse neural actor repository for high-fidelity human-centric rendering.
\newblock In \emph{International Conference on Computer Vision (ICCV)} (2023), pp.~19982--19993.

\bibitem[CCH{\etalchar{*}}25]{Chen2025doubly}
\textsc{Chen Y., Carrasco F.~V., H\"{a}ne C., Nam G., Bazin J.-C., De~la Torre F.}:
\newblock Doubly hierarchical geometric representations for strand-based human hairstyle generation.
\newblock In \emph{Advances in Neural Information Processing Systems (NeurIPS)} (2025).

\bibitem[CDA{\etalchar{*}}24]{chhatre2024amuse}
\textsc{Chhatre K., Daněček R., Athanasiou N., Becherini G., Peters C., Black M.~J., Bolkart T.}:
\newblock {AMUSE}: Emotional speech-driven {3D} body animation via disentangled latent diffusion.
\newblock In \emph{Conference on Computer Vision and Pattern Recognition (CVPR)} (2024), pp.~1942--1953.

\bibitem[CFS{\etalchar{*}}18]{Christensen2018RenderMan}
\textsc{Christensen P., Fong J., Shade J., Wooten W., Schubert B., Kensler A., Friedman S., Kilpatrick C., Ramshaw C., Bannister M., Rayner B., Brouillat J., Liani M.}:
\newblock Renderman: An advanced path-tracing architecture for movie rendering.
\newblock \emph{Transactions on Graphics (TOG) 37}, 3 (2018).

\bibitem[CGH{\etalchar{*}}25]{cheng2025wananimate}
\textsc{Cheng G., Gao X., Hu L., Hu S., Huang M., Ji C., Li J., Meng D., Qi J., Qiao P., Shen Z., Song Y., Sun K., Tian L., Wang F., Wang G., Wang Q., Wang Z., Xiao J., Xu S., Zhang B., Zhang P., Zhang X., Zhang Z., Zhou J., Zhuo L.}:
\newblock Wan-animate: Unified character animation and replacement with holistic replication, 2025.

\bibitem[CGZE19]{chan2019everybody}
\textsc{Chan C., Ginosar S., Zhou T., Efros A.~A.}:
\newblock Everybody dance now.
\newblock In \emph{International Conference on Computer Vision (ICCV)} (2019), pp.~5933--5942.

\bibitem[CH24]{Chu2024generalizable}
\textsc{Chu X., Harada T.}:
\newblock Generalizable and animatable gaussian head avatar.
\newblock \emph{Advances in Neural Information Processing Systems (NeurIPS) 37} (2024), 57642--57670.

\bibitem[CHM{\etalchar{*}}23]{chen2023primdiffusion}
\textsc{Chen Z., Hong F., Mei H., Wang G., Yang L., Liu Z.}:
\newblock Primdiffusion: Volumetric primitives diffusion for 3d human generation.
\newblock \emph{Advances in Neural Information Processing Systems (NeurIPS) 36} (2023), 13664--13677.

\bibitem[CHV{\etalchar{*}}22]{corona2022lisa}
\textsc{Corona E., Hodan T., Vo M., Moreno-Noguer F., Sweeney C., Newcombe R., Ma L.}:
\newblock Lisa: Learning implicit shape and appearance of hands.
\newblock In \emph{Conference on Computer Vision and Pattern Recognition (CVPR)} (2022), pp.~20533--20543.

\bibitem[CHW{\etalchar{*}}25]{chen2025taoavatar}
\textsc{Chen J., Hu J., Wang G., Jiang Z., Zhou T., Chen Z., Lv C.}:
\newblock Taoavatar: Real-time lifelike full-body talking avatars for augmented reality via 3d gaussian splatting.
\newblock In \emph{Conference on Computer Vision and Pattern Recognition (CVPR)} (2025).

\bibitem[CJS{\etalchar{*}}22]{Chen2022gdna}
\textsc{Chen X., Jiang T., Song J., Yang J., Black M.~J., Geiger A., Hilliges O.}:
\newblock gdna: Towards generative detailed neural avatars.
\newblock In \emph{Conference on Computer Vision and Pattern Recognition (CVPR)} (2022), pp.~20427--20437.

\bibitem[CKD{\etalchar{*}}25]{celarek2025does}
\textsc{Celarek A., Kopanas G., Drettakis G., Wimmer M., Kerbl B.}:
\newblock Does 3d gaussian splatting need accurate volumetric rendering?
\newblock In \emph{Computer Graphics Forum} (2025), Wiley Online Library, p.~e70032.

\bibitem[CKK98]{colton1998inverse}
\textsc{Colton D.~L., Kress R., Kress R.}:
\newblock \emph{Inverse acoustic and electromagnetic scattering theory}, vol.~93.
\newblock Springer, 1998.

\bibitem[CL22]{chen2022relighting4d}
\textsc{Chen Z., Liu Z.}:
\newblock Relighting4d: Neural relightable human from videos.
\newblock In \emph{European Conference on Computer Vision (ECCV)} (2022), Springer, pp.~606--623.

\bibitem[CLC{\etalchar{*}}22]{chan2022eg3d}
\textsc{Chan E.~R., Lin C.~Z., Chan M.~A., Nagano K., Pan B., De~Mello S., Gallo O., Guibas L.~J., Tremblay J., Khamis S., et~al.}:
\newblock Efficient geometry-aware 3d generative adversarial networks.
\newblock In \emph{Conference on Computer Vision and Pattern Recognition (CVPR)} (2022), Computer Vision Foundation / {IEEE}, pp.~16123--16133.

\bibitem[CLGK{\etalchar{*}}24]{carbonera2024relightable}
\textsc{Carbonera~Luvizon D., Golyanik V., Kortylewski A., Habermann M., Theobalt C.}:
\newblock Relightable neural actor with intrinsic decomposition and pose control.
\newblock In \emph{European Conference on Computer Vision (ECCV)} (2024), pp.~465--483.

\bibitem[CLS{\etalchar{*}}15]{Chai2015high}
\textsc{Chai M., Luo L., Sunkavalli K., Carr N., Hadap S., Zhou K.}:
\newblock High-quality hair modeling from a single portrait photo.
\newblock \emph{Transactions on Graphics, (Proc. SIGGRAPH) 34}, 6 (2015).

\bibitem[CLYY24]{Chen2024VertexBlockDescent}
\textsc{Chen A.~H., Liu Z., Yang Y., Yuksel C.}:
\newblock Vertex block descent.
\newblock \emph{Transactions on Graphics, (Proc. SIGGRAPH) 43}, 4 (2024), 1--16.

\bibitem[CLZ{\etalchar{*}}25]{chen2025hunyuanvideoavatarhighfidelityaudiodrivenhuman}
\textsc{Chen Y., Liang S., Zhou Z., Huang Z., Ma Y., Tang J., Lin Q., Zhou Y., Lu Q.}:
\newblock Hunyuanvideo-avatar: High-fidelity audio-driven human animation for multiple characters, 2025.

\bibitem[CMG{\etalchar{*}}24]{Chen2024urhand}
\textsc{Chen Z., Moon G., Guo K., Cao C., Pidhorskyi S., Simon T., Joshi R., Dong Y., Xu Y., Pires B., et~al.}:
\newblock Urhand: Universal relightable hands.
\newblock In \emph{Conference on Computer Vision and Pattern Recognition (CVPR)} (2024), pp.~119--129.

\bibitem[CMH{\etalchar{*}}22]{choutas2022accurate}
\textsc{Choutas V., M{\"u}ller L., Huang C.-H.~P., Tang S., Tzionas D., Black M.~J.}:
\newblock Accurate 3d body shape regression using metric and semantic attributes.
\newblock In \emph{Conference on Computer Vision and Pattern Recognition (CVPR)} (2022), pp.~2718--2728.

\bibitem[CMM{\etalchar{*}}20]{chentanez2020cloth}
\textsc{Chentanez N., Macklin M., M{"u}ller M., Jeschke S., Kim T.-Y.}:
\newblock Cloth and skin deformation with a triangle mesh based convolutional neural network.
\newblock \emph{Computer Graphics Forum (CGF) 39}, 8 (2020), 123--134.

\bibitem[CMT{\etalchar{*}}25]{chen2025interactavatar}
\textsc{Chen K., Mohan S., Theiss J., Oprea S., Sridhar S., Prakash A.}:
\newblock Interactavatar: Modeling hand-face interaction in photorealistic avatars with deformable gaussians.
\newblock In \emph{International Conference on Computer Vision (ICCV)} (2025), pp.~10410--10420.

\bibitem[CMZ{\etalchar{*}}25]{chen2025foundhand}
\textsc{Chen K., Min C., Zhang L., Hampali S., Keskin C., Sridhar S.}:
\newblock Foundhand: Large-scale domain-specific learning for controllable hand image generation.
\newblock In \emph{Conference on Computer Vision and Pattern Recognition (CVPR)} (2025), pp.~17448--17460.

\bibitem[Com18]{Blender}
\textsc{Community B.~O.}:
\newblock Blender - a 3d modelling and rendering package, 2018.

\bibitem[CPA{\etalchar{*}}21]{corona2021smplicit}
\textsc{Corona E., Pumarola A., Alenya G., Pons-Moll G., Moreno-Noguer F.}:
\newblock Smplicit: Topology-aware generative model for clothed people.
\newblock In \emph{Conference on Computer Vision and Pattern Recognition (CVPR)} (2021), pp.~11875--11885.

\bibitem[CPLC21]{choi2021viton}
\textsc{Choi S., Park S., Lee M., Choo J.}:
\newblock Viton-hd: High-resolution virtual try-on via misalignment-aware normalization.
\newblock In \emph{Conference on Computer Vision and Pattern Recognition (CVPR)} (2021), pp.~14131--14140.

\bibitem[CPNV24]{cozzolino2024zeroshotdetectionaigeneratedimages}
\textsc{Cozzolino D., Poggi G., NieÃŸner M., Verdoliva L.}:
\newblock Zero-shot detection of ai-generated images, 2024.

\bibitem[CPTZ25]{chen2025d}
\textsc{Chen H., Peng B., Tao Y., Zhang J.}:
\newblock D\^{} 3-human: Dynamic disentangled digital human from monocular video.
\newblock In \emph{Conference on Computer Vision and Pattern Recognition (CVPR)} (2025), pp.~10836--10846.

\bibitem[CRT{\etalchar{*}}20]{cozzolino2020idreveal}
\textsc{Cozzolino D., R{\"o}ssler A., Thies J., Nie{\ss}ner M., Verdoliva L.}:
\newblock Id-reveal: Identity-aware deepfake video detection.
\newblock \emph{arXiv} (2020).

\bibitem[CSK{\etalchar{*}}22]{cao2022authentic}
\textsc{Cao C., Simon T., Kim J.~K., Schwartz G., Zollhoefer M., Saito S., Lombardi S., Wei S.-E., Belko D., Yu S.-I., et~al.}:
\newblock Authentic volumetric avatars from a phone scan.
\newblock \emph{Transactions on Graphics, (Proc. SIGGRAPH) 41}, 4 (2022), 1--19.

\bibitem[CSW{\etalchar{*}}16]{Chai2016autohair}
\textsc{Chai M., Shao T., Wu H., Weng Y., Zhou K.}:
\newblock Autohair: Fully automatic hair modeling from a single image.
\newblock \emph{Transactions on Graphics, (Proc. SIGGRAPH) 35} (2016), 1--12.

\bibitem[CT82]{cook1982reflectance}
\textsc{Cook R.~L., Torrance K.~E.}:
\newblock A reflectance model for computer graphics.
\newblock \emph{Transactions on Graphics, (Proc. SIGGRAPH) 1}, 1 (1982), 7--24.

\bibitem[CTR{\etalchar{*}}18]{cozzolino2018forensictransfer}
\textsc{Cozzolino D., Thies J., R{\"o}ssler A., Riess C., Nie{\ss}ner M., Verdoliva L.}:
\newblock Forensictransfer: Weakly-supervised domain adaptation for forgery detection.
\newblock \emph{arXiv} (2018).

\bibitem[CTT17]{clyde2017modeling}
\textsc{Clyde D., Teran J., Tamstorf R.}:
\newblock Modeling and data-driven parameter estimation for woven fabrics.
\newblock In \emph{Symposium on Computer Animation (SCA)} (2017), pp.~1--11.

\bibitem[CWS23]{chen2023handavatar}
\textsc{Chen X., Wang B., Shum H.-Y.}:
\newblock Hand avatar: Free-pose hand animation and rendering from monocular video.
\newblock In \emph{Conference on Computer Vision and Pattern Recognition (CVPR)} (2023), pp.~8683--8693.

\bibitem[CWW{\etalchar{*}}12]{Chai2012single}
\textsc{Chai M., Wang L., Weng Y., Yu Y., Guo B., Zhou K.}:
\newblock Single-view hair modeling for portrait manipulation.
\newblock \emph{Transactions on Graphics, (Proc. SIGGRAPH) 31}, 4 (2012).

\bibitem[CWW{\etalchar{*}}13]{Chai2013dynamic}
\textsc{Chai M., Wang L., Weng Y., Jin X., Zhou K.}:
\newblock Dynamic hair manipulation in images and videos.
\newblock \emph{Transactions on Graphics, (Proc. SIGGRAPH) 32}, 4 (2013).

\bibitem[CWZ{\etalchar{*}}22]{chen2022structure}
\textsc{Chen X., Wang G., Zhu D., Liang X., Torr P., Lin L.}:
\newblock Structure-preserving 3d garment modeling with neural sewing machines.
\newblock \emph{Advances in Neural Information Processing Systems (NeurIPS) 35} (2022), 15147--15159.

\bibitem[CWZ{\etalchar{*}}24]{chen2024egoavatar}
\textsc{Chen J., Wang J., Zhang Y., Pandey R., Beeler T., Habermann M., Theobalt C.}:
\newblock {EgoAvatar}: Egocentric view-driven and photorealistic full-body avatars.
\newblock In \emph{SIGGRAPH Asia Conference Papers (SA)} (2024), pp.~1--11.

\bibitem[CYL{\etalchar{*}}24]{chen2024diffusionfake}
\textsc{Chen S., Yao T., Liu H., Sun X., Ding S., Ji R., et~al.}:
\newblock Diffusionfake: Enhancing generalization in deepfake detection via guided stable diffusion.
\newblock \emph{Advances in Neural Information Processing Systems 37} (2024), 101474--101497.

\bibitem[CZB{\etalchar{*}}21]{Chen2021snarf}
\textsc{Chen X., Zheng Y., Black M.~J., Hilliges O., Geiger A.}:
\newblock Snarf: Differentiable forward skinning for animating non-rigid neural implicit shapes.
\newblock In \emph{International Conference on Computer Vision (ICCV)} (2021), pp.~11594--11604.

\bibitem[CZL{\etalchar{*}}24]{chen2024meshavatar}
\textsc{Chen Y., Zheng Z., Li Z., Xu C., Liu Y.}:
\newblock Meshavatar: Learning high-quality triangular human avatars from multi-view videos, 2024.

\bibitem[CZL{\etalchar{*}}25]{cao2025uni3c}
\textsc{Cao C., Zhou J., Li S., Liang J., Yu C., Wang F., Xue X., Fu Y.}:
\newblock Uni3c: Unifying precisely 3d-enhanced camera and human motion controls for video generation.
\newblock In \emph{Proceedings of the SIGGRAPH Asia 2025 Conference Papers} (2025), pp.~1--12.

\bibitem[CZZ17]{Chai2017adaptive}
\textsc{Chai M., Zheng C., Zhou K.}:
\newblock Adaptive skinning for interactive hair-solid simulation.
\newblock \emph{Transactions on Visualization and Computer Graphics (TVCG) 23}, 7 (2017), 1725--1738.

\bibitem[Dav23]{daviet2023interactive}
\textsc{Daviet G.}:
\newblock Interactive hair simulation on the gpu using admm.
\newblock In \emph{SIGGRAPH Conference Papers (SA)} (New York, NY, USA, 2023), SIGGRAPH '23, Association for Computing Machinery.

\bibitem[DCK{\etalchar{*}}22]{Drobyshev2022megaportraits}
\textsc{Drobyshev N., Chelishev J., Khakhulin T., Ivakhnenko A., Lempitsky V., Zakharov E.}:
\newblock Megaportraits: One-shot megapixel neural head avatars.
\newblock In \emph{International Conference on Multimedia and Expo (ICME)} (2022), pp.~2663--2671.

\bibitem[DCT{\etalchar{*}}23]{danecek2023emote}
\textsc{Daněček R., Chhatre K., Tripathi S., Wen Y., Black M., Bolkart T.}:
\newblock Emotional speech-driven animation with content-emotion disentanglement.
\newblock In \emph{SIGGRAPH Asia Conference Papers (SA)} (2023), ACM.

\bibitem[DCY{\etalchar{*}}23]{dong2023ag3d}
\textsc{Dong Z., Chen X., Yang J., Black M.~J., Hilliges O., Geiger A.}:
\newblock {AG3D:} {L}earning to generate {3D} avatars from {2D} image collections.
\newblock In \emph{International Conference on Computer Vision (ICCV)} (2023), {IEEE}, pp.~14870--14881.

\bibitem[DDS{\etalchar{*}}25]{dong2025moga}
\textsc{Dong Z., Duan L., Song J., Black M.~J., Geiger A.}:
\newblock Moga: 3d generative avatar prior for monocular gaussian avatar reconstruction.
\newblock In \emph{International Conference on Computer Vision (ICCV)} (2025), pp.~1--11.

\bibitem[Deb08]{debevec2008rendering}
\textsc{Debevec P.}:
\newblock Rendering synthetic objects into real scenes: Bridging traditional and image-based graphics with global illumination and high dynamic range photography.
\newblock In \emph{Transactions on Graphics, (Proc. SIGGRAPH)}. {ACM}, 2008, pp.~189--198.

\bibitem[dFH{\etalchar{*}}11]{Deon2011energy}
\textsc{d'Eon E., Francois G., Hill M., Letteri J., Aubry J.-M.}:
\newblock An energy-conserving hair reflectance model.
\newblock In \emph{Computer Graphics Forum (CGF)} (2011), pp.~1181--1187.

\bibitem[DHT{\etalchar{*}}00]{debevec2000acquiring}
\textsc{Debevec P., Hawkins T., Tchou C., Duiker H.-P., Sarokin W., Sagar M.}:
\newblock Acquiring the reflectance field of a human face.
\newblock In \emph{Proceedings of the 27th annual conference on Computer graphics and interactive techniques} (2000), pp.~145--156.

\bibitem[dLGHT{\etalchar{*}}25]{delagorce2025volume}
\textsc{de~La~Gorce M., Hewitt C., Takacs T., Gerdisch R., Hosenie Z., Meishvili G., Kowalski M., Cashman T.~J., Criminisi A.}:
\newblock {VoluMe} -- authentic 3d video calls from live gaussian splat prediction.
\newblock In \emph{International Conference on Computer Vision (ICCV)} (2025).

\bibitem[DLLG{\etalchar{*}}23]{de2023drapenet}
\textsc{De~Luigi L., Li R., Guillard B., Salzmann M., Fua P.}:
\newblock Drapenet: Garment generation and self-supervised draping.
\newblock In \emph{Conference on Computer Vision and Pattern Recognition (CVPR)} (2023), pp.~1451--1460.

\bibitem[DLS{\etalchar{*}}19]{dong2019towards}
\textsc{Dong H., Liang X., Shen X., Wang B., Lai H., Zhu J., Hu Z., Yin J.}:
\newblock Towards multi-pose guided virtual try-on network.
\newblock In \emph{International Conference on Computer Vision (ICCV)} (2019), pp.~9026--9035.

\bibitem[dMH14]{Deon2014fiber}
\textsc{d'Eon E., Marschner S., Hanika J.}:
\newblock A fiber scattering model with non-separable lobes.
\newblock In \emph{SIGGRAPH Conference Papers (SA)} (2014).

\bibitem[DWR{\etalchar{*}}24]{Deng2024portrait4d}
\textsc{Deng Y., Wang D., Ren X., Chen X., Wang B.}:
\newblock Portrait4d: Learning one-shot 4d head avatar synthesis using synthetic data.
\newblock In \emph{Conference on Computer Vision and Pattern Recognition (CVPR)} (2024), pp.~7119--7130.

\bibitem[DZX{\etalchar{*}}23]{ding2023diffusionrig}
\textsc{Ding Z., Zhang X., Xia Z., Jebe L., Tu Z., Zhang X.}:
\newblock Diffusionrig: Learning personalized priors for facial appearance editing.
\newblock In \emph{Conference on Computer Vision and Pattern Recognition (CVPR)} (2023), pp.~12736--12746.

\bibitem[DZZ{\etalchar{*}}24]{deng2024ram}
\textsc{Deng X., Zheng Z., Zhang Y., Sun J., Xu C., Yang X., Wang L., Liu Y.}:
\newblock Ram-avatar: Real-time photo-realistic avatar from monocular videos with full-body control.
\newblock In \emph{Conference on Computer Vision and Pattern Recognition (CVPR)} (2024), pp.~1996--2007.

\bibitem[EST{\etalchar{*}}20]{Egger20203d}
\textsc{Egger B., Smith W.~A., Tewari A., Wuhrer S., Zollhoefer M., Beeler T., Bernard F., Bolkart T., Kortylewski A., Romdhani S., et~al.}:
\newblock 3d morphable face models—past, present, and future.
\newblock \emph{Transactions on Graphics, (Proc. SIGGRAPH) 39}, 5 (2020), 1--38.

\bibitem[FBC{\etalchar{*}}25]{fung2025embodied}
\textsc{Fung P., Bachrach Y., Celikyilmaz A., Chaudhuri K., Chen D., Chung W., Dupoux E., Gong H., J{\'e}gou H., Lazaric A., et~al.}:
\newblock Embodied ai agents: Modeling the world.
\newblock \emph{arXiv preprint arXiv:2506.22355} (2025).

\bibitem[FCB{\etalchar{*}}21]{feng2021pixie}
\textsc{Feng Y., Choutas V., Bolkart T., Tzionas D., Black M.}:
\newblock Collaborative regression of expressive bodies using moderation.
\newblock In \emph{International Conference on 3D Vision (3DV)} (2021), pp.~792--804.

\bibitem[FFBB21]{Feng2021deca}
\textsc{Feng Y., Feng H., Black M.~J., Bolkart T.}:
\newblock Learning an animatable detailed 3d face model from in-the-wild images.
\newblock \emph{Transactions on Graphics, (Proc. SIGGRAPH) 40}, 4 (2021), 1--13.

\bibitem[FP09]{furukawa2009accurate}
\textsc{Furukawa Y., Ponce J.}:
\newblock Accurate, dense, and robust multiview stereopsis.
\newblock \emph{Transactions on Pattern Analysis and Machine Intelligence (TPAMI) 32}, 8 (2009), 1362--1376.

\bibitem[FPK{\etalchar{*}}24]{fan2024hold}
\textsc{Fan Z., Parelli M., Kadoglou M.~E., Chen X., Kocabas M., Black M.~J., Hilliges O.}:
\newblock Hold: Category-agnostic 3d reconstruction of interacting hands and objects from video.
\newblock In \emph{Conference on Computer Vision and Pattern Recognition (CVPR)} (2024), pp.~494--504.

\bibitem[FYA{\etalchar{*}}24]{fu2024adaptive}
\textsc{Fu Q., Yang X., Asad M., Oh C., Yuan S., Slabaugh G.}:
\newblock Adaptive multi-modal control of digital human hand synthesis using a region-aware cycle loss.
\newblock In \emph{European Conference on Computer Vision (ECCV)} (2024), Springer, pp.~68--84.

\bibitem[FYP{\etalchar{*}}22]{Feng2022scarf}
\textsc{Feng Y., Yang J., Pollefeys M., Black M.~J., Bolkart T.}:
\newblock Capturing and animation of body and clothing from monocular video.
\newblock In \emph{SIGGRAPH Asia Conference Papers (SA)} (2022), pp.~1--9.

\bibitem[GBB{\etalchar{*}}24]{Grigorev2024contourcraft}
\textsc{Grigorev A., Becherini G., Black M., Hilliges O., Thomaszewski B.}:
\newblock Contourcraft: Learning to resolve intersections in neural multi-garment simulations.
\newblock In \emph{SIGGRAPH Conference Papers (SA)} (2024), pp.~1--10.

\bibitem[GBH23]{grigorev2023hood}
\textsc{Grigorev A., Black M.~J., Hilliges O.}:
\newblock Hood: Hierarchical graphs for generalized modelling of clothing dynamics.
\newblock In \emph{Conference on Computer Vision and Pattern Recognition (CVPR)} (2023), pp.~16965--16974.

\bibitem[GCS{\etalchar{*}}19]{gundogdu2019garnet}
\textsc{Gundogdu E., Constantin V., Seifoddini A., Dang M., Salzmann M., Fua P.}:
\newblock Garnet: A two-stream network for fast and accurate 3d cloth draping.
\newblock In \emph{International Conference on Computer Vision (ICCV)} (2019), pp.~8739--8748.

\bibitem[GCS{\etalchar{*}}25]{Guo2025pgc}
\textsc{Guo M., Chiang M. J.-Y., Santesteban I., Sarafianos N., Chen H.-y., Halimi O., Bo{\v{z}}i{\v{c}} A., Saito S., Wu J., Liu C.~K., et~al.}:
\newblock Pgc: Physics-based gaussian cloth from a single pose.
\newblock In \emph{Conference on Computer Vision and Pattern Recognition (CVPR)} (2025).

\bibitem[GIF{\etalchar{*}}18]{Georgiev2018arnold}
\textsc{Georgiev I., Ize T., Farnsworth M., Montoya-Vozmediano R., King A., Lommel B.~V., Jimenez A., Anson O., Ogaki S., Johnston E., Herubel A., Russell D., Servant F., Fajardo M.}:
\newblock Arnold: A brute-force production path tracer.
\newblock \emph{Transactions on Graphics (TOG) 37}, 3 (2018).

\bibitem[GJC{\etalchar{*}}23]{Guo2023vid2avatar}
\textsc{Guo C., Jiang T., Chen X., Song J., Hilliges O.}:
\newblock Vid2avatar: 3d avatar reconstruction from videos in the wild via self-supervised scene decomposition.
\newblock In \emph{Conference on Computer Vision and Pattern Recognition (CVPR)} (2023), pp.~12858--12868.

\bibitem[GJK{\etalchar{*}}24]{guo2024reloo}
\textsc{Guo C., Jiang T., Kaufmann M., Zheng C., Valentin J., Song J., Hilliges O.}:
\newblock Reloo: Reconstructing humans dressed in loose garments from monocular video in the wild.
\newblock In \emph{European Conference on Computer Vision (ECCV)} (2024), pp.~21--38.

\bibitem[GKG{\etalchar{*}}23]{giebenhain2023nphm}
\textsc{Giebenhain S., Kirschstein T., Georgopoulos M., R{\"{u}}nz M., Agapito L., Nie{\ss}ner M.}:
\newblock Learning neural parametric head models.
\newblock In \emph{Conference on Computer Vision and Pattern Recognition (CVPR)} (2023).

\bibitem[GKG{\etalchar{*}}24]{Giebenhain2024mononphm}
\textsc{Giebenhain S., Kirschstein T., Georgopoulos M., R{\"u}nz M., Agapito L., Nie{\ss}ner M.}:
\newblock Mononphm: Dynamic head reconstruction from monocular videos.
\newblock In \emph{Conference on Computer Vision and Pattern Recognition (CVPR)} (2024), pp.~10747--10758.

\bibitem[GKR{\etalchar{*}}24]{giebenhain2024npga}
\textsc{Giebenhain S., Kirschstein T., R{\"u}nz M., Agapito L., Nie{\ss}ner M.}:
\newblock Npga: Neural parametric gaussian avatars.
\newblock In \emph{SIGGRAPH Asia Conference Papers (SA)} (2024), pp.~1--11.

\bibitem[GLD{\etalchar{*}}19]{guo2019relightables}
\textsc{Guo K., Lincoln P., Davidson P., Busch J., Yu X., Whalen M., Harvey G., Orts-Escolano S., Pandey R., Dourgarian J., et~al.}:
\newblock The relightables: Volumetric performance capture of humans with realistic relighting.
\newblock \emph{Transactions on Graphics (TOG) 38}, 6 (2019), 1--19.

\bibitem[GLK{\etalchar{*}}25]{Guo2025vid2avatarpro}
\textsc{Guo C., Li J., Kant Y., Sheikh Y., Saito S., Cao C.}:
\newblock Vid2avatar-pro: Authentic avatar from videos in the wild via universal prior.
\newblock In \emph{Conference on Computer Vision and Pattern Recognition (CVPR)} (2025).

\bibitem[GLW{\etalchar{*}}25]{gu20253dhumans}
\textsc{Gu M., Li J., Wu Y., Luo H., Zheng J., Bai X.}:
\newblock 3d human avatar reconstruction with neural fields: A recent survey.
\newblock \emph{Image Vision Comput. 154}, C (2025).

\bibitem[GPL{\etalchar{*}}22]{Grassal2022nha}
\textsc{Grassal P.-W., Prinzler M., Leistner T., Rother C., Nie{\ss}ner M., Thies J.}:
\newblock Neural head avatars from monocular rgb videos.
\newblock In \emph{Conference on Computer Vision and Pattern Recognition (CVPR)} (2022), pp.~18653--18664.

\bibitem[GPX{\etalchar{*}}23]{geng2023learning}
\textsc{Geng C., Peng S., Xu Z., Bao H., Zhou X.}:
\newblock Learning neural volumetric representations of dynamic humans in minutes.
\newblock In \emph{Conference on Computer Vision and Pattern Recognition (CVPR)} (2023), pp.~8759--8770.

\bibitem[GPX{\etalchar{*}}24]{guo2024diffusion}
\textsc{Guo J., Prada F., Xiang D., Romero J., Wu C., Park H.~S., Shiratori T., Saito S.}:
\newblock Diffusion shape prior for wrinkle-accurate cloth registration.
\newblock In \emph{International Conference on 3D Vision (3DV)} (2024), IEEE, pp.~790--799.

\bibitem[GTZN21]{Gafni2021dynamic}
\textsc{Gafni G., Thies J., Zollhofer M., Nie{\ss}ner M.}:
\newblock Dynamic neural radiance fields for monocular 4d facial avatar reconstruction.
\newblock In \emph{Conference on Computer Vision and Pattern Recognition (CVPR)} (2021), pp.~8649--8658.

\bibitem[gua]{guardian2024}
The guardian -- deepfake video conference.
\newblock \href{https://www.theguardian.com/world/2024/feb/05/hong-kong-company-deepfake-video-conference-call-scam}{Deepfake video conference}.
\newblock Accessed: 2025-10-10.

\bibitem[GZL{\etalchar{*}}24]{guo2024liveportrait}
\textsc{Guo J., Zhang D., Liu X., Zhong Z., Zhang Y., Wan P., Zhang D.}:
\newblock Liveportrait: Efficient portrait animation with stitching and retargeting control.
\newblock \emph{arXiv} (2024).

\bibitem[GZL{\etalchar{*}}25]{learn2control}
\textsc{Gao X., Zhou J., Liu D., Zhou Y., Zhang J.}:
\newblock Controlling avatar diffusion with learnable gaussian embedding, 2025.

\bibitem[GZW{\etalchar{*}}23]{guo2023handnerf}
\textsc{Guo Z., Zhou W., Wang M., Li L., Li H.}:
\newblock Handnerf: Neural radiance fields for animatable interacting hands.
\newblock In \emph{Conference on Computer Vision and Pattern Recognition (CVPR)} (2023), pp.~21078--21087.

\bibitem[Haa14]{Haas2014unity}
\textsc{Haas J.~K.}:
\newblock A history of the unity game engine, 2014.

\bibitem[HAH{\etalchar{*}}24]{hema2024famous}
\textsc{Hema V.~M., Aich S., Haene C., Bazin J.-C., De~la Torre F.}:
\newblock Famous: High-fidelity monocular 3d human digitization using view synthesis.
\newblock In \emph{European Conference on Computer Vision (ECCV)} (2024), pp.~57--73.

\bibitem[HAHZ{\etalchar{*}}24]{He2024digitalsalon}
\textsc{He C., Amador~Herrera J.~A., Zhou Y., Shu Z., Sun X., Feng Y., Pirk S., Michels D.~L., Zhang M., Wang T.~Y., Rushmeier H.}:
\newblock Digital salon: An ai and physics-driven tool for 3d hair grooming and simulation.
\newblock In \emph{SIGGRAPH Asia 2024 Real-Time Live!} (2024), SA Real-Time Live! '24.

\bibitem[HBLB17]{Hu2017simulation}
\textsc{Hu L., Bradley D., Li H., Beeler T.}:
\newblock Simulation-ready hair capture.
\newblock \emph{Computer Graphics Forum (CGF)} (2017), 281--294.

\bibitem[HCB{\etalchar{*}}24]{hacohen2024ltx}
\textsc{HaCohen Y., Chiprut N., Brazowski B., Shalem D., Moshe D., Richardson E., Levin E., Shiran G., Zabari N., Gordon O., et~al.}:
\newblock Ltx-video: Realtime video latent diffusion.
\newblock \emph{arXiv preprint arXiv:2501.00103} (2024).

\bibitem[HCL{\etalchar{*}}23]{hong2023eva3d}
\textsc{Hong F., Chen Z., Lan Y., Pan L., Liu Z.}:
\newblock {EVA3D:} {C}ompositional {3D} human generation from {2D} image collections.
\newblock In \emph{International Conference on Learning Representations (ICLR)} (2023), OpenReview.net.

\bibitem[HCT{\etalchar{*}}24]{he2024diffrelight}
\textsc{He M., Clausen P., Ta\c{s}el A.~L., Ma L., Pilarski O., Xian W., Rikker L., Yu X., Burgert R., Yu N., Debevec P.}:
\newblock Diffrelight: Diffusion-based facial performance relighting.
\newblock In \emph{SIGGRAPH Asia Conference Papers (SA)} (2024).

\bibitem[HCW{\etalchar{*}}24]{he2024wildvidfit}
\textsc{He Z., Chen P., Wang G., Li G., Torr P.~H., Lin L.}:
\newblock Wildvidfit: Video virtual try-on in the wild via image-based controlled diffusion models.
\newblock In \emph{European Conference on Computer Vision (ECCV)} (2024), Springer, pp.~123--139.

\bibitem[HDZ{\etalchar{*}}22]{hong2022cogvideo}
\textsc{Hong W., Ding M., Zheng W., Liu X., Tang J.}:
\newblock Cogvideo: Large-scale pretraining for text-to-video generation via transformers.
\newblock \emph{arXiv preprint arXiv:2205.15868} (2022).

\bibitem[HFW{\etalchar{*}}24]{hu2024expressive}
\textsc{Hu H., Fan Z., Wu T., Xi Y., Lee S., Pavlakos G., Wang Z., et~al.}:
\newblock Expressive gaussian human avatars from monocular rgb video.
\newblock \emph{Advances in Neural Information Processing Systems (NeurIPS) 37} (2024), 5646--5660.

\bibitem[HGY{\etalchar{*}}25]{he2025lam}
\textsc{He Y., Gu X., Ye X., Xu C., Zhao Z., Dong Y., Yuan W., Dong Z., Bo L.}:
\newblock Lam: Large avatar model for one-shot animatable gaussian head.
\newblock In \emph{SIGGRAPH Conference Papers (SA)} (2025), pp.~1--13.

\bibitem[HGZ{\etalchar{*}}23]{hu2023animateanyone}
\textsc{Hu L., Gao X., Zhang P., Sun K., Zhang B., Bo L.}:
\newblock Animate anyone: Consistent and controllable image-to-video synthesis for character animation.
\newblock \emph{arXiv preprint arXiv:2311.17117} (2023).

\bibitem[HHH22]{Huang2022microfacet}
\textsc{Huang W., Hullin M., Hanika J.}:
\newblock A microfacet-based hair scattering model.
\newblock \emph{Computer Graphics Forum (CGF) 41} (2022), 79--91.

\bibitem[HHL24a]{hu2024gauhuman}
\textsc{Hu S., Hu T., Liu Z.}:
\newblock Gauhuman: Articulated gaussian splatting from monocular human videos.
\newblock In \emph{Conference on Computer Vision and Pattern Recognition (CVPR)} (2024), pp.~20418--20431.

\bibitem[HHL24b]{hu2024structldm}
\textsc{Hu T., Hong F., Liu Z.}:
\newblock {StructLDM}: {S}tructured latent diffusion for {3D} human generation.
\newblock In \emph{European Conference on Computer Vision (ECCV)} (2024), vol.~15109, Springer, pp.~363--381.

\bibitem[HJA20]{ho2020denoising}
\textsc{Ho J., Jain A., Abbeel P.}:
\newblock Denoising diffusion probabilistic models.
\newblock \emph{Advances in neural information processing systems 33} (2020), 6840--6851.

\bibitem[HK93]{hanrahan1993reflection}
\textsc{Hanrahan P., Krueger W.}:
\newblock Reflection from layered surfaces due to subsurface scattering.
\newblock In \emph{SIGGRAPH}. ACM, 1993, pp.~165--174.

\bibitem[HLK{\etalchar{*}}25]{he2025head}
\textsc{He C., Li J., Kirschstein T., Sevastopolsky A., Saito S., Tan Q., Romero J., Cao C., Rushmeier H., Nam G.}:
\newblock 3dgh: 3d head generation with composable hair and face.
\newblock \emph{Transactions on Graphics, (Proc. SIGGRAPH) 44}, 4 (2025), 1--12.

\bibitem[HLL{\etalchar{*}}24]{huang2024learning}
\textsc{Huang X., Li H., Liu W., Liang X., Yan Y., Cheng Y., Gao C.}:
\newblock Learning interaction-aware 3d gaussian splatting for one-shot hand avatars.
\newblock \emph{Advances in Neural Information Processing Systems 37} (2024), 14127--14147.

\bibitem[HLX{\etalchar{*}}21]{habermann2021real}
\textsc{Habermann M., Liu L., Xu W., Zollhoefer M., Pons-Moll G., Theobalt C.}:
\newblock Real-time deep dynamic characters.
\newblock \emph{Transactions on Graphics, (Proc. SIGGRAPH) 40}, 4 (2021), 1--16.

\bibitem[HLX{\etalchar{*}}23]{habermann2023hdhumans}
\textsc{Habermann M., Liu L., Xu W., Pons-Moll G., Zollhoefer M., Theobalt C.}:
\newblock Hdhumans: A hybrid approach for high-fidelity digital humans.
\newblock \emph{Proceedings of the ACM on Computer Graphics and Interactive Techniques 6}, 3 (2023), 1--23.

\bibitem[HML{\etalchar{*}}14]{Hu2014capturing}
\textsc{Hu L., Ma C., Luo L., Wei L.-Y., Li H.}:
\newblock Capturing braided hairstyles.
\newblock In \emph{Transactions on Graphics, (Proc. SIGGRAPH Asia)} (2014), pp.~225:1--225:9.

\bibitem[HMLL14]{Hu2014robust}
\textsc{Hu L., Ma C., Luo L., Li H.}:
\newblock Robust hair capture using simulated examples.
\newblock \emph{Transactions on Graphics, (Proc. SIGGRAPH) 33}, 4 (2014).

\bibitem[HMLL15]{Hu2015single}
\textsc{Hu L., Ma C., Luo L., Li H.}:
\newblock Single-view hair modeling using a hairstyle database.
\newblock \emph{Transactions on Graphics, (Proc. SIGGRAPH)} (2015), 125:1--125:9.

\bibitem[HPX{\etalchar{*}}22]{hong2022headnerf}
\textsc{Hong Y., Peng B., Xiao H., Liu L., Zhang J.}:
\newblock Headnerf: A real-time nerf-based parametric head model.
\newblock In \emph{Conference on Computer Vision and Pattern Recognition (CVPR)} (2022), pp.~20374--20384.

\bibitem[HSS{\etalchar{*}}25]{He2025perm}
\textsc{He C., Sun X., Shu Z., Luan F., Pirk S., Herrera J. A.~A., Michels D.~L., Wang T.~Y., Zhang M., Rushmeier H., Zhou Y.}:
\newblock Perm: A parametric representation for multi-style 3d hair modeling.
\newblock In \emph{International Conference on Learning Representations (ICLR)} (2025).

\bibitem[HSW{\etalchar{*}}17]{Hu2017avatar}
\textsc{Hu L., Saito S., Wei L., Nagano K., Seo J., Fursund J., Sadeghi I., Sun C., Chen Y.-C., Li H.}:
\newblock Avatar digitization from a single image for real-time rendering.
\newblock \emph{Transactions on Graphics, (Proc. SIGGRAPH) 36}, 6 (2017).

\bibitem[HSX{\etalchar{*}}22]{halimi2022pattern}
\textsc{Halimi O., Stuyck T., Xiang D., Bagautdinov T.~M., Wen H., Kimmel R., Shiratori T., Wu C., Sheikh Y., Prada F.}:
\newblock Pattern-based cloth registration and sparse-view animation.
\newblock \emph{Transactions on Graphics, (Proc. SIGGRAPH Asia) 41}, 6 (2022), 196--1.

\bibitem[Hu24]{hu2024animate}
\textsc{Hu L.}:
\newblock Animate anyone: Consistent and controllable image-to-video synthesis for character animation.
\newblock In \emph{Conference on Computer Vision and Pattern Recognition (CVPR)} (2024), pp.~8153--8163.

\bibitem[HWP{\etalchar{*}}23]{Hsu2023sag}
\textsc{Hsu J., Wang T., Pan Z., Gao X., Yuksel C., Wu K.}:
\newblock Sag-free initialization for strand-based hybrid hair simulation.
\newblock \emph{Transactions on Graphics, (Proc. SIGGRAPH) 42}, 4 (2023).

\bibitem[HWP{\etalchar{*}}24]{Hsu2024real}
\textsc{Hsu J., Wang T., Pan Z., Gao X., Yuksel C., Wu K.}:
\newblock Real-time physically guided hair interpolation.
\newblock \emph{Transactions on Graphics, (Proc. SIGGRAPH) 43}, 4 (2024).

\bibitem[HWS{\etalchar{*}}25]{hu2025animate}
\textsc{Hu L., Wang G., Shen Z., Gao X., Meng D., Zhuo L., Zhang P., Zhang B., Bo L.}:
\newblock Animate anyone 2: High-fidelity character image animation with environment affordance.
\newblock In \emph{International Conference on Computer Vision (ICCV)} (2025), pp.~1--11.

\bibitem[HWW{\etalchar{*}}18]{han2018viton}
\textsc{Han X., Wu Z., Wu Z., Yu R., Davis L.~S.}:
\newblock Viton: An image-based virtual try-on network.
\newblock In \emph{Conference on Computer Vision and Pattern Recognition (CVPR)} (2018), pp.~7543--7552.

\bibitem[HXL{\etalchar{*}}25]{hong2025free}
\textsc{Hong F.-T., Xu Z., Liu H., Lin Q., Song L., Shu Z., Zhou Y., Ceylan D., Xu D.}:
\newblock Free-viewpoint human animation with pose-correlated reference selection.
\newblock In \emph{Conference on Computer Vision and Pattern Recognition (CVPR)} (2025), pp.~26253--26262.

\bibitem[HXSH23]{ho2023learning}
\textsc{Ho H.-I., Xue L., Song J., Hilliges O.}:
\newblock Learning locally editable virtual humans.
\newblock In \emph{Conference on Computer Vision and Pattern Recognition (CVPR)} (2023), pp.~21024--21035.

\bibitem[HYL{\etalchar{*}}25]{huang2025adahuman}
\textsc{Huang Y., Yuan Y., Li X., Kautz J., Iqbal U.}:
\newblock Adahuman: Animatable detailed 3d human generation with compositional multiview diffusion.
\newblock In \emph{International Conference on Computer Vision (ICCV)} (2025), pp.~1--11.

\bibitem[HYW{\etalchar{*}}23]{Huang2023towards}
\textsc{Huang L., Yang F., Wei C., Chen Y. J.~E., Yuan C., Gao M.}:
\newblock Towards realtime: A hybrid physics-based method for hair animation on gpu.
\newblock \emph{Computer Graphics and Interactive Techniques (PACMCGIT) 6}, 3 (2023).

\bibitem[HYZ{\etalchar{*}}24]{he2024dresscode}
\textsc{He K., Yao K., Zhang Q., Yu J., Liu L., Xu L.}:
\newblock Dresscode: Autoregressively sewing and generating garments from text guidance.
\newblock \emph{Transactions on Graphics, (Proc. SIGGRAPH) 43}, 4 (2024), 1--13.

\bibitem[HZG{\etalchar{*}}24]{hong2024lrm}
\textsc{Hong Y., Zhang K., Gu J., Bi S., Zhou Y., Liu D., Liu F., Sunkavalli K., Bui T., Tan H.}:
\newblock Lrm: Large reconstruction model for single image to 3d.
\newblock In \emph{International Conference on Learning Representations (ICLR)} (2024).

\bibitem[HZS{\etalchar{*}}24]{Herrera2024augmented}
\textsc{Herrera J. A.~A., Zhou Y., Sun X., Shu Z., He C., Pirk S., Michels D.~L.}:
\newblock Augmented mass-spring model for real-time dense hair simulation.
\newblock In \emph{International Conference on Computer Vision (ICCV)} (2024).

\bibitem[HZZ{\etalchar{*}}24]{hu2024gaussianavatar}
\textsc{Hu L., Zhang H., Zhang Y., Zhou B., Liu B., Zhang S., Nie L.}:
\newblock Gaussianavatar: Towards realistic human avatar modeling from a single video via animatable 3d gaussians.
\newblock In \emph{Conference on Computer Vision and Pattern Recognition (CVPR)} (2024), pp.~634--644.

\bibitem[ICN{\etalchar{*}}23]{iqbal2023rana}
\textsc{Iqbal U., Caliskan A., Nagano K., Khamis S., Molchanov P., Kautz J.}:
\newblock Rana: Relightable articulated neural avatars.
\newblock In \emph{International Conference on Computer Vision (ICCV)} (2023), pp.~23142--23153.

\bibitem[IPS{\etalchar{*}}25]{iandola2025squeezeme}
\textsc{Iandola F., Pidhorskyi S., Santesteban I., Gupta D., Pahuja A., Bartolovic N., Yu F., Garbin E., Simon T., Saito S.}:
\newblock Squeezeme: Mobile-ready distillation of gaussian full-body avatars.
\newblock In \emph{Transactions on Graphics, (Proc. SIGGRAPH)} (2025), pp.~1--11.

\bibitem[IRG{\etalchar{*}}23]{Icsik2023humanrf}
\textsc{I{\c{s}}{\i}k M., R{\"u}nz M., Georgopoulos M., Khakhulin T., Starck J., Agapito L., Nie{\ss}ner M.}:
\newblock Humanrf: High-fidelity neural radiance fields for humans in motion.
\newblock \emph{Transactions on Graphics, (Proc. SIGGRAPH) 42}, 4 (2023), 1--12.

\bibitem[ISS{\etalchar{*}}23]{Iwase2023relightablehands}
\textsc{Iwase S., Saito S., Simon T., Lombardi S., Bagautdinov T., Joshi R., Prada F., Shiratori T., Sheikh Y., Saragih J.}:
\newblock Relightablehands: Efficient neural relighting of articulated hand models.
\newblock In \emph{Conference on Computer Vision and Pattern Recognition (CVPR)} (2023), pp.~16663--16673.

\bibitem[JCSH23]{jiang2023instantavatar}
\textsc{Jiang T., Chen X., Song J., Hilliges O.}:
\newblock Instantavatar: Learning avatars from monocular video in 60 seconds.
\newblock In \emph{Conference on Computer Vision and Pattern Recognition (CVPR)} (2023), pp.~16922--16932.

\bibitem[JGK{\etalchar{*}}24]{jiang2024multiply}
\textsc{Jiang Z., Guo C., Kaufmann M., Jiang T., Valentin J., Hilliges O., Song J.}:
\newblock Multiply: Reconstruction of multiple people from monocular video in the wild.
\newblock In \emph{Conference on Computer Vision and Pattern Recognition (CVPR)} (2024), pp.~109--118.

\bibitem[JLL{\etalchar{*}}24]{jin2024neural_graffer}
\textsc{Jin H., Li Y., Luan F., Xiangli Y., Bi S., Zhang K., Xu Z., Sun J., Snavely N.}:
\newblock Neural gaffer: Relighting any object via diffusion.
\newblock \emph{Advances in Neural Information Processing Systems (NeurIPS) 37} (2024), 141129--141152.

\bibitem[JPW{\etalchar{*}}25]{jin2025diffuman4d}
\textsc{Jin Y., Peng S., Wang X., Xie T., Xu Z., Yang Y., Shen Y., Bao H., Zhou X.}:
\newblock Diffuman4d: 4d consistent human view synthesis from sparse-view videos with spatio-temporal diffusion models.
\newblock In \emph{International Conference on Computer Vision (ICCV)} (2025), pp.~1--11.

\bibitem[JSR{\etalchar{*}}22]{Wenzel2022Mitsuba3}
\textsc{Jakob W., Speierer S., Roussel N., Nimier-David M., Vicini D., Zeltner T., Nicolet B., Crespo M., Leroy V., Zhang Z.}:
\newblock Mitsuba 3 renderer, 2022.
\newblock https://mitsuba-renderer.org.

\bibitem[JSZ{\etalchar{*}}25]{junkawitsch2025eva}
\textsc{Junkawitsch H., Sun G., Zhu H., Theobalt C., Habermann M.}:
\newblock Eva: Expressive virtual avatars from multi-view videos.
\newblock In \emph{SIGGRAPH Conference Papers (SA)} (2025), pp.~1--11.

\bibitem[JYG{\etalchar{*}}22]{ji2022geometry}
\textsc{Ji C., Yu T., Guo K., Liu J., Liu Y.}:
\newblock Geometry-aware single-image full-body human relighting.
\newblock In \emph{European Conference on Computer Vision (ECCV)} (2022), Springer, pp.~388--405.

\bibitem[JZH{\etalchar{*}}20]{jiang2020bcnet}
\textsc{Jiang B., Zhang J., Hong Y., Luo J., Liu L., Bao H.}:
\newblock Bcnet: Learning body and cloth shape from a single image.
\newblock In \emph{European Conference on Computer Vision (ECCV)} (2020), pp.~18--35.

\bibitem[JZZ{\etalchar{*}}25]{jiang2025omnihuman}
\textsc{Jiang J., Zeng W., Zheng Z., Yang J., Liang C., Liao W., Liang H., Zhang Y., Gao M.}:
\newblock Omnihuman-1.5: Instilling an active mind in avatars via cognitive simulation.
\newblock \emph{arXiv} (2025).

\bibitem[KAC{\etalchar{*}}24]{kolotouros2024instant}
\textsc{Kolotouros N., Alldieck T., Corona E., Bazavan E.~G., Sminchisescu C.}:
\newblock Instant 3d human avatar generation using image diffusion models.
\newblock In \emph{European Conference on Computer Vision (ECCV)} (2024), Springer, pp.~177--195.

\bibitem[Kaj86]{kajiya1986rendering}
\textsc{Kajiya J.~T.}:
\newblock The rendering equation.
\newblock In \emph{Proceedings of the 13th annual conference on Computer graphics and interactive techniques} (1986), pp.~143--150.

\bibitem[KAL{\etalchar{*}}21]{karras2021alias}
\textsc{Karras T., Aittala M., Laine S., H{\"a}rk{\"o}nen E., Hellsten J., Lehtinen J., Aila T.}:
\newblock Alias-free generative adversarial networks.
\newblock \emph{Advances in Neural Information Processing Systems (NeurIPS) 34} (2021), 852--863.

\bibitem[KAM24]{Kerbiriou2024reconstruction}
\textsc{Kerbiriou G., Avril Q., Marchal M.}:
\newblock {3D Reconstruction and Semantic Modeling of Eyelashes}.
\newblock \emph{Computer Graphics Forum (CGF)} (2024).

\bibitem[KAZ{\etalchar{*}}23]{kolotouros2023dreamhuman}
\textsc{Kolotouros N., Alldieck T., Zanfir A., Bazavan E., Fieraru M., Sminchisescu C.}:
\newblock Dreamhuman: Animatable 3d avatars from text.
\newblock \emph{Advances in Neural Information Processing Systems (NeurIPS) 36} (2023), 10516--10529.

\bibitem[KBM{\etalchar{*}}24]{Khirodkar2024sapiens}
\textsc{Khirodkar R., Bagautdinov T., Martinez J., Zhaoen S., James A., Selednik P., Anderson S., Saito S.}:
\newblock Sapiens: Foundation for human vision models.
\newblock In \emph{European Conference on Computer Vision (ECCV)} (2024), Springer, pp.~206--228.

\bibitem[KCF{\etalchar{*}}22]{Kuang2022deepmvshair}
\textsc{Kuang Z., Chen Y., Fu H., Zhou K., Zheng Y.}:
\newblock {DeepMVSHair}: Deep hair modeling from sparse views.
\newblock In \emph{SIGGRAPH Asia Conference Papers (SA)} (2022).

\bibitem[KE19]{kanamori2019relighting}
\textsc{Kanamori Y., Endo Y.}:
\newblock Relighting humans: occlusion-aware inverse rendering for full-body human images.
\newblock \emph{Transactions on Graphics, (Proc. SIGGRAPH)} (2019).

\bibitem[KGE{\etalchar{*}}21]{kappel2021high}
\textsc{Kappel M., Golyanik V., Elgharib M., Henningson J.-O., Seidel H.-P., Castillo S., Theobalt C., Magnor M.}:
\newblock High-fidelity neural human motion transfer from monocular video.
\newblock In \emph{Conference on Computer Vision and Pattern Recognition (CVPR)} (2021), pp.~1541--1550.

\bibitem[KGN24]{kirschstein2024diffusionavatars}
\textsc{Kirschstein T., Giebenhain S., Nie{\ss}ner M.}:
\newblock Diffusionavatars: Deferred diffusion for high-fidelity 3d head avatars.
\newblock In \emph{Conference on Computer Vision and Pattern Recognition (CVPR)} (2024), pp.~5481--5492.

\bibitem[KGT{\etalchar{*}}24]{kirschstein2024gghead}
\textsc{Kirschstein T., Giebenhain S., Tang J., Georgopoulos M., Nie\ss{}ner M.}:
\newblock {GGHead: Fast and Generalizable 3D Gaussian Heads}.
\newblock In \emph{SIGGRAPH Asia Conference Papers (SA)} (2024), SA '24, Association for Computing Machinery.

\bibitem[KHHB21]{kocabas2021pare}
\textsc{Kocabas M., Huang C.-H.~P., Hilliges O., Black M.~J.}:
\newblock Pare: Part attention regressor for 3d human body estimation.
\newblock In \emph{International Conference on Computer Vision (ICCV)} (2021), pp.~11127--11137.

\bibitem[KHTG24]{kairanda2024neuralclothsim}
\textsc{Kairanda N., Habermann M., Theobalt C., Golyanik V.}:
\newblock Neuralclothsim: Neural deformation fields meet the thin shell theory.
\newblock \emph{Advances in Neural Information Processing Systems (NeurIPS) 37} (2024), 110533--110572.

\bibitem[KJJ{\etalchar{*}}26]{ki2026avatar}
\textsc{Ki T., Jang S., Jo J., Yoon J., Hwang S.~J.}:
\newblock Avatar forcing: Real-time interactive head avatar generation for natural conversation.
\newblock \emph{arXiv preprint arXiv:2601.00664} (2026).

\bibitem[KK89]{Kajiya1989Rendering}
\textsc{Kajiya J.~T., Kay T.~L.}:
\newblock Rendering fur with three dimensional textures.
\newblock \emph{SIGGRAPH Conference Papers (SA)} (1989).

\bibitem[KKLD23]{Kerbl20233d}
\textsc{Kerbl B., Kopanas G., Leimk{\"u}hler T., Drettakis G.}:
\newblock 3d gaussian splatting for real-time radiance field rendering.
\newblock \emph{Transactions on Graphics, (Proc. SIGGRAPH) 42}, 4 (2023), 139--1.

\bibitem[KKSJ24]{kim2024gala}
\textsc{Kim T., Kim B., Saito S., Joo H.}:
\newblock Gala: Generating animatable layered assets from a single scan.
\newblock In \emph{Conference on Computer Vision and Pattern Recognition (CVPR)} (2024), pp.~1535--1545.

\bibitem[KL21]{korosteleva2021sewingpatterns}
\textsc{Korosteleva M., Lee S.-H.}:
\newblock Generating datasets of 3d garments with sewing patterns.
\newblock \emph{arXiv} (2021).

\bibitem[KL22]{korosteleva2022neuraltailor}
\textsc{Korosteleva M., Lee S.-H.}:
\newblock Neuraltailor: Reconstructing sewing pattern structures from 3d point clouds of garments.
\newblock \emph{Transactions on Graphics, (Proc. SIGGRAPH) 41}, 4 (2022), 1--16.

\bibitem[KLA19]{Karras2019style}
\textsc{Karras T., Laine S., Aila T.}:
\newblock A style-based generator architecture for generative adversarial networks.
\newblock In \emph{Conference on Computer Vision and Pattern Recognition (CVPR)} (2019), pp.~4401--4410.

\bibitem[KLF{\etalchar{*}}23]{kwon2023deliffas}
\textsc{Kwon Y., Liu L., Fuchs H., Habermann M., Theobalt C.}:
\newblock Deliffas: Deformable light fields for fast avatar synthesis.
\newblock \emph{Advances in Neural Information Processing Systems (NeurIPS) 36} (2023), 40944--40962.

\bibitem[KLL{\etalchar{*}}24]{karras2024fashion}
\textsc{Karras J., Li Y., Liu N., Zhu L., Yoo I., Lugmayr A., Lee C., Kemelmacher-Shlizerman I.}:
\newblock Fashion-vdm: Video diffusion model for virtual try-on.
\newblock In \emph{SIGGRAPH Asia Conference Papers (SA)} (2024), pp.~1--11.

\bibitem[KM17]{Khungurn2017azimuthal}
\textsc{Khungurn P., Marschner S.}:
\newblock Azimuthal scattering from elliptical hair fibers.
\newblock \emph{Transactions on Graphics, (Proc. SIGGRAPH) 36}, 2 (2017).

\bibitem[KMC24]{ki2024export3d}
\textsc{Ki T., Min D., Chae G.}:
\newblock Learning to generate conditional tri-plane for 3d-aware expression controllable portrait animation.
\newblock In \emph{European Conference on Computer Vision (ECCV)} (2024), Springer, pp.~476--493.

\bibitem[KNS{\etalchar{*}}25]{kansy2025reenact}
\textsc{Kansy M., Naruniec J., Schroers C., Gross M., Weber R.~M.}:
\newblock Reenact anything: Semantic video motion transfer using motion-textual inversion.
\newblock In \emph{Proceedings of the Special Interest Group on Computer Graphics and Interactive Techniques Conference Conference Papers} (2025), pp.~1--12.

\bibitem[KPBD19]{kolotouros2019learning}
\textsc{Kolotouros N., Pavlakos G., Black M.~J., Daniilidis K.}:
\newblock Learning to reconstruct 3d human pose and shape via model-fitting in the loop.
\newblock In \emph{International Conference on Computer Vision (ICCV)} (2019), pp.~2252--2261.

\bibitem[KPHT23]{karunratanakul2023harp}
\textsc{Karunratanakul K., Prokudin S., Hilliges O., Tang S.}:
\newblock Harp: Personalized hand reconstruction from a monocular rgb video.
\newblock In \emph{Conference on Computer Vision and Pattern Recognition (CVPR)} (2023), pp.~12802--12813.

\bibitem[KPJD21]{kolotouros2021probabilistic}
\textsc{Kolotouros N., Pavlakos G., Jayaraman D., Daniilidis K.}:
\newblock Probabilistic modeling for human mesh recovery.
\newblock In \emph{International Conference on Computer Vision (ICCV)} (2021), pp.~11605--11614.

\bibitem[KQG{\etalchar{*}}23]{Kirschstein2023nersemble}
\textsc{Kirschstein T., Qian S., Giebenhain S., Walter T., Nie\ss{}ner M.}:
\newblock Nersemble: Multi-view radiance field reconstruction of human heads.
\newblock \emph{Transactions on Graphics, (Proc. SIGGRAPH) 42}, 4 (2023).

\bibitem[KRS{\etalchar{*}}25]{kirschstein2025avat3r}
\textsc{Kirschstein T., Romero J., Sevastopolsky A., Nie\ss{}ner M., Saito S.}:
\newblock Avat3r: Large animatable gaussian reconstruction model for high-fidelity 3d head avatars.
\newblock In \emph{International Conference on Computer Vision (ICCV)} (2025).

\bibitem[KSLZ22]{khakhulin2022rome}
\textsc{Khakhulin T., Sklyarova V., Lempitsky V., Zakharov E.}:
\newblock Realistic one-shot mesh-based head avatars.
\newblock In \emph{European Conference on Computer Vision (ECCV)} (2022), Springer, pp.~345--362.

\bibitem[KSN{\etalchar{*}}25]{kim2025haircup}
\textsc{Kim B., Saito S., Nam G., Simon T., Saragih J., Joo H., Li J.}:
\newblock Haircup: Hair compositional universal prior for 3d gaussian avatars.
\newblock In \emph{International Conference on Computer Vision (ICCV)} (2025).

\bibitem[KTS{\etalchar{*}}15]{khamis2015learning}
\textsc{Khamis S., Taylor J., Shotton J., Keskin C., Izadi S., Fitzgibbon A.}:
\newblock Learning an efficient model of hand shape variation from depth images.
\newblock In \emph{Conference on Computer Vision and Pattern Recognition (CVPR)} (2015), pp.~2540--2548.

\bibitem[KTZ{\etalchar{*}}24]{kong2024hunyuanvideo}
\textsc{Kong W., Tian Q., Zhang Z., Min R., Dai Z., Zhou J., Xiong J., Li X., Wu B., Zhang J., et~al.}:
\newblock Hunyuanvideo: A systematic framework for large video generative models.
\newblock \emph{arXiv preprint arXiv:2412.03603} (2024).

\bibitem[KWG{\etalchar{*}}19]{dkulon2019rec}
\textsc{Kulon D., Wang H., G{\"{u}}ler R.~A., Bronstein M.~M., Zafeiriou S.}:
\newblock Single image 3d hand reconstruction with mesh convolutions.
\newblock In \emph{British Machine Vision Conference (BMVC)} (2019).

\bibitem[KWK{\etalchar{*}}25]{kant2025pippo}
\textsc{Kant Y., Weber E., Kim J.~K., Khirodkar R., Zhaoen S., Martinez J., Gilitschenski I., Saito S., Bagautdinov T.}:
\newblock Pippo: High-resolution multi-view humans from a single image.
\newblock In \emph{Conference on Computer Vision and Pattern Recognition (CVPR)} (2025), pp.~16418--16429.

\bibitem[KZB{\etalchar{*}}24]{kabadayi24ganavatar}
\textsc{Kabadayi B., Zielonka W., Bhatnagar B.~L., Pons-Moll G., Thies J.}:
\newblock Gan-avatar: Controllable personalized gan-based human head avatar.
\newblock In \emph{International Conference on 3D Vision (3DV)} (2024).

\bibitem[Lam60]{lambert1760photometria}
\textsc{Lambert J.~H.}:
\newblock \emph{Photometria sive de mensura et gradibus luminis, colorum et umbrae}.
\newblock sumptibus vidvae E. Klett, typis CP Detleffsen, 1760.

\bibitem[LBB{\etalchar{*}}17]{Li2017flame}
\textsc{Li T., Bolkart T., Black M.~J., Li H., Romero J.}:
\newblock Learning a model of facial shape and expression from 4d scans.
\newblock \emph{Transactions on Graphics, (Proc. SIGGRAPH Asia) 36}, 6 (2017), 194--1.

\bibitem[LBB{\etalchar{*}}25]{flux2024}
\textsc{Labs B.~F., Batifol S., Blattmann A., Boesel F., Consul S., Diagne C., Dockhorn T., English J., English Z., Esser P., et~al.}:
\newblock Flux. 1 kontext: Flow matching for in-context image generation and editing in latent space, 2025.

\bibitem[LCCZ22]{Lyu2022real}
\textsc{Lyu Q., Chai M., Chen X., Zhou K.}:
\newblock Real-time hair simulation with neural interpolation.
\newblock \emph{Transactions on Visualization and Computer Graphics (TVCG) 28}, 4 (2022), 1894--1905.

\bibitem[LCD{\etalchar{*}}25]{li2025dmap}
\textsc{Li R., Cao C., Dumery C., You Y., Li H., Fua P.}:
\newblock Single view garment reconstruction using diffusion mapping via pattern coordinates.
\newblock In \emph{SIGGRAPH Conference Papers (SA)} (2025), pp.~1--11.

\bibitem[LCL{\etalchar{*}}24]{li2024diffavatar}
\textsc{Li Y., Chen H.-y., Larionov E., Sarafianos N., Matusik W., Stuyck T.}:
\newblock Diffavatar: Simulation-ready garment optimization with differentiable simulation.
\newblock In \emph{Conference on Computer Vision and Pattern Recognition (CVPR)} (2024), pp.~4368--4378.

\bibitem[LCR24]{mast3r_eccv24}
\textsc{Leroy V., Cabon Y., Revaud J.}:
\newblock Grounding image matching in 3d with mast3r, 2024.

\bibitem[LCS{\etalchar{*}}24a]{Li2024urvatar}
\textsc{Li J., Cao C., Schwartz G., Khirodkar R., Richardt C., Simon T., Sheikh Y., Saito S.}:
\newblock {URAvatar}: {U}niversal relightable gaussian codec avatars.
\newblock In \emph{SIGGRAPH Asia Conference Papers (SA)} (2024), {ACM}, pp.~128:1--128:11.

\bibitem[LCS{\etalchar{*}}24b]{li2024uravatar}
\textsc{Li J., Cao C., Schwartz G., Khirodkar R., Richardt C., Simon T., Sheikh Y., Saito S.}:
\newblock Uravatar: Universal relightable gaussian codec avatars.
\newblock In \emph{SIGGRAPH Asia Conference Papers (SA)} (2024), pp.~1--11.

\bibitem[LDK{\etalchar{*}}25]{lu2025gas}
\textsc{Lu Y., Dong J., Kwon Y., Zhao Q., Dai B., De~la Torre F.}:
\newblock Gas: Generative avatar synthesis from a single image.
\newblock In \emph{International Conference on Computer Vision (ICCV)} (2025), pp.~1--11.

\bibitem[LDN{\etalchar{*}}25]{liu2025lucas}
\textsc{Liu D., Deng T., Nam G., Rong Y., Pidhorskyi S., Li J., Saragih J., Metaxas D.~N., Cao C.}:
\newblock Lucas: Layered universal codec avatars.
\newblock In \emph{Conference on Computer Vision and Pattern Recognition (CVPR)} (2025).

\bibitem[LDR14]{lagarde2014moving}
\textsc{Lagarde S., De~Rousiers C.}:
\newblock Moving frostbite to physically based rendering 3.0.
\newblock \emph{SIGGRAPH Course: Physically Based Shading in Theory and Practice 3} (2014).

\bibitem[LFL{\etalchar{*}}23]{li2023subspace}
\textsc{Li X., Fang Y., Lan L., Wang H., Yang Y., Li M., Jiang C.}:
\newblock Subspace-preconditioned gpu projective dynamics with contact for cloth simulation.
\newblock In \emph{SIGGRAPH Asia Conference Papers (SA)} (2023), pp.~1--12.

\bibitem[LHR{\etalchar{*}}21]{liu2021neural}
\textsc{Liu L., Habermann M., Rudnev V., Sarkar K., Gu J., Theobalt C.}:
\newblock Neural actor: Neural free-view synthesis of human actors with pose control.
\newblock \emph{Transactions on Graphics, (Proc. SIGGRAPH) 40}, 6 (2021), 1--16.

\bibitem[LHX{\etalchar{*}}25]{lin2025interanimate}
\textsc{Lin Y., Hong Y., Xu Z., Li X., Xu C., Song C., Li R., Chen H., Lan J., Zhu H., et~al.}:
\newblock Interanimate: Taming region-aware diffusion model for realistic human interaction animation.
\newblock \emph{arXiv} (2025).

\bibitem[LJX{\etalchar{*}}25]{li2025magicid}
\textsc{Li H., Jiang L., Xiao X., Wang T., Yi H., Wu B., Cai D.}:
\newblock Magicid: Hybrid preference optimization for id-consistent and dynamic-preserved video customization.
\newblock In \emph{ICCV} (2025), pp.~1--8.

\bibitem[LJY{\etalchar{*}}25]{lin2025omnihuman1}
\textsc{Lin G., Jiang J., Yang J., Zheng Z., Liang C.}:
\newblock Omnihuman-1: Rethinking the scaling-up of one-stage conditioned human animation models.
\newblock In \emph{International Conference on Computer Vision (ICCV)} (2025), pp.~1--11.

\bibitem[LJZ{\etalchar{*}}23]{Li2023ems}
\textsc{Li C., Jin L., Zheng Y., Yu Y., Han X.}:
\newblock Ems: 3d eyebrow modeling from single-view images.
\newblock \emph{Transactions on Graphics, (Proc. SIGGRAPH Asia) 42}, 6 (2023).

\bibitem[LKJ21]{li2021codimensional}
\textsc{Li M., Kaufman D.~M., Jiang C.}:
\newblock Codimensional incremental potential contact.
\newblock \emph{Transactions on Graphics, (Proc. SIGGRAPH) 40}, 4 (2021), 1--24.

\bibitem[LKK24]{lee2024learning}
\textsc{Lee J., Kim H., Kwon T.}:
\newblock Learning-based self-collision avoidance in retargeting using body part-specific signed distance fields.

\bibitem[LKL{\etalchar{*}}25]{lee2025geoavatar}
\textsc{Lee S., Kim S., Lee H., Jeong W.-S., Lee J.~H.}:
\newblock Geoavatar: Geometrically-consistent multi-person avatar reconstruction from sparse multi-view videos.
\newblock In \emph{Conference on Computer Vision and Pattern Recognition (CVPR)} (2025), pp.~21138--21147.

\bibitem[LL24]{Li2024strand}
\textsc{Li H., Liu X.}:
\newblock Strand-accurate multi-view facial hair reconstruction and tracking.
\newblock \emph{The Visual Computer (VC) 40}, 7 (2024), 4713--4724.

\bibitem[LLJ{\etalchar{*}}25]{li2025garmentdreamer}
\textsc{Li B., Li X., Jiang Y., Xie T., Gao F., Wang H., Yang Y., Jiang C.}:
\newblock Garmentdreamer: 3dgs guided garment synthesis with diverse geometry and texture details.
\newblock In \emph{International Conference on 3D Vision (3DV)} (2025).

\bibitem[LLLL24]{liu2024improvedllava}
\textsc{Liu H., Li C., Li Y., Lee Y.~J.}:
\newblock Improved baselines with visual instruction tuning.
\newblock In \emph{Conference on Computer Vision and Pattern Recognition (CVPR)} (2024), {IEEE}, pp.~26286--26296.

\bibitem[LLR13]{Luo2013structure}
\textsc{Luo L., Li H., Rusinkiewicz S.}:
\newblock Structure-aware hair capture.
\newblock \emph{Transactions on Graphics, (Proc. SIGGRAPH) 32}, 4 (2013).

\bibitem[LLS{\etalchar{*}}24]{lin2024layga}
\textsc{Lin S., Li Z., Su Z., Zheng Z., Zhang H., Liu Y.}:
\newblock Layga: Layered gaussian avatars for animatable clothing transfer.
\newblock In \emph{SIGGRAPH Conference Papers (SA)} (2024), pp.~1--11.

\bibitem[LLWL23]{liu2023llava}
\textsc{Liu H., Li C., Wu Q., Lee Y.~J.}:
\newblock Visual instruction tuning, 2023.

\bibitem[LLZ25]{lin2025controlhairphysicallybasedvideodiffusion}
\textsc{Lin W., Li H., Zhu Y.}:
\newblock Controlhair: Physically-based video diffusion for controllable dynamic hair rendering, 2025.

\bibitem[LMR{\etalchar{*}}15]{loper2015smpl}
\textsc{Loper M., Mahmood N., Romero J., Pons-Moll G., Black M.~J.}:
\newblock {SMPL}: A skinned multi-person linear model.
\newblock \emph{Transactions on Graphics, (Proc. SIGGRAPH Asia) 34}, 6 (2015), 248:1--248:16.

\bibitem[LOZ{\etalchar{*}}24]{Luo2024gaussianhair}
\textsc{Luo H., Ouyang M., Zhao Z., Jiang S., Zhang L., Zhang Q., Yang W., Xu L., Yu J.}:
\newblock {GaussianHair}: Hair modeling and rendering with light-aware gaussians.
\newblock \emph{arXiv} (2024).

\bibitem[LQZ{\etalchar{*}}25]{li2025personalvideo}
\textsc{Li H., Qiu H., Zhang S., Wang X., Wei Y., Li Z., Zhang Y., Wu B., Cai D.}:
\newblock Personalvideo: High id-fidelity video customization without dynamic and semantic degradation.
\newblock In \emph{International Conference on Computer Vision (ICCV)} (2025), pp.~19406--19416.

\bibitem[LSS{\etalchar{*}}19]{lombardi2019neuvol}
\textsc{Lombardi S., Simon T., Saragih J., Schwartz G., Lehrmann A., Sheikh Y.}:
\newblock Neural volumes: Learning dynamic renderable volumes from images.
\newblock \emph{Transactions on Graphics, (Proc. SIGGRAPH) 38}, 4 (2019), 65:1--65:14.

\bibitem[LSS{\etalchar{*}}21]{Lombardi21mvp}
\textsc{Lombardi S., Simon T., Schwartz G., Zollhoefer M., Sheikh Y., Saragih J.}:
\newblock Mixture of volumetric primitives for efficient neural rendering.
\newblock \emph{Transactions on Graphics, (Proc. SIGGRAPH) 40}, 4 (2021).

\bibitem[LSS{\etalchar{*}}23]{li2023megane}
\textsc{Li J., Saito S., Simon T., Lombardi S., Li H., Saragih J.}:
\newblock Megane: Morphable eyeglass and avatar network.
\newblock In \emph{Conference on Computer Vision and Pattern Recognition (CVPR)} (2023), pp.~12769--12779.

\bibitem[LSSS18]{lombardi2018deep}
\textsc{Lombardi S., Saragih J., Simon T., Sheikh Y.}:
\newblock Deep appearance models for face rendering.
\newblock \emph{Transactions on Graphics, (Proc. SIGGRAPH) 37}, 4 (2018), 1--13.

\bibitem[LSZ{\etalchar{*}}23]{li2023animatable}
\textsc{Li Z., Sun Y., Zheng Z., Wang L., Zhang S., Liu Y.}:
\newblock Animatable and relightable gaussians for high-fidelity human avatar modeling.
\newblock \emph{arXiv} (2023).

\bibitem[LSZ{\etalchar{*}}25]{li2025anydressing}
\textsc{Li X., Sun Q., Zhang P., Ye F., Liao Z., Feng W., Zhao S., He Q.}:
\newblock Anydressing: Customizable multi-garment virtual dressing via latent diffusion models.
\newblock In \emph{Conference on Computer Vision and Pattern Recognition (CVPR)} (2025), IEEE, pp.~23723--23733.

\bibitem[LVKS21]{lewis2021tryongan}
\textsc{Lewis K.~M., Varadharajan S., Kemelmacher-Shlizerman I.}:
\newblock Tryongan: Body-aware try-on via layered interpolation.
\newblock \emph{Transactions on Graphics (TOG) 40}, 4 (2021), 1--10.

\bibitem[LWH{\etalchar{*}}25]{liu2025beyond}
\textsc{Liu J., Wang J., Hou S., Ren M., Wu H., Ma L., Pei R., He Z.}:
\newblock Beyond face swapping: A diffusion-based digital human benchmark for multimodal deepfake detection.
\newblock \emph{arXiv} (2025).

\bibitem[LXH{\etalchar{*}}20]{liu2020Neural}
\textsc{Liu L., Xu W., Habermann M., Zollh\"ofer M., Bernard F., Kim H., Wang W., Theobalt C.}:
\newblock Neural human video rendering by learning dynamic textures and rendering-to-video translation.
\newblock \emph{TVCG} (2020), 1--10.

\bibitem[LXL{\etalchar{*}}23]{liu2023sewformer}
\textsc{Liu L., Xu X., Lin Z., Liang J., Yan S.}:
\newblock Towards garment sewing pattern reconstruction from a single image.
\newblock \emph{Transactions on Graphics, (Proc. SIGGRAPH Asia) 42}, 6 (2023), 1--15.

\bibitem[LXL{\etalchar{*}}25a]{li2025adaviewplanner}
\textsc{Li Y., Xia M., Liu G., Bai J., Wang X., Zhang C., Lin Y., Chu R., Wan P., Yang Y.}:
\newblock Adaviewplanner: Adapting video diffusion models for viewpoint planning in 4d scenes.
\newblock \emph{arXiv preprint arXiv:2510.10670} (2025).

\bibitem[LXL{\etalchar{*}}25b]{Liao2025hhavatar}
\textsc{Liao Z., Xu Y., Li Z., Li Q., Zhou B., Bai R., Xu D., Zhang H., Liu Y.}:
\newblock Hhavatar: Gaussian head avatar with dynamic hairs.
\newblock \emph{Transactions on Pattern Analysis and Machine Intelligence (TPAMI) PP} (2025).

\bibitem[LXS{\etalchar{*}}25]{li2025tokenmotion}
\textsc{Li R., Xing D., Sun H., Ha Y., Shen J., Ho C.}:
\newblock Tokenmotion: Decoupled motion control via token disentanglement for human-centric video generation.
\newblock In \emph{CVPR} (2025), pp.~1951--1961.

\bibitem[LXZ{\etalchar{*}}19]{liu2019neural}
\textsc{Liu L., Xu W., Zollhoefer M., Kim H., Bernard F., Habermann M., Wang W., Theobalt C.}:
\newblock Neural rendering and reenactment of human actor videos.
\newblock \emph{TOG} (2019), 1--14.

\bibitem[LYD{\etalchar{*}}25]{li2025dress1to3}
\textsc{Li X., Yu C., Du W., Jiang Y., Xie T., Chen Y., Yang Y., Jiang C.}:
\newblock Dress-1-to-3: Single image to simulation-ready 3d outfit with diffusion prior and differentiable physics.
\newblock \emph{Transactions on Graphics, (Proc. SIGGRAPH) 44}, 4 (2025), 1--16.

\bibitem[LYDM{\etalchar{*}}25]{li2025simavatar}
\textsc{Li X., Yuan Y., De~Mello S., Daviet G., Leaf J., Macklin M., Kautz J., Iqbal U.}:
\newblock Simavatar: Simulation-ready avatars with layered hair and clothing.
\newblock In \emph{Conference on Computer Vision and Pattern Recognition (CVPR)} (2025), pp.~26320--26330.

\bibitem[LZL{\etalchar{*}}23]{li2023posevocab}
\textsc{Li Z., Zheng Z., Liu Y., Zhou B., Liu Y.}:
\newblock Posevocab: Learning joint-structured pose embeddings for human avatar modeling.
\newblock In \emph{SIGGRAPH Conference Papers (SA)} (2023), pp.~1--11.

\bibitem[LZL{\etalchar{*}}25a]{li2025pshuman}
\textsc{Li P., Zheng W., Liu Y., Yu T., Li Y., Qi X., Chi X., Xia S., Cao Y.-P., Xue W., et~al.}:
\newblock Pshuman: Photorealistic single-image 3d human reconstruction using cross-scale multiview diffusion and explicit remeshing.
\newblock In \emph{Conference on Computer Vision and Pattern Recognition (CVPR)} (2025), pp.~16008--16018.

\bibitem[LZL{\etalchar{*}}25b]{liang2025realismotion}
\textsc{Liang J., Zhou J., Li S., Cao C., Sun L., Qian Y., Chen W., Wang F.}:
\newblock Realismotion: Decomposed human motion control and video generation in the world space.
\newblock \emph{arXiv preprint arXiv:2508.08588} (2025).

\bibitem[LZQ{\etalchar{*}}22]{li2022nimble}
\textsc{Li Y., Zhang L., Qiu Z., Jiang Y., Li N., Ma Y., Zhang Y., Xu L., Yu J.}:
\newblock Nimble: a non-rigid hand model with bones and muscles.
\newblock \emph{Transactions on Graphics, (Proc. SIGGRAPH) 41}, 4 (2022), 1--16.

\bibitem[LZWL24]{Li2024animatable}
\textsc{Li Z., Zheng Z., Wang L., Liu Y.}:
\newblock Animatable gaussians: Learning pose-dependent gaussian maps for high-fidelity human avatar modeling.
\newblock In \emph{Conference on Computer Vision and Pattern Recognition (CVPR)} (2024), pp.~19711--19722.

\bibitem[LZY{\etalchar{*}}25]{li2025pursuing}
\textsc{Li D., Zhong W., Yu W., Pan Y., Zhang D., Yao T., Han J., Mei T.}:
\newblock Pursuing temporal-consistent video virtual try-on via dynamic pose interaction.
\newblock In \emph{Conference on Computer Vision and Pattern Recognition (CVPR)} (2025), pp.~22648--22657.

\bibitem[LZZ{\etalchar{*}}22]{lin2022learning}
\textsc{Lin S., Zhang H., Zheng Z., Shao R., Liu Y.}:
\newblock Learning implicit templates for point-based clothed human modeling.
\newblock In \emph{European Conference on Computer Vision (ECCV)} (2022), Springer, pp.~210--228.

\bibitem[may]{maya}
{Autodesk, INC.} maya.
\newblock \href{https://autodesk.com/maya}{Maya}.
\newblock Accessed: 2025-10-10.

\bibitem[MBT{\etalchar{*}}12]{miguel2012data}
\textsc{Miguel E., Bradley D., Thomaszewski B., Bickel B., Matusik W., Otaduy M.~A., Marschner S.}:
\newblock Data-driven estimation of cloth simulation models.
\newblock \emph{Computer Graphics Forum (CGF) 31}, 2pt2 (2012), 519--528.

\bibitem[MBZ{\etalchar{*}}22]{Mihajlovic2022keypointnerf}
\textsc{Mihajlovic M., Bansal A., Zollhoefer M., Tang S., Saito S.}:
\newblock Keypointnerf: Generalizing image-based volumetric avatars using relative spatial encoding of keypoints.
\newblock In \emph{European Conference on Computer Vision (ECCV)} (2022), Springer, pp.~179--197.

\bibitem[MCH{\etalchar{*}}24]{Meishvili2024hairmony}
\textsc{Meishvili G., Clemoes J., Hewitt C., Hosenie Z., Xiao X., de~La~Gorce M., Takacs T., Baltrusaitis T., Criminisi A., McRae C., Jablonski N., Wilczkowiak M.}:
\newblock Hairmony: Fairness-aware hairstyle classification.
\newblock In \emph{SIGGRAPH Asia Conference Papers (SA)} (2024).

\bibitem[MDT{\etalchar{*}}25]{mendiratta2025grmm}
\textsc{Mendiratta M., Deshmukh M., Teotia K., Golyanik V., Kortylewski A., Theobalt C.}:
\newblock Grmm: Real-time high-fidelity gaussian morphable head model with learned residuals.
\newblock \emph{arXiv} (2025).

\bibitem[MFH{\etalchar{*}}25]{Ma2025diff_star}
\textsc{Ma Y., Feng K., Hu Z., Wang X., Wang Y., Zheng M., He X., Zhu C., Liu H., He Y., Wang Z., Li Z., Li X., Liu W., Xu D., Zhang L., Chen Q.}:
\newblock Controllable video generation: A survey, 2025.

\bibitem[MHHR07]{muller2007position}
\textsc{M{"u}ller M., Heidelberger B., Hennix M., Ratcliff J.}:
\newblock Position based dynamics.
\newblock \emph{Journal of Visual Communication and Image Representation 18}, 2 (2007), 109--118.

\bibitem[MHM{\etalchar{*}}25]{mei2025lux}
\textsc{Mei Y., He M., Ma L., Philip J., Xian W., George D.~M., Yu X., Dedic G., Taşel A.~L., Yu N., Patel V.~M., Debevec P.}:
\newblock Lux post facto: Learning portrait performance relighting with conditional video diffusion and a hybrid dataset.
\newblock \emph{Conference on Computer Vision and Pattern Recognition (CVPR)} (2025).

\bibitem[MJC{\etalchar{*}}03]{marschner2003light}
\textsc{Marschner S.~R., Jensen H.~W., Cammarano M., Worley S., Hanrahan P.}:
\newblock Light scattering from human hair fibers.
\newblock \emph{Transactions on Graphics, (Proc. SIGGRAPH) 22}, 3 (2003), 780--791.

\bibitem[MKR{\etalchar{*}}24]{Martinez2024codec}
\textsc{Martinez J., Kim E., Romero J., Bagautdinov T., Saito S., Yu S.-I., Anderson S., Zollh{\"o}fer M., Wang T.-L., Bai S., et~al.}:
\newblock Codec avatar studio: Paired human captures for complete, driveable, and generalizable avatars.
\newblock \emph{Advances in Neural Information Processing Systems (NeurIPS) 37} (2024), 83008--83023.

\bibitem[MLW{\etalchar{*}}24]{ma2024followyouremoji}
\textsc{Ma Y., Liu H., Wang H., Pan H., He Y., Yuan J., Zeng A., Cai C., Shum H.-Y., Liu W., et~al.}:
\newblock Follow-your-emoji: Fine-controllable and expressive freestyle portrait animation.
\newblock \emph{arXiv} (2024).

\bibitem[MMC16]{Macklin2016XPBD}
\textsc{Macklin M., M\"uller M., Chentanez N.}:
\newblock Xpbd: Position-based simulation of compliant constrained dynamics.
\newblock In \emph{Proceedings of the 9th International Conference on Motion in Games (MiG)} (2016), pp.~49--54.

\bibitem[MNSL22]{moon20223d}
\textsc{Moon G., Nam H., Shiratori T., Lee K.~M.}:
\newblock 3d clothed human reconstruction in the wild.
\newblock In \emph{European Conference on Computer Vision (ECCV)} (2022), pp.~184--200.

\bibitem[MPS21]{Morales2021survey}
\textsc{Morales A., Piella G., Sukno F.~M.}:
\newblock Survey on 3d face reconstruction from uncalibrated images.
\newblock \emph{Computer Science Review 40} (2021), 100400.

\bibitem[MSD{\etalchar{*}}24]{moreau2024human}
\textsc{Moreau A., Song J., Dhamo H., Shaw R., Zhou Y., P{\'e}rez-Pellitero E.}:
\newblock Human gaussian splatting: Real-time rendering of animatable avatars.
\newblock In \emph{Conference on Computer Vision and Pattern Recognition (CVPR)} (2024), pp.~788--798.

\bibitem[MSL20]{moon2020deephandmesh}
\textsc{Moon G., Shiratori T., Lee K.~M.}:
\newblock Deephandmesh: A weakly-supervised deep encoder-decoder framework for high-fidelity hand mesh modeling.
\newblock In \emph{European Conference on Computer Vision (ECCV)} (2020), Springer, pp.~440--455.

\bibitem[MST{\etalchar{*}}20]{Mildenhall2020nerf}
\textsc{Mildenhall B., Srinivasan P.~P., Tancik M., Barron J.~T., Ramamoorthi R., Ng R.}:
\newblock Nerf: Representing scenes as neural radiance fields for view synthesis.
\newblock In \emph{European Conference on Computer Vision (ECCV)} (2020), Springer.

\bibitem[MSY{\etalchar{*}}21]{Ma2021scale}
\textsc{Ma Q., Saito S., Yang J., Tang S., Black M.~J.}:
\newblock Scale: Modeling clothed humans with a surface codec of articulated local elements.
\newblock In \emph{Conference on Computer Vision and Pattern Recognition (CVPR)} (2021), pp.~16082--16093.

\bibitem[M{\"u}l06]{muller2006spherical}
\textsc{M{\"u}ller C.}:
\newblock \emph{Spherical harmonics}, vol.~17.
\newblock Springer, 2006.

\bibitem[MWH{\etalchar{*}}23]{mundra2023livehand}
\textsc{Mundra A., Wang J., Habermann M., Theobalt C., Elgharib M., et~al.}:
\newblock Livehand: Real-time and photorealistic neural hand rendering.
\newblock In \emph{International Conference on Computer Vision (ICCV)} (2023), pp.~18035--18045.

\bibitem[MXJ{\etalchar{*}}24]{moon2024authentic}
\textsc{Moon G., Xu W., Joshi R., Wu C., Shiratori T.}:
\newblock Authentic hand avatar from a phone scan via universal hand model.
\newblock In \emph{Conference on Computer Vision and Pattern Recognition (CVPR)} (2024), pp.~2029--2038.

\bibitem[MYCB25]{men2025mimo}
\textsc{Men Y., Yao Y., Cui M., Bo L.}:
\newblock Mimo: Controllable character video synthesis with spatial decomposed modeling.
\newblock In \emph{Conference on Computer Vision and Pattern Recognition (CVPR)} (2025), pp.~21181--21191.

\bibitem[MYR{\etalchar{*}}20]{Ma2020cape}
\textsc{Ma Q., Yang J., Ranjan A., Pujades S., Pons-Moll G., Tang S., Black M.~J.}:
\newblock Learning to dress 3d people in generative clothing.
\newblock In \emph{Conference on Computer Vision and Pattern Recognition (CVPR)} (2020), pp.~6469--6478.

\bibitem[MYTB21]{ma2021power}
\textsc{Ma Q., Yang J., Tang S., Black M.~J.}:
\newblock The power of points for modeling humans in clothing.
\newblock In \emph{International Conference on Computer Vision (ICCV)} (2021), pp.~10974--10984.

\bibitem[MYW{\etalchar{*}}20]{moon2020interhand2}
\textsc{Moon G., Yu S.-I., Wen H., Shiratori T., Lee K.~M.}:
\newblock {InterHand2.6M}: A dataset and baseline for {3D} interacting hand pose estimation from a single {RGB} image.
\newblock In \emph{European Conference on Computer Vision (ECCV)} (2020).

\bibitem[MZBT21]{mihajlovic2021leap}
\textsc{Mihajlovic M., Zhang Y., Black M.~J., Tang S.}:
\newblock Leap: Learning articulated occupancy of people.
\newblock In \emph{Conference on Computer Vision and Pattern Recognition (CVPR)} (2021), pp.~10461--10471.

\bibitem[MZQ{\etalchar{*}}23]{ma2023otavatar}
\textsc{Ma Z., Zhu X., Qi G.-J., Lei Z., Zhang L.}:
\newblock Otavatar: One-shot talking face avatar with controllable tri-plane rendering.
\newblock In \emph{Conference on Computer Vision and Pattern Recognition (CVPR)} (2023), pp.~16901--16910.

\bibitem[NAK{\etalchar{*}}25]{nakayama2025aipparel}
\textsc{Nakayama K., Ackermann J., Kesdogan T.~L., Zheng Y., Korosteleva M., Sorkine-Hornung O., Guibas L.~J., Yang G., Wetzstein G.}:
\newblock Aipparel: A multimodal foundation model for digital garments.
\newblock In \emph{Conference on Computer Vision and Pattern Recognition (CVPR)} (2025), pp.~8138--8149.

\bibitem[NBC{\etalchar{*}}24]{narasimhaswamy2024handiffuser}
\textsc{Narasimhaswamy S., Bhattacharya U., Chen X., Dasgupta I., Mitra S., Hoai M.}:
\newblock Handiffuser: Text-to-image generation with realistic hand appearances.
\newblock In \emph{Conference on Computer Vision and Pattern Recognition (CVPR)} (2024), pp.~2468--2479.

\bibitem[NCZ17]{nagrani2017voxceleb}
\textsc{Nagrani A., Chung J.~S., Zisserman A.}:
\newblock Voxceleb: a large-scale speaker identification dataset.
\newblock \emph{CoRR} (2017).

\bibitem[NHZ{\etalchar{*}}22]{neekhara2022facesigns}
\textsc{Neekhara P., Hussain S., Zhang X., Huang K., McAuley J., Koushanfar F.}:
\newblock Facesigns: semi-fragile neural watermarks for media authentication and countering deepfakes.
\newblock \emph{arXiv} (2022).

\bibitem[NKML25]{nam2025parte}
\textsc{Nam H., Kim D., Moon G., Lee K.~M.}:
\newblock Parte: Part-guided texturing for 3d human reconstruction from a single image.
\newblock In \emph{International Conference on Computer Vision (ICCV)} (2025), pp.~1--11.

\bibitem[NKOL25]{nam2025decloth}
\textsc{Nam H., Kim D., Oh J., Lee K.~M.}:
\newblock Decloth: Decomposable 3d cloth and human body reconstruction from a single image.
\newblock In \emph{Conference on Computer Vision and Pattern Recognition (CVPR)} (2025), pp.~5636--5645.

\bibitem[NWKS19]{Nam2019strandaccurate}
\textsc{Nam G., Wu C., Kim M.~H., Sheikh Y.}:
\newblock Strand-accurate multi-view hair capture.
\newblock In \emph{Conference on Computer Vision and Pattern Recognition (CVPR)} (2019), pp.~155--164.

\bibitem[OBT25]{HAIRFREE:NEURIPS:25}
\textsc{Ostrek M., Black M.~J., Thies J.}:
\newblock Hairfree: Compositional 2d head prior for text-driven 360° bald texture synthesis.
\newblock In \emph{Advances in Neural Information Processing Systems (NeurIPS)} (2025).

\bibitem[ODM{\etalchar{*}}23]{Oquab2023DINOv2LR}
\textsc{Oquab M., Darcet T., Moutakanni T., Vo H.~Q., Szafraniec M., Khalidov V., Fernandez P., Haziza D., Massa F., El-Nouby A., Assran M., Ballas N., Galuba W., Howes R., Huang P.-Y.~B., Li S.-W., Misra I., Rabbat M.~G., Sharma V., Synnaeve G., Xu H., J{\'e}gou H., Mairal J., Labatut P., Joulin A., Bojanowski P.}:
\newblock Dinov2: Learning robust visual features without supervision.
\newblock \emph{ArXiv} (2023).

\bibitem[OKA11a]{oikonomidis2011efficient}
\textsc{Oikonomidis I., Kyriazis N., Argyros A.}:
\newblock Efficient model-based 3d tracking of hand articulations using kinect.
\newblock In \emph{British Machine Vision Conference (BMVC)} (2011).

\bibitem[OKA11b]{oikonomidis2011full}
\textsc{Oikonomidis I., Kyriazis N., Argyros A.~A.}:
\newblock Full dof tracking of a hand interacting with an object by modeling occlusions and physical constraints.
\newblock In \emph{International Conference on Computer Vision (ICCV)} (2011), IEEE, pp.~2088--2095.

\bibitem[OLS{\etalchar{*}}25]{ohkawa2025generative}
\textsc{Ohkawa T., Lee J., Saito S., Saragih J., Prada F., Xu Y., Yu S.-I., Furuta R., Sato Y., Shiratori T.}:
\newblock Generative modeling of shape-dependent self-contact human poses.
\newblock In \emph{International Conference on Computer Vision (ICCV)} (2025), pp.~5426--5436.

\bibitem[oSN77]{united1977geometrical}
\textsc{of~Standards U. S. N.~B., Nicodemus F.~E.}:
\newblock \emph{Geometrical considerations and nomenclature for reflectance}, vol.~160.
\newblock US Department of Commerce, National Bureau of Standards Washington, DC, USA, 1977.

\bibitem[OT24]{SVP:ECCV:24}
\textsc{Ostrek M., Thies J.}:
\newblock Stable video portraits.
\newblock In \emph{European Conference on Computer Vision (ECCV)} (2024).

\bibitem[PBD{\etalchar{*}}10]{Parker2010Optix}
\textsc{Parker S.~G., Bigler J., Dietrich A., Friedrich H., Hoberock J., Luebke D., McAllister D., McGuire M., Morley K., Robison A., Stich M.}:
\newblock Optix: a general purpose ray tracing engine.
\newblock \emph{Transactions on Graphics (TOG) 29}, 4 (2010).

\bibitem[PBV23]{Petrovich2023tmr}
\textsc{Petrovich M., Black M.~J., Varol G.}:
\newblock {TMR:} text-to-motion retrieval using contrastive {3D} human motion synthesis.
\newblock In \emph{International Conference on Computer Vision (ICCV)} (2023), pp.~9488--9497.

\bibitem[PCG{\etalchar{*}}19]{pavlakos2019expressive}
\textsc{Pavlakos G., Choutas V., Ghorbani N., Bolkart T., Osman A.~A., Tzionas D., Black M.~J.}:
\newblock Expressive body capture: 3d hands, face, and body from a single image.
\newblock In \emph{Conference on Computer Vision and Pattern Recognition (CVPR)} (2019), pp.~10975--10985.

\bibitem[PDW{\etalchar{*}}21]{peng2021animatable}
\textsc{Peng S., Dong J., Wang Q., Zhang S., Shuai Q., Zhou X., Bao H.}:
\newblock Animatable neural radiance fields for modeling dynamic human bodies.
\newblock In \emph{International Conference on Computer Vision (ICCV)} (2021), pp.~14314--14323.

\bibitem[PFS{\etalchar{*}}19]{park2019deepsdf}
\textsc{Park J.~J., Florence P., Straub J., Newcombe R., Lovegrove S.}:
\newblock Deepsdf: Learning continuous signed distance functions for shape representation.
\newblock In \emph{Conference on Computer Vision and Pattern Recognition (CVPR)} (2019), pp.~165--174.

\bibitem[PHBB{\etalchar{*}}24]{parker2024genie}
\textsc{Parker-Holder J., Ball P., Bruce J., Dasagi V., Holsheimer K., Kaplanis C., Moufarek A., Scully G., Shar J., Shi J., et~al.}:
\newblock Genie 2: A large-scale foundation world model.
\newblock \emph{URL: https://deepmind. google/discover/blog/genie-2-a-large-scale-foundation-world-model} (2024).

\bibitem[PJBM23]{poole2023dreamfusion}
\textsc{Poole B., Jain A., Barron J.~T., Mildenhall B.}:
\newblock Dreamfusion: Text-to-3d using 2d diffusion.
\newblock In \emph{International Conference on Learning Representations (ICLR)} (2023).

\bibitem[PJH23]{pharr2023physically}
\textsc{Pharr M., Jakob W., Humphreys G.}:
\newblock \emph{Physically based rendering: From theory to implementation}.
\newblock MIT Press, 2023.

\bibitem[PKA{\etalchar{*}}09]{paysan2009bfm}
\textsc{Paysan P., Knothe R., Amberg B., Romdhani S., Vetter T.}:
\newblock A 3d face model for pose and illumination invariant face recognition.
\newblock In \emph{International Conference on Advanced Video and Signal based Surveillance (AAAI)} (2009), Ieee, pp.~296--301.

\bibitem[PKP{\etalchar{*}}24]{paudel2024ihuman}
\textsc{Paudel P., Khanal A., Paudel D.~P., Tandukar J., Chhatkuli A.}:
\newblock ihuman: Instant animatable digital humans from monocular videos.
\newblock In \emph{European Conference on Computer Vision (ECCV)} (2024), Springer, pp.~304--323.

\bibitem[PLPM20]{patel2020tailornet}
\textsc{Patel C., Liao Z., Pons-Moll G.}:
\newblock Tailornet: Predicting clothing in 3d as a function of human pose, shape and garment style.
\newblock In \emph{Conference on Computer Vision and Pattern Recognition (CVPR)} (2020), pp.~7365--7375.

\bibitem[PLT{\etalchar{*}}25]{pang2025mavid}
\textsc{Pang Y., Liu J., Tan L., Zhang Y., Gao F., Deng X., Kang Z., Wei X., Liu Y.}:
\newblock Mavid: A multimodal framework for audio-visual dialogue understanding and generation.
\newblock \emph{arXiv} (2025).

\bibitem[PMJ{\etalchar{*}}22]{pan2022predicting}
\textsc{Pan X., Mai J., Jiang X., Tang D., Li J., Shao T., Zhou K., Jin X., Manocha D.}:
\newblock Predicting loose-fitting garment deformations using bone-driven motion networks.
\newblock In \emph{SIGGRAPH Conference Papers (SA)} (2022), pp.~1--10.

\bibitem[PMPHB17]{pons2017clothcap}
\textsc{Pons-Moll G., Pujades S., Hu S., Black M.~J.}:
\newblock Clothcap: Seamless 4d clothing capture and retargeting.
\newblock \emph{Transactions on Graphics, (Proc. SIGGRAPH) 36}, 4 (2017), 1--15.

\bibitem[PNK25]{pan2025hairgs}
\textsc{Pan Y., Niessner M., Kirschstein T.}:
\newblock Hairgs: Hair strand reconstruction based on 3d gaussian splatting.
\newblock \emph{arXiv} (2025).

\bibitem[PPM{\etalchar{*}}23]{potamias2023handy}
\textsc{Potamias R.~A., Ploumpis S., Moschoglou S., Triantafyllou V., Zafeiriou S.}:
\newblock Handy: Towards a high fidelity 3d hand shape and appearance model.
\newblock In \emph{Conference on Computer Vision and Pattern Recognition (CVPR)} (2023), pp.~4670--4680.

\bibitem[PSB{\etalchar{*}}21]{Park2021nerfies}
\textsc{Park K., Sinha U., Barron J.~T., Bouaziz S., Goldman D.~B., Seitz S.~M., Martin-Brualla R.}:
\newblock Nerfies: Deformable neural radiance fields.
\newblock In \emph{International Conference on Computer Vision (ICCV)} (2021), pp.~5865--5874.

\bibitem[PSH{\etalchar{*}}21]{Park2021hypernerf}
\textsc{Park K., Sinha U., Hedman P., Barron J.~T., Bouaziz S., Goldman D.~B., Martin-Brualla R., Seitz S.~M.}:
\newblock Hypernerf: A higher-dimensional representation for topologically varying neural radiance fields.
\newblock \emph{Transactions on Graphics, (Proc. SIGGRAPH) 40} (2021).

\bibitem[PSZ{\etalchar{*}}25]{pang2025manivideo}
\textsc{Pang Y., Shao R., Zhang J., Tu H., Liu Y., Zhou B., Zhang H., Liu Y.}:
\newblock Manivideo: Generating hand-object manipulation video with dexterous and generalizable grasping.
\newblock In \emph{Conference on Computer Vision and Pattern Recognition (CVPR)} (2025), pp.~12209--12219.

\bibitem[PWS{\etalchar{*}}23]{peng2023emotalk}
\textsc{Peng Z., Wu H., Song Z., Xu H., Zhu X., He J., Liu H., Fan Z.}:
\newblock {EmoTalk}: {S}peech-driven emotional disentanglement for {3D} face animation.
\newblock In \emph{International Conference on Computer Vision (ICCV)} (2023), {IEEE}, pp.~20630--20640.

\bibitem[PX23]{peebles2023scalable}
\textsc{Peebles W., Xie S.}:
\newblock Scalable diffusion models with transformers.
\newblock In \emph{Proceedings of the IEEE/CVF international conference on computer vision} (2023), pp.~4195--4205.

\bibitem[PYG{\etalchar{*}}24]{po2024state}
\textsc{Po R., Yifan W., Golyanik V., Aberman K., Barron J.~T., Bermano A., Chan E., Dekel T., Holynski A., Kanazawa A., et~al.}:
\newblock State of the art on diffusion models for visual computing.
\newblock \emph{Computer Graphics Forum (CGF) 43}, 2 (2024), e15063.

\bibitem[PZK{\etalchar{*}}24]{Pang2024ash}
\textsc{Pang H., Zhu H., Kortylewski A., Theobalt C., Habermann M.}:
\newblock Ash: Animatable gaussian splats for efficient and photoreal human rendering.
\newblock In \emph{Conference on Computer Vision and Pattern Recognition (CVPR)} (2024), pp.~1165--1175.

\bibitem[PZS{\etalchar{*}}25]{prinzler2024joker}
\textsc{Prinzler M., Zakharov E., Sklyarova V., Kabadayi B., Thies J.}:
\newblock Joker: Conditional 3d head synthesis with extreme facial expressions.
\newblock In \emph{International Conference on 3D Vision (3DV)} (2025), {IEEE}, pp.~1583--1593.

\bibitem[PZX{\etalchar{*}}21]{peng2021neural}
\textsc{Peng S., Zhang Y., Xu Y., Wang Q., Shuai Q., Bao H., Zhou X.}:
\newblock Neural body: Implicit neural representations with structured latent codes for novel view synthesis of dynamic humans.
\newblock In \emph{Conference on Computer Vision and Pattern Recognition (CVPR)} (2021), pp.~9054--9063.

\bibitem[QCZ{\etalchar{*}}23]{qiu2023rec}
\textsc{Qiu L., Chen G., Zhou J., Xu M., Wang J., Han X.}:
\newblock Rec-mv: Reconstructing 3d dynamic cloth from monocular videos.
\newblock In \emph{Conference on Computer Vision and Pattern Recognition (CVPR)} (2023), pp.~4637--4646.

\bibitem[QGL{\etalchar{*}}25]{qiu2025lhm}
\textsc{Qiu L., Gu X., Li P., Zuo Q., Shen W., Zhang J., Qiu K., Yuan W., Chen G., Dong Z., et~al.}:
\newblock Lhm: Large animatable human reconstruction model from a single image in seconds.
\newblock \emph{arXiv} (2025).

\bibitem[QKS{\etalchar{*}}24]{Qian2024gaussianavatars}
\textsc{Qian S., Kirschstein T., Schoneveld L., Davoli D., Giebenhain S., Nie{\ss}ner M.}:
\newblock Gaussianavatars: Photorealistic head avatars with rigged 3d gaussians.
\newblock In \emph{Conference on Computer Vision and Pattern Recognition (CVPR)} (2024), pp.~20299--20309.

\bibitem[QLZ{\etalchar{*}}25]{qiu2025pf}
\textsc{Qiu L., Li P., Zuo Q., Gu X., Dong Y., Yuan W., Zhu S., Han X., Chen G., Dong Z.}:
\newblock Pf-lhm: 3d animatable avatar reconstruction from pose-free articulated human images.
\newblock In \emph{International Conference on Computer Vision (ICCV)} (2025), pp.~1--11.

\bibitem[QWM{\etalchar{*}}20]{qian2020html}
\textsc{Qian N., Wang J., Mueller F., Bernard F., Golyanik V., Theobalt C.}:
\newblock Html: A parametric hand texture model for 3d hand reconstruction and personalization.
\newblock In \emph{European Conference on Computer Vision (ECCV)} (2020), Springer, pp.~54--71.

\bibitem[QXL{\etalchar{*}}22]{qian2022unif}
\textsc{Qian S., Xu J., Liu Z., Ma L., Gao S.}:
\newblock Unif: United neural implicit functions for clothed human reconstruction and animation.
\newblock In \emph{European Conference on Computer Vision (ECCV)} (2022), Springer, pp.~121--137.

\bibitem[QZZ{\etalchar{*}}25]{qiu2025anigs}
\textsc{Qiu L., Zhu S., Zuo Q., Gu X., Dong Y., Zhang J., Xu C., Li Z., Yuan W., Bo L., et~al.}:
\newblock Anigs: Animatable gaussian avatar from a single image with inconsistent gaussian reconstruction.
\newblock In \emph{Conference on Computer Vision and Pattern Recognition (CVPR)} (2025), pp.~21148--21158.

\bibitem[RAFA{\etalchar{*}}25]{ressler2025dismo}
\textsc{Ressler-Antal T., Fundel F., Alaya M.~B., Baumann S.~A., Krause F., Gui M., Ommer B.}:
\newblock Dismo: Disentangled motion representations for open-world motion transfer.
\newblock \emph{arXiv preprint arXiv:2511.23428} (2025).

\bibitem[Ram02]{ramamoorthi2002analytic}
\textsc{Ramamoorthi R.}:
\newblock Analytic pca construction for theoretical analysis of lighting variability in images of a lambertian object.
\newblock \emph{Transactions on Pattern Analysis and Machine Intelligence (TPAMI) 24}, 10 (2002), 1322--1333.

\bibitem[RBS{\etalchar{*}}22]{Remelli2022drivable}
\textsc{Remelli E., Bagautdinov T., Saito S., Wu C., Simon T., Wei S.-E., Guo K., Cao Z., Prada F., Saragih J., et~al.}:
\newblock Drivable volumetric avatars using texel-aligned features.
\newblock In \emph{SIGGRAPH Conference Papers (SA)} (2022), pp.~1--9.

\bibitem[RCV{\etalchar{*}}19]{rossler2019faceforensics++}
\textsc{Rossler A., Cozzolino D., Verdoliva L., Riess C., Thies J., Nie{\ss}ner M.}:
\newblock Faceforensics++: Learning to detect manipulated facial images.
\newblock In \emph{International Conference on Computer Vision (ICCV)} (2019), pp.~1--11.

\bibitem[RGB{\etalchar{*}}20]{riviere2020single}
\textsc{Riviere J., Gotardo P.~F., Bradley D., Ghosh A., Beeler T.}:
\newblock Single-shot high-quality facial geometry and skin appearance capture.
\newblock \emph{Transactions on Graphics, (Proc. SIGGRAPH) 39}, 4 (2020), 81.

\bibitem[RGW{\etalchar{*}}25]{Rong2024gaussgarment}
\textsc{Rong B., Grigorev A., Wang W., Black M.~J., Thomaszewski B., Tsalicoglou C., Hilliges O.}:
\newblock Gaussian garments: Reconstructing simulation-ready clothing with photorealistic appearance from multi-view video.
\newblock \emph{International Conference on 3D Vision (3DV)} (2025).

\bibitem[RH01]{ramamoorthi2001efficient}
\textsc{Ramamoorthi R., Hanrahan P.}:
\newblock An efficient representation for irradiance environment maps.
\newblock In \emph{Proceedings of the 28th annual conference on Computer graphics and interactive techniques} (2001), pp.~497--500.

\bibitem[RK94]{rehg1994visual}
\textsc{Rehg J.~M., Kanade T.}:
\newblock Visual tracking of high dof articulated structures: an application to human hand tracking.
\newblock In \emph{European Conference on Computer Vision (ECCV)} (1994), Springer, pp.~35--46.

\bibitem[RMBCO22]{Roich2022pivotal}
\textsc{Roich D., Mokady R., Bermano A.~H., Cohen-Or D.}:
\newblock Pivotal tuning for latent-based editing of real images.
\newblock \emph{Transactions on Graphics, (Proc. SIGGRAPH) 42}, 1 (2022), 1--13.

\bibitem[RMZ{\etalchar{*}}25]{rao20253dpr}
\textsc{Rao P., Meka A., Zhou X., Fox G., B~R M., Zhan F., Weyrich T., Bickel B., Pfister H., Matusik W., Beeler T., Elgharib M., Habermann M., Theobalt C.}:
\newblock {3DPR}: Single image {3D} portrait relighting with generative priors.
\newblock In \emph{SIGGRAPH Asia Conference Papers (SA)} (2025), pp.~1--11.

\bibitem[RSW{\etalchar{*}}22]{Radu2022neuralstrands}
\textsc{Rosu R.~A., Saito S., Wang Z., Wu C., Behnke S., Nam G.}:
\newblock Neural strands: Learning hair geometry and appearance from multi-view images.
\newblock In \emph{European Conference on Computer Vision (ECCV)} (2022).

\bibitem[RTB17]{romero2017mano}
\textsc{Romero J., Tzionas D., Black M.~J.}:
\newblock Embodied hands: Modeling and capturing hands and bodies together.
\newblock \emph{Transactions on Graphics, (Proc. SIGGRAPH Asia) 36}, 6 (Nov. 2017).

\bibitem[RWF{\etalchar{*}}25]{Radu2025difflocks}
\textsc{Rosu R.~A., Wu K., Feng Y., Zheng Y., Black M.~J.}:
\newblock {DiffLocks}: Generating 3d hair from a single image using diffusion models.
\newblock In \emph{Conference on Computer Vision and Pattern Recognition (CVPR)} (2025).

\bibitem[RYCT23]{ranjan2023facelit}
\textsc{Ranjan A., Yi K.~M., Chang J.-H.~R., Tuzel O.}:
\newblock Facelit: Neural 3d relightable faces.
\newblock In \emph{Conference on Computer Vision and Pattern Recognition (CVPR)} (2023), pp.~8619--8628.

\bibitem[RYV{\etalchar{*}}25]{svdp}
\textsc{R. M.~B., Yin F., Voleti V., Drobyshev N., Lapin M., Vasishta A., Jampani V.}:
\newblock Stable video-driven portraits, 2025.

\bibitem[SAH{\etalchar{*}}25]{Saleh2025david}
\textsc{Saleh F., Aliakbarian S., Hewitt C., Petikam L., Xiao-Xian, Criminisi A., Cashman T.~J., Baltrušaitis T.}:
\newblock {DAViD}: Data-efficient and accurate vision models from synthetic data.
\newblock In \emph{International Conference on Computer Vision (ICCV)} (2025).

\bibitem[SBR23a]{Su2023npc}
\textsc{Su S.-Y., Bagautdinov T., Rhodin H.}:
\newblock Npc: Neural point characters from video.
\newblock In \emph{International Conference on Computer Vision (ICCV)} (2023), pp.~14795--14805.

\bibitem[SBR23b]{su2023iccv}
\textsc{Su S.-Y., Bagautdinov T., Rhodin H.}:
\newblock Npc: Neural point characters from video.
\newblock In \emph{International Conference on Computer Vision (ICCV)} (2023), pp.~1--8.

\bibitem[SC23]{stuyck2023diffxpbd}
\textsc{Stuyck T., Chen H.-y.}:
\newblock Diffxpbd: Differentiable position-based simulation of compliant constraint dynamics.
\newblock \emph{Symposium on Computer Animation (SCA) 6}, 3 (2023), 1--14.

\bibitem[SCD{\etalchar{*}}23]{Sklyarova2023neural}
\textsc{Sklyarova V., Chelishev J., Dogaru A., Medvedev I., Lempitsky V., Zakharov E.}:
\newblock Neural haircut: Prior-guided strand-based hair reconstruction.
\newblock In \emph{International Conference on Computer Vision (ICCV)} (2023), pp.~19762--19773.

\bibitem[SCT{\etalchar{*}}25]{Sun2025strandhead}
\textsc{Sun X., Cai Z., Tai Y., Yang J., Zhang Z.}:
\newblock Strandhead: Text to hair-disentangled 3d head avatars using human-centric priors.
\newblock In \emph{International Conference on Computer Vision (ICCV)} (2025).

\bibitem[SDF{\etalchar{*}}24]{sun2024metacap}
\textsc{Sun G., Dabral R., Fua P., Theobalt C., Habermann M.}:
\newblock {MetaCap}: Meta-learning priors from multi-view imagery for sparse-view human performance capture and rendering.
\newblock In \emph{European Conference on Computer Vision (ECCV)} (2024), pp.~341--361.

\bibitem[SDZ{\etalchar{*}}25]{sun2025dut}
\textsc{Sun G., Dabral R., Zhu H., Fua P., Theobalt C., Habermann M.}:
\newblock Real-time free-view human rendering from sparse-view {RGB} videos using double unprojected textures.
\newblock In \emph{Conference on Computer Vision and Pattern Recognition (CVPR)} (2025), pp.~562--573.

\bibitem[SF16]{schoenberger2016sfm}
\textsc{Sch\"{o}nberger J.~L., Frahm J.-M.}:
\newblock Structure-from-motion revisited.
\newblock In \emph{Conference on Computer Vision and Pattern Recognition (CVPR)} (2016).

\bibitem[SGBL23]{svitov2023dinar}
\textsc{Svitov D., Gudkov D., Bashirov R., Lempitsky V.}:
\newblock Dinar: Diffusion inpainting of neural textures for one-shot human avatars.
\newblock In \emph{International Conference on Computer Vision (ICCV)} (2023), pp.~7062--7072.

\bibitem[SGK{\etalchar{*}}23]{shen2023x}
\textsc{Shen K., Guo C., Kaufmann M., Zarate J.~J., Valentin J., Song J., Hilliges O.}:
\newblock X-avatar: Expressive human avatars.
\newblock In \emph{Conference on Computer Vision and Pattern Recognition (CVPR)} (2023), pp.~16911--16921.

\bibitem[SGN25]{schmidt2025becominglit}
\textsc{Schmidt J., Giebenhain S., Niessner M.}:
\newblock Becominglit: Relightable gaussian avatars with hybrid neural shading, 2025.

\bibitem[SGPT23]{shimada2023decaf}
\textsc{Shimada S., Golyanik V., P{\'e}rez P., Theobalt C.}:
\newblock Decaf: Monocular deformation capture for face and hand interactions.
\newblock \emph{Transactions on Graphics (TOG) 42}, 6 (2023), 1--16.

\bibitem[SGY{\etalchar{*}}24]{sanyal2024sculpt}
\textsc{Sanyal S., Ghosh P., Yang J., Black M.~J., Thies J., Bolkart T.}:
\newblock {SCULPT}: Shape-conditioned unpaired learning of pose-dependent clothed and textured human meshes.
\newblock In \emph{Conference on Computer Vision and Pattern Recognition (CVPR)} (2024), pp.~2362--2371.

\bibitem[SHL{\etalchar{*}}23]{su2023caphy}
\textsc{Su Z., Hu L., Lin S., Zhang H., Zhang S., Thies J., Liu Y.}:
\newblock Caphy: Capturing physical properties for animatable human avatars.
\newblock In \emph{International Conference on Computer Vision (ICCV)} (2023), pp.~14150--14160.

\bibitem[SHM{\etalchar{*}}18]{Shunsuke2018hair}
\textsc{Saito S., Hu L., Ma C., Ibayashi H., Luo L., Li H.}:
\newblock 3d hair synthesis using volumetric variational autoencoders.
\newblock \emph{Transactions on Graphics, (Proc. SIGGRAPH) 37}, 6 (2018).

\bibitem[SHM{\etalchar{*}}25]{shan2025populate}
\textsc{Shan M., He Z., Ma H., Juefei-Xu F., Zhang P., Hou T., Chuang C.-Y.}:
\newblock Populate-a-scene: Affordance-aware human video generation.
\newblock \emph{arXiv} (2025).

\bibitem[SHN{\etalchar{*}}19]{Saito2019pifu}
\textsc{Saito S., Huang Z., Natsume R., Morishima S., Kanazawa A., Li H.}:
\newblock Pifu: Pixel-aligned implicit function for high-resolution clothed human digitization.
\newblock In \emph{International Conference on Computer Vision (ICCV)} (2019), pp.~2304--2314.

\bibitem[SHS{\etalchar{*}}24]{shetty2024holoported}
\textsc{Shetty A., Habermann M., Sun G., Luvizon D., Golyanik V., Theobalt C.}:
\newblock Holoported characters: Real-time free-viewpoint rendering of humans from sparse {RGB} cameras.
\newblock In \emph{Conference on Computer Vision and Pattern Recognition (CVPR)} (2024), pp.~1206--1215.

\bibitem[SKS02]{sloan2002precomputed}
\textsc{Sloan P.~J., Kautz J., Snyder J.~M.}:
\newblock Precomputed radiance transfer for real-time rendering in dynamic, low-frequency lighting environments.
\newblock \emph{Transactions on Graphics (TOG) 21}, 3 (2002), 527--536.

\bibitem[SLB{\etalchar{*}}22]{sun2022putting}
\textsc{Sun Y., Liu W., Bao Q., Fu Y., Mei T., Black M.~J.}:
\newblock Putting people in their place: Monocular regression of 3d people in depth.
\newblock In \emph{Conference on Computer Vision and Pattern Recognition (CVPR)} (2022), pp.~13243--13252.

\bibitem[SLD23]{shao2023towards}
\textsc{Shao Y., Loy C.~C., Dai B.}:
\newblock Towards multi-layered 3d garments animation.
\newblock In \emph{International Conference on Computer Vision (ICCV)} (2023), pp.~14361--14370.

\bibitem[SLD24]{shao2024learning}
\textsc{Shao Y., Loy C.~C., Dai B.}:
\newblock Learning 3d garment animation from trajectories of a piece of cloth.
\newblock In \emph{Advances in Neural Information Processing Systems (NeurIPS)} (2024), pp.~41803--41825.

\bibitem[SLH{\etalchar{*}}19]{smith2019facsimile}
\textsc{Smith D., Loper M., Hu X., Mavroidis P., Romero J.}:
\newblock Facsimile: Fast and accurate scans from an image in less than a second.
\newblock In \emph{International Conference on Computer Vision (ICCV)} (2019), pp.~5330--5339.

\bibitem[SLL{\etalchar{*}}24a]{shin2024canonicalfusion}
\textsc{Shin J., Lee J., Lee S., Park M.-G., Kang J.-M., Yoon J.~H., Jeon H.-G.}:
\newblock Canonicalfusion: Generating drivable 3d human avatars from multiple images.
\newblock In \emph{European Conference on Computer Vision (ECCV)} (2024), Springer, pp.~38--56.

\bibitem[SLL{\etalchar{*}}24b]{Stuyck2024quaffure}
\textsc{Stuyck T., Lin G. W.-C., Larionov E., Chen H.-y., Bozic A., Sarafianos N., Roble D.}:
\newblock Quaffure: Real-time quasi-static neural hair simulation.
\newblock In \emph{Conference on Computer Vision and Pattern Recognition (CVPR)} (2024).

\bibitem[SM25]{sim2025persona}
\textsc{Sim G., Moon G.}:
\newblock Persona: Personalized whole-body 3d avatar with pose-driven deformations from a single image.
\newblock In \emph{International Conference on Computer Vision (ICCV)} (2025), pp.~1--11.

\bibitem[SMOT15]{sridhar2015fast}
\textsc{Sridhar S., Mueller F., Oulasvirta A., Theobalt C.}:
\newblock Fast and robust hand tracking using detection-guided optimization.
\newblock In \emph{Conference on Computer Vision and Pattern Recognition (CVPR)} (2015), pp.~3213--3221.

\bibitem[SNA{\etalchar{*}}21]{Sun2021humanhair}
\textsc{Sun T., Nam G., Aliaga C., Hery C., Ramamoorthi R.}:
\newblock {Human Hair Inverse Rendering using Multi-View Photometric data}.
\newblock In \emph{Computer Graphics Forum (CGF)} (2021).

\bibitem[SNF14]{schmidt2014dart}
\textsc{Schmidt T., Newcombe R.~A., Fox D.}:
\newblock Dart: Dense articulated real-time tracking.
\newblock \emph{Robotics: Science and Systems (RSS) 2}, 1 (2014), 1--9.

\bibitem[SOC19]{santesteban2019learning}
\textsc{Santesteban I., Otaduy M.~A., Casas D.}:
\newblock Learning-based animation of clothing for virtual try-on.
\newblock \emph{Computer Graphics Forum (CGF) 38}, 2 (2019), 355--366.

\bibitem[SOC22]{santesteban2022snug}
\textsc{Santesteban I., Otaduy M.~A., Casas D.}:
\newblock Snug: Self-supervised neural dynamic garments.
\newblock In \emph{Conference on Computer Vision and Pattern Recognition (CVPR)} (2022), pp.~8140--8150.

\bibitem[SOT13]{sridhar2013interactive}
\textsc{Sridhar S., Oulasvirta A., Theobalt C.}:
\newblock Interactive markerless articulated hand motion tracking using rgb and depth data.
\newblock In \emph{Conference on Computer Vision and Pattern Recognition (CVPR)} (2013), pp.~2456--2463.

\bibitem[SPJT10]{Sadeghi2010artist}
\textsc{Sadeghi I., Pritchett H., Jensen H.~W., Tamstorf R.}:
\newblock An artist friendly hair shading system.
\newblock \emph{Transactions on Graphics, (Proc. SIGGRAPH) 29}, 4 (2010).

\bibitem[SPZ{\etalchar{*}}24]{shao2024human4dit}
\textsc{Shao R., Pang Y., Zheng Z., Sun J., Liu Y.}:
\newblock 360-degree human video generation with 4d diffusion transformer.
\newblock \emph{Transactions on Graphics, (Proc. SIGGRAPH) 43}, 6 (2024), 1--13.

\bibitem[SRRW21]{sztrajman2021neural}
\textsc{Sztrajman A., Rainer G., Ritschel T., Weyrich T.}:
\newblock Neural brdf representation and importance sampling.
\newblock \emph{Computer Graphics Forum (CGF) 40}, 6 (2021), 332--346.

\bibitem[SS99]{stockman1999cone}
\textsc{Stockman A., Sharpe L.~T.}:
\newblock Cone spectral sensitivities and color matching.
\newblock \emph{Color vision: From genes to perception} (1999), 53--88.

\bibitem[SSK{\etalchar{*}}05]{scholz2005garment}
\textsc{Scholz V., Stich T., Keckeisen M., Wacker M., Magnor M.}:
\newblock Garment motion capture using color-coded patterns.
\newblock \emph{Computer Graphics Forum (CGF) 24}, 3 (2005), 439--448.

\bibitem[SSS{\etalchar{*}}24]{Saito2024relightable}
\textsc{Saito S., Schwartz G., Simon T., Li J., Nam G.}:
\newblock Relightable gaussian codec avatars.
\newblock In \emph{Conference on Computer Vision and Pattern Recognition (CVPR)} (2024), pp.~130--141.

\bibitem[SSSJ20]{Saito2020pifuhd}
\textsc{Saito S., Simon T., Saragih J., Joo H.}:
\newblock Pifuhd: Multi-level pixel-aligned implicit function for high-resolution 3d human digitization.
\newblock In \emph{Conference on Computer Vision and Pattern Recognition (CVPR)} (2020), pp.~84--93.

\bibitem[SSW{\etalchar{*}}23]{shen2023CT2Hair}
\textsc{Shen Y., Saito S., Wang Z., Maury O., Wu C., Hodgins J., Zheng Y., Nam G.}:
\newblock Ct2hair: High-fidelity 3d hair modeling using computed tomography.
\newblock \emph{Transactions on Graphics, (Proc. SIGGRAPH) 42}, 4 (2023), 1--13.

\bibitem[SSX{\etalchar{*}}25]{sarafianos2025garment3dgen}
\textsc{Sarafianos N., Stuyck T., Xiang X., Li Y., Popovic J., Ranjan R.}:
\newblock Garment3dgen: 3d garment stylization and texture generation.
\newblock In \emph{International Conference on 3D Vision (3DV)} (2025), IEEE, pp.~1382--1393.

\bibitem[STH{\etalchar{*}}24]{cage-based-deformation-survey}
\textsc{Ströter D., Thiery J.~M., Hormann K., Chen J., Chang Q., Besler S., Mueller-Roemer J.~S., Boubekeur T., Stork A., Fellner D.~W.}:
\newblock A survey on cage-based deformation of 3d models.
\newblock \emph{Computer Graphics Forum 43}, 2 (2024), e15060.

\bibitem[SWL{\etalchar{*}}24]{Shao2024splattingavatar}
\textsc{Shao Z., Wang Z., Li Z., Wang D., Lin X., Zhang Y., Fan M., Wang Z.}:
\newblock Splattingavatar: Realistic real-time human avatars with mesh-embedded gaussian splatting.
\newblock In \emph{Conference on Computer Vision and Pattern Recognition (CVPR)} (2024), pp.~1606--1616.

\bibitem[SWW{\etalchar{*}}20]{smith2020constraining}
\textsc{Smith B., Wu C., Wen H., Peluse P., Sheikh Y., Hodgins J.~K., Shiratori T.}:
\newblock Constraining dense hand surface tracking with elasticity.
\newblock \emph{Transactions on Graphics (TOG) 39}, 6 (2020), 1--14.

\bibitem[SWW{\etalchar{*}}23]{Sun2023next3d}
\textsc{Sun J., Wang X., Wang L., Li X., Zhang Y., Zhang H., Liu Y.}:
\newblock Next3d: Generative neural texture rasterization for 3d-aware head avatars.
\newblock In \emph{Conference on Computer Vision and Pattern Recognition (CVPR)} (2023), pp.~20991--21002.

\bibitem[SXS{\etalchar{*}}25]{shao2024isa4d}
\textsc{Shao R., Xu Y., Shen Y., Yang C., Zheng Y., Chen C., Liu Y., Wetzstein G.}:
\newblock Isa4d: Interspatial attention for efficient 4d human video generation.
\newblock \emph{SIGGRAPH} (2025).

\bibitem[SYH{\etalchar{*}}17]{shu2017neural}
\textsc{Shu Z., Yumer E., Hadap S., Sunkavalli K., Shechtman E., Samaras D.}:
\newblock Neural face editing with intrinsic image disentangling.
\newblock In \emph{Conference on Computer Vision and Pattern Recognition (CVPR)} (2017), pp.~5541--5550.

\bibitem[SYMB21]{Saito2021scanimate}
\textsc{Saito S., Yang J., Ma Q., Black M.~J.}:
\newblock Scanimate: Weakly supervised learning of skinned clothed avatar networks.
\newblock In \emph{Conference on Computer Vision and Pattern Recognition (CVPR)} (2021), pp.~2886--2897.

\bibitem[SYZR21]{su2021anerf}
\textsc{Su S.-Y., Yu F., Zollh{\"o}fer M., Rhodin H.}:
\newblock A-nerf: Articulated neural radiance fields for learning human shape, appearance, and pose.
\newblock In \emph{Advances in Neural Information Processing Systems (NIPS)} (2021).

\bibitem[SZG{\etalchar{*}}25]{sang2025lynxhighfidelitypersonalizedvideo}
\textsc{Sang S., Zhi T., Gu T., Liu J., Luo L.}:
\newblock Lynx: Towards high-fidelity personalized video generation, 2025.

\bibitem[SZH{\etalchar{*}}23]{Sklyarova2024haar}
\textsc{Sklyarova V., Zakharov E., Hilliges O., Black M.~J., Thies J.}:
\newblock Text-conditioned generative model of 3d strand-based human hairstyles.
\newblock \emph{Conference on Computer Vision and Pattern Recognition (CVPR)} (2023), 4703--4712.

\bibitem[SZP{\etalchar{*}}25]{sklyarova2025im2haircut}
\textsc{Sklyarova V., Zakharov E., Prinzler M., Becherini G., Black M., Thies J.}:
\newblock Im2haircut: Single-view strand-based hair reconstruction for human avatars.
\newblock In \emph{International Conference on Computer Vision (ICCV)} (2025).

\bibitem[SZPF16]{schoenberger2016mvs}
\textsc{Sch\"{o}nberger J.~L., Zheng E., Pollefeys M., Frahm J.-M.}:
\newblock Pixelwise view selection for unstructured multi-view stereo.
\newblock In \emph{European Conference on Computer Vision (ECCV)} (2016).

\bibitem[SZZ{\etalchar{*}}22]{shao2022doublefield}
\textsc{Shao R., Zhang H., Zhang H., Chen M., Cao Y.-P., Yu T., Liu Y.}:
\newblock Doublefield: Bridging the neural surface and radiance fields for high-fidelity human reconstruction and rendering.
\newblock In \emph{Conference on Computer Vision and Pattern Recognition (CVPR)} (2022), pp.~15872--15882.

\bibitem[TCL23]{tseng2023edge}
\textsc{Tseng J., Castellon R., Liu C.~K.}:
\newblock {EDGE:} editable dance generation from music.
\newblock In \emph{Conference on Computer Vision and Pattern Recognition (CVPR)} (2023), {IEEE}, pp.~448--458.

\bibitem[TDK{\etalchar{*}}25]{tang2025gaf}
\textsc{Tang J., Davoli D., Kirschstein T., Schoneveld L., Niessner M.}:
\newblock Gaf: Gaussian avatar reconstruction from monocular videos via multi-view diffusion.
\newblock In \emph{Conference on Computer Vision and Pattern Recognition (CVPR)} (2025), pp.~5546--5558.

\bibitem[TFT{\etalchar{*}}20]{Tewari2020state}
\textsc{Tewari A., Fried O., Thies J., Sitzmann V., Lombardi S., Sunkavalli K., Martin-Brualla R., Simon T., Saragih J., Nie{\ss}ner M., et~al.}:
\newblock State of the art on neural rendering.
\newblock \emph{Computer Graphics Forum (CGF) 39}, 2 (2020), 701--727.

\bibitem[TKB{\etalchar{*}}23]{TretschkNonRigidSurvey2023}
\textsc{Tretschk E., Kairanda N., {B R} M., Dabral R., Kortylewski A., Egger B., Habermann M., Fua P., Theobalt C., Golyanik V.}:
\newblock State of the art in dense monocular non-rigid 3d reconstruction.
\newblock \emph{Computer Graphics Forum (Eurographics State of the Art Reports)} (2023).

\bibitem[TKG{\etalchar{*}}24]{teotia2024gaussianheads}
\textsc{Teotia K., Kim H., Garrido P., Habermann M., Elgharib M., Theobalt C.}:
\newblock Gaussianheads: End-to-end learning of drivable gaussian head avatars from coarse-to-fine representations.
\newblock \emph{Transactions on Graphics, (Proc. SIGGRAPH) 43}, 6 (2024), 1--12.

\bibitem[TPT16]{tkach2016sphere}
\textsc{Tkach A., Pauly M., Tagliasacchi A.}:
\newblock Sphere-meshes for real-time hand modeling and tracking.
\newblock \emph{Transactions on Graphics (TOG) 35}, 6 (2016), 1--11.

\bibitem[TRM{\etalchar{*}}25]{teotia2025audiodrivenuniversalgaussianhead}
\textsc{Teotia K., Rhodin H., Mendiratta M., Kim H., Habermann M., Theobalt C.}:
\newblock Audio-driven universal gaussian head avatars, 2025.

\bibitem[TSD{\etalchar{*}}24]{tu2024tele}
\textsc{Tu H., Shao R., Dong X., Zheng S., Zhang H., Chen L., Wang M., Li W., Ma S., Zhang S., et~al.}:
\newblock Tele-aloha: A low-budget and high-authenticity telepresence system using sparse rgb cameras.
\newblock In \emph{SIGGRAPH Conference Papers (SA)} (2024), pp.~145--162.

\bibitem[TST{\etalchar{*}}15]{tagliasacchi2015robust}
\textsc{Tagliasacchi A., Schr{\"o}der M., Tkach A., Bouaziz S., Botsch M., Pauly M.}:
\newblock Robust articulated-icp for real-time hand tracking.
\newblock \emph{Computer Graphics Forum (CGF) 34}, 5 (2015), 101--114.

\bibitem[TSTPM21]{tiwari2021neural}
\textsc{Tiwari G., Sarafianos N., Tung T., Pons-Moll G.}:
\newblock Neural-gif: Neural generalized implicit functions for animating people in clothing.
\newblock In \emph{International Conference on Computer Vision (ICCV)} (2021), pp.~11708--11718.

\bibitem[TTM{\etalchar{*}}22]{Tewari2022advances}
\textsc{Tewari A., Thies J., Mildenhall B., Srinivasan P., Tretschk E., Yifan W., Lassner C., Sitzmann V., Martin-Brualla R., Lombardi S., et~al.}:
\newblock Advances in neural rendering.
\newblock \emph{Computer Graphics Forum (CGF) 41}, 2 (2022), 703--735.

\bibitem[TTS{\etalchar{*}}24]{Takimoto2024drhair}
\textsc{Takimoto Y., Takehara H., Sato H., Zhu Z., Zheng B.}:
\newblock Dr.hair: Reconstructing scalp-connected hair strands without pre-training via differentiable rendering of line segments.
\newblock In \emph{Conference on Computer Vision and Pattern Recognition (CVPR)} (2024).

\bibitem[TXT{\etalchar{*}}25]{tan2025dressrecon}
\textsc{Tan J., Xiang D., Tulsiani S., Ramanan D., Yang G.}:
\newblock Dressrecon: Freeform 4d human reconstruction from monocular video.
\newblock In \emph{International Conference on 3D Vision (3DV)} (2025), pp.~250--260.

\bibitem[TZH{\etalchar{*}}24a]{tran2024voodooxp}
\textsc{Tran P., Zakharov E., Ho L.-N., Karmanov A., Bermudez~Venegas A., Goldwhite M., Agarwal A., Hu L., Tran A., Li H.}:
\newblock Voodoo xp: Expressive one-shot head reenactment for vr telepresence.
\newblock \emph{Transactions on Graphics, (Proc. SIGGRAPH) 43}, 6 (2024), 1--26.

\bibitem[TZH{\etalchar{*}}24b]{tran2024voodoo3d}
\textsc{Tran P., Zakharov E., Ho L.-N., Tran A.~T., Hu L., Li H.}:
\newblock Voodoo 3d: Volumetric portrait disentanglement for one-shot 3d head reenactment.
\newblock In \emph{Conference on Computer Vision and Pattern Recognition (CVPR)} (2024), pp.~10336--10348.

\bibitem[TZN19]{thies2019deferred}
\textsc{Thies J., Zollh{\"o}fer M., Nie{\ss}ner M.}:
\newblock Deferred neural rendering: Image synthesis using neural textures.
\newblock \emph{Transactions on Graphics (TOG) 38}, 4 (2019), 1--12.

\bibitem[TZS{\etalchar{*}}16]{Thies2016face}
\textsc{Thies J., Zollh{\"o}fer M., Stamminger M., Theobalt C., Nie{\ss}ner M.}:
\newblock Face2face: Real-time face capture and reenactment of rgb videos.
\newblock In \emph{Conference on Computer Vision and Pattern Recognition (CVPR)} (2016).

\bibitem[TZT{\etalchar{*}}25]{taubner2025mvp4d}
\textsc{Taubner F., Zhang R., Tuli M., Bahmani S., Lindell D.~B.}:
\newblock Mvp4d: Multi-view portrait video diffusion for animatable 4d avatars.
\newblock In \emph{SIGGRAPH Asia Conference Papers (SA)} (2025), pp.~1--11.

\bibitem[TZTL25]{Taubner2025cap4d}
\textsc{Taubner F., Zhang R., Tuli M., Lindell D.~B.}:
\newblock Cap4d: Creating animatable 4d portrait avatars with morphable multi-view diffusion models.
\newblock In \emph{Conference on Computer Vision and Pattern Recognition (CVPR)} (2025), pp.~5318--5330.

\bibitem[unr]{unrealengine}
{Epic Games} unreal engine.
\newblock \href{https://www.unrealengine.com}{Unreal Engine}.
\newblock Accessed: 2025-10-10.

\bibitem[VCH{\etalchar{*}}21]{villegas2021contact}
\textsc{Villegas R., Ceylan D., Hertzmann A., Yang J., Saito J.}:
\newblock Contact-aware retargeting of skinned motion.
\newblock In \emph{Proceedings of the IEEE/CVF International Conference on Computer Vision} (2021), pp.~9720--9729.

\bibitem[VCR{\etalchar{*}}18]{varol2018bodynet}
\textsc{Varol G., Ceylan D., Russell B., Yang J., Yumer E., Laptev I., Schmid C.}:
\newblock Bodynet: Volumetric inference of 3d human body shapes.
\newblock In \emph{European Conference on Computer Vision (ECCV)} (2018), pp.~20--36.

\bibitem[VGC24]{vidaurre2025diffusedwrinkles}
\textsc{Vidaurre R., Garces E., Casas D.}:
\newblock Diffusedwrinkles: A diffusion-based model for data-driven garment animation.
\newblock \emph{British Machine Vision Conference (BMVC)} (2024).

\bibitem[VRM{\etalchar{*}}17]{varol2017surreal}
\textsc{Varol G., Romero J., Martin X., Mahmood N., Black M.~J., Laptev I., Schmid C.}:
\newblock Learning from synthetic humans.
\newblock In \emph{Conference on Computer Vision and Pattern Recognition (CVPR)} (2017), pp.~109--117.

\bibitem[VSGC20]{vidaurre2020fully}
\textsc{Vidaurre R., Santesteban I., Garces E., Casas D.}:
\newblock Fully convolutional graph neural networks for parametric virtual try-on.
\newblock \emph{Computer Graphics Forum (CGF) 39}, 8 (2020), 145--156.

\bibitem[WAGT24]{wang2024intrinsicavatar}
\textsc{Wang S., Antic B., Geiger A., Tang S.}:
\newblock Intrinsicavatar: Physically based inverse rendering of dynamic humans from monocular videos via explicit ray tracing.
\newblock In \emph{Conference on Computer Vision and Pattern Recognition (CVPR)} (2024), pp.~1877--1888.

\bibitem[War92]{ward1992measuring}
\textsc{Ward G.~J.}:
\newblock Measuring and modeling anisotropic reflection.
\newblock In \emph{Proceedings of the 19th annual conference on Computer graphics and interactive techniques} (1992), pp.~265--272.

\bibitem[WCHW24]{wang2024survey}
\textsc{Wang R., Cao Y., Han K., Wong K.-Y.~K.}:
\newblock A survey on 3d human avatar modeling--from reconstruction to generation.
\newblock \emph{arXiv} (2024).

\bibitem[WCK{\etalchar{*}}25]{wang2025vggt}
\textsc{Wang J., Chen M., Karaev N., Vedaldi A., Rupprecht C., Novotny D.}:
\newblock Vggt: Visual geometry grounded transformer.
\newblock In \emph{Conference on Computer Vision and Pattern Recognition (CVPR)} (2025), Computer Vision Foundation / {IEEE}.

\bibitem[WCL{\etalchar{*}}23]{wang2023fast}
\textsc{Wang T., Chen J., Li D., Liu X., Wang H., Zhou K.}:
\newblock Fast gpu-based two-way continuous collision handling.
\newblock \emph{Transactions on Graphics (TOG) 42}, 5 (2023), 1--15.

\bibitem[WCY{\etalchar{*}}22]{wang2022faceverse}
\textsc{Wang L., Chen Z., Yu T., Ma C., Li L., Liu Y.}:
\newblock Faceverse: a fine-grained and detail-controllable 3d face morphable model from a hybrid dataset.
\newblock In \emph{Conference on Computer Vision and Pattern Recognition (CVPR)} (2022).

\bibitem[WJMLH01]{wann2001practical}
\textsc{Wann~Jensen H., Marschner S.~R., Levoy M., Hanrahan P.}:
\newblock A practical model for subsurface light transport.
\newblock In \emph{SIGGRAPH} (2001), {ACM}, pp.~511--518.

\bibitem[WLC{\etalchar{*}}24a]{dust3r_cvpr24}
\textsc{Wang S., Leroy V., Cabon Y., Chidlovskii B., Revaud J.}:
\newblock Dust3r: Geometric 3d vision made easy.
\newblock In \emph{CVPR} (2024).

\bibitem[WLC{\etalchar{*}}24b]{wang2024dust3r}
\textsc{Wang S., Leroy V., Cabon Y., Chidlovskii B., Revaud J.}:
\newblock Dust3r: Geometric 3d vision made easy.
\newblock In \emph{Conference on Computer Vision and Pattern Recognition (CVPR)} (2024), Computer Vision Foundation / {IEEE}, pp.~20697--20709.

\bibitem[WLD{\etalchar{*}}24]{wang2024disentangled}
\textsc{Wang J., Liu Y., Dou Z., Yu Z., Liang Y., Lin C., Xie R., Song L., Li X., Wang W.}:
\newblock Disentangled clothed avatar generation from text descriptions.
\newblock In \emph{European Conference on Computer Vision (ECCV)} (2024), pp.~381--401.

\bibitem[WLZ{\etalchar{*}}25]{wang2025multi}
\textsc{Wang Z., Li Y., Zeng Y., Guo Y., Lin D., Xue T., Dai B.}:
\newblock Multi-identity human image animation with structural video diffusion.
\newblock \emph{arXiv} (2025).

\bibitem[WMB19]{wang2019hand}
\textsc{Wang B., Matcuk G., Barbi{\v{c}} J.}:
\newblock Hand modeling and simulation using stabilized magnetic resonance imaging.
\newblock \emph{Transactions on Graphics (TOG) 38}, 4 (2019), 1--14.

\bibitem[WML21]{wang2021facevid2vid}
\textsc{Wang T.-C., Mallya A., Liu M.-Y.}:
\newblock One-shot free-view neural talking-head synthesis for video conferencing.
\newblock In \emph{Conference on Computer Vision and Pattern Recognition (CVPR)} (2021), pp.~10039--10049.

\bibitem[WMM{\etalchar{*}}21]{Wang2021metaavatar}
\textsc{Wang S., Mihajlovic M., Ma Q., Geiger A., Tang S.}:
\newblock Metaavatar: Learning animatable clothed human models from few depth images.
\newblock \emph{Advances in Neural Information Processing Systems (NeurIPS) 34} (2021), 2810--2822.

\bibitem[WNS{\etalchar{*}}22]{Wang2022hvh}
\textsc{Wang Z., Nam G., Stuyck T., Lombardi S., Zollhoefer M., Hodgins J., Lassner C.}:
\newblock Hvh: Learning a hybrid neural volumetric representation for dynamic hair performance capture.
\newblock In \emph{Conference on Computer Vision and Pattern Recognition (CVPR)} (2022).

\bibitem[WNS{\etalchar{*}}23]{Wang2023neuwigs}
\textsc{Wang Z., Nam G., Stuyck T., Lombardi S., Cao C., Saragih J., Zollhoefer M., Hodgins J., Lassner C.}:
\newblock Neuwigs: A neural dynamic model for volumetric hair capture and animation.
\newblock In \emph{Conference on Computer Vision and Pattern Recognition (CVPR)} (2023).

\bibitem[Woo80]{woodham1980photometric}
\textsc{Woodham R.~J.}:
\newblock Photometric method for determining surface orientation from multiple images.
\newblock \emph{Optical engineering 19}, 1 (1980), 139--144.

\bibitem[WOR11]{wang2011data}
\textsc{Wang H., O'Brien J.~F., Ramamoorthi R.}:
\newblock Data-driven elastic models for cloth: modeling and measurement.
\newblock \emph{Transactions on Graphics, (Proc. SIGGRAPH) 30}, 4 (2011), 1--12.

\bibitem[WPJT25]{wu2025animportrait3d}
\textsc{Wu Y., Prinzler M., Jin X., Tang S.}:
\newblock Text-based animatable 3d avatars with morphable model alignment.
\newblock In \emph{SIGGRAPH Conference Papers (SA)} (2025), pp.~1--11.

\bibitem[WPW{\etalchar{*}}25]{Wang2025fresa}
\textsc{Wang R., Prada F., Wang Z., Jiang Z., Yin C., Li J., Saito S., Santesteban I., Romero J., Joshi R., et~al.}:
\newblock Fresa: Feedforward reconstruction of personalized skinned avatars from few images.
\newblock In \emph{Conference on Computer Vision and Pattern Recognition (CVPR)} (2025).

\bibitem[WRG{\etalchar{*}}09]{wang2009all}
\textsc{Wang J., Ren P., Gong M., Snyder J., Guo B.}:
\newblock All-frequency rendering of dynamic, spatially-varying reflectance.
\newblock \emph{Transactions on Graphics, (Proc. SIGGRAPH Asia) 28}, 5 (2009), 133.

\bibitem[WS00]{wyszecki2000color}
\textsc{Wyszecki G., Stiles W.~S.}:
\newblock \emph{Color science: concepts and methods, quantitative data and formulae}.
\newblock John wiley \& sons, 2000.

\bibitem[WSCKS23]{weng2023personnerf}
\textsc{Weng C.-Y., Srinivasan P.~P., Curless B., Kemelmacher-Shlizerman I.}:
\newblock Personnerf: Personalized reconstruction from photo collections.
\newblock In \emph{Conference on Computer Vision and Pattern Recognition (CVPR)} (2023), pp.~524--533.

\bibitem[WSS{\etalchar{*}}25]{wang2025relightable}
\textsc{Wang S., Simon T., Santesteban I., Bagautdinov T., Li J., Agrawal V., Prada F., Yu S.-I., Nalbone P., Gramlich M., et~al.}:
\newblock Relightable full-body gaussian codec avatars.
\newblock In \emph{SIGGRAPH Conference Papers (SA)} (2025), pp.~1--12.

\bibitem[WTD{\etalchar{*}}25]{wang2025mosa}
\textsc{Wang H., Tang H., Di D., Zhang Z., Zuo W., Gao F., Ma S., Zhang S.}:
\newblock Mosa: Motion-coherent human video generation via structure-appearance decoupling.
\newblock \emph{arXiv} (2025).

\bibitem[WWA{\etalchar{*}}25]{wan2025wan}
\textsc{Wan T., Wang A., Ai B., Wen B., Mao C., Xie C.-W., Chen D., Yu F., Zhao H., Yang J., et~al.}:
\newblock Wan: Open and advanced large-scale video generative models.
\newblock \emph{arXiv} (2025).

\bibitem[WWJ{\etalchar{*}}25]{wang2025fantasytalking}
\textsc{Wang M., Wang Q., Jiang F., Fan Y., Zhang Y., Qi Y., Zhao K., Xu M.}:
\newblock Fantasytalking: Realistic talking portrait generation via coherent motion synthesis, 2025.

\bibitem[WWO{\etalchar{*}}19]{wang2019detecting}
\textsc{Wang S.-Y., Wang O., Owens A., Zhang R., Efros A.~A.}:
\newblock Detecting photoshopped faces by scripting photoshop.
\newblock In \emph{International Conference on Computer Vision (ICCV)} (2019), pp.~10072--10081.

\bibitem[WWS{\etalchar{*}}20]{wang2020mead}
\textsc{Wang K., Wu Q., Song L., Yang Z., Wu W., Qian C., He R., Qiao Y., Loy C.~C.}:
\newblock Mead: A large-scale audio-visual dataset for emotional talking-face generation.
\newblock In \emph{European Conference on Computer Vision (ECCV)} (2020), Springer, pp.~700--717.

\bibitem[WXT{\etalchar{*}}25]{wang2025dgh}
\textsc{Wang J., Xu Y., Tretschk E., Wang Z., Ianina A., Bozic A., Neumann U., Tung T.}:
\newblock Dgh: Dynamic gaussian hair.
\newblock In \emph{Advances in Neural Information Processing Systems (NeurIPS)} (2025).

\bibitem[WYK{\etalchar{*}}24]{Wu2024monohair}
\textsc{Wu K., Yang L., Kuang Z., Feng Y., Han X., Shen Y., Fu H., Zhou K., Zheng Y.}:
\newblock Monohair: High-fidelity hair modeling from a monocular video.
\newblock In \emph{Conference on Computer Vision and Pattern Recognition (CVPR)} (2024).

\bibitem[WYL{\etalchar{*}}25]{wei2025echovideo}
\textsc{Wei J., Yan S., Lin W., Liu B., Chen R., Guo M.}:
\newblock Echovideo: Identity-preserving human video generation by multimodal feature fusion.
\newblock \emph{arXiv} (2025).

\bibitem[WYY{\etalchar{*}}22]{Wu2022neuralhdhair}
\textsc{Wu K., Ye Y., Yang L., Fu H., Zhou K., Zhengl Y.}:
\newblock {NeuralHDHair}: Automatic high-fidelity hair modeling from a single image using implicit neural representations.
\newblock In \emph{Conference on Computer Vision and Pattern Recognition (CVPR)} (2022), pp.~1516--1525.

\bibitem[WZG{\etalchar{*}}26]{wang2026flowact}
\textsc{Wang L., Zhu Y., Ge Z., Zheng Y., Zhang L., Hu T., Qin S., Luo M., Zhang J., Chen X., et~al.}:
\newblock Flowact-r1: Towards interactive humanoid video generation.
\newblock \emph{arXiv} (2026).

\bibitem[WZL{\etalchar{*}}18]{wang2018toward}
\textsc{Wang B., Zheng H., Liang X., Chen Y., Lin L., Yang M.}:
\newblock Toward characteristic-preserving image-based virtual try-on network.
\newblock In \emph{European Conference on Computer Vision (ECCV)} (2018), pp.~589--604.

\bibitem[WZN{\etalchar{*}}25]{wang20254real}
\textsc{Wang C., Zhuang P., Ngo T.~D., Menapace W., Siarohin A., Vasilkovsky M., Skorokhodov I., Tulyakov S., Wonka P., Lee H.-Y.}:
\newblock 4real-video: Learning generalizable photo-realistic 4d video diffusion.
\newblock In \emph{Conference on Computer Vision and Pattern Recognition (CVPR)} (2025), pp.~17723--17732.

\bibitem[WZR{\etalchar{*}}24]{wen2024gomavatar}
\textsc{Wen J., Zhao X., Ren Z., Schwing A.~G., Wang S.}:
\newblock Gomavatar: Efficient animatable human modeling from monocular video using gaussians-on-mesh.
\newblock In \emph{Conference on Computer Vision and Pattern Recognition (CVPR)} (2024), pp.~2059--2069.

\bibitem[XAS21]{Xu2021hnerf}
\textsc{Xu H., Alldieck T., Sminchisescu C.}:
\newblock H-nerf: Neural radiance fields for rendering and temporal reconstruction of humans in motion.
\newblock \emph{Advances in Neural Information Processing Systems (NeurIPS) 34} (2021), 14955--14966.

\bibitem[XBS{\etalchar{*}}22]{xiang2022DressingAvatars}
\textsc{Xiang D., Bagautdinov T., Stuyck T., Prada F., Romero J., Xu W., Saito S., Guo J., Smith B., Shiratori T., et~al.}:
\newblock Dressing avatars: Deep photorealistic appearance for physically simulated clothing.
\newblock \emph{Transactions on Graphics, (Proc. SIGGRAPH Asia) 41}, 6 (2022), 1--15.

\bibitem[XCG{\etalchar{*}}24]{xu2024vasa1}
\textsc{Xu S., Chen G., Guo Y., Yang J., Li C., Zang Z., Zhang Y., Tong X., Guo B.}:
\newblock {VASA-1:} lifelike audio-driven talking faces generated in real time.
\newblock In \emph{Advances in Neural Information Processing Systems (NeurIPS)} (2024).

\bibitem[XCL{\etalchar{*}}24]{Xu2024gaussian}
\textsc{Xu Y., Chen B., Li Z., Zhang H., Wang L., Zheng Z., Liu Y.}:
\newblock Gaussian head avatar: Ultra high-fidelity head avatar via dynamic gaussians.
\newblock In \emph{Conference on Computer Vision and Pattern Recognition (CVPR)} (2024), pp.~1931--1941.

\bibitem[XGGZ24]{Xiang2024flashavatar}
\textsc{Xiang J., Gao X., Guo Y., Zhang J.}:
\newblock Flashavatar: High-fidelity head avatar with efficient gaussian embedding.
\newblock In \emph{Conference on Computer Vision and Pattern Recognition (CVPR)} (2024), pp.~1802--1812.

\bibitem[XGL{\etalchar{*}}25]{xie2025x}
\textsc{Xie Y., Gu T., Li Z., Zhang C., Song G., Zhao X., Liang C., Jiang J., Xu H., Luo L.}:
\newblock X-streamer: Unified human world modeling with audiovisual interaction.
\newblock \emph{arXiv} (2025).

\bibitem[XMR{\etalchar{*}}11]{Xu2011interactive}
\textsc{Xu K., Ma L.-Q., Ren B., Wang R., Hu S.-M.}:
\newblock Interactive hair rendering and appearance editing under environment lighting.
\newblock \emph{Transactions on Graphics (TOG) 30}, 6 (2011), 173:1--173:10.

\bibitem[XPB{\etalchar{*}}21]{xiang2021ClothingCodecAvatar}
\textsc{Xiang D., Prada F., Bagautdinov T., Xu W., Dong Y., Wen H., Hodgins J., Wu C.}:
\newblock Modeling clothing as a separate layer for an animatable human avatar.
\newblock \emph{Transactions on Graphics, (Proc. SIGGRAPH Asia) 40}, 6 (2021), 1--15.

\bibitem[XPC{\etalchar{*}}23]{Xiang2023drivable}
\textsc{Xiang D., Prada F., Cao Z., Guo K., Wu C., Hodgins J., Bagautdinov T.}:
\newblock Drivable avatar clothing: Faithful full-body telepresence with dynamic clothing driven by sparse rgb-d input.
\newblock In \emph{SIGGRAPH Asia Conference Papers (SA)} (2023), pp.~1--11.

\bibitem[XPG{\etalchar{*}}24]{xu2024relightable}
\textsc{Xu Z., Peng S., Geng C., Mou L., Yan Z., Sun J., Bao H., Zhou X.}:
\newblock Relightable and animatable neural avatar from sparse-view video.
\newblock In \emph{Conference on Computer Vision and Pattern Recognition (CVPR)} (2024), pp.~990--1000.

\bibitem[XPWH20]{xiang2020monoclothcap}
\textsc{Xiang D., Prada F., Wu C., Hodgins J.}:
\newblock Monoclothcap: Towards temporally coherent clothing capture from monocular rgb video.
\newblock In \emph{International Conference on 3D Vision (3DV)} (2020), pp.~322--332.

\bibitem[XTS{\etalchar{*}}22]{xie2022neural}
\textsc{Xie Y., Takikawa T., Saito S., Litany O., Yan S., Khan N., Tombari F., Tompkin J., Sitzmann V., Sridhar S.}:
\newblock Neural fields in visual computing and beyond.
\newblock \emph{Computer Graphics Forum (CGF) 41}, 2 (2022), 641--676.

\bibitem[XWW{\etalchar{*}}14]{Zexiang2014dynamic}
\textsc{Xu Z., Wu H.-T., Wang L., Zheng C., Tong X., Qi Y.}:
\newblock Dynamic hair capture using spacetime optimization.
\newblock \emph{Transactions on Graphics, (Proc. SIGGRAPH) 33}, 6 (2014).

\bibitem[XWZ{\etalchar{*}}22]{xie2022vfhq}
\textsc{Xie L., Wang X., Zhang H., Dong C., Shan Y.}:
\newblock Vfhq: A high-quality dataset and benchmark for video face super-resolution.
\newblock In \emph{Conference on Computer Vision and Pattern Recognition (CVPR)} (2022), pp.~657--666.

\bibitem[XWZ{\etalchar{*}}24]{Xu2024gphm}
\textsc{Xu Y., Wang L., Zheng Z., Su Z., Liu Y.}:
\newblock 3d gaussian parametric head model.
\newblock In \emph{European Conference on Computer Vision (ECCV)} (2024), Springer, pp.~129--147.

\bibitem[XYC{\etalchar{*}}23]{Xiu2023econ}
\textsc{Xiu Y., Yang J., Cao X., Tzionas D., Black M.~J.}:
\newblock Econ: Explicit clothed humans optimized via normal integration.
\newblock In \emph{Conference on Computer Vision and Pattern Recognition (CVPR)} (2023), pp.~512--523.

\bibitem[XYL{\etalchar{*}}24]{xiu2024puzzleavatar}
\textsc{Xiu Y., Ye Y., Liu Z., Tzionas D., Black M.~J.}:
\newblock Puzzleavatar: Assembling 3d avatars from personal albums.
\newblock \emph{Transactions on Graphics, (Proc. SIGGRAPH) 43}, 6 (2024), 1--15.

\bibitem[XYTB22]{Xiu2022icon}
\textsc{Xiu Y., Yang J., Tzionas D., Black M.~J.}:
\newblock Icon: Implicit clothed humans obtained from normals.
\newblock In \emph{Conference on Computer Vision and Pattern Recognition (CVPR)} (2022), pp.~13286--13296.

\bibitem[XZJJ25]{xie2025physanimatorphysicsguidedgenerativecartoon}
\textsc{Xie T., Zhao Y., Jiang Y., Jiang C.}:
\newblock Physanimator: Physics-guided generative cartoon animation, 2025.

\bibitem[XZL{\etalchar{*}}23]{xu2023xagen}
\textsc{Xu Z., Zhang J., Liew J.~H., Feng J., Shou M.~Z.}:
\newblock {XAGen}: {3D} expressive human avatars generation.
\newblock In \emph{Advances in Neural Information Processing Systems (NeurIPS)} (2023).

\bibitem[YA06]{Yuksel2006rendering}
\textsc{Yuksel C., Akleman E.}:
\newblock Rendering hair with global illumination.
\newblock In \emph{ACM SIGGRAPH 2006 Research Posters} (New York, NY, USA, 2006), ACM Press, p.~124.

\bibitem[YBM{\etalchar{*}}24]{yu2024one2avatar}
\textsc{Yu Z., Bai Z., Meka A., Tan F., Xu Q., Pandey R., Fanello S., Park H.~S., Zhang Y.}:
\newblock One2avatar: Generative implicit head avatar for few-shot user adaptation.
\newblock \emph{arXiv} (2024).

\bibitem[YCG{\etalchar{*}}24]{yang2024have}
\textsc{Yang X., Chen X., Gao D., Wang S., Han X., Wang B.}:
\newblock Have-fun: Human avatar reconstruction from few-shot unconstrained images.
\newblock In \emph{Conference on Computer Vision and Pattern Recognition (CVPR)} (2024), pp.~742--752.

\bibitem[YCW{\etalchar{*}}22]{yoon2022learning}
\textsc{Yoon J.~S., Ceylan D., Wang T.~Y., Lu J., Yang J., Shu Z., Park H.~S.}:
\newblock Learning motion-dependent appearance for high-fidelity rendering of dynamic humans from a single camera.
\newblock In \emph{Conference on Computer Vision and Pattern Recognition (CVPR)} (2022), pp.~3407--3417.

\bibitem[YFZ{\etalchar{*}}23]{yu2023nofa}
\textsc{Yu W., Fan Y., Zhang Y., Wang X., Yin F., Bai Y., Cao Y.-P., Shan Y., Wu Y., Sun Z., et~al.}:
\newblock Nofa: Nerf-based one-shot facial avatar reconstruction.
\newblock In \emph{SIGGRAPH Conference Papers (SA)} (2023), pp.~1--12.

\bibitem[YK08]{Yuksel2008deep}
\textsc{Yuksel C., Keyser J.}:
\newblock Deep opacity maps.
\newblock \emph{Computer Graphics Forum (CGF) 27}, 2 (2008), 675--680.

\bibitem[YLB{\etalchar{*}}25]{yu2025humanram}
\textsc{Yu Z., Li Z., Bao H., Yang C., Zhou X.}:
\newblock Humanram: Feed-forward human reconstruction and animation model using transformers.
\newblock In \emph{SIGGRAPH Conference Papers (SA)} (2025).

\bibitem[YLL{\etalchar{*}}25]{yang2025sigman}
\textsc{Yang Y., Liu F., Lu Y., Zhao Q., Wu P., Zhai W., Yi R., Cao Y., Ma L., Zha Z.-J., et~al.}:
\newblock Sigman: Scaling 3d human gaussian generation with millions of assets.
\newblock In \emph{International Conference on Computer Vision (ICCV)} (2025), pp.~1--11.

\bibitem[YLS{\etalchar{*}}25]{yu2025gaia}
\textsc{Yu Z., Li T., Sun J., Shapira O., Park S., Stengel M., Chan M., Li X., Wang W., Nagano K., et~al.}:
\newblock Gaia: Generative animatable interactive avatars with expression-conditioned gaussians.
\newblock In \emph{SIGGRAPH Conference Papers (SA)} (2025), pp.~1--10.

\bibitem[YSC{\etalchar{*}}25]{yang2025echomotion}
\textsc{Yang Y., Sheng H., Cai S., Lin J., Wang J., Deng B., Lu J., Wang H., Ye J.}:
\newblock Echomotion: Unified human video and motion generation via dual-modality diffusion transformer.
\newblock \emph{arXiv} (2025).

\bibitem[YSZZ19]{Yang2019dynamic}
\textsc{Yang L., Shi Z., Zheng Y., Zhou K.}:
\newblock Dynamic hair modeling from monocular videos using deep neural networks.
\newblock \emph{Transactions on Graphics, (Proc. SIGGRAPH) 38}, 6 (2019).

\bibitem[YTB{\etalchar{*}}21]{yenamandra2021i3dmm}
\textsc{Yenamandra T., Tewari A., Bernard F., Seidel H.-P., Elgharib M., Cremers D., Theobalt C.}:
\newblock i3dmm: Deep implicit 3d morphable model of human heads.
\newblock In \emph{Conference on Computer Vision and Pattern Recognition (CVPR)} (2021), pp.~12803--12813.

\bibitem[YVN{\etalchar{*}}22]{yang2022banmo}
\textsc{Yang G., Vo M., Neverova N., Ramanan D., Vedaldi A., Joo H.}:
\newblock Banmo: Building animatable 3d neural models from many casual videos.
\newblock In \emph{Conference on Computer Vision and Pattern Recognition (CVPR)} (2022), pp.~2863--2873.

\bibitem[YWO09]{Yamaguchi2009video}
\textsc{Yamaguchi T., Wilburn B., Ofek E.}:
\newblock Video-based modeling of dynamic hair.
\newblock In \emph{Pacific-Rim Symposium on Image and Video Technology (PSIVT)} (2009), pp.~585--596.

\bibitem[YZG{\etalchar{*}}21]{yu2021function4d}
\textsc{Yu T., Zheng Z., Guo K., Liu P., Dai Q., Liu Y.}:
\newblock Function4d: Real-time human volumetric capture from very sparse consumer rgbd sensors.
\newblock In \emph{Conference on Computer Vision and Pattern Recognition (CVPR)} (2021), pp.~5746--5756.

\bibitem[YZM{\etalchar{*}}24]{yang2024vrmm}
\textsc{Yang H., Zheng M., Ma C., Lai Y.-K., Wan P., Huang H.}:
\newblock Vrmm: A volumetric relightable morphable head model.
\newblock In \emph{SIGGRAPH Conference Papers (SA)} (2024), pp.~1--11.

\bibitem[YZR{\etalchar{*}}24]{ye2024real3d}
\textsc{Ye Z., Zhong T., Ren Y., Yang J., Li W., Huang J., Jiang Z., He J., Huang R., Liu J., et~al.}:
\newblock Real3d-portrait: One-shot realistic 3d talking portrait synthesis.
\newblock In \emph{International Conference on Learning Representations (ICLR)} (2024).

\bibitem[ZAB{\etalchar{*}}22]{Zheng2022avatar}
\textsc{Zheng Y., Abrevaya V.~F., B{\"u}hler M.~C., Chen X., Black M.~J., Hilliges O.}:
\newblock Im avatar: Implicit morphable head avatars from videos.
\newblock In \emph{Conference on Computer Vision and Pattern Recognition (CVPR)} (2022), pp.~13545--13555.

\bibitem[ZBBT25]{Zielonka2025gem}
\textsc{Zielonka W., Bolkart T., Beeler T., Thies J.}:
\newblock Gaussian eigen models for human heads.
\newblock \emph{Conference on Computer Vision and Pattern Recognition (CVPR)} (2025).

\bibitem[ZBS{\etalchar{*}}25]{Zielonka2023d3ga}
\textsc{Zielonka W., Bagautdinov T., Saito S., Zollh{\"o}fer M., Thies J., Romero J.}:
\newblock Drivable 3d gaussian avatars.
\newblock \emph{International Conference on 3D Vision (3DV)} (2025).

\bibitem[ZBT22]{Zielonka2022mica}
\textsc{Zielonka W., Bolkart T., Thies J.}:
\newblock Towards metrical reconstruction of human faces.
\newblock In \emph{European Conference on Computer Vision (ECCV)} (2022), Springer, pp.~250--269.

\bibitem[ZBT23]{Zielonka2023insta}
\textsc{Zielonka W., Bolkart T., Thies J.}:
\newblock Instant volumetric head avatars.
\newblock In \emph{Conference on Computer Vision and Pattern Recognition (CVPR)} (2023), pp.~4574--4584.

\bibitem[ZBT{\etalchar{*}}24]{Zhang2024gslrm}
\textsc{Zhang K., Bi S., Tan H., Xiangli Y., Zhao N., Sunkavalli K., Xu Z.}:
\newblock Gs-lrm: Large reconstruction model for 3d gaussian splatting.
\newblock In \emph{European Conference on Computer Vision (ECCV)} (2024), Springer, pp.~1--19.

\bibitem[ZCD{\etalchar{*}}24]{zhu2024champ}
\textsc{Zhu S., Chen J.~L., Dai Z., Dong Z., Xu Y., Cao X., Yao Y., Zhu H., Zhu S.}:
\newblock Champ: Controllable and consistent human image animation with 3d parametric guidance.
\newblock In \emph{European Conference on Computer Vision (ECCV)} (2024), pp.~145--162.

\bibitem[ZCM22]{zhang2022motion}
\textsc{Zhang M., Ceylan D., Mitra N.~J.}:
\newblock Motion guided deep dynamic 3d garments.
\newblock \emph{Transactions on Graphics, (Proc. SIGGRAPH Asia) 41}, 6 (2022), 1--12.

\bibitem[ZCP{\etalchar{*}}23]{Zhou2023groomgen}
\textsc{Zhou Y., Chai M., Pepe A., Gross M., Beeler T.}:
\newblock Groomgen: A high-quality generative hair model using hierarchical latent representations.
\newblock \emph{Transactions on Graphics, (Proc. SIGGRAPH)} (2023).

\bibitem[ZCV{\etalchar{*}}25]{Zheng2025GroomLight}
\textsc{Zheng Y., Chai M., Vicini D., Zhou Y., Xu Y., Guibas L., Wetzstein G., Beeler T.}:
\newblock Groomlight: Hybrid inverse rendering for relightable human hair appearance modeling.
\newblock In \emph{Conference on Computer Vision and Pattern Recognition (CVPR)} (2025), pp.~16040--16050.

\bibitem[ZCW{\etalchar{*}}17]{Zhang2017data}
\textsc{Zhang M., Chai M., Wu H., Yang H., Zhou K.}:
\newblock A data-driven approach to four-view image-based hair modeling.
\newblock \emph{Transactions on Graphics, (Proc. SIGGRAPH) 36}, 4 (2017).

\bibitem[ZCW{\etalchar{*}}24a]{Zhang2024rodinhd}
\textsc{Zhang B., Cheng Y., Wang C., Zhang T., Yang J., Tang Y., Zhao F., Chen D., Guo B.}:
\newblock Rodinhd: High-fidelity 3d avatar generation with diffusion models.
\newblock In \emph{European Conference on Computer Vision (ECCV)} (2024), Springer, pp.~465--483.

\bibitem[ZCW{\etalchar{*}}24b]{Zhou2024groomcap}
\textsc{Zhou Y., Chai M., Wang D., Winberg S., Wood E., Sarkar K., Gross M., Beeler T.}:
\newblock Groomcap: High-fidelity prior-free hair capture.
\newblock \emph{Transactions on Graphics, (Proc. SIGGRAPH)} (2024).

\bibitem[ZCW{\etalchar{*}}25]{zhai2025taga}
\textsc{Zhai Z., Chen G., Wang W., Zheng D., Xiao J.}:
\newblock Taga: Self-supervised learning for template-free animatable gaussian articulated model.
\newblock In \emph{Conference on Computer Vision and Pattern Recognition (CVPR)} (2025), pp.~21159--21169.

\bibitem[ZFK{\etalchar{*}}24]{zhang2024teca}
\textsc{Zhang H., Feng Y., Kulits P., Wen Y., Thies J., Black M.~J.}:
\newblock Teca: Text-guided generation and editing of compositional 3d avatars.
\newblock In \emph{International Conference on 3D Vision (3DV)} (2024), IEEE, pp.~1520--1530.

\bibitem[ZGL{\etalchar{*}}25]{Zielonka2025synshot}
\textsc{Zielonka W., Garbin S.~J., Lattas A., Kopanas G., Gotardo P., Beeler T., Thies J., Bolkart T.}:
\newblock Synthetic prior for few-shot drivable head avatar inversion.
\newblock \emph{Conference on Computer Vision and Pattern Recognition (CVPR)} (2025).

\bibitem[ZGW{\etalchar{*}}25]{zhang2025mimicmotion}
\textsc{Zhang Y., Gu J., Wang L.-W., Wang H., Zhu Y., Zou F., et~al.}:
\newblock Mimicmotion: High-quality human motion video generation with confidence-aware pose guidance.
\newblock In \emph{International Conference on Machine Learning (ICML)} (2025), pp.~1--10.

\bibitem[ZHX{\etalchar{*}}18]{Zhou2018hairnet}
\textsc{Zhou Y., Hu L., Xing J., Chen W., Kung H.-W., Tong X., Li H.}:
\newblock {HairNet}: Single-view hair reconstruction using convolutional neural networks.
\newblock In \emph{European Conference on Computer Vision (ECCV)} (2018), pp.~249--265.

\bibitem[ZHY{\etalchar{*}}22]{zheng2022structured}
\textsc{Zheng Z., Huang H., Yu T., Zhang H., Guo Y., Liu Y.}:
\newblock Structured local radiance fields for human avatar modeling.
\newblock In \emph{Conference on Computer Vision and Pattern Recognition (CVPR)} (2022), pp.~15893--15903.

\bibitem[ZJL{\etalchar{*}}23]{zheng2023hairstep}
\textsc{Zheng Y., Jin Z., Li M., Huang H., Ma C., Cui S., Han X.}:
\newblock Hairstep: Transfer synthetic to real using strand and depth maps for single-view 3d hair modeling.
\newblock In \emph{Conference on Computer Vision and Pattern Recognition (CVPR)} (2023), pp.~12726--12735.

\bibitem[ZKB{\etalchar{*}}25]{zhuang2025dagsm}
\textsc{Zhuang J., Kang D., Bao L., Lin L., Li G.}:
\newblock Dagsm: Disentangled avatar generation with gs-enhanced mesh.
\newblock In \emph{Conference on Computer Vision and Pattern Recognition (CVPR)} (2025), pp.~292--303.

\bibitem[ZLB{\etalchar{*}}24]{zhang2024estimating}
\textsc{Zhang J.~X., Lin G. W.-C., Bode L., Chen H.-Y., Stuyck T., Larionov E.}:
\newblock Estimating cloth elasticity parameters from homogenized yarn-level models.
\newblock In \emph{Proceedings of the 17th ACM SIGGRAPH Conference on Motion, Interaction, and Games} (2024), pp.~1--12.

\bibitem[ZLDF21]{zhang2021hdtf}
\textsc{Zhang Z., Li L., Ding Y., Fan C.}:
\newblock Flow-guided one-shot talking face generation with a high-resolution audio-visual dataset.
\newblock In \emph{Proceedings of the IEEE/CVF conference on computer vision and pattern recognition} (2021), pp.~3661--3670.

\bibitem[ZLL{\etalchar{*}}24]{zhu2024m}
\textsc{Zhu L., Li Y., Liu N., Peng H., Yang D., Kemelmacher-Shlizerman I.}:
\newblock M\&m vto: Multi-garment virtual try-on and editing.
\newblock In \emph{Conference on Computer Vision and Pattern Recognition (CVPR)} (2024), pp.~1346--1356.

\bibitem[ZLS{\etalchar{*}}23]{zhang2023closet}
\textsc{Zhang H., Lin S., Shao R., Zhang Y., Zheng Z., Huang H., Guo Y., Liu Y.}:
\newblock Closet: Modeling clothed humans on continuous surface with explicit template decomposition.
\newblock In \emph{Conference on Computer Vision and Pattern Recognition (CVPR)} (2023), pp.~501--511.

\bibitem[ZLW{\etalchar{*}}21]{zhang2021physg}
\textsc{Zhang K., Luan F., Wang Q., Bala K., Snavely N.}:
\newblock Physg: Inverse rendering with spherical gaussians for physics-based material editing and relighting.
\newblock In \emph{Conference on Computer Vision and Pattern Recognition (CVPR)} (2021), Computer Vision Foundation / {IEEE}, pp.~5453--5462.

\bibitem[ZLW{\etalchar{*}}25]{zhuang2025idol}
\textsc{Zhuang Y., Lv J., Wen H., Shuai Q., Zeng A., Zhu H., Chen S., Yang Y., Cao X., Liu W.}:
\newblock Idol: Instant photorealistic 3d human creation from a single image.
\newblock In \emph{Conference on Computer Vision and Pattern Recognition (CVPR)} (2025), pp.~26308--26319.

\bibitem[ZMF{\etalchar{*}}24]{zhou2024headstudio}
\textsc{Zhou Z., Ma F., Fan H., Yang Z., Yang Y.}:
\newblock Headstudio: Text to animatable head avatars with 3d gaussian splatting.
\newblock In \emph{European Conference on Computer Vision (ECCV)} (2024).

\bibitem[ZMFC25]{zhou2025zero1toA}
\textsc{Zhou Z., Ma F., Fan H., Chua T.-S.}:
\newblock Zero-1-to-a: Zero-shot one image to animatable head avatars using video diffusion.
\newblock In \emph{Conference on Computer Vision and Pattern Recognition (CVPR)} (2025), pp.~15941--15952.

\bibitem[ZMZ{\etalchar{*}}23]{zhang2023probabilistic}
\textsc{Zhang S., Ma Q., Zhang Y., Aliakbarian S., Cosker D., Tang S.}:
\newblock Probabilistic human mesh recovery in 3d scenes from egocentric views.
\newblock In \emph{International Conference on Computer Vision (ICCV)} (2023), pp.~7989--8000.

\bibitem[ZPBPM17]{zhang2017detailed}
\textsc{Zhang C., Pujades S., Black M.~J., Pons-Moll G.}:
\newblock Detailed, accurate, human shape estimation from clothed 3d scan sequences.
\newblock In \emph{Conference on Computer Vision and Pattern Recognition (CVPR)} (2017), pp.~4191--4200.

\bibitem[ZPX{\etalchar{*}}24]{zhou2024animatable}
\textsc{Zhou X., Peng S., Xu Z., Dong J., Wang Q., Zhang S., Shuai Q., Bao H.}:
\newblock Animatable implicit neural representations for creating realistic avatars from videos.
\newblock \emph{Transactions on Pattern Analysis and Machine Intelligence (TPAMI) 46}, 6 (2024), 4147--4159.

\bibitem[ZQQH22]{zhu2022registering}
\textsc{Zhu H., Qiu L., Qiu Y., Han X.}:
\newblock Registering explicit to implicit: Towards high-fidelity garment mesh reconstruction from single images.
\newblock In \emph{Conference on Computer Vision and Pattern Recognition (CVPR)} (2022), pp.~3845--3854.

\bibitem[ZRA23]{zhang2023adding}
\textsc{Zhang L., Rao A., Agrawala M.}:
\newblock Adding conditional control to text-to-image diffusion models.
\newblock In \emph{International Conference on Computer Vision (ICCV)} (2023), pp.~3836--3847.

\bibitem[ZRL{\etalchar{*}}09]{Zinke2009practical}
\textsc{Zinke A., Rump M., Lay T., Weber A., Andriyenko A., Klein R.}:
\newblock A practical approach for photometric acquisition of hair color.
\newblock \emph{Transactions on Graphics, (Proc. SIGGRAPH) 28}, 5 (2009), 1--9.

\bibitem[ZSB{\etalchar{*}}24]{Zakharov2024haircut}
\textsc{Zakharov E., Sklyarova V., Black M., Nam G., Thies J., Hilliges O.}:
\newblock Human hair reconstruction with strand-aligned 3d gaussians.
\newblock In \emph{European Conference on Computer Vision (ECCV)} (2024), Springer, pp.~409--425.

\bibitem[ZSW{\etalchar{*}}24]{Zhao2024invertavatar}
\textsc{Zhao X., Sun J., Wang L., Suo J., Liu Y.}:
\newblock Invertavatar: Incremental gan inversion for generalized head avatars.
\newblock In \emph{SIGGRAPH Conference Papers (SA)} (2024), pp.~1--10.

\bibitem[ZSYZ25]{zhan2025real}
\textsc{Zhan Y., Shao T., Yang Y., Zhou K.}:
\newblock Real-time high-fidelity gaussian human avatars with position-based interpolation of spatially distributed mlps.
\newblock In \emph{Conference on Computer Vision and Pattern Recognition (CVPR)} (2025), pp.~26297--26307.

\bibitem[ZTG{\etalchar{*}}18]{Zollhofer2018state}
\textsc{Zollh{\"o}fer M., Thies J., Garrido P., Bradley D., Beeler T., P{\'e}rez P., Stamminger M., Nie{\ss}ner M., Theobalt C.}:
\newblock State of the art on monocular 3d face reconstruction, tracking, and applications.
\newblock \emph{Computer Graphics Forum (CGF) 37}, 2 (2018), 523--550.

\bibitem[ZWHB22]{zheng2022simulation}
\textsc{Zheng M., Wang B., Huang J., Barbi{\v{c}} J.}:
\newblock Simulation of hand anatomy using medical imaging.
\newblock \emph{Transactions on Graphics (TOG) 41}, 6 (2022), 1--20.

\bibitem[ZWL{\etalchar{*}}22]{zhuo2022fast}
\textsc{Zhuo L., Wang G., Li S., Wu W., Liu Z.}:
\newblock Fast-vid2vid: Spatial-temporal compression for video-to-video synthesis.
\newblock In \emph{European Conference on Computer Vision (ECCV)} (2022), pp.~289--305.

\bibitem[ZWL{\etalchar{*}}25]{Xheng2024headgap}
\textsc{Zheng X., Wen C., Li Z., Zhang W., Su Z., Chang X., Zhao Y., Lv Z., Zhang X., Zhang Y., et~al.}:
\newblock Headgap: Few-shot 3d head avatar via generalizable gaussian priors.
\newblock In \emph{International Conference on 3D Vision (3DV)} (2025), {IEEE}, pp.~946--957.

\bibitem[ZWLY25]{zhang2025disentangled}
\textsc{Zhang W., Wu S., Liao M., Yan Y.}:
\newblock Disentangled clothed avatar generation with layered representation.
\newblock In \emph{International Conference on Computer Vision (ICCV)} (2025).

\bibitem[ZWS{\etalchar{*}}24]{zheng2024ohta}
\textsc{Zheng X., Wen C., Su Z., Xu Z., Li Z., Zhao Y., Xue Z.}:
\newblock Ohta: One-shot hand avatar via data-driven implicit priors.
\newblock In \emph{Conference on Computer Vision and Pattern Recognition (CVPR)} (2024), pp.~799--810.

\bibitem[ZWW{\etalchar{*}}18]{Zhang2018modeling}
\textsc{Zhang M., Wu P., Wu H., Weng Y., Zheng Y., Zhou K.}:
\newblock Modeling hair from an rgb-d camera.
\newblock \emph{Transactions on Graphics, (Proc. SIGGRAPH) 37}, 6 (2018), 205:1--205:10.

\bibitem[ZWZ{\etalchar{*}}22]{zhu2022celebvhq}
\textsc{Zhu H., Wu W., Zhu W., Jiang L., Tang S., Zhang L., Liu Z., Loy C.~C.}:
\newblock Celebv-hq: A large-scale video facial attributes dataset.
\newblock In \emph{European Conference on Computer Vision (ECCV)} (2022), Springer, pp.~650--667.

\bibitem[ZYG{\etalchar{*}}23]{Zhu2023facescape}
\textsc{Zhu H., Yang H., Guo L., Zhang Y., Wang Y., Huang M., Wu M., Shen Q., Yang R., Cao X.}:
\newblock Facescape: 3d facial dataset and benchmark for single-view 3d face reconstruction.
\newblock \emph{Transactions on Pattern Analysis and Machine Intelligence (TPAMI) 45}, 12 (2023), 14528--14545.

\bibitem[ZYHC22]{zheng2022imface}
\textsc{Zheng M., Yang H., Huang D., Chen L.}:
\newblock Imface: A nonlinear 3d morphable face model with implicit neural representations.
\newblock In \emph{Conference on Computer Vision and Pattern Recognition (CVPR)} (2022), pp.~20343--20352.

\bibitem[ZYL{\etalchar{*}}24]{yang2024e3gen}
\textsc{Zhang W., Yan Y., Liu Y., Sheng X., Yang X.}:
\newblock \emph{E}\({}^{\mbox{3}}\)gen: Efficient, expressive and editable avatars generation.
\newblock In \emph{International Conference on Multimedia, {MM}} (2024), {ACM}, pp.~6860--6869.

\bibitem[ZYW{\etalchar{*}}23]{zheng2023pointavatar}
\textsc{Zheng Y., Yifan W., Wetzstein G., Black M.~J., Hilliges O.}:
\newblock Pointavatar: Deformable point-based head avatars from videos.
\newblock In \emph{Conference on Computer Vision and Pattern Recognition (CVPR)} (2023), pp.~21057--21067.

\bibitem[ZYWK08]{zinke2008dual}
\textsc{Zinke A., Yuksel C., Weber A., Keyser J.}:
\newblock Dual scattering approximation for fast multiple scattering in hair.
\newblock \emph{Transactions on Graphics, (Proc. SIGGRAPH) 27}, 3 (2008), 32:1--32:10.

\bibitem[ZYX{\etalchar{*}}24]{zhang2024dual}
\textsc{Zhang Y., Ye D., Xie C., Tang L., Liao X., Liu Z., Chen C., Deng J.}:
\newblock Dual defense: Adversarial, traceable, and invisible robust watermarking against face swapping.
\newblock \emph{IEEE Transactions on Information Forensics and Security 19} (2024), 4628--4641.

\bibitem[ZYZ{\etalchar{*}}23]{zhu2023tryondiffusion}
\textsc{Zhu L., Yang D., Zhu T., Reda F., Chan W., Saharia C., Norouzi M., Kemelmacher-Shlizerman I.}:
\newblock Tryondiffusion: A tale of two unets.
\newblock In \emph{Conference on Computer Vision and Pattern Recognition (CVPR)} (2023), pp.~4606--4615.

\bibitem[ZZ19]{Zhang2019hair}
\textsc{Zhang M., Zheng Y.}:
\newblock Hair-gan: Recovering 3d hair structure from a single image using generative adversarial networks.
\newblock \emph{Visual Informatics (VI) 3}, 2 (2019), 102--112.

\bibitem[ZZC{\etalchar{*}}23]{zhang2023getavatar}
\textsc{Zhang X., Zhang J., Chacko R., Xu H., Song G., Yang Y., Feng J.}:
\newblock Getavatar: Generative textured meshes for animatable human avatars.
\newblock In \emph{International Conference on Computer Vision (ICCV)} (2023), pp.~2273--2282.

\bibitem[ZZR{\etalchar{*}}25]{zhu2025infp}
\textsc{Zhu Y., Zhang L., Rong Z., Hu T., Liang S., Ge Z.}:
\newblock Infp: Audio-driven interactive head generation in dyadic conversations.
\newblock In \emph{Proceedings of the Computer Vision and Pattern Recognition Conference} (2025), pp.~10667--10677.

\bibitem[ZZS{\etalchar{*}}24]{Zheng2024gps}
\textsc{Zheng S., Zhou B., Shao R., Liu B., Zhang S., Nie L., Liu Y.}:
\newblock Gps-gaussian: Generalizable pixel-wise 3d gaussian splatting for real-time human novel view synthesis.
\newblock In \emph{Conference on Computer Vision and Pattern Recognition (CVPR)} (2024), pp.~19680--19690.

\bibitem[ZZSC22]{zhuang2022mofanerf}
\textsc{Zhuang Y., Zhu H., Sun X., Cao X.}:
\newblock Mofanerf: Morphable facial neural radiance field.
\newblock In \emph{European Conference on Computer Vision (ECCV)} (2022), Springer, pp.~268--285.

\bibitem[ZZTH24]{zhu2024trihuman}
\textsc{Zhu H., Zhan F., Theobalt C., Habermann M.}:
\newblock Trihuman: a real-time and controllable tri-plane representation for detailed human geometry and appearance synthesis.
\newblock \emph{Transactions on Graphics, (Proc. SIGGRAPH) 44}, 1 (2024), 1--17.

\bibitem[ZZY{\etalchar{*}}24]{zheng2024physavatar}
\textsc{Zheng Y., Zhao Q., Yang G., Yifan W., Xiang D., Dubost F., Lagun D., Beeler T., Tombari F., Guibas L., et~al.}:
\newblock Physavatar: Learning the physics of dressed 3d avatars from visual observations.
\newblock In \emph{European Conference on Computer Vision (ECCV)} (2024), Springer, pp.~262--284.

\bibitem[ZZZ{\etalchar{*}}23]{zheng2023avatarrex}
\textsc{Zheng Z., Zhao X., Zhang H., Liu B., Liu Y.}:
\newblock Avatarrex: Real-time expressive full-body avatars.
\newblock \emph{Transactions on Graphics, (Proc. SIGGRAPH) 42}, 4 (2023), 1--19.

\end{thebibliography}
